%% file: PionForward.tex
\documentclass[fleqn,twoside]{article}
\usepackage{epsfig}
\usepackage{xspace}
\usepackage{longtable}
\usepackage{graphicx}
\usepackage[figuresright]{rotating}

\newcommand{\GeVc}{\ensuremath{\mbox{GeV}/c}\xspace}
\newcommand{\MeVc}{\ensuremath{\mbox{MeV}/c}\xspace}

\newcommand{\mm}{\ensuremath{\mbox{mm}}\xspace}

\newcommand{\mrad}{\ensuremath{\mbox{mrad}}\xspace}
\newcommand{\rad}{\ensuremath{\mbox{rad}}\xspace}

\newcommand{\ps}{\ensuremath{\mbox{ps}}\xspace}

\newcommand{\pip}{\ensuremath{\pi^+}\xspace}
\newcommand{\pim}{\ensuremath{\pi^-}\xspace}

\newcommand{\bfpip}{\ensuremath{\mathbf {\pi^+}}\xspace}
\newcommand{\bfpim}{\ensuremath{\mathbf {\pi^-}}\xspace}

\def\be{\begin{equation}}
\def\ee{\end{equation}}
\def\bea{\begin{eqnarray}}
\def\eea{\end{eqnarray}}

\newcommand{\bfGeVc}{\ensuremath{\mathbf {\mbox{\bf GeV}/c}}\xspace}
\begin{document}
\newpage
\title{\bf Forward production of  charged pions
with incident $\pi^{\pm}$ \\ on  
  nuclear targets  measured  at the CERN PS.}

\author{HARP Collaboration}

\maketitle

\begin{abstract}
  Measurements of the  double-differential $\pi^{\pm}$ production
  cross-section    in the range of momentum $0.5~\GeVc \leq p \le 8.0~\GeVc$ 
  and angle $0.025~\rad \leq \theta  \le 0.25~\rad$
  in interactions of charged pions on
  beryllium, carbon, aluminium, copper,
  tin, tantalum and lead are presented. 
  These data represent the first experimental campaign to systematically
  measure forward pion hadroproduction.

  The data were taken with the large acceptance HARP detector in the T9 beam
  line of the CERN PS.
  %
  Incident particles, impinging on a 5\% nuclear interaction length target, 
   were identified by an elaborate system of beam
  detectors.
  The tracking and identification of the
  produced particles was performed using the forward spectrometer of the 
  HARP detector.
  Results are obtained for the double-differential cross-sections 
  $
  {{\mathrm{d}^2 \sigma}}/{{\mathrm{d}p\mathrm{d}\Omega }}
  $
  mainly at four incident pion beam momenta (3~\GeVc, 5~\GeVc, 8~\GeVc and 
  12~\GeVc). 
  The measurements are compared with the  GEANT4 and MARS Monte Carlo
  simulation.
\end{abstract}

\clearpage

\thispagestyle{plain}
\begin{center}
\vspace{1cm}
{\small
M.~Apollonio$^{x}$, 
A.~Artamonov$^{f,4}$,   
A. Bagulya$^{m}$, 
G.~Barr$^{p}$, 
A.~Blondel$^{g}$, 
F.~Bobisut$^{q,19}$, 
M.~Bogomilov$^{w}$, 
M.~Bonesini$^{l,*}$, 
C.~Booth$^{u}$, 
S.~Borghi$^{g,12}$,  
S.~Bunyatov$^{d}$, 
J.~Burguet--Castell$^{z}$, 
M.G.~Catanesi$^{a}$, 
A.~Cervera--Villanueva$^{z}$, 
P.~Chimenti$^{x}$,  
L.~Coney$^{15,18}$, 
E.~Di~Capua$^{e}$, 
U.~Dore$^{s}$,
J.~Dumarchez$^{r}$,
R.~Edgecock$^{b}$, 
M.~Ellis$^{b,1}$,          
F.~Ferri$^{l}$,           
U.~Gastaldi$^{i}$, 
S.~Giani$^{f}$, 
G.~Giannini$^{x}$, 
D.~Gibin$^{q,19}$,
S.~Gilardoni$^{f}$,       
P.~Gorbunov$^{f,4}$,  
C.~G\"{o}\ss ling $^{c}$,
J.J.~G\'{o}mez--Cadenas$^{z}$, 
A.~Grant$^{f}$,  
J.S.~Graulich$^{k,16}$, 
G.~Gr\'{e}goire$^{k}$ 
V.~Grichine$^{n}$,  
A.~Grossheim$^{f,6}$, 
A.~Guglielmi$^{q}$, 
L.~Howlett$^{t}$,
A.~Ivanchenko$^{f,7}$,
V.~Ivanchenko$^{f,8}$,  
\newcommand{\afkyot}{{19}\xspace}
A.~Kayis-Topaksu$^{f,9}$,
M.~Kirsanov$^{m}$,
D.~Kolev$^{w}$, 
A.~Krasnoperov$^{d}$, 
J. Mart\'{i}n--Albo$^{z}$,
C.~Meurer$^{h}$,
M.~Mezzetto$^{q}$,
G.~B.~Mills$^{17,20}$  
M.C.~Morone$^{g,13}$, 
P.~Novella$^{z}$,
D.~Orestano$^{t,20}$, 
V.~Palladino$^{o}$
J.~Panman$^{f}$, 
I.~Papadopoulos$^{f}$,  
F.~Pastore$^{t,20}$, 
S.~Piperov$^{w}$, 
N.~Polukhina$^{n}$, 
B.~Popov$^{d,2}$, 
G.~Prior$^{g,14}$,   
E.~Radicioni$^{a}$,
D.~Schmitz$^{15,18}$,
R.~Schroeter$^{g}$,
G~Skoro$^{u}$,
M.~Sorel$^{z}$,
E.~Tcherniaev$^{f}$, 
P.~Temnikov$^{y}$,
V.~Tereschenko$^{d}$,  
A.~Tonazzo$^{t,20}$, 
L.~Tortora$^{t}$,
R.~Tsenov$^{w}$
I.~Tsukerman$^{f,4}$,   
G.~Vidal--Sitjes$^{e,3}$,  
C.~Wiebusch$^{f,10}$,    
P.~Zucchelli$^{f,5,11}$ 
}  
\vskip .5cm
{\large (HARP collaboration)}\\
\newcommand{\afdoct}{{3}\xspace}
\end{center}
\thispagestyle{plain}
\vfill
\rule{0.3\textwidth}{0.4mm}
\newline
{ (a)  Sezione INFN, Bari, Italy} 
\newline
{ (b) Rutherford Appleton Laboratory, Chilton, Didcot, UK} 
\newline
{ (c) Institut f\"{u}r Physik, Universit\"{a}t Dortmund, Germany} 
\newline
{ (d) Joint Institute for Nuclear Research, JINR Dubna, Russia} 
\newline
{ (e) Universit\`{a} degli Studi e Sezione INFN, Ferrara, Italy}  
\newline
{ (f) CERN, Geneva, Switzerland} 
\newline
{ (g) Section de Physique, Universit\'{e} de Gen\`{e}ve, Switzerland} 
\newline
{ (h) Institut f\"{u}r Physik, Forschungszentrum Karlsruhe, Germany}
\newline
{ (i) Laboratori Nazionali di Legnaro dell' INFN, Legnaro, Italy} 
\newline
{ (k) Institut de Physique Nucl\'{e}aire, UCL, Louvain-la-Neuve,
  Belgium} 
\newline
{ (l)  Sezione INFN Milano Bicocca, Milano, Italy} 
\newline
{ (m) Institute for Nuclear Research, Moscow, Russia} 
\newline
{ (n) P. N. Lebedev Institute of Physics (FIAN), Russian Academy of
Sciences, Moscow, Russia} 
\newline
{ (o) Universit\`{a} ``Federico II'' e Sezione INFN, Napoli, Italy} 
\newline
{ (p) Nuclear and Astrophysics Laboratory, University of Oxford, UK} 
\newline
{ (q) Sezione INFN, Padova, Italy} 
\newline
{ (r) LPNHE, Universit\'{e}s de Paris VI et VII, Paris, France} 
\newline
{ (s) Universit\`{a} ``La Sapienza'' e Sezione INFN Roma I, Roma,
  Italy} 
\newline
{ (t) Sezione INFN Roma III, Roma, Italy}
\newline
{ (u) Dept. of Physics, University of Sheffield, UK} 
\newline
{ (w) Faculty of Physics, St. Kliment Ohridski University, Sofia,
  Bulgaria} 
\newline
{ (y) Institute for Nuclear Research and Nuclear Energy, 
Academy of Sciences, Sofia, Bulgaria} 
\newline
{ (x) Universit\`{a} degli Studi e Sezione INFN, Trieste, Italy} 
\newline
{  (z) Instituto de F\'{i}sica Corpuscular, IFIC, CSIC and Universidad de Valencia,
Spain} 
\newpage
\vskip 1cm
$^{*}$ Corresponding author.
\newline
 E-mail address: maurizio.bonesini@mib.infn.it.
\newline
$^{~1}${Now at FNAL, Batavia, Illinois, USA.}
\newline
$^{~2}${Also supported by LPNHE, Paris, France.}
\newline
%
$^{~3}${Now at Imperial College, University of London, UK.}
\newline
$^{~4}${ITEP, Moscow, Russian Federation.}
\newline
$^{~5}${Now at SpinX Technologies, Geneva, Switzerland.}
\newline
$^{6}${Now at TRIUMF, Vancouver, Canada.}
\newline
$^{7}${On leave of absence from Novosibirsk University, Russia.}
\newline
$^{8}${On leave of absence from Ecoanalitica, Moscow State University,
Moscow, Russia.}
\newline
$^{9}${Now at \c{C}ukurova University, Adana, Turkey.}
\newline
$^{10}${Now at III Phys. Inst. B, RWTH Aachen, Aachen, Germany.}
\newline
$^{11}$On leave of absence from INFN, Sezione di Ferrara, Italy.
\newline
$^{12}${Now at CERN, Geneva, Switzerland.}
\newline
$^{13}${Now at Univerity of Rome Tor Vergata, Italy.}
\newline
$^{14}${Now at Lawrence Berkeley National Laboratory, Berkeley, California, USA.}
\newline
$^{15}${MiniBooNE Collaboration.}
\newline
$^{16}${Now at Section de Physique, Universit\'{e} de Gen\`{e}ve, Switzerland, Switzerland.}
\newline
$^{17}${Los Alamos National Laboratory, Los Alamos, USA}
\newline
$^{18}${Columbia University, New York, USA
\newline
$^{19}${also at Universit\`a di Padova, Padova, Italy}
\newline
$^{20}${also at Universit\`a di Roma III, Roma, Italy}

\clearpage

\section{Introduction}


The HARP experiment~\cite{ref:harp-prop} at the CERN PS
was designed to measure hadron yields from a large range
of nuclear targets and for incident particle momenta from
 1.5~\GeVc to 15~\GeVc.
With incident protons, this corresponds to a  momentum region of great interest for
neutrino beams and far from the coverage by earlier dedicated hadroproduction
experiments~\cite{ref:na56,ref:atherton,ref:physrep}.
Measurements with incident charged pions are relevant
for the calculation of cosmic-ray muon and neutrino fluxes using
extended air shower simulations.
These data are also important to simulate re-interactions in particle
detectors and neutrino beam production targets. 
At present such calculations rely on models with large uncertainties
given the lack of experimental data in this energy region.  

Covering an extended range of solid targets in
the same experiment, it is also possible to perform systematic
comparison of hadron production models with measurements 
at different incoming beam momenta over a large range 
of target atomic number $A$.

This paper presents our final  
measurements of the double-differential cross-section, 
$
{{\mathrm{d}^2 \sigma^{\pi}}}/{{\mathrm{d}p\mathrm{d}\Omega }}
$
for $\pi^{\pm}$ forward production by
incident charged pions of 3~\GeVc, 5~\GeVc, 8~\GeVc and  12~\GeVc 
momentum impinging
on a thin beryllium, carbon, aluminium, copper, tin, tantalum  
or lead target of 5\% nuclear interaction length ($\lambda_I$). 
In addition, high statistics data at 8.9~\GeVc 
(12.9~\GeVc) on a thin beryllium (aluminium) target
are presented.

Some HARP results on pion production in the forward region 
with incident protons have already been published in
papers~\cite{ref:alPaper,ref:bePaper,ref:carbonfw,ref:cnofw}.

Pion production data at low momenta and in the forward region   
from incident pions are extremely scarce~\cite{ref:Blieden}. HARP
is the first experiment to provide a high statistics data set, taken with many 
different targets, full particle identification and large-acceptance detector.
The collected statistics,
for the different nuclear targets, are reported in Table~\ref{tab:events}. 
\input{All5_table1.tex}


Other data with an incident pion beam, up to higher momenta, has been
collected by the Fermilab E970-MIPP experiment \cite{MIPP} 
and more data at other momenta would be available with its proposed upgrade
\cite{ref:mippupgr}.

The paper is organized as follows. In subsection~\ref{subsec:harp_det} we briefly 
describe the HARP detector, while section~\ref{sec:analysis} is devoted to 
the main features of the analysis procedure. Section~\ref{sec:results} presents results. 
It is followed by a summary presented in section~\ref{sec:conclusions}.

\subsection{Experimental apparatus}
\label{subsec:harp_det}

 The HARP experiment
 makes use of a large-acceptance spectrometer consisting of a
 forward and large-angle detection system.
 The HARP detector is shown in Fig.~\ref{fig:harp}.
 A detailed
 description of the experimental apparatus can be found in Ref.~\cite{ref:harpTech}.
 The forward spectrometer -- 
 based on five modules of large area drift chambers
 (NDC1-5)~\cite{ref:NOMAD_NIM_DC} and a dipole magnet
 complemented by a set of detectors for particle identification (PID): 
 a time-of-flight wall (TOFW)~\cite{ref:tofPaper}, a large Cherenkov detector (CHE) 
 and an electromagnetic calorimeter (ECAL) --
 covers polar angles up to 250~mrad. 
 The muon contamination of the beam is measured with a muon identifier 
 consisting of thick iron absorbers and scintillation counters.
 The large-angle spectrometer -- based on a Time Projection Chamber (TPC) 
 and Resistive Plate Chambers (RPCs)
 located inside a solenoidal magnet --
 has a large acceptance in  momentum
 and angular range for the pions relevant to the production of the
 muons in a neutrino factory. 
 For the analysis described here  \underline{only}  the forward spectrometer and
 the beam instrumentation are used.

\begin{figure}[tb]]
\centering
\includegraphics[width=0.8\textwidth]{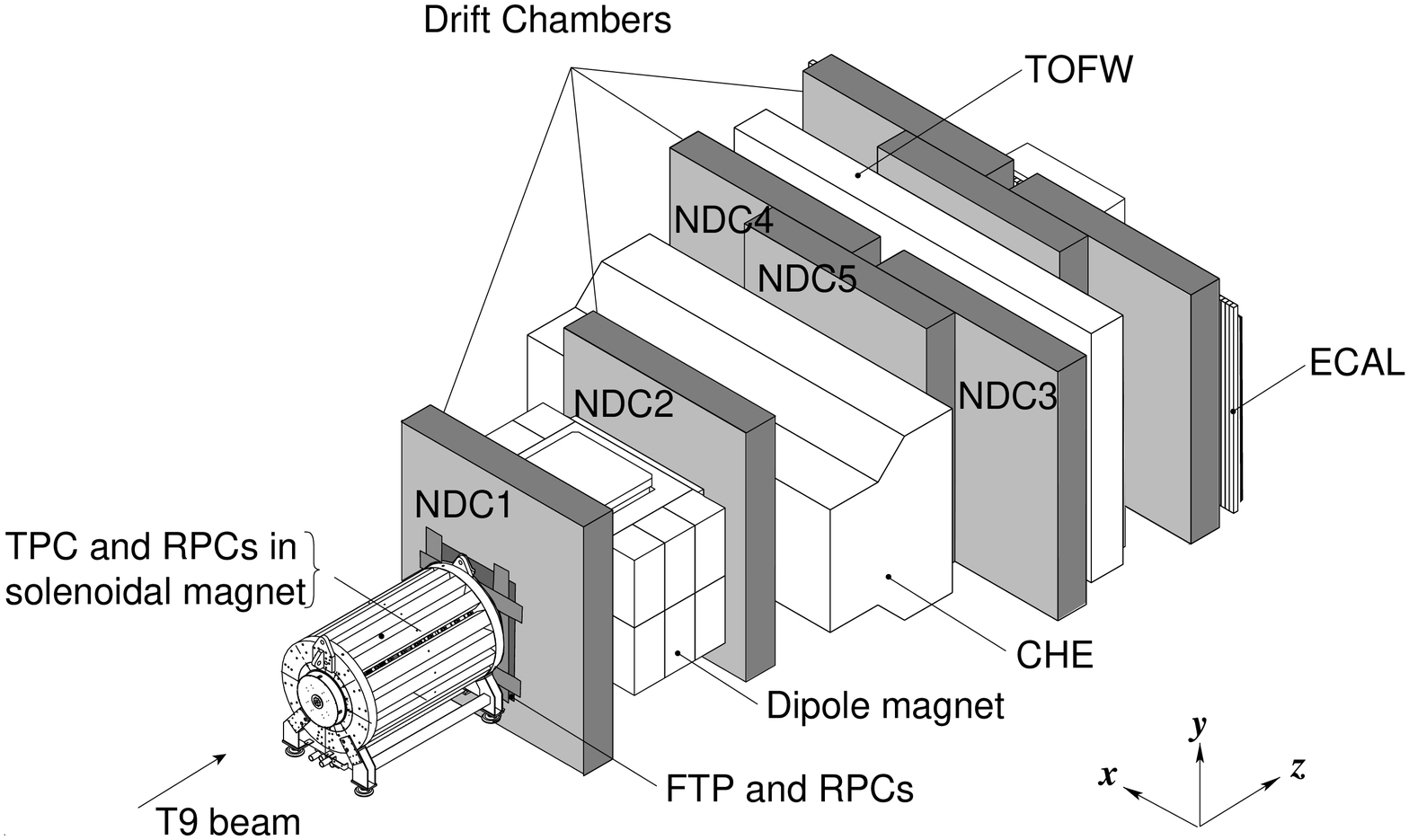} 
\caption{\label{fig:harp} 
Schematic layout of the HARP detector.
The convention for the coordinate system is shown in the lower-right
corner.
}
\end{figure}


The HARP experiment, located in the T9 beam of the CERN PS, took data in 2001
and 2002.
The momentum definition of the T9 beam 
is known with a precision of the order of 1\%~\cite{ref:t9}. 

The target is placed inside the inner field cage (IFC) of the TPC,
in an assembly that can be moved in and out of the solenoid magnet.
The targets used for the measurements reported here
have a cylindrical shape with a nominal diameter of about 30~\mm.
Their thickness is equivalent to about 5\% $\lambda_I$.

A set of four multi-wire
proportional chambers (MWPCs) measures the position and direction of
the incoming beam particles with an accuracy of $\approx$1~\mm in
position and $\approx$0.2~\mrad in angle per projection.
A beam time-of-flight system (BTOF)
measures the time difference of particles over a $21.4$~m path-length. 
It is made of two
identical scintillation hodoscopes, TOFA and TOFB (originally built
for the NA52 experiment~\cite{ref:NA52}),
which, together with a small target-defining trigger counter, TDS
(also used for the trigger), provide particle
identification at low momenta. This provides separation of pions, kaons
and protons up to 5~\GeVc and determines the initial time at the
interaction vertex ($t_0$). 
The timing resolution of the combined BTOF system is about 70~\ps.
A system of two N$_2$-filled Cherenkov detectors (BCA and BCB) is
used to tag electrons at low energies and pions at higher energies. 
The electron and pion tagging efficiency is found to be close to
100\%.
At the beam energy used for this analysis the Cherenkov counters select
all particles lighter than protons, while the BTOF is used to reject ions. 
A set of trigger detectors completes the beam instrumentation.

The selection of beam pions is performed as described in \cite{ref:harpTech}.
At 3~\GeVc BCB gas pressure is set to tag $e^\pm$ (BCA is evacuated to reduce multiple 
scattering of the beam) while the TOFs are capable of resolving pions from protons. 
At 5~\GeVc pions are separated from protons by using joint information from the
TOFs and one of the Cherenkovs (usually BCB), while the
gas pressure in the other Cherenkov (BCA) is
set to tag $e^\pm$. At higher momenta the $e^\pm$ contamination
drops below 1\% and the TOF system becomes unable to separate pions from protons 
efficiently: the beam Cherenkov detectors are used for $\pi$-p separation. 
At the highest beam energy the gas pressure in one of the Cherenkov
detectors has been set to distinguish protons from kaons.
This allows us to estimate the kaon component to be negligible in the
pion beam selection at all momenta.

A downstream trigger in the forward scintillator trigger plane (FTP) 
was required to record the event,
accepting only tracks with a trajectory outside the central hole
(60~mm) which allows beam particles to pass. 
The trigger counter covers the spectrometer acceptance fully and has a
high efficiency ($>$99.8\%).

The length of the accelerator spill was 400~ms with a typical intensity
of 15~000 beam particles per spill.
The average number of events recorded by the data acquisition ranged
from 300 to 350 per spill.

The absolute normalization of the number of incident pions was
performed using `incident-particle' triggers. 
These are triggers where the same selection on the beam particle was
applied but no selection on the interaction was performed.
The rate of this trigger was down-scaled by a factor 64.

\section{Data Analysis}
\label{sec:analysis}
\subsection{Event and particle selection}

A detailed description of the experimental techniques 
used
for data analysis in the HARP forward spectrometer 
can be found in Ref.~\cite{ref:alPaper,ref:pidPaper}.
%
%
Further details of the improved analysis techniques 
can be found in~\cite{ref:bePaper,ref:carbonfw}.
For the current analysis we have used identical reconstruction and PID
algorithms, while at the final stage of the analysis 
the unfolding technique introduced as UFO in~\cite{ref:alPaper} has been
applied. 
This technique has already been described in details in Ref.~\cite{ref:harp:tantalum}.

At the first stage of the analysis a beam pion is selected
using the beam time of flight system (TOF-A, TOF-B) and the Cherenkov
counters (BCA, BCB) as described in section~\ref{subsec:harp_det}.
We always require time measurements in TOF-A, TOF-B and/or TDS to be
present which are needed for calculating 
the arrival time of the beam pion at the target. 



Secondary track selection criteria,
described in~\cite{ref:carbonfw},  are optimized to ensure the quality
of momentum reconstruction and a clean time-of-flight measurement
while maintaining a high reconstruction efficiency. 

The background induced by
interactions of beam particles in the materials outside the target
is measured  by taking data without a
target in the target holder (``empty target data'').  
These data are  subject to the same  event and track
selection criteria as the standard  data sets. 

To take into account this background the number of particles of the
observed 
type ($\pi^+$, $\pi^-$) in the ``empty target data''
are subtracted bin-by-bin (momentum and angular bins) from the number
of particles of the same type. The uncertainty
induced by
this method is discussed in section~\ref{errorest} and
labeled ``empty target subtraction''. 
The event statistics is summarized in Table~\ref{tab:events}.
 
The negative beam consists of $e^-$ and $\pi^-$ (with a dominant
fraction of $ \pi^{-} $), while the positive beam is dominated by
protons at high momenta and by $ \pi^{+} $ at low
momenta~\footnote{the proton fraction in the incoming beam goes from
35\% at 3~\GeVc to about 92\% at 12~\GeVc}. 
The kaon background is estimated to be $<$0.5\% and is neglected in the
analysis, while the muon contamination (tagged by the muon identifier)
is around 3\% and is subtracted.

The small fraction of pions in the positively charged beam explains the 
significantly different statistics of the 
data-sets with incident $\pi^{-}$ and $\pi^{+}$. 
The relatively small fraction of accepted final state pions compared to 
the accepted beam pions is due to the stringent quality cuts applied 
in the analysis.
The major loss is due to the small acceptance of the dipole magnet in 
the vertical plane, and the fact that only focused particles are 
accepted for analyses.
The latter requirement reduces the statistics by a factor two but
improves the quality of reconstruction and reduces the 
background~\cite{ref:bePaper}.

\subsection{Cross-section calculation}

The cross-section is calculated as follows
\begin{eqnarray}
\frac{d^2 \sigma^{\alpha}}{dp d\Omega}(p_i,\theta_j) & = & 
\frac{A}{N_A \rho t} \cdot \frac{1}{N_{\rm pot}} \cdot \frac{1}{\Delta p_i \Delta \Omega_j} \cdot 
\sum_{p'_i,\theta'_j,\alpha'} \mathcal{M}^{\rm cor}_{p_i\theta_j\alpha p'_i\theta'_j\alpha'} \cdot 
N^{\alpha'}(p'_i,\theta'_j)\hspace{0.1cm},
\end{eqnarray} 
where 
\begin{itemize}
\item $\frac{d^2 \sigma^{\alpha}}{dp d\Omega}(p_i,\theta_j)$ is the
  cross-section in mb/(\GeVc sr) for the particle type $\alpha$ (p,
  $\pi^+$ or $\pi^-$) for each true momentum and angle bin ($p_i,\theta_j$)
  covered in this analysis;
\item $N^{\alpha'}(p'_i,\theta'_j)$  is the number of particles of
  type $\alpha'$ in bins of reconstructed momentum $p'_i$ and angle
  $\theta_j'$ in the raw data;
\item $\mathcal{M}^{\rm cor}_{p\theta\alpha p'\theta'\alpha'}$ is the
  correction matrix which accounts for efficiency and resolution of
  the detector;
\item $\frac{A}{N_A \rho t}$, $\frac{1}{N_{\rm pot}}$ and
  $\frac{1}{\Delta p_i \Delta \Omega_j}$ are normalization factors,
  namely:
\subitem $\frac{N_A \rho t}{A}$ is the number of target nuclei per unit area 
\footnote{$A$ - atomic  mass, $N_A$ - Avogadro number, $\rho$ - target
  density and $t$ - target thickness};
\subitem $N_{\rm pot}$ is the number of incident beam particles on
  target (particles on target);
\subitem $\Delta p_i $ and $\Delta \Omega_j $ are the bin sizes in
  momentum and solid angle, respectively 
\footnote{$\Delta p_i = p^{\rm max}_i-p^{\rm min}_i$,\hspace{0.2cm}
  $\Delta \Omega_j = 2 \pi (\cos(\theta^{\rm min}_j)- 
  \cos(\theta^{\rm max}_j))$}.
\end{itemize}
We do not make a correction for the attenuation
of the incoming beam in the target, so that strictly speaking the
cross-sections are valid for $\lambda_{\mathrm{I}}=5\%$
targets.

The  calculation of the
correction matrix $M^{\rm cor}_{p_i\theta_j\alpha
  p'_i\theta'_j\alpha'}$ is 
a rather difficult task.
Various techniques are
described in the literature to obtain this matrix. As 
discussed 
in Ref.~\cite{ref:alPaper} for the p-Al analysis of HARP data at 12.9~{\GeVc}, two
complementary analyses have been performed to cross-check internal
consistency and possible biases in the respective procedures.
A comparison of both analyses shows that the results are consistent
within the overall systematic error~\cite{ref:alPaper}.

In the first method -- called ``Atlantic'' in~\cite{ref:alPaper} -- 
the correction matrix $M^{\rm
  cor}_{p_i\theta_j\alpha p'_i\theta'_j\alpha'}$ is decomposed into
distinct independent contributions, which are computed mostly using
the data themselves.
The second method -- called ``UFO'' in~\cite{ref:alPaper} -- 
is the unfolding method introduced 
by D'Agostini~\cite{ref:DAgostini}~\footnote{
The  unfolding method tries to put in correspondence the
vector of measured observables (such as particle momentum, polar
angle and particle type) $x_{\rm meas}$ with the vector of true values
$x_{\rm true}$ using a migration matrix: $x_{meas} = {\sl M}_{migr} \times x_{true}$.
The goal of the method is to compute a transformation 
(correction matrix) to obtain the expected
values for $x_{true}$ from the measured ones. The most simple
and obvious solution, based on simple matrix inversion 
${\sl M}^{-1}_{\rm migr}$, 
is usually unstable and is dominated by large variances and strong negative
correlations between neighbouring bins.
In the method of D' Agostini, 
the correction matrix ${\sl M}^{\rm UFO}$ tries to connect
the measurement space (effects) with the space of the true values (causes) 
using an iterative Bayesian approach, based on Monte Carlo simulations to
estimate the probability for a given effect to be produced by a certain 
cause.}. 
This method has been used in the recent HARP 
publications~\cite{ref:cnofw,ref:harp:tantalum,ref:harp:carboncoppertin,ref:harp:bealpb,
ref:finalproton} 
and it is also applied in the analysis described here 
(see~\cite{ref:carbonfw} for additional information).

The Monte Carlo simulation of the HARP setup is based on 
the GEANT4 package~\cite{ref:geant4,ref:ivan}. 
The detector
materials are accurately 
described
in this simulation as well as the
relevant features of the detector response and the digitization
process. All relevant physics processes are considered, including
multiple scattering, energy loss, absorption and
re-interactions. 
The simulation is independent of the beam momentum and particle type
because it only generates for each event
exactly one secondary particle of a specific particle type inside the
target material and propagates it through the 
complete detector. 
A small difference (at the few percent level) is observed between the
efficiency calculated for 
events simulated with the single-particle Monte Carlo and with a
simulation using a multi-particle hadron-production model.
A similar difference is seen between the single-particle Monte Carlo and
the efficiencies measured directly from the data.
The single-particle Monte Carlo predicts an efficiency more in agreement 
with the efficiency measured with the data.
A momentum-dependent correction factor determined using the efficiency
measured with the  data is applied to take this into account. 
The track reconstruction algorithms used in this analysis and the simulation are
identical to the ones used for the $\pi^+$ production in p-Be
collisions~\cite{ref:bePaper}. 
A detailed description of the corrections and their magnitude can be
found there. 

The reconstruction efficiency (inside the geometrical acceptance) is
larger than 95\% above 1.5~\GeVc and drops to 80\% at 0.5~\GeVc. 
The requirement of a match with a TOFW hit has an efficiency between
90\% and 95\% above 1.0~\GeVc and is responsible for the drop in 
efficiency towards lower momenta.
The electron veto rejects about 1\% of the pions and protons below
3~\GeVc with a remaining electron background of less than 0.5\%.
Below Cherenkov threshold the TOFW separates pions and protons with
negligible background and an efficiency of $\approx$98\% for pions.
Above Cherenkov threshold the efficiency for pions is greater than 99\%
with only 1.5\% of the protons mis-identified as a pion.
The kaon background in the pion spectra is smaller than 1\%.

The absorption and decay of particles is simulated by the Monte Carlo.
The generated single particle can re-interact and produce background
particles by hadronic or electromagnetic processes, thus giving rise to
tracks in the forward spectrometer (``tertiaries'').
In such cases also the additional measurements are entered into the
migration matrix thereby taking into account the combined effect of the
generated particle and any secondaries it creates.
The absorption correction is on average 20\%, approximately independent
of momentum.
Uncertainties in the absorption of secondaries in the dipole
spectrometer material are taken into account by
a variation of 10\% of this effect in the simulation. 
The effect of pion decay is treated in the same way as the absorption
and is 20\% at 500~\MeVc and negligible at 3~\GeVc. 

The uncertainty in the production of background due to tertiary
particles is larger. 
The average correction is $\approx$10\% and up to 20\% at
1~\GeVc. 
The correction includes re-interactions in the detector material as well
as a small component coming from re-interactions in the target.
The validity of the generators used in the simulation was checked by an
analysis of HARP data with incoming protons and charged pions on
aluminium and carbon targets at lower momenta (3~\GeVc and 5~\GeVc).
A 30\% variation of the secondary production was applied.
The average empty-target subtraction amounts to $\approx$20\%.


Owing to the redundancy of the tracking system downstream of the
target the reconstruction efficiency is very robust under the usual
variations of the detector performance during the long data taking
periods. 
Since the momentum is reconstructed without making use of the upstream
drift chamber module (which is more sensitive in its performance to the beam
intensity) the reconstruction efficiency is uniquely determined by the
downstream system.
No variation of the overall efficiency has been observed.
The performance of the TOFW and CHE system have been monitored to be
constant for the data taking periods used in this analysis.
The calibration of the detectors was performed on a day-by-day basis.

\subsection{Error estimation}
\label{errorest}


Different types of sources 
induce
systematic errors for the analysis
described here: 
track yield corrections ($\sim 5 \%$), particle identification ($\sim 0.1 \%$),
momentum and angular reconstruction ($\sim 0.5 \%$)~\footnote{
The quoted error in parenthesis refers to fractional error of the integrated cross-section
in the kinematic range covered by the HARP experiment}.
The dominant error is due to track yield correction, in particular
due to the subtraction of tertiary particles, with smaller contributions from particle
absorption, empty target subtraction and reconstruction efficiency. 
The strategy to calculate these systematic errors and the different
methods used for their evaluation are described in~\cite{ref:carbonfw} 
where the numerical values are tabulated.
An additional source of error is due to misidentified secondary kaons, 
which are not considered in the particle identification method used for
this analysis and are subtracted on the
basis of a Monte Carlo simulation, as in~\cite{ref:carbonfw}. 
No explicit correction is made for pions coming from decays of other 
particles created in the target, as they give a very small contribution
according to the selection criteria applied in the analysis.

Examples of experimental uncertainties are shown in Figure~\ref{fig:syst1}  
for Be target with incident $\pi^{-}$
for $\pi^{\pm}$  production at two incident beam momenta (5 and 8~\GeVc).
They are very similar for the other beam momenta  and targets. 
Going from lighter (Be, C) to heavier targets (Ta, Pb)
the corrections for $\pi^{0}$ conversion and absorption/tertiaries
increase slightly.
\begin{figure*}[tbp]
  \begin{center}
\includegraphics[width=0.49\textwidth]{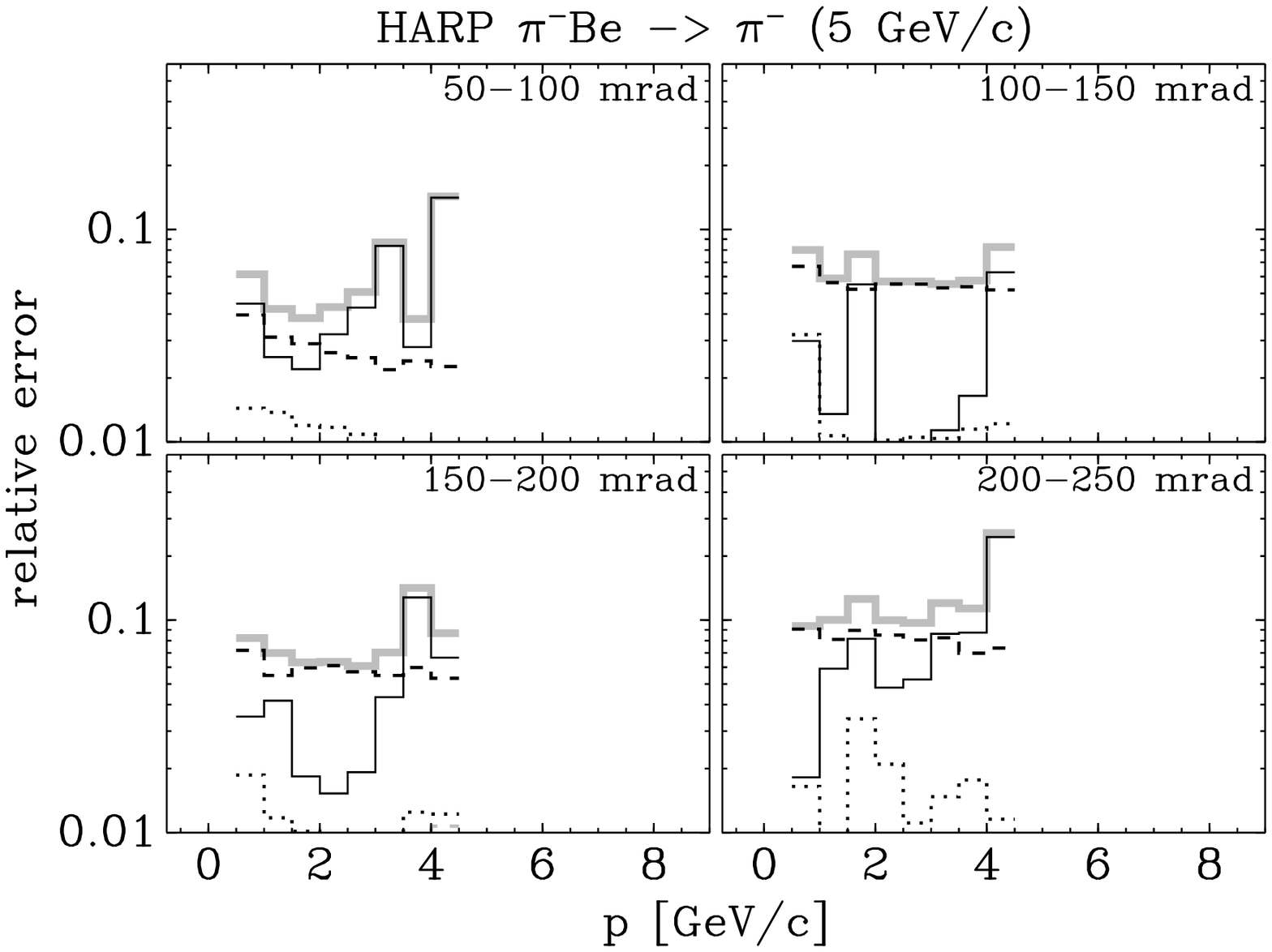}
\includegraphics[width=0.49\textwidth]{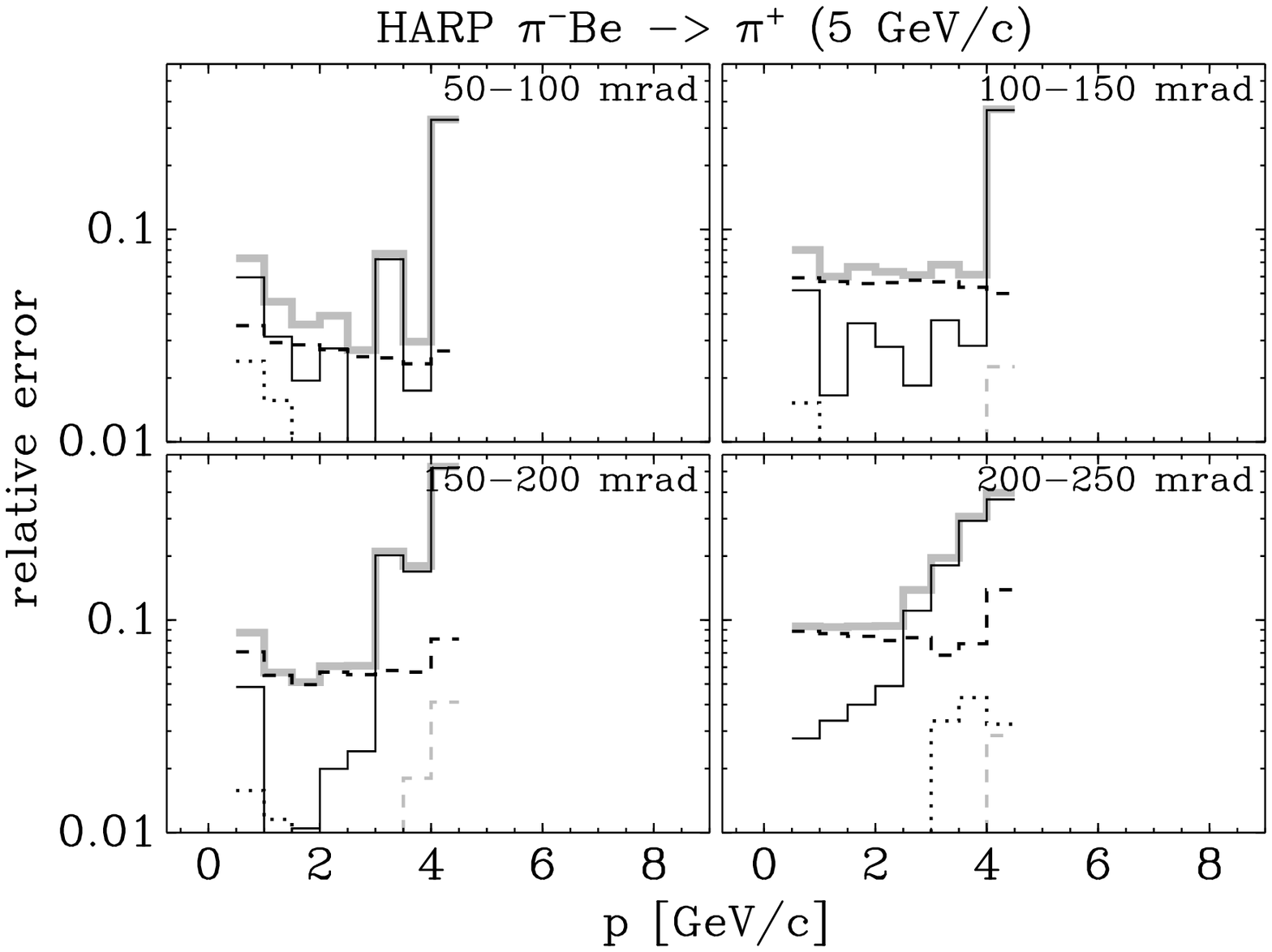}
\\
\includegraphics[width=0.49\textwidth]{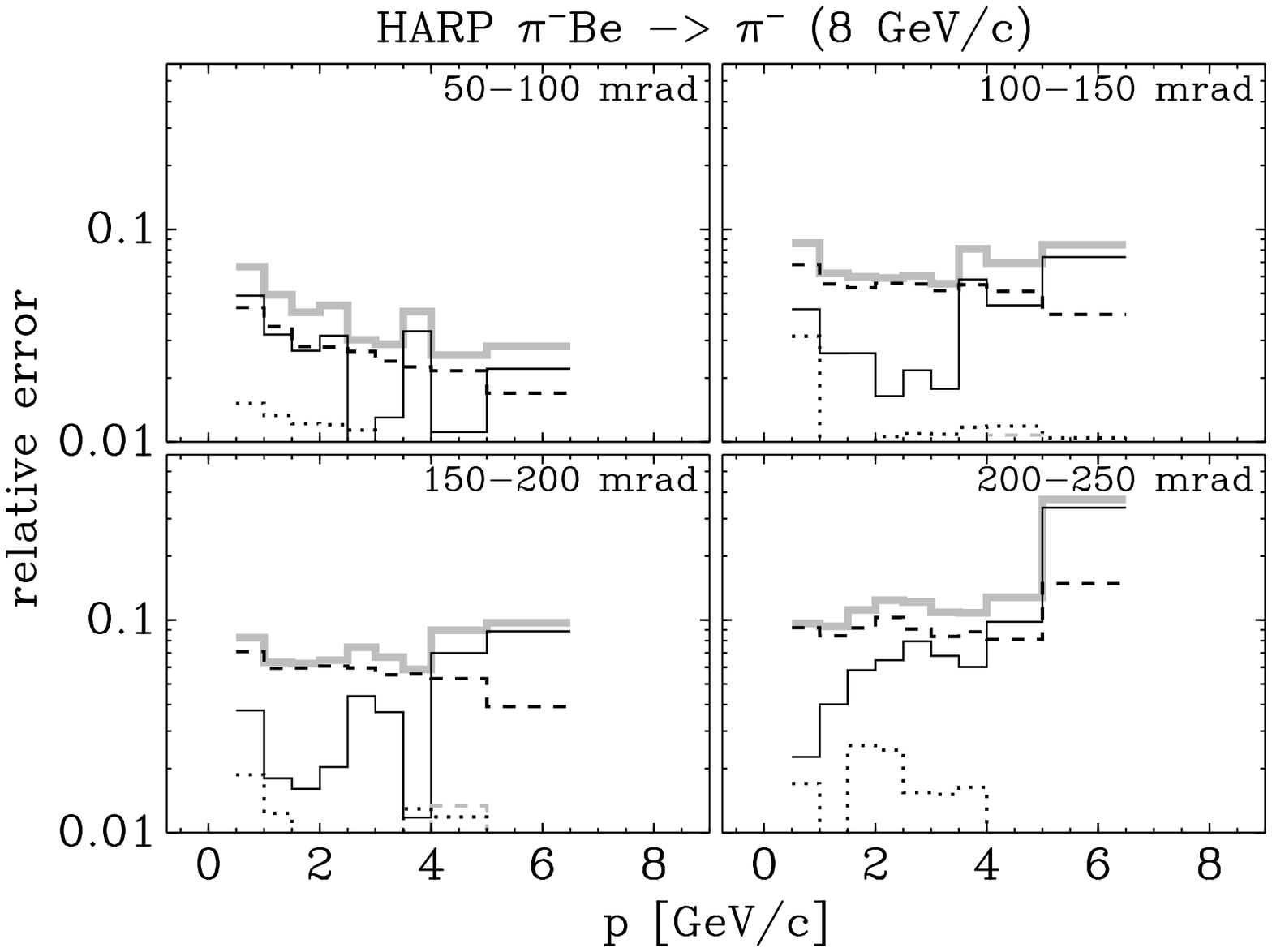}
\includegraphics[width=0.49\textwidth]{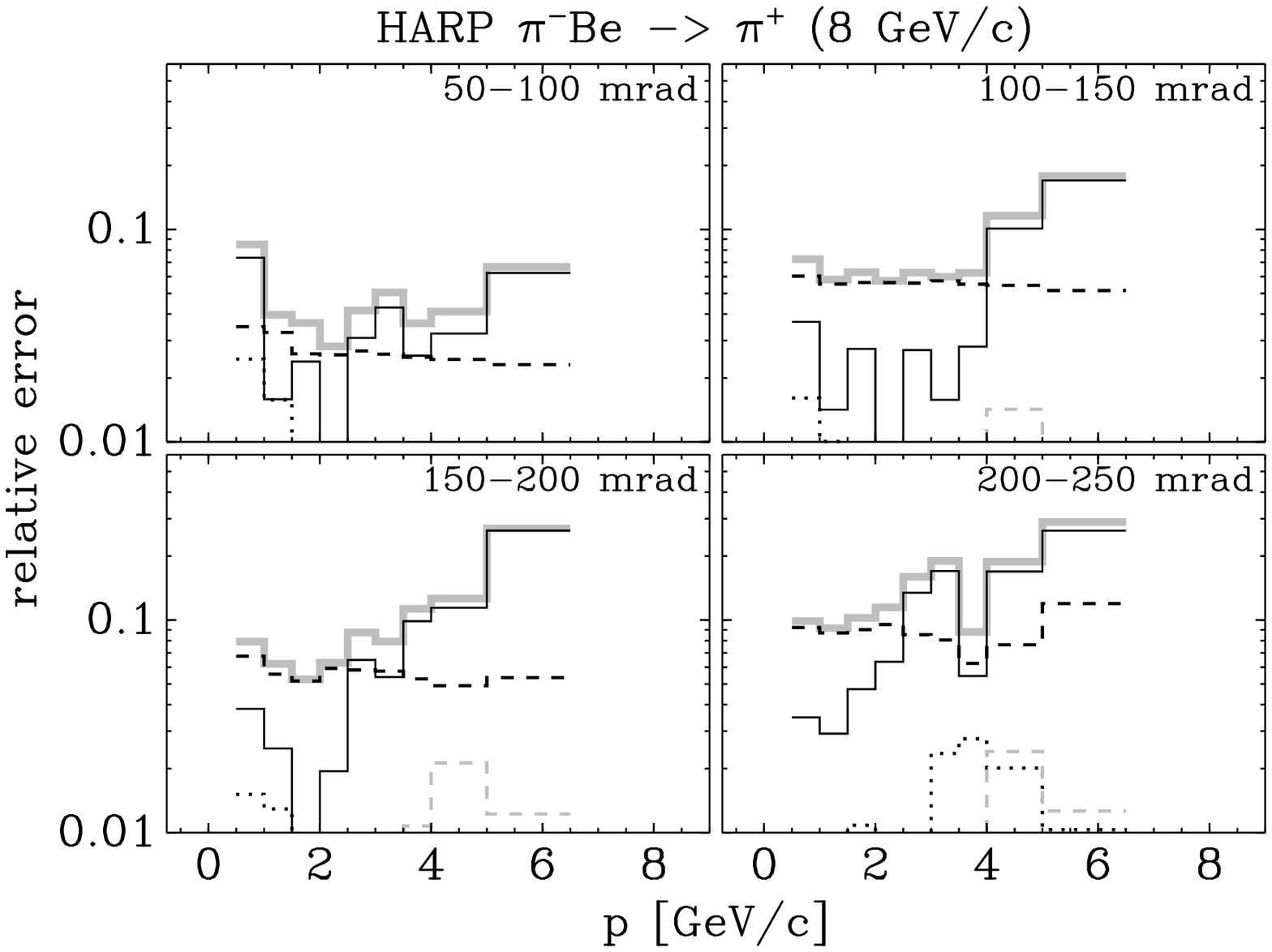}
\end{center}
\caption{Systematic errors as a function of momentum of the outgoing
pions for the particular case of 5 and 8 \GeVc \pim  interacting on a
 beryllium target.
 Plots in the upper left panel are  for incident 5 \GeVc $\pim$ producing $\pim$,
plots in the upper right panel are for 5 \GeVc $\pim$ producing $\pip$, plots in the lower left panels
are for  8 \GeVc $\pim$ producing $\pim$ and 
plots in the lower right panel are for 8 \GeVc $\pim$ producing $\pip$. 
Total systematic error (grey solid line) are displayed together with the
 main components:
black short-dashed line for absorption+tertiaries interactions,
black dotted line for track efficiency and target pointing efficiency,
black dot-dashed line for $\pi^{0}$ subtraction, black solid line for
momentum scale+resolution and angle scale, grey short-dashed line
for PID.
}
\label{fig:syst1}
\end{figure*}

The overall normalization has an uncertainty of $\sim 2\%$ and is
mainly due to the uncertainty in the efficiency that beam pions counted
in the normalization actually hit the target, with smaller components
from the target density and the beam particle counting procedure. 
On average the total integrated systematic error is around $5-6\%$,
with a differential bin-to-bin systematic error of the order of
$10-11 \%$, to be compared with a statistical integrated (bin-to-bin
differential) error of $\sim 2-3 \%$ ($\sim 10-13 \%$).
Systematic and statistical errors are roughly of the same order.

\section{Results}
\label{sec:results}

The measured double-differential cross-sections for the
production of \pip and \pim in the laboratory system as a function of
the momentum and the polar angle for each incident beam momentum are
shown in Figures \ref{fig:Be} to \ref{fig:Pb} for
targets from Be to Pb.
The error bars  shown are the
square-roots of the diagonal elements in the covariance matrix,
where statistical and systematic uncertainties are combined
in quadrature.
The correlation of the statistical errors (introduced by the unfolding
procedure) are typically smaller than 20\% for adjacent momentum bins and
even smaller for adjacent angular bins.
The correlations of the systematic errors are larger, typically 80\% for
adjacent bins.
The overall scale error ($<2\%$) is not shown.
The results of this analysis are also tabulated in Appendix A.
We have not included the scale errors in the tables to make it possible to 
calculate e.g. integrated particle ratios taking the scale errors into 
account only when applicable, i.e. when different beams are compared.

\begin{figure}[tb]
\centering
\includegraphics[width=.43\textwidth]{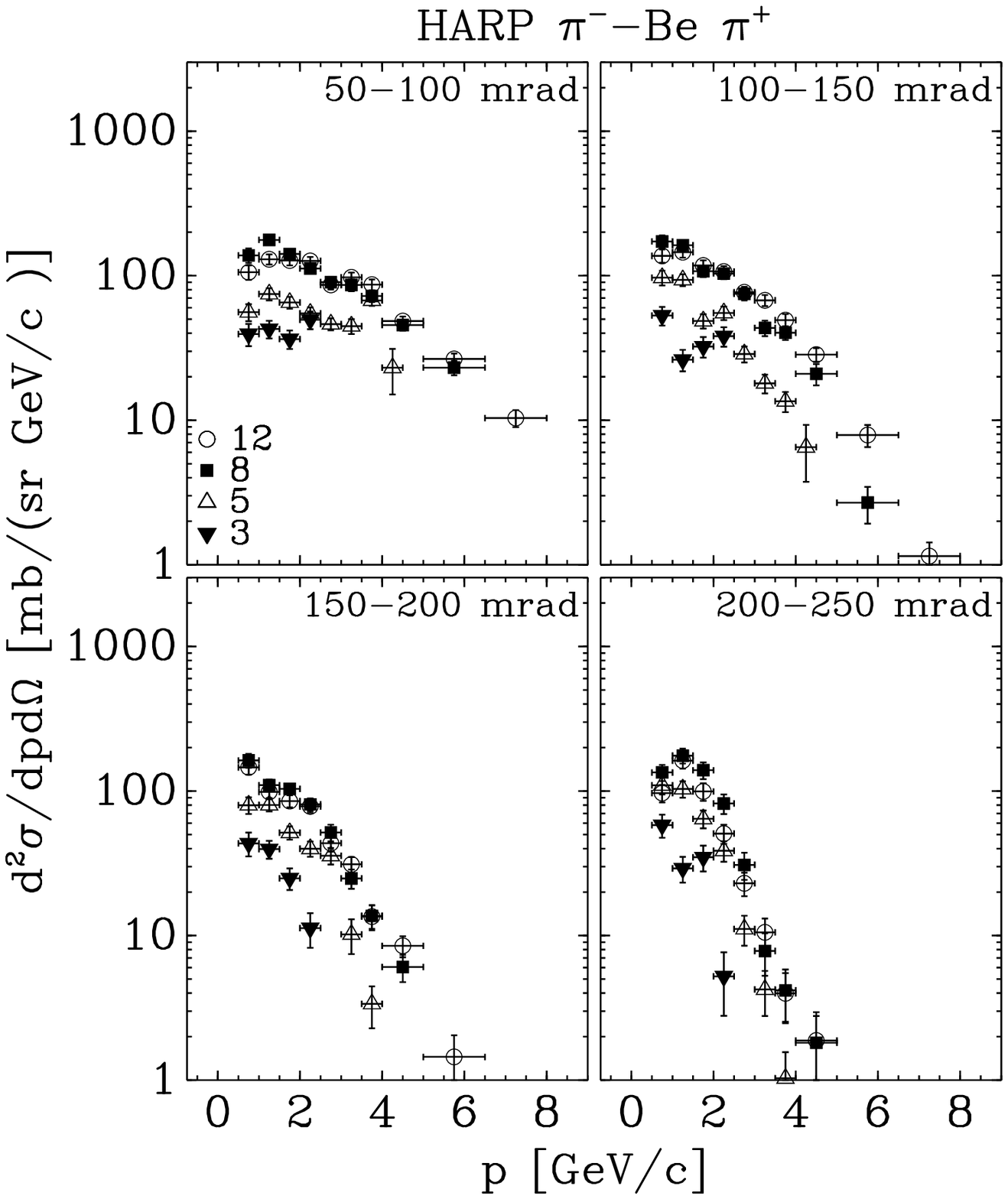}
\includegraphics[width=.43\textwidth]{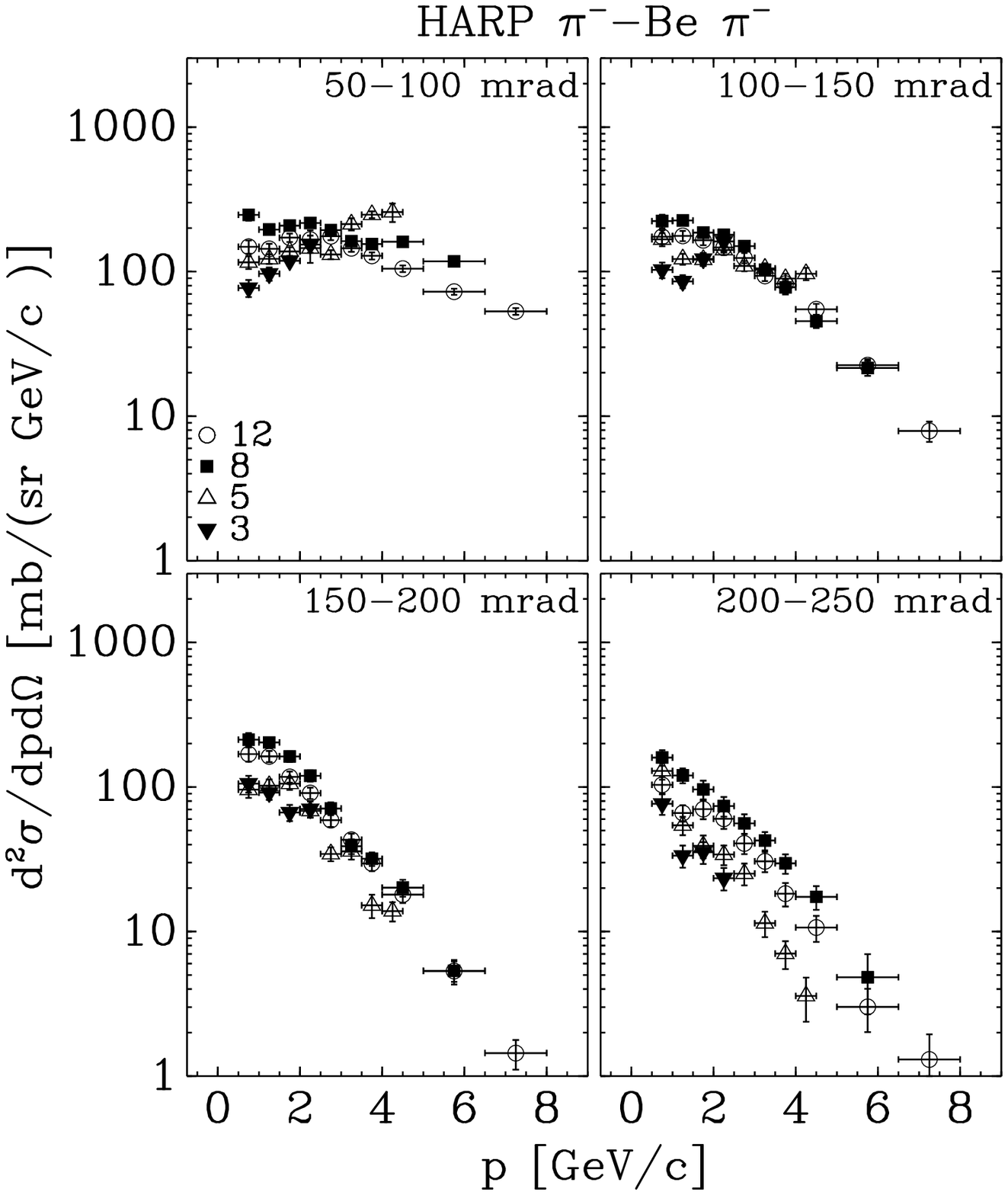}
\includegraphics[width=.43\textwidth]{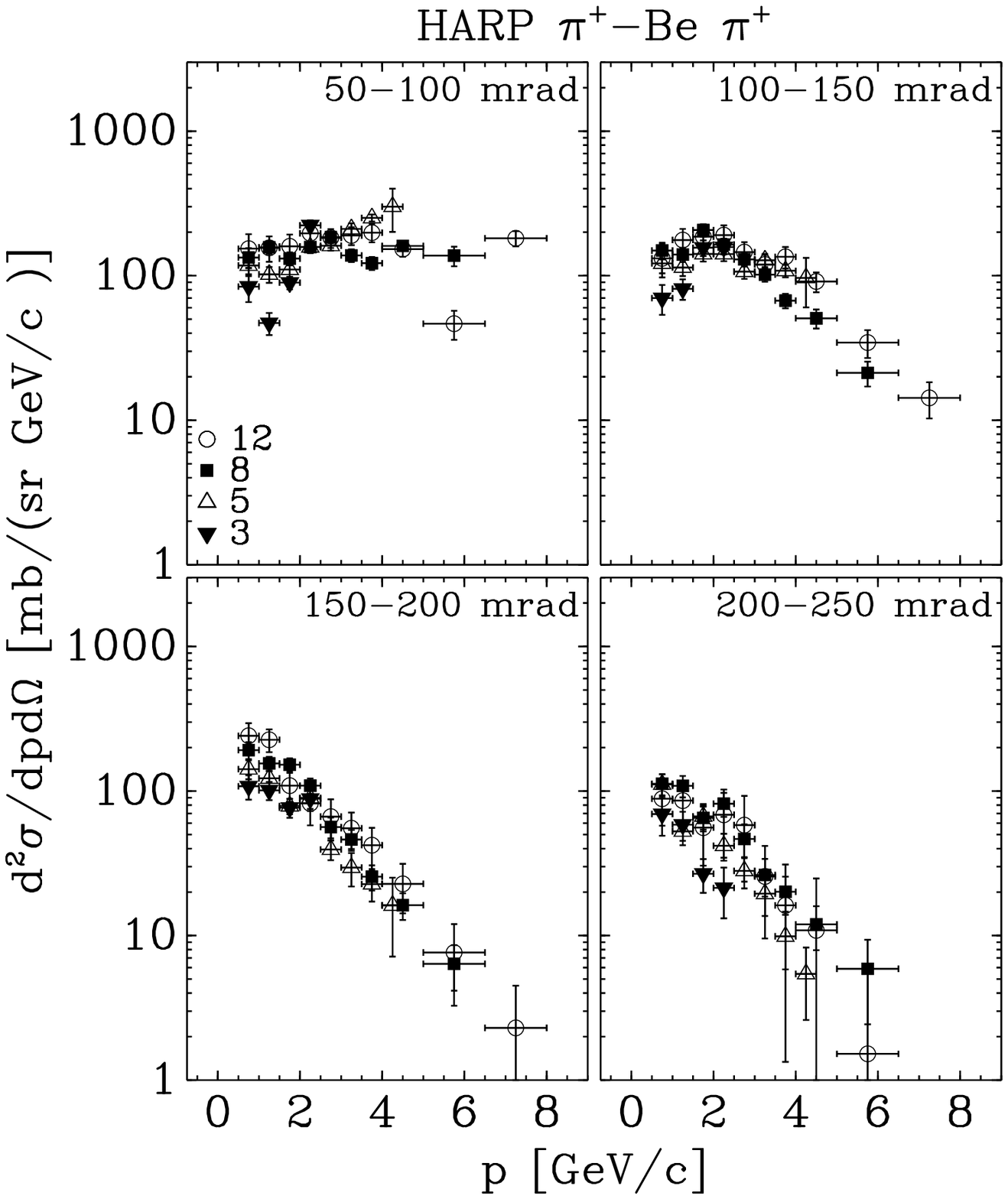}
\includegraphics[width=.43\textwidth]{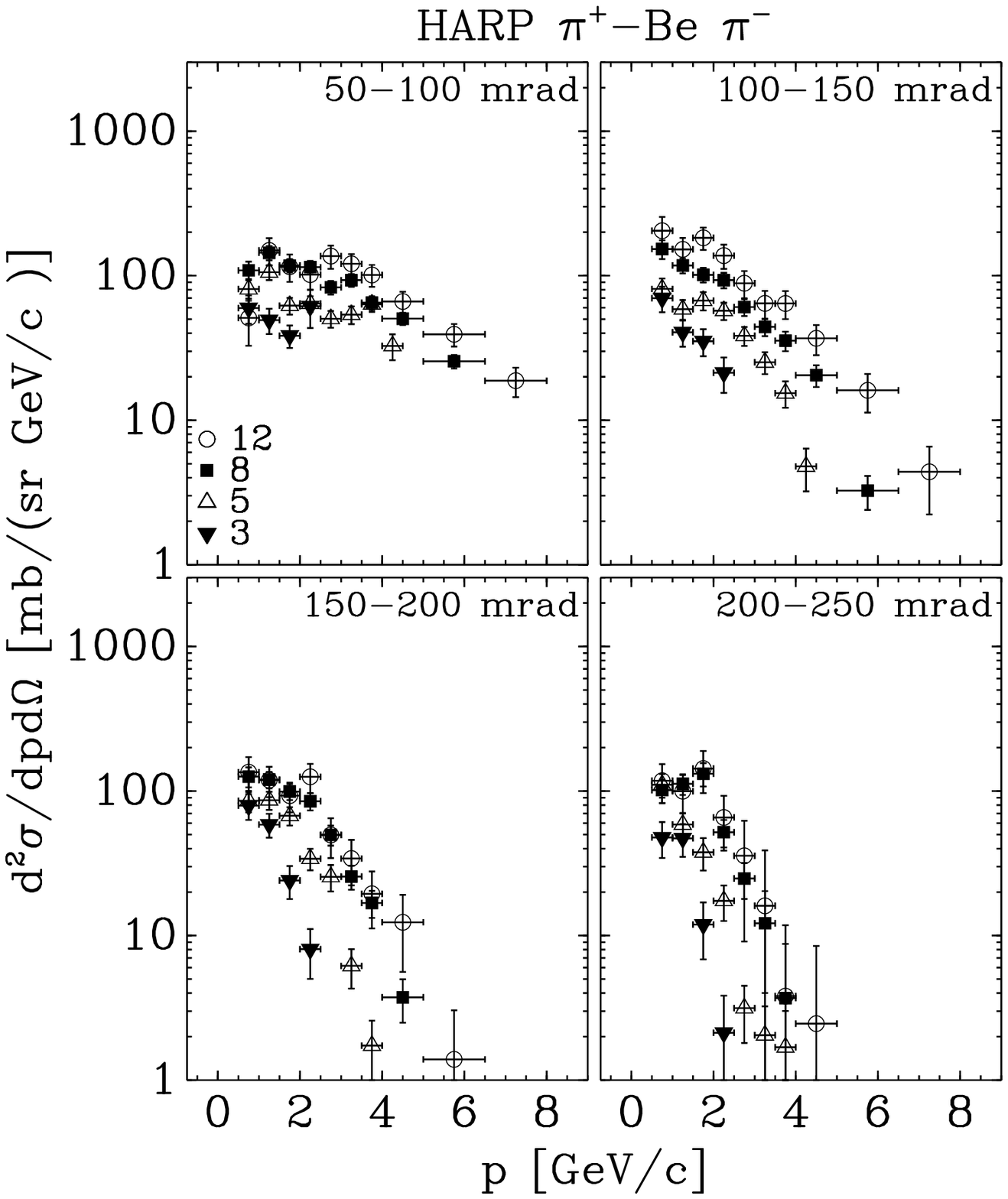}
\caption{$\pi^{-}-$Be (top) and $\pi^{+}-$Be (bottom) differential cross-sections
for different incoming beam momenta (from 3 to 12~\GeVc).
Left panels show the
$\pi^{+}$ production, while right panels show the $\pi^{-}$ production.   
In the top right corner of each plot the covered angular range is shown in mrad.}
\label{fig:Be}
\end{figure}

\begin{figure}[tb]
\centering
\includegraphics[width=.49\textwidth]{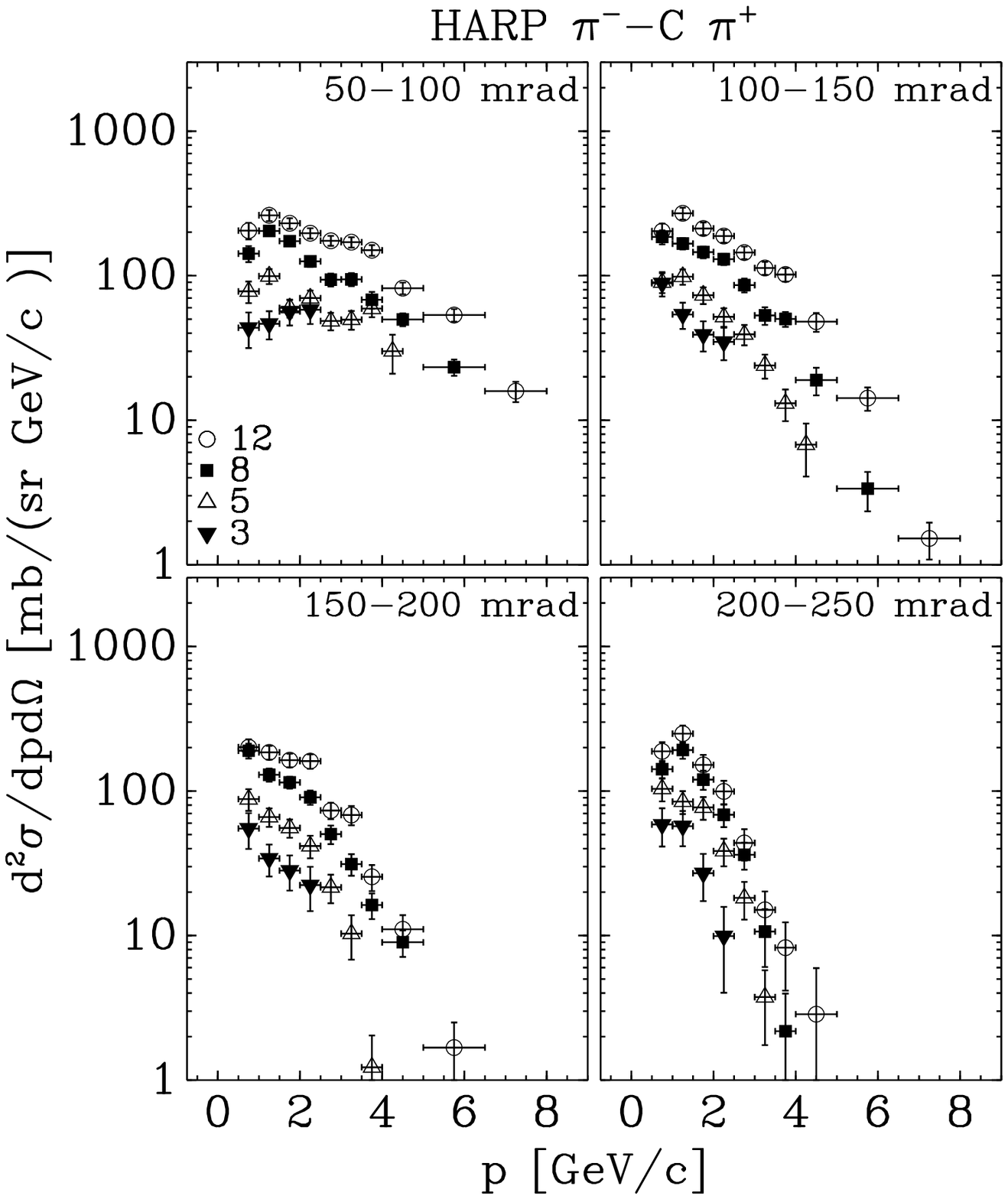}
\includegraphics[width=.49\textwidth]{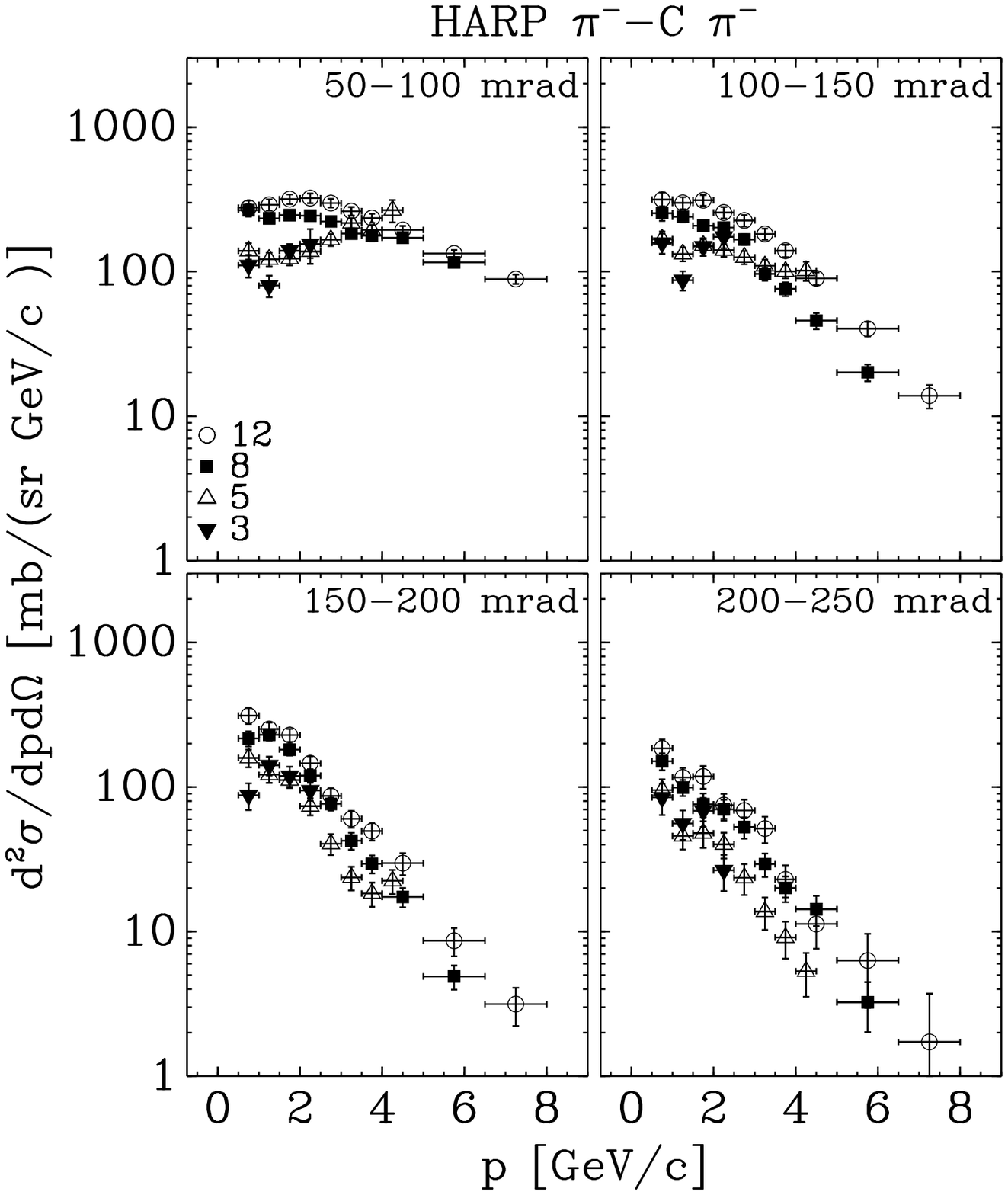}
\includegraphics[width=.49\textwidth]{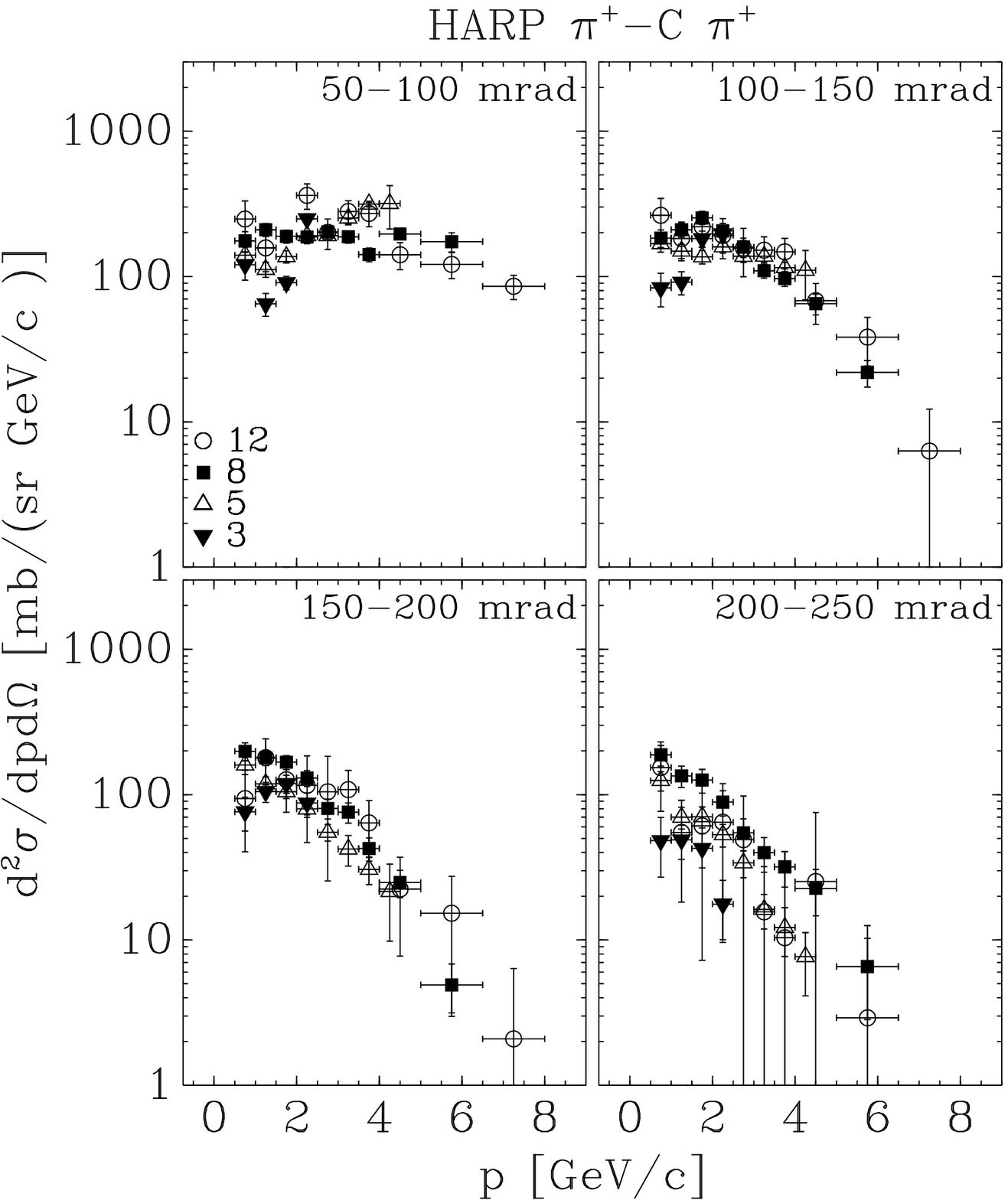}
\includegraphics[width=.49\textwidth]{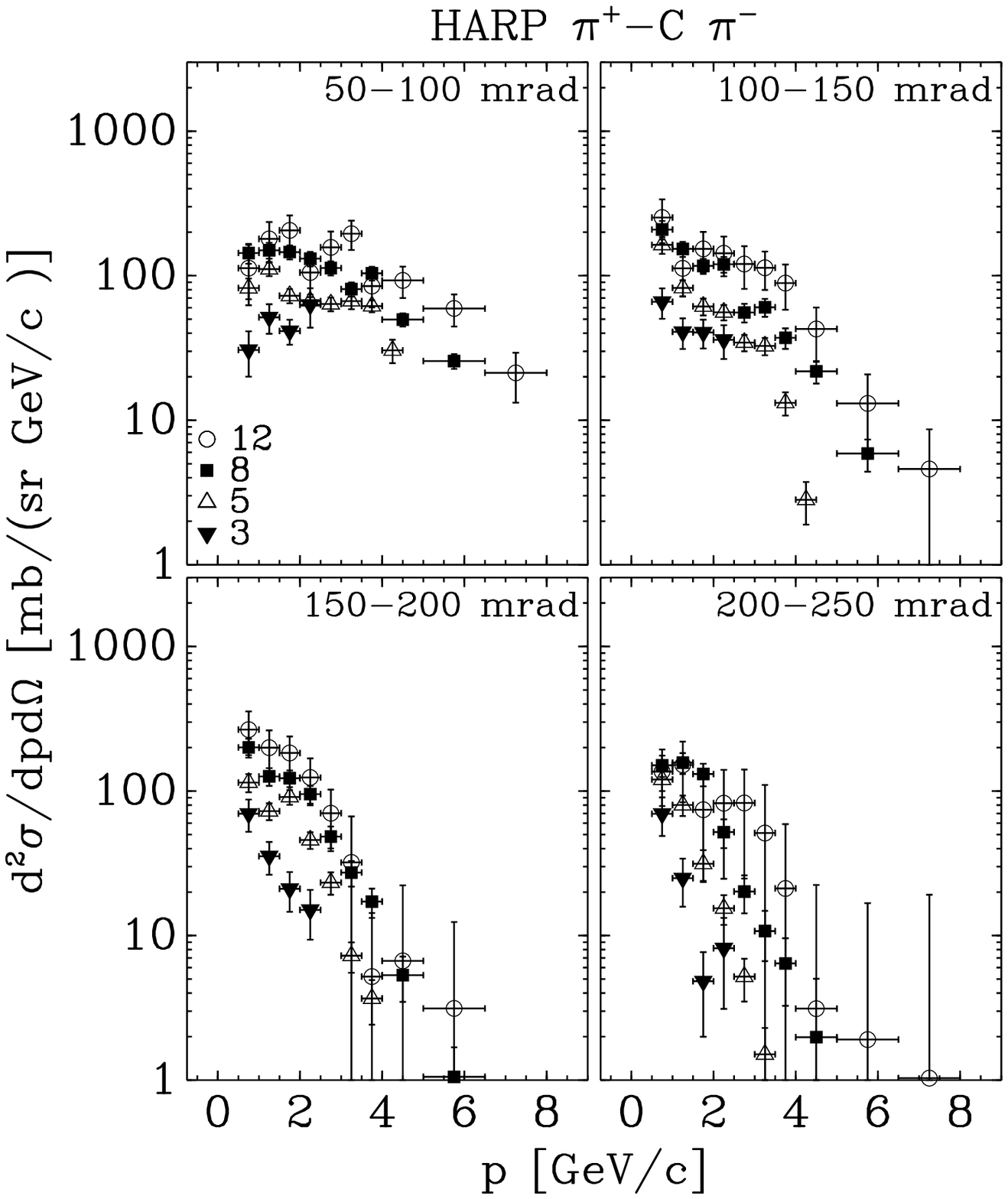}
\caption{$\pi^{-}-$C (top) and $\pi^{+}-$C (bottom) differential cross-sections
for different incoming beam momenta (from 3 to 12~\GeVc).
Left panels show the
$\pi^{+}$ production, while right panels show the $\pi^{-}$ production.   
In the top right corner of each plot the covered angular range is shown in mrad.}
\label{fig:C}
\end{figure}

\begin{figure}[tb]
\centering
\includegraphics[width=.49\textwidth]{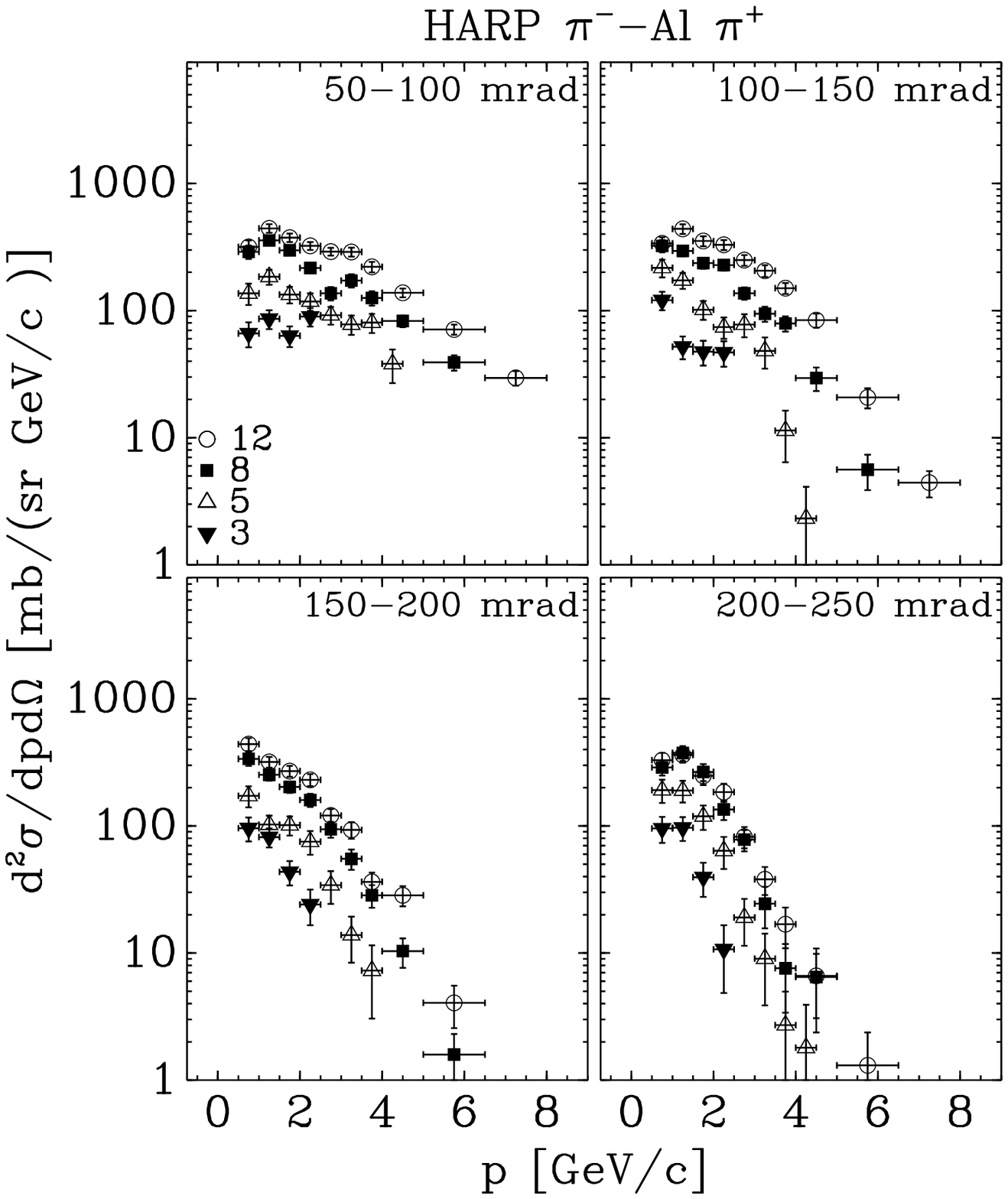}
\includegraphics[width=.49\textwidth]{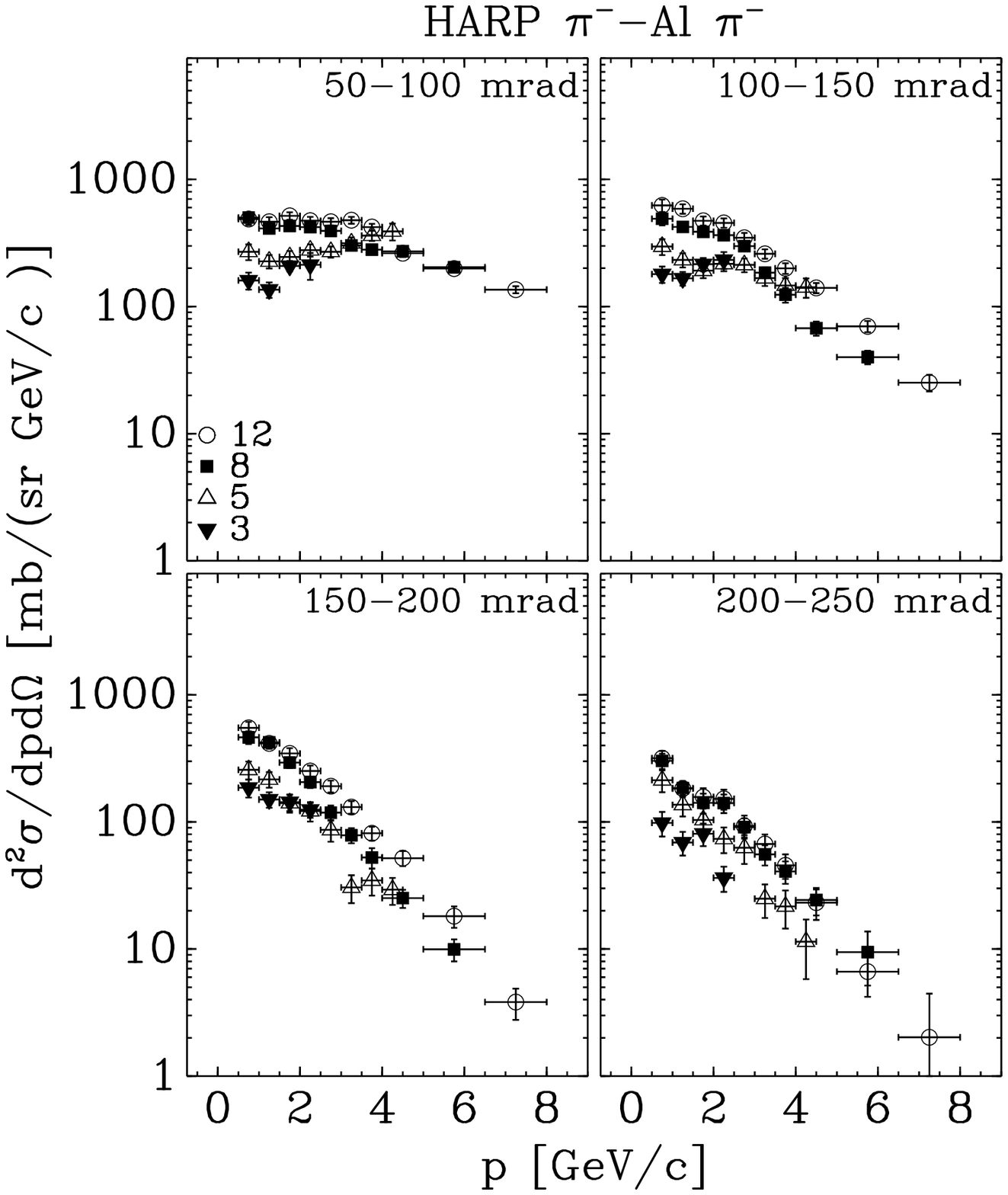}
\includegraphics[width=.49\textwidth]{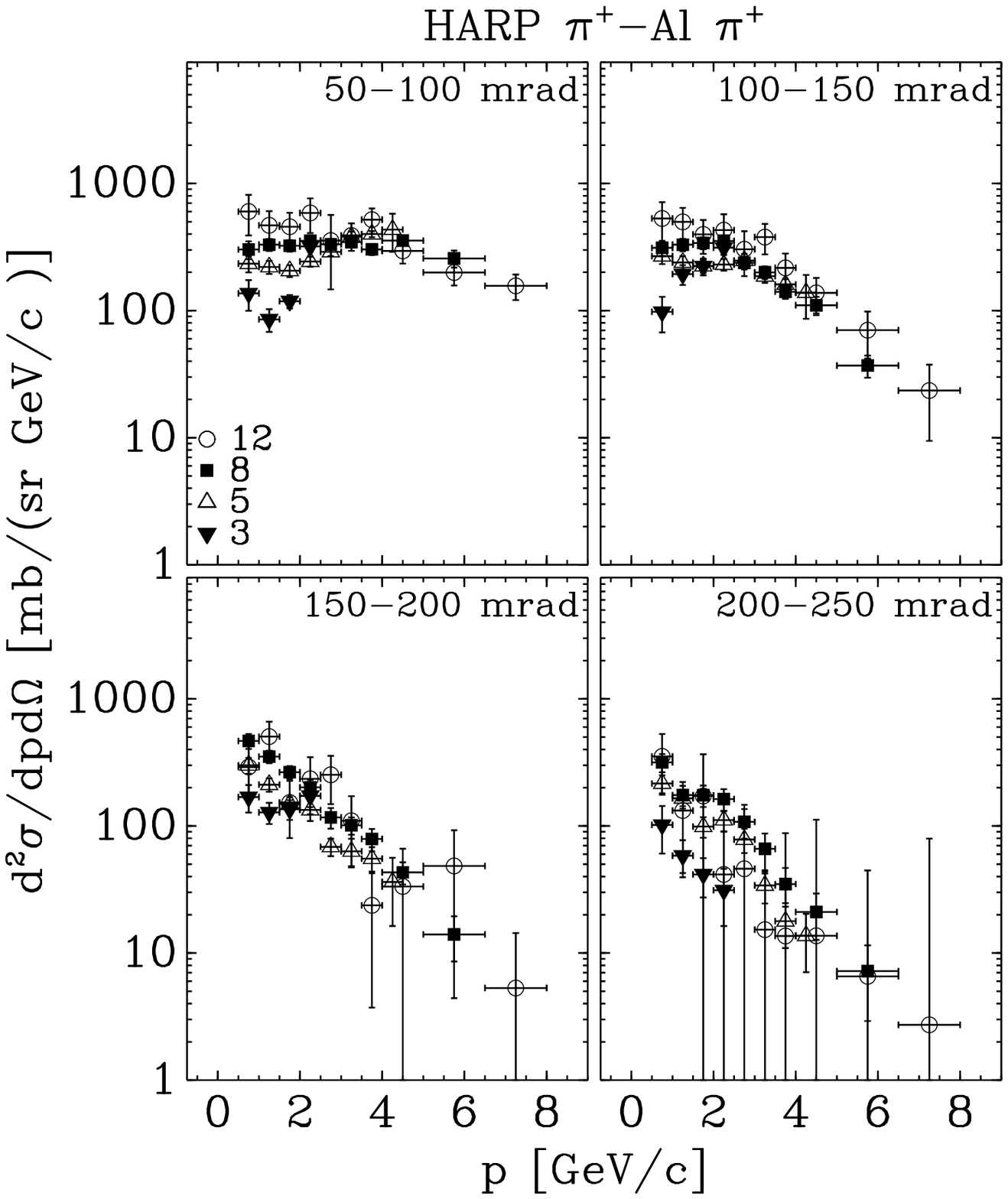}
\includegraphics[width=.49\textwidth]{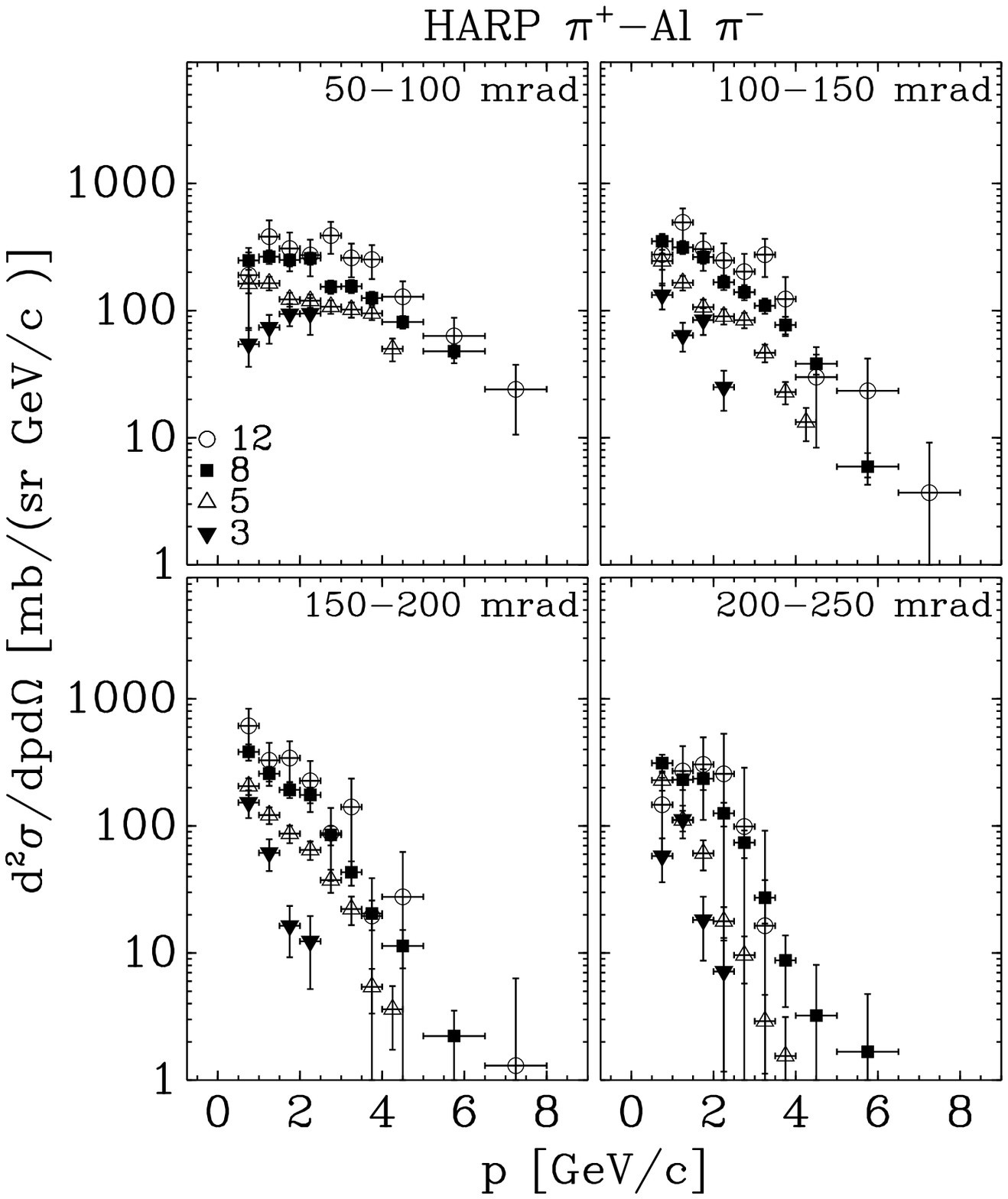}
\caption{$\pi^{-}-$Al (top) and $\pi^{+}-$Al (bottom) differential cross-sections
for different incoming beam momenta (from 3 to 12~\GeVc).
Left panels show the
$\pi^{+}$ production, while right panels show the $\pi^{-}$ production.   
In the top right corner of each plot the covered angular range is shown in mrad.}
\label{fig:Al}
\end{figure}

\begin{figure}[tb]
\centering
\includegraphics[width=.49\textwidth]{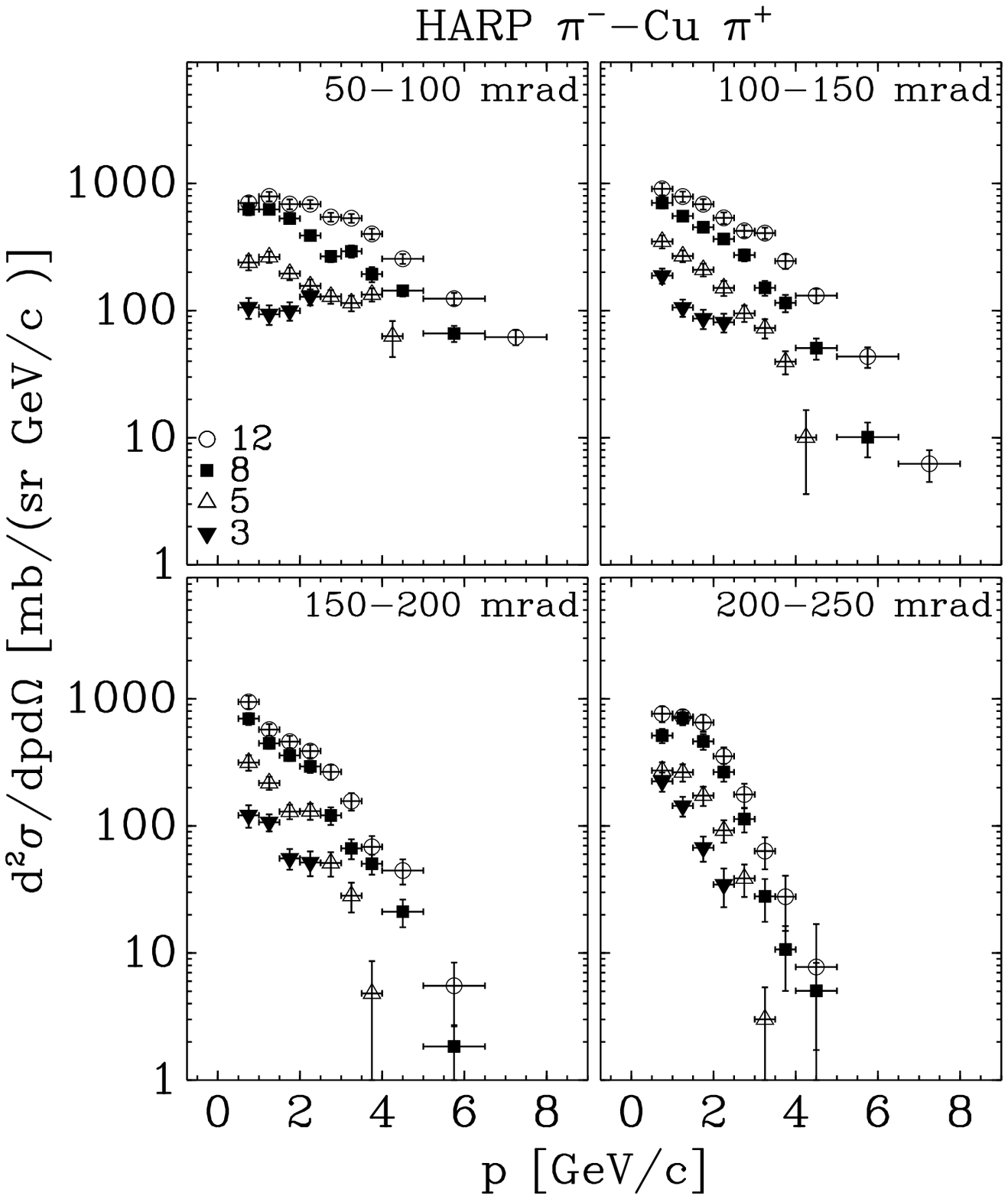}
\includegraphics[width=.49\textwidth]{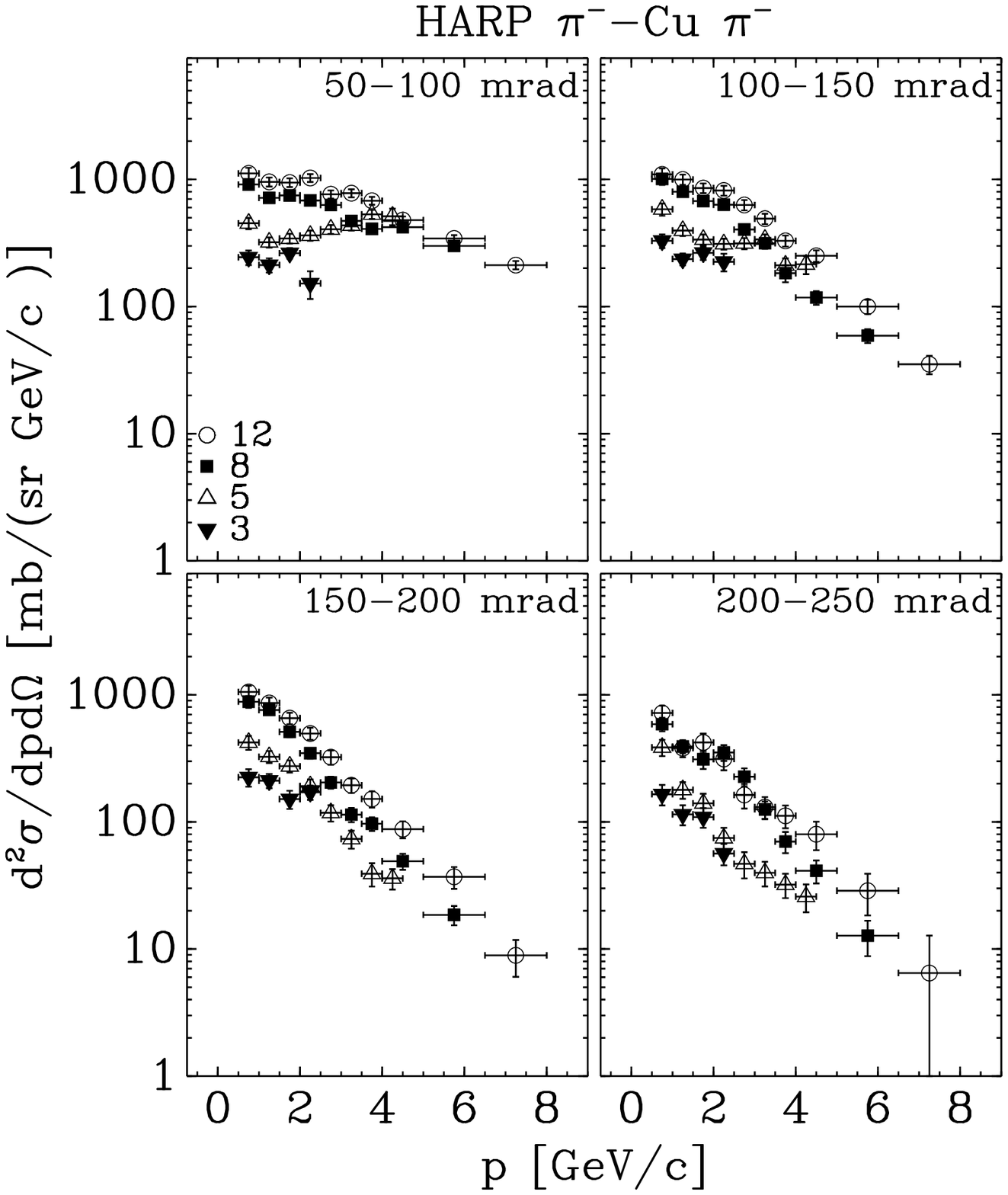}
\includegraphics[width=.49\textwidth]{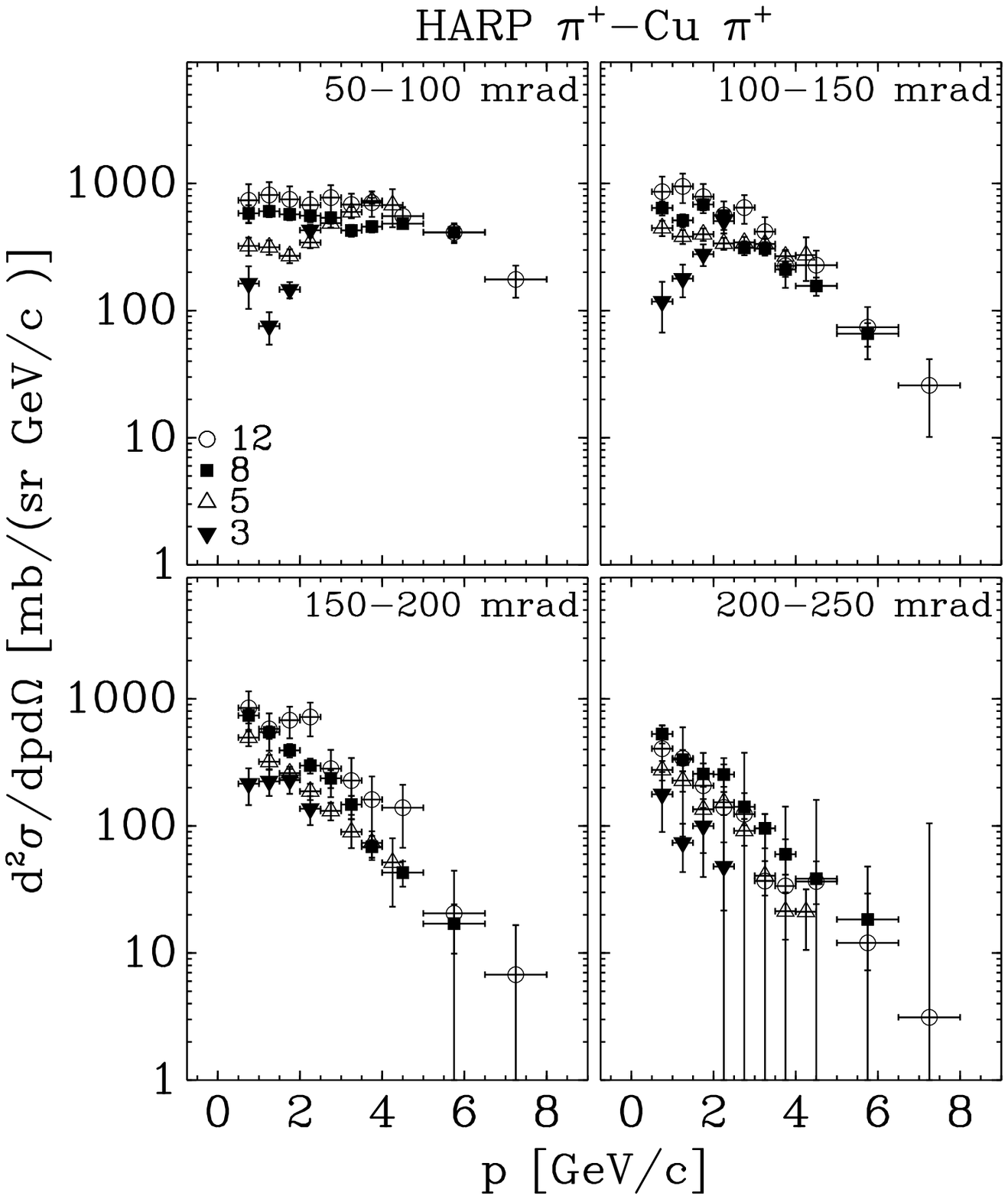}
\includegraphics[width=.49\textwidth]{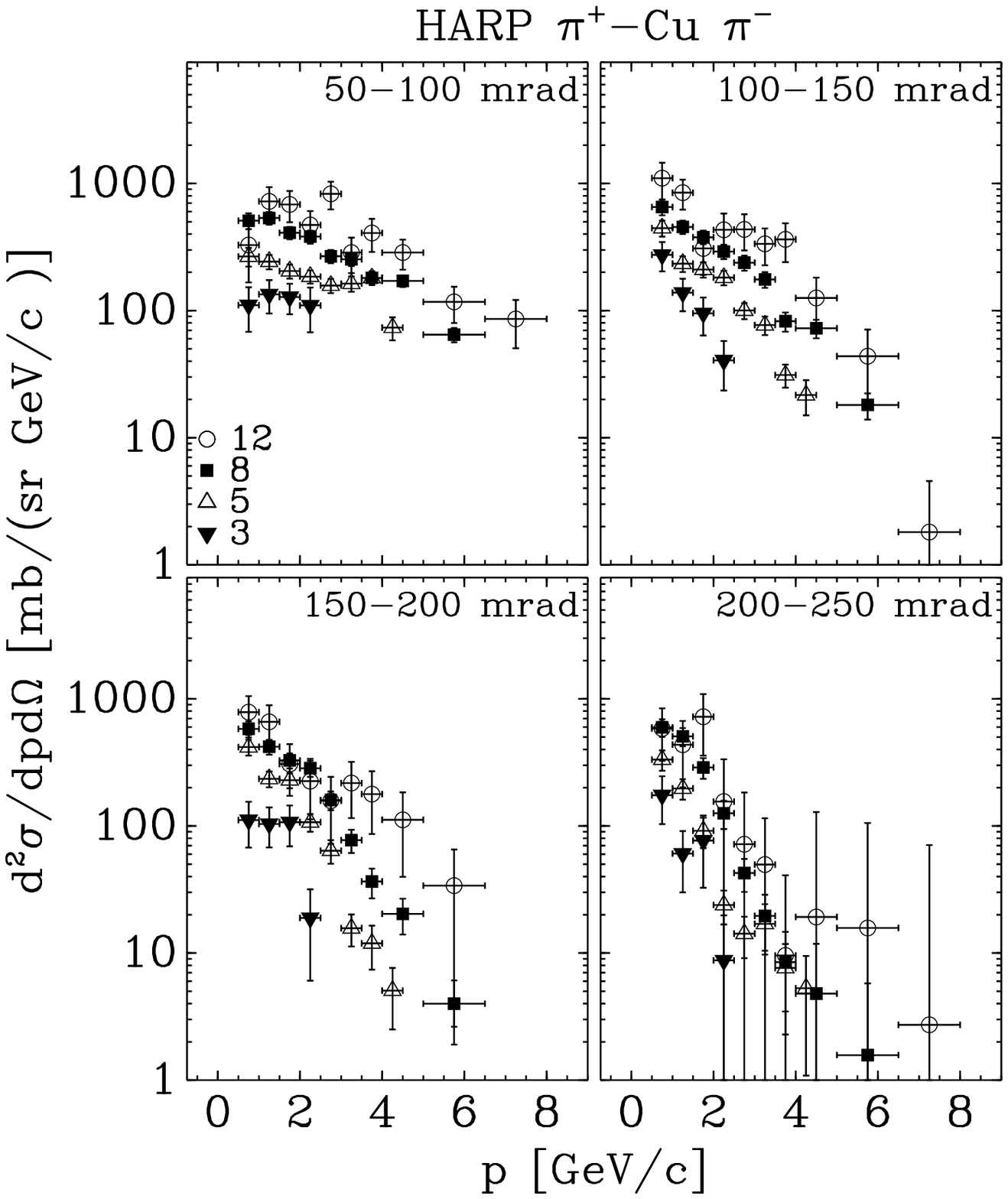}
\caption{$\pi^{-}-$Cu (top) and $\pi^{+}-$Cu (bottom) differential cross-sections
for different incoming beam momenta (from 3 to 12~\GeVc).
Left panels show the
$\pi^{+}$ production, while right panels show the $\pi^{-}$ production.   
In the top right corner of each plot the covered angular range is shown in mrad.}
\label{fig:Cu}
\end{figure}

\begin{figure}[tb]
\centering
\includegraphics[width=.49\textwidth]{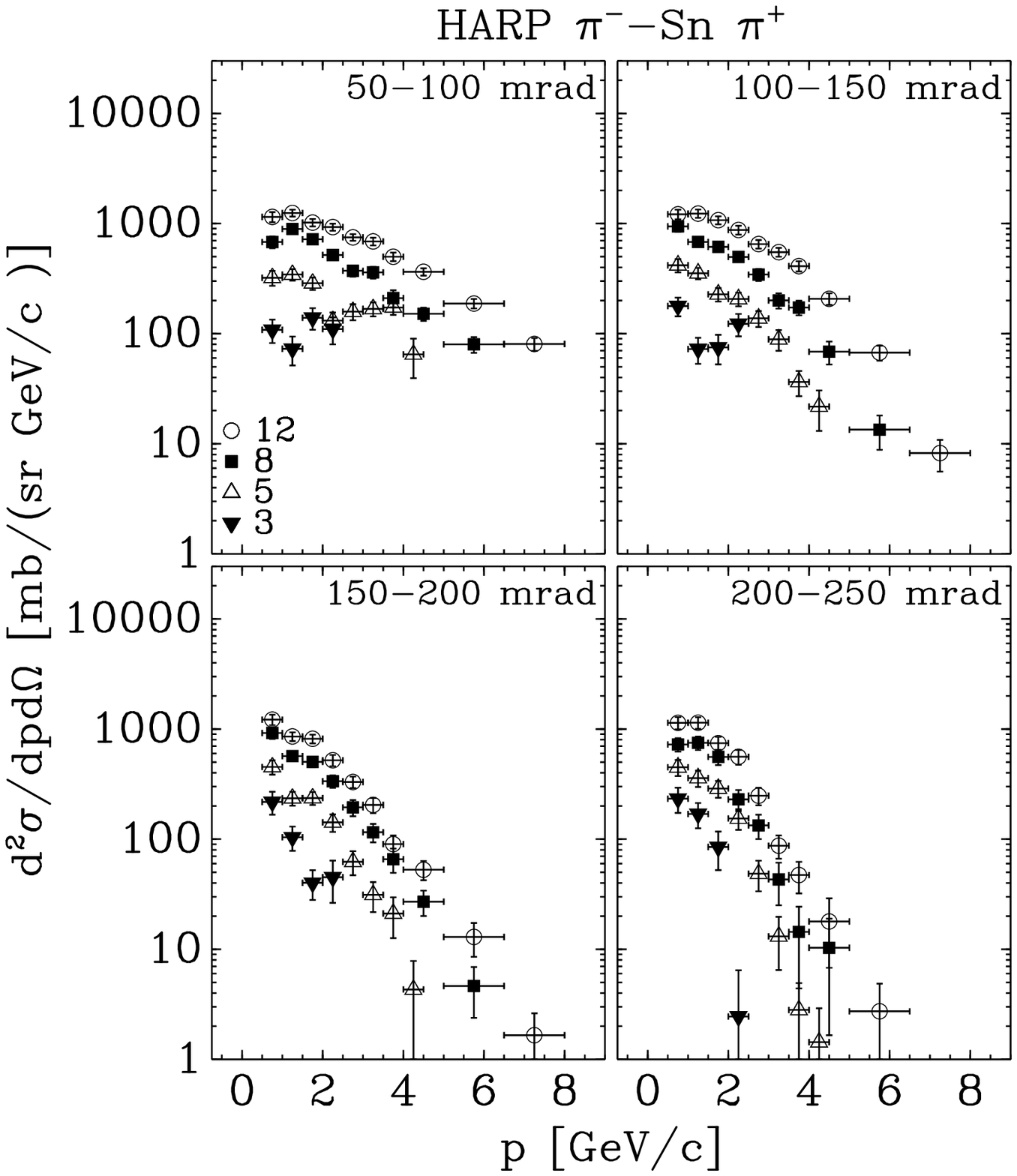}
\includegraphics[width=.49\textwidth]{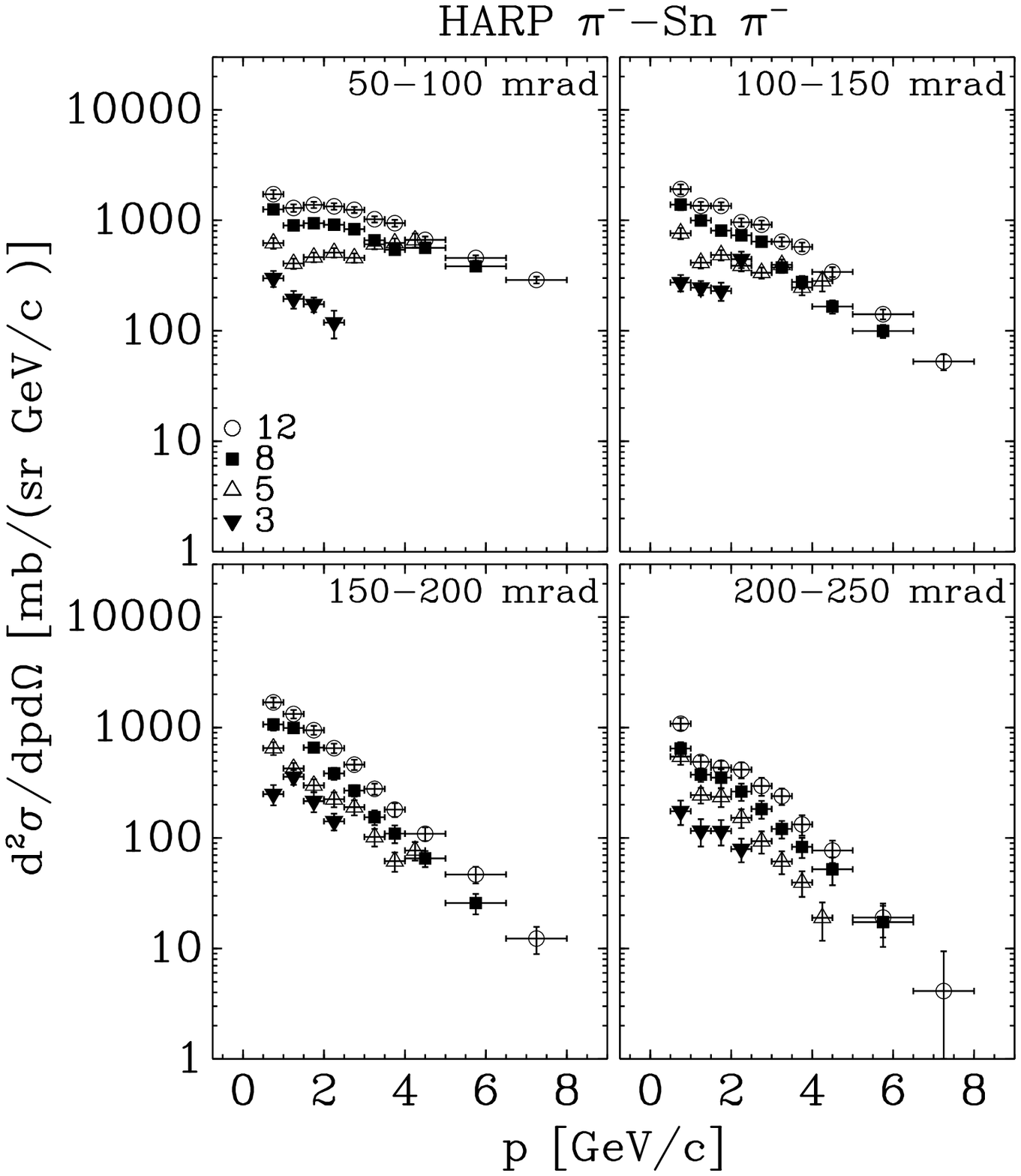}
\includegraphics[width=.49\textwidth]{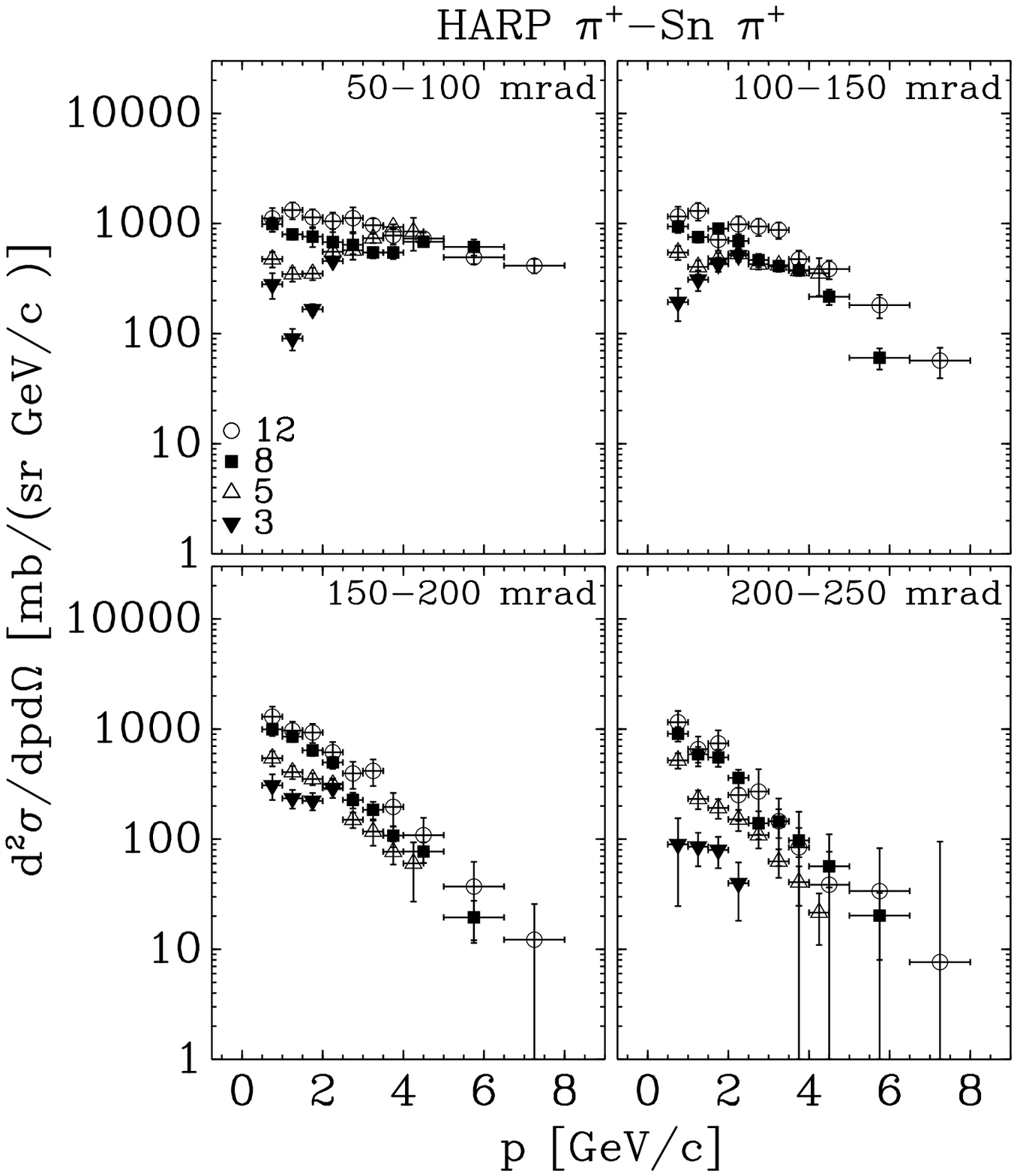}
\includegraphics[width=.49\textwidth]{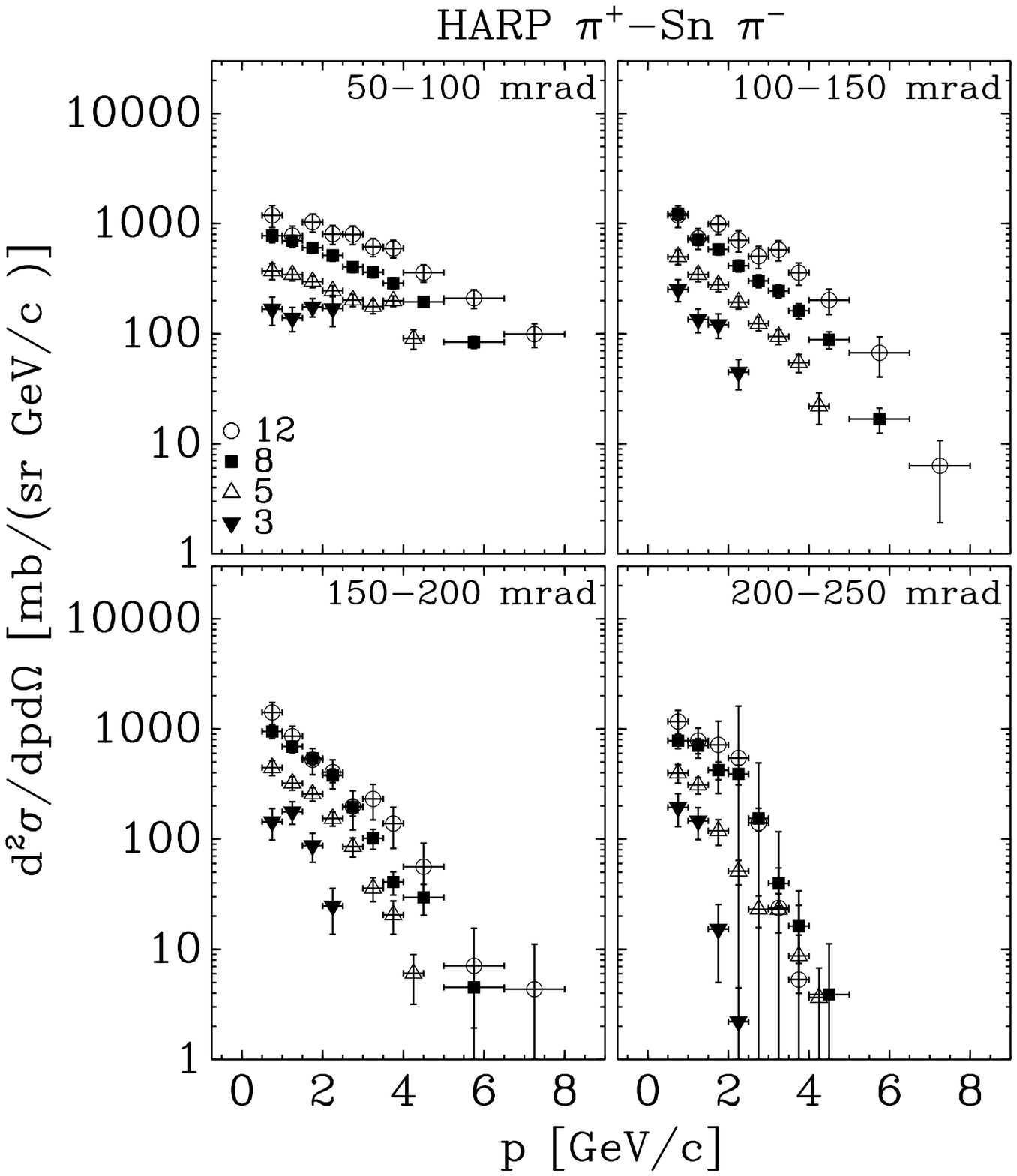}
\caption{$\pi^{-}-$Sn (top) and $\pi^{+}-$Sn (bottom) differential cross-sections
for different incoming beam momenta (from 3 to 12~\GeVc).
Left panels show the
$\pi^{+}$ production, while right panels show the $\pi^{-}$ production.   
In the top right corner of each plot the covered angular range is shown in mrad.}
\label{fig:Sn}
\end{figure}
\begin{figure}[tb]
\centering
\includegraphics[width=.49\textwidth]{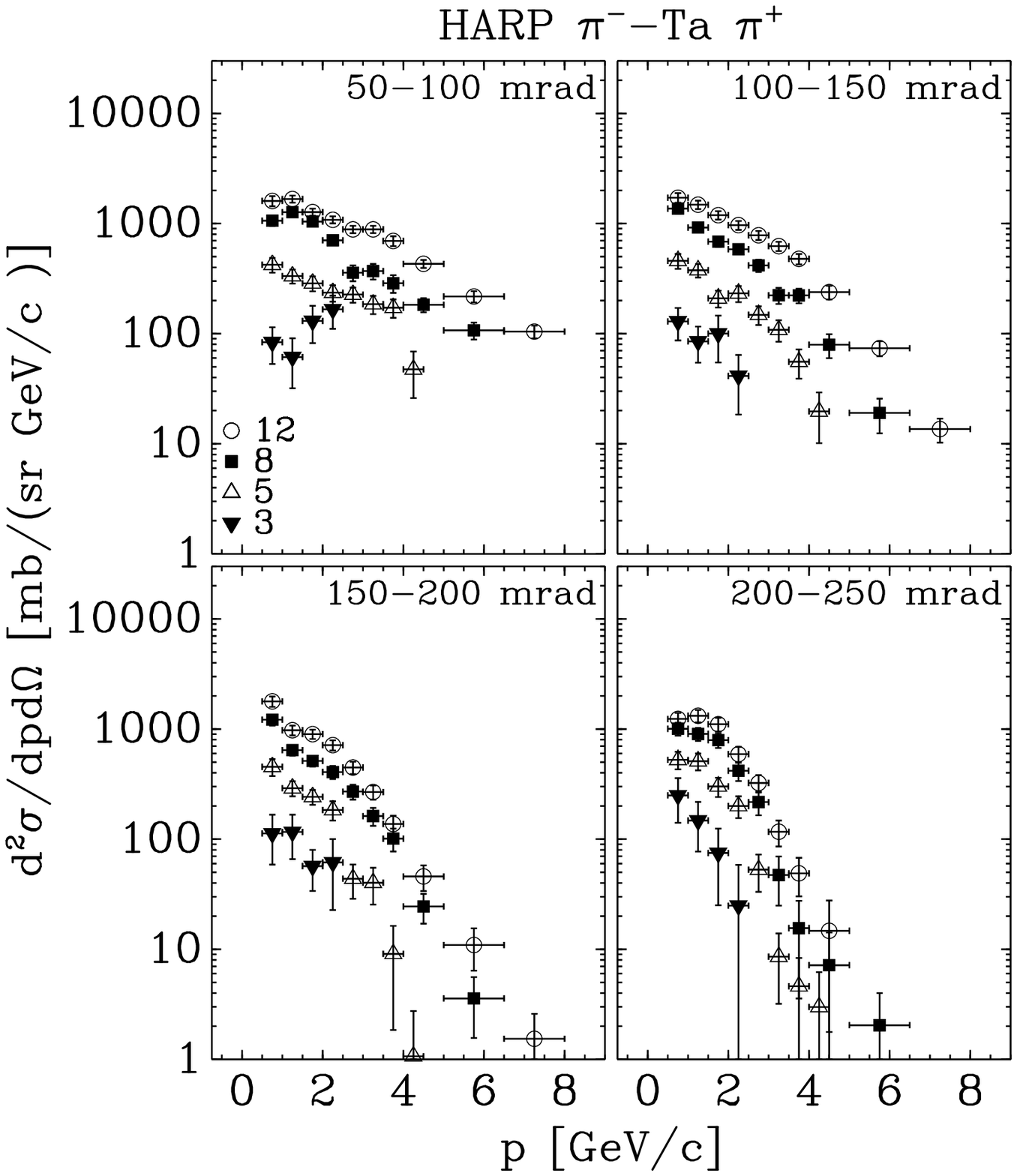}
\includegraphics[width=.49\textwidth]{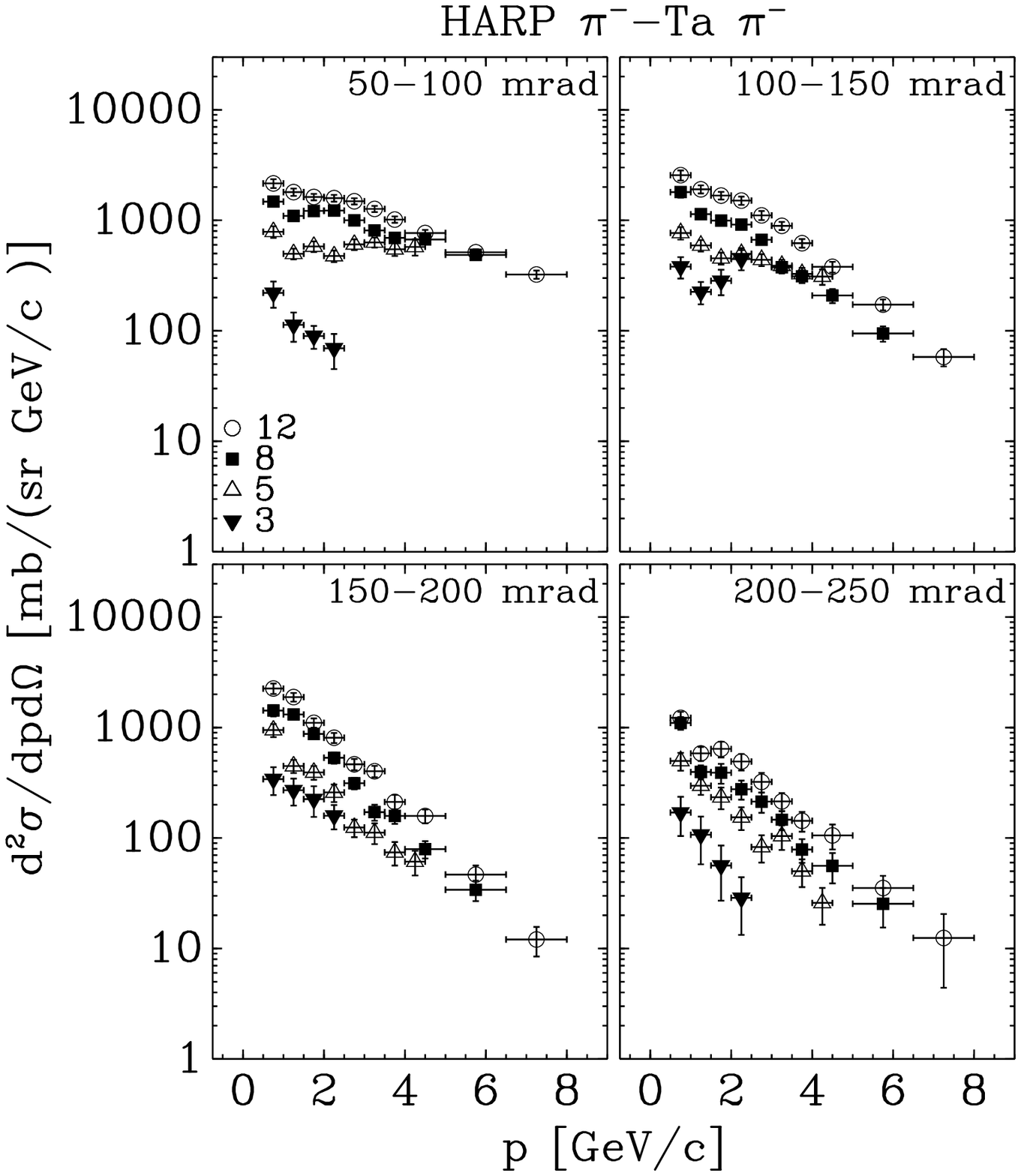}
\includegraphics[width=.49\textwidth]{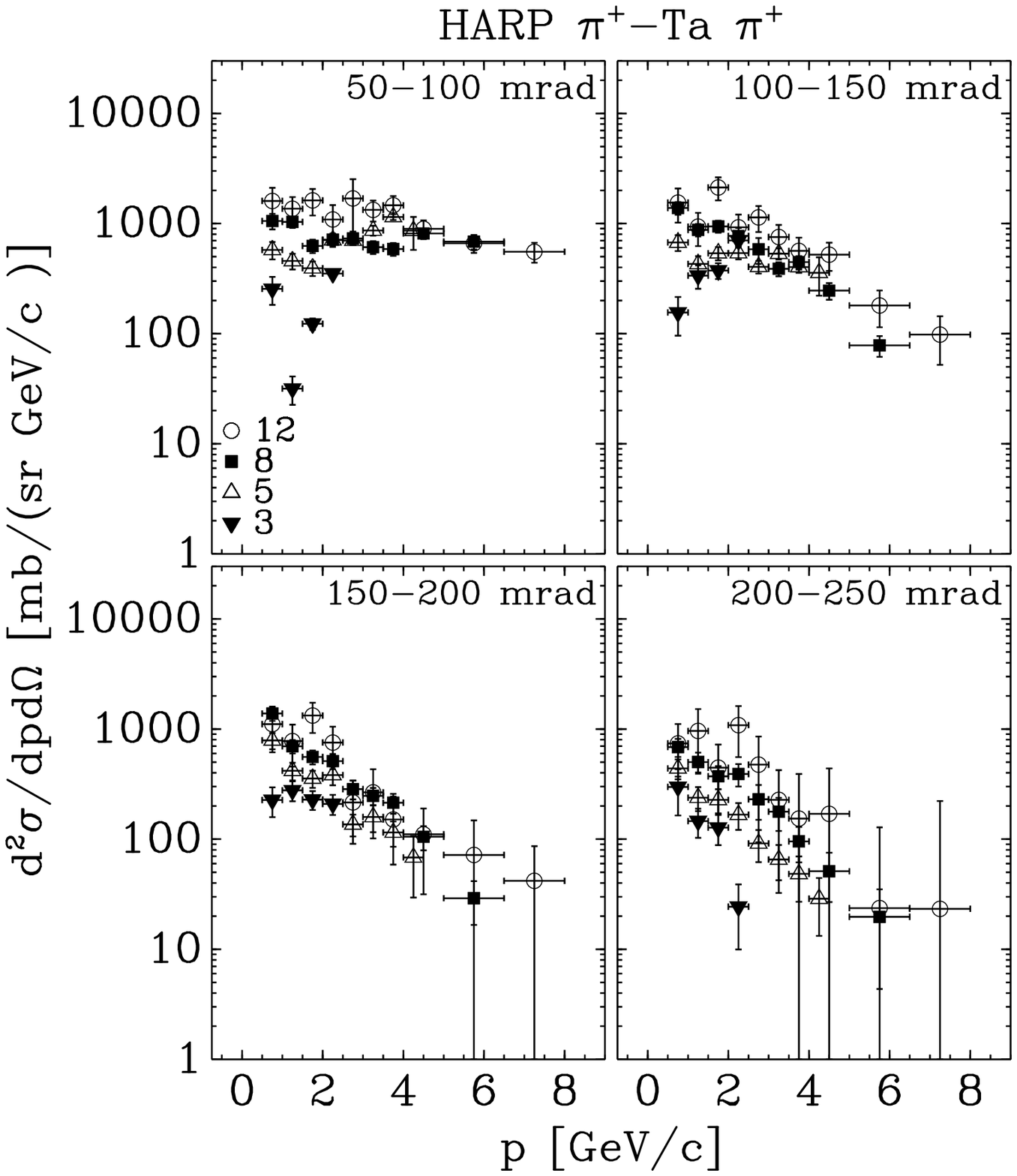}
\includegraphics[width=.49\textwidth]{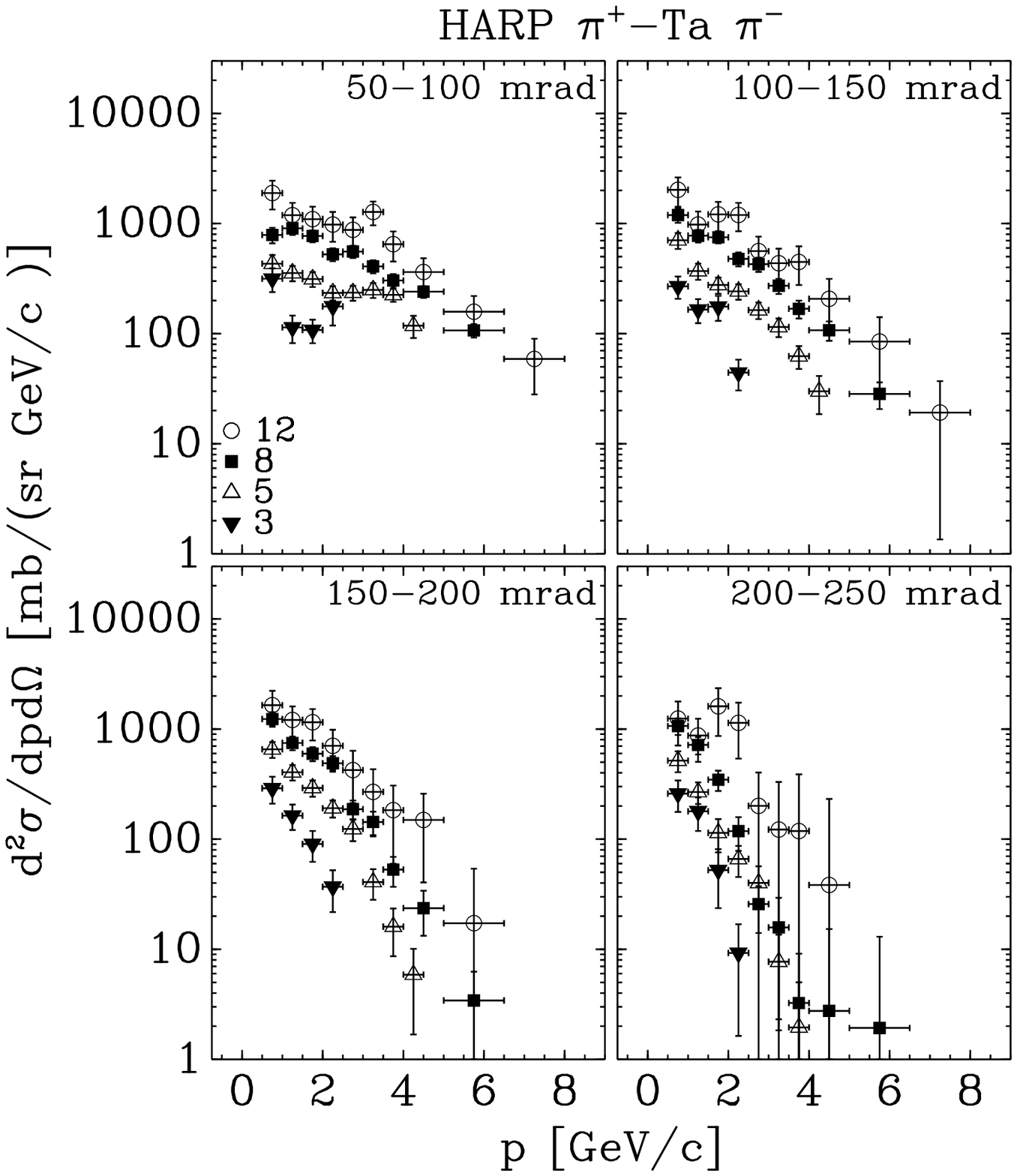}
\caption{$\pi^{-}-$Ta (top) and $\pi^{+}-$Ta (bottom) differential cross-sections
for different incoming beam momenta (from 3 to 12~\GeVc).
Left panels show the
$\pi^{+}$ production, while right panels show the $\pi^{-}$ production.   
In the top right corner of each plot the covered angular range is shown in mrad.}
\label{fig:Ta}
\end{figure}

\begin{figure}[tb]
\centering
\includegraphics[width=.49\textwidth]{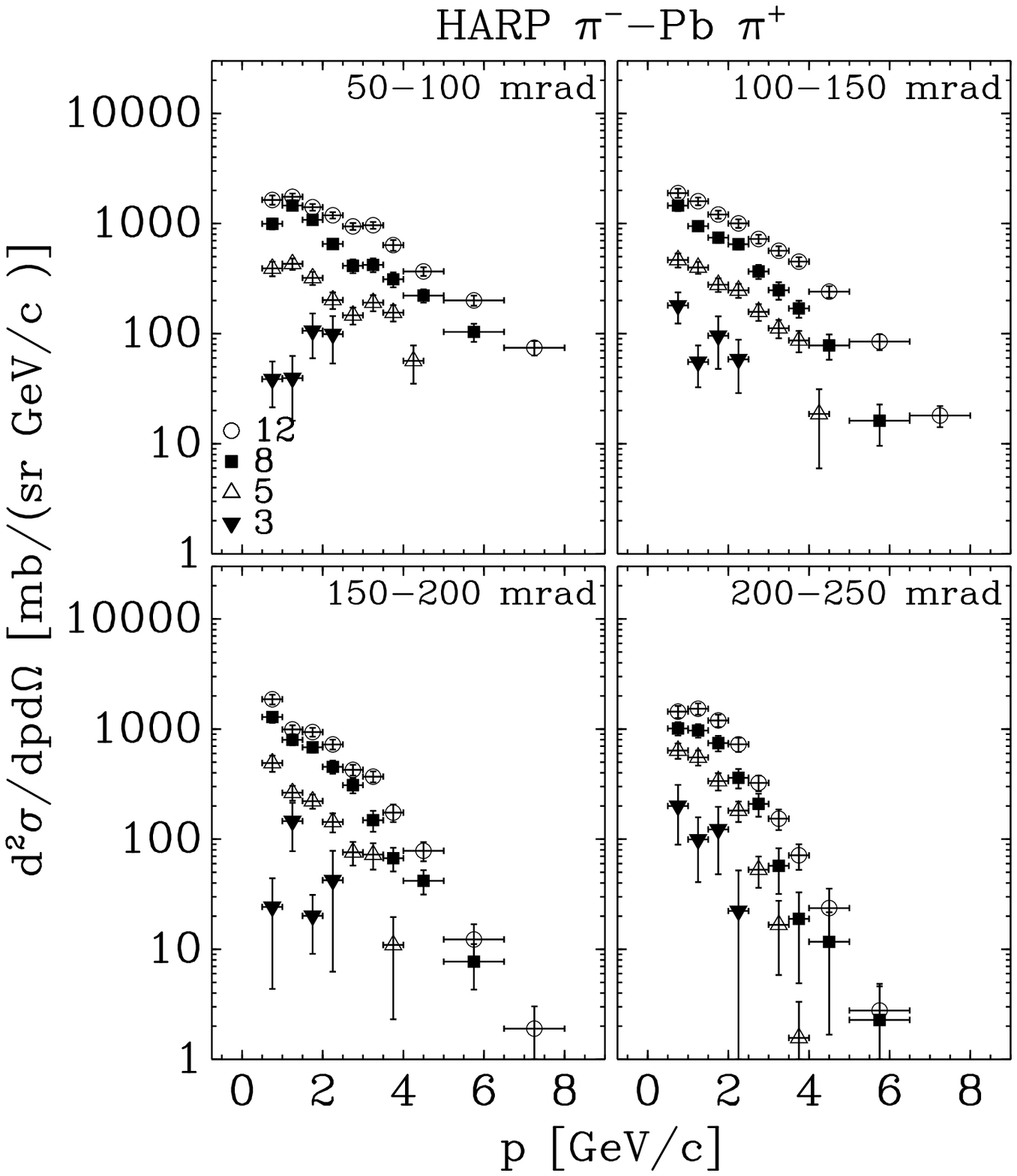}
\includegraphics[width=.49\textwidth]{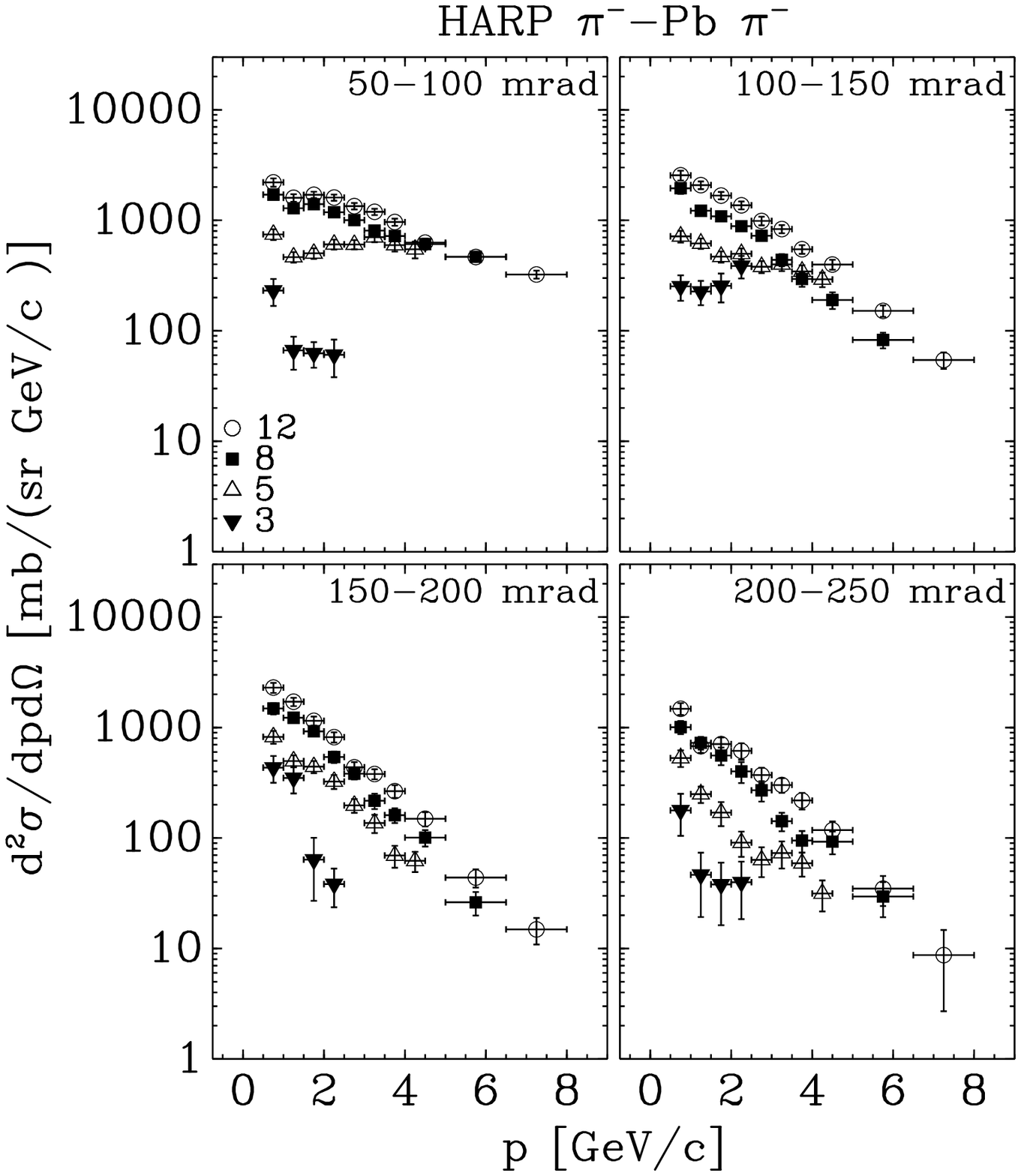}
\includegraphics[width=.49\textwidth]{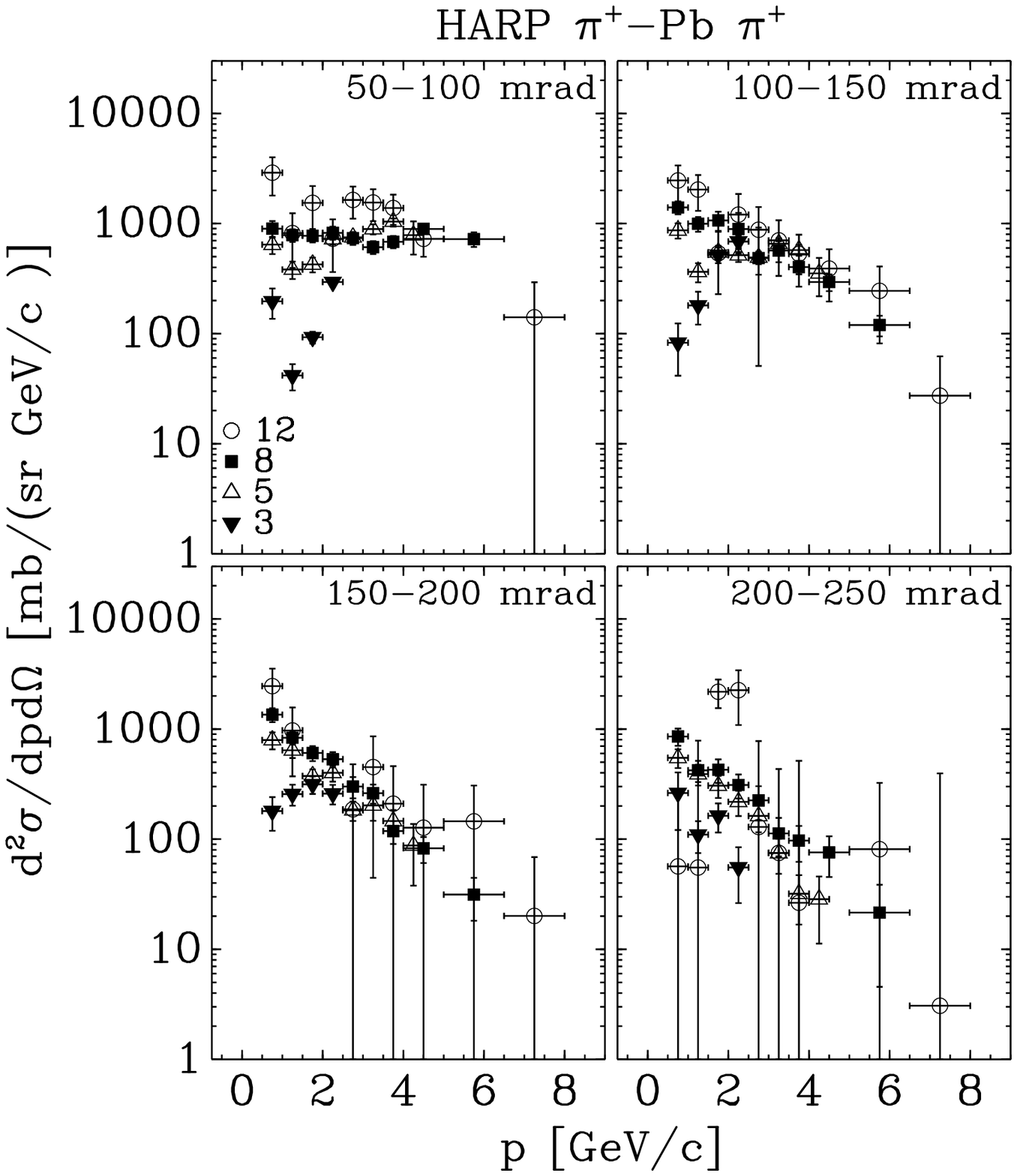}
\includegraphics[width=.49\textwidth]{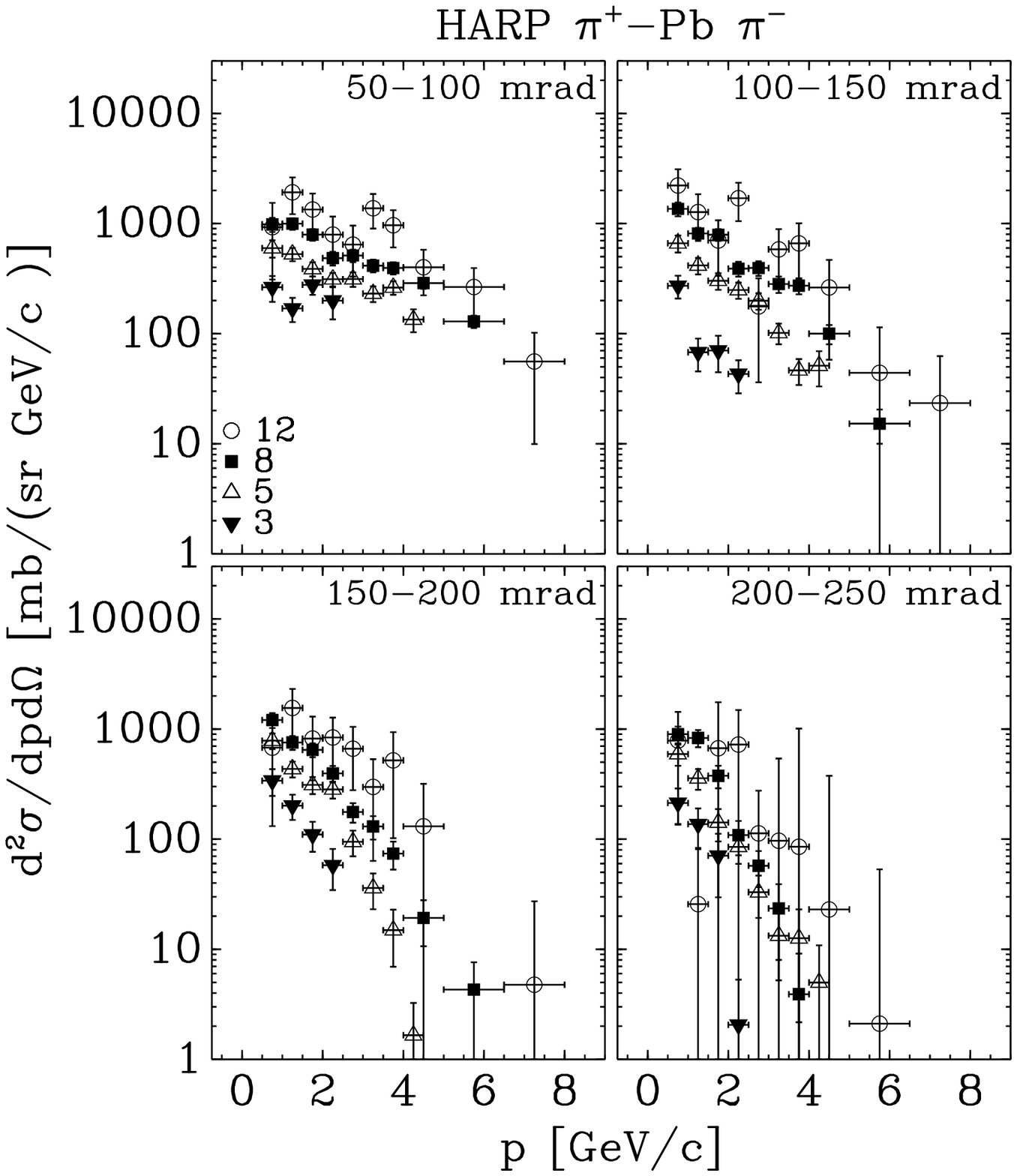}
\caption{$\pi^{-}-$Pb (top) and $\pi^{+}-$Pb (bottom) differential cross-sections
for different incoming beam momenta (from 3 to 12~\GeVc).
Left panels show the
$\pi^{+}$ production, while right panels show the $\pi^{-}$ production.   
In the top right corner of each plot the covered angular range is shown in mrad.}
\label{fig:Pb}
\end{figure}

The dependence of the averaged 
pion yields on the atomic number $A$ is
shown from Fig.~\ref{fig:xs-a-dep-1}  to Fig.~\ref{fig:xs-a-dep-4}.
The pion  yields, averaged over two angular regions
 ($0.05~\rad \leq \theta < 0.15~\rad$ and 
  $0.15~\rad \leq \theta < 0.25~\rad$)
 and four momentum regions 
  ($0.5~\GeVc \leq p < 1.5~\GeVc$,
   $1.5~\GeVc \leq p < 2.5~\GeVc$,
   $2.5~\GeVc \leq p < 3.5~\GeVc$ and
   $3.5~\GeVc \leq p < 4.5~\GeVc$),
are shown  for four different beam momenta.
One observes a smooth behaviour of the averaged yields.
The most striking feature is the difference between same-sign and opposite-sign 
pion production compared to the beam particle.
Here one observes a stronger beam energy dependence for opposite-sign production.
The $A$-dependence is similar for \pim and \pip production if one compares 
opposite-sign and same-sign production, although one observes that 
especially at lower beam momenta the opposite-sign pion production saturates 
earlier towards higher $A$.

\begin{figure*}[tbp]
\begin{center}
  \includegraphics[width=0.79\textwidth]{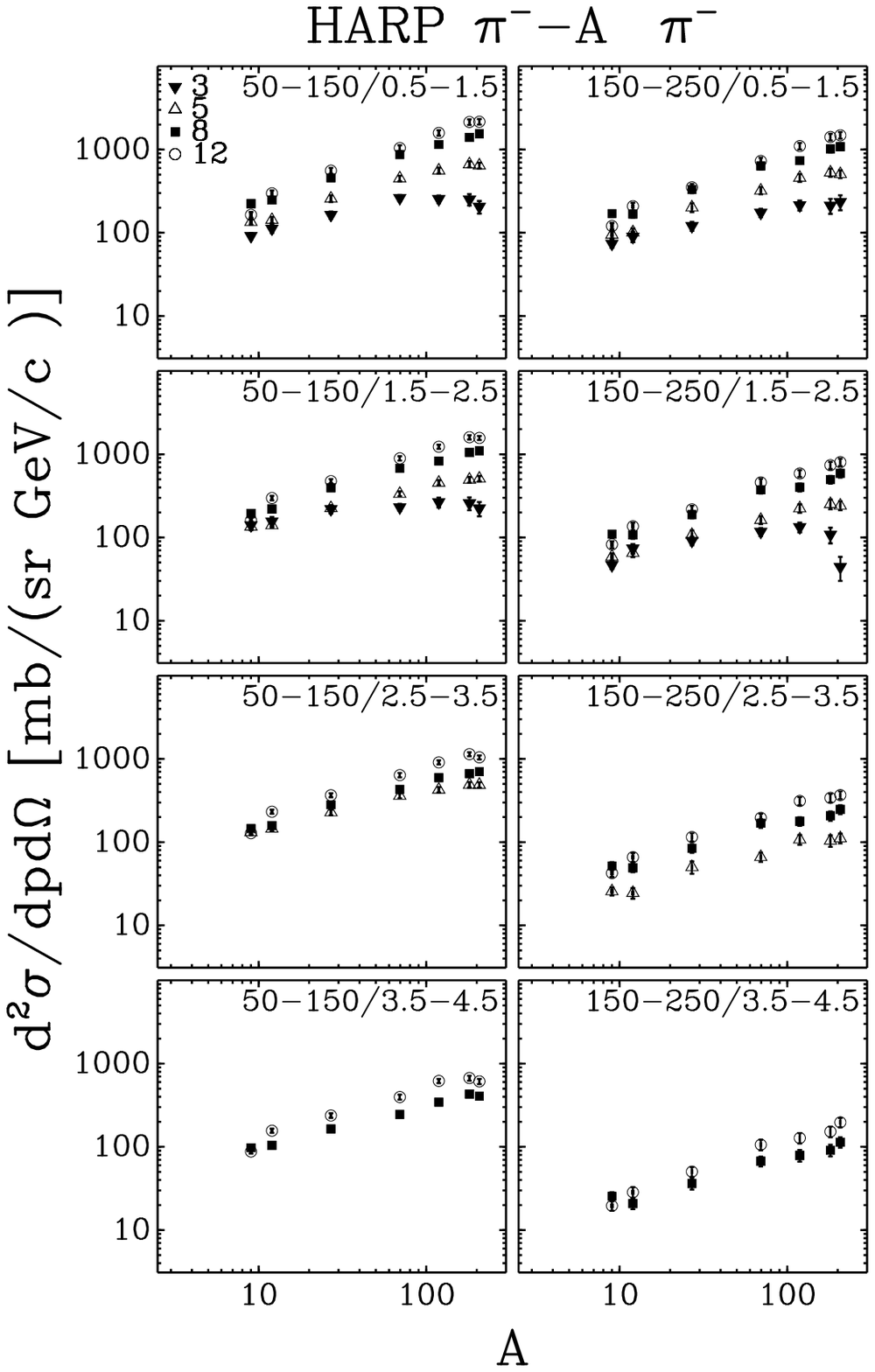}
\end{center}
\caption{
 The dependence on the atomic number $A$ of the \pim  production yields
 in $\pi^-$--Be, $\pi^-$--C, $\pi^-$--Al, $\pi^-$--Cu, $\pi^-$--Sn, $\pi^-$--Ta, 
  $\pi^-$--Pb
 interactions averaged over two forward angular region
 ($0.05~\rad \leq \theta < 0.15~\rad$ and 
  $0.15~\rad \leq \theta < 0.25~\rad$)
 and four momentum regions 
  ($0.5~\GeVc \leq p < 1.5~\GeVc$,
   $1.5~\GeVc \leq p < 2.5~\GeVc$,
   $2.5~\GeVc \leq p < 3.5~\GeVc$ and
   $3.5~\GeVc \leq p < 4.5~\GeVc$), for the four different
  incoming beam momenta (from 3~\GeVc to 12~\GeVc).
}
\label{fig:xs-a-dep-1}
\end{figure*}
\begin{figure*}[tbp]
\begin{center}
  \includegraphics[width=0.79\textwidth]{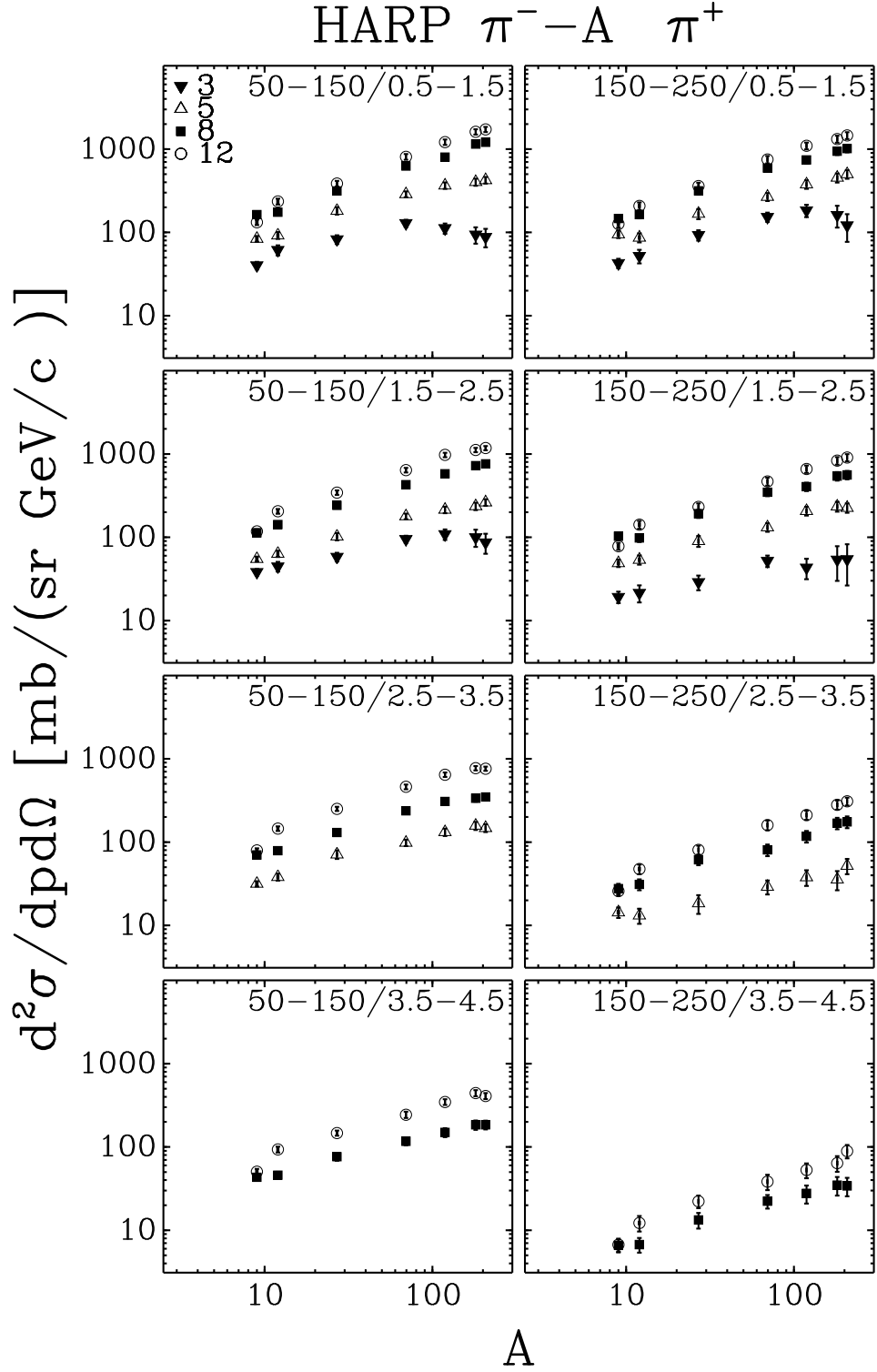}
\end{center}
\caption{
 The dependence on the atomic number $A$ of the \pip  production yields
 in $\pi^-$--Be, $\pi^-$--C, $\pi^-$--Al, $\pi^-$--Cu, $\pi^-$--Sn, $\pi^-$--Ta, 
  $\pi^-$--Pb
 interactions averaged over two forward angular region
 ($0.05~\rad \leq \theta < 0.15~\rad$ and 
  $0.15~\rad \leq \theta < 0.25~\rad$)
 and four momentum regions 
  ($0.5~\GeVc \leq p < 1.5~\GeVc$,
   $1.5~\GeVc \leq p < 2.5~\GeVc$,
   $2.5~\GeVc \leq p < 3.5~\GeVc$ and
   $3.5~\GeVc \leq p < 4.5~\GeVc$), for the four different
  incoming beam momenta (from 3~\GeVc to 12~\GeVc).
}
\label{fig:xs-a-dep-2}
\end{figure*}

\begin{figure*}[tbp]
\begin{center}
  \includegraphics[width=0.79\textwidth]{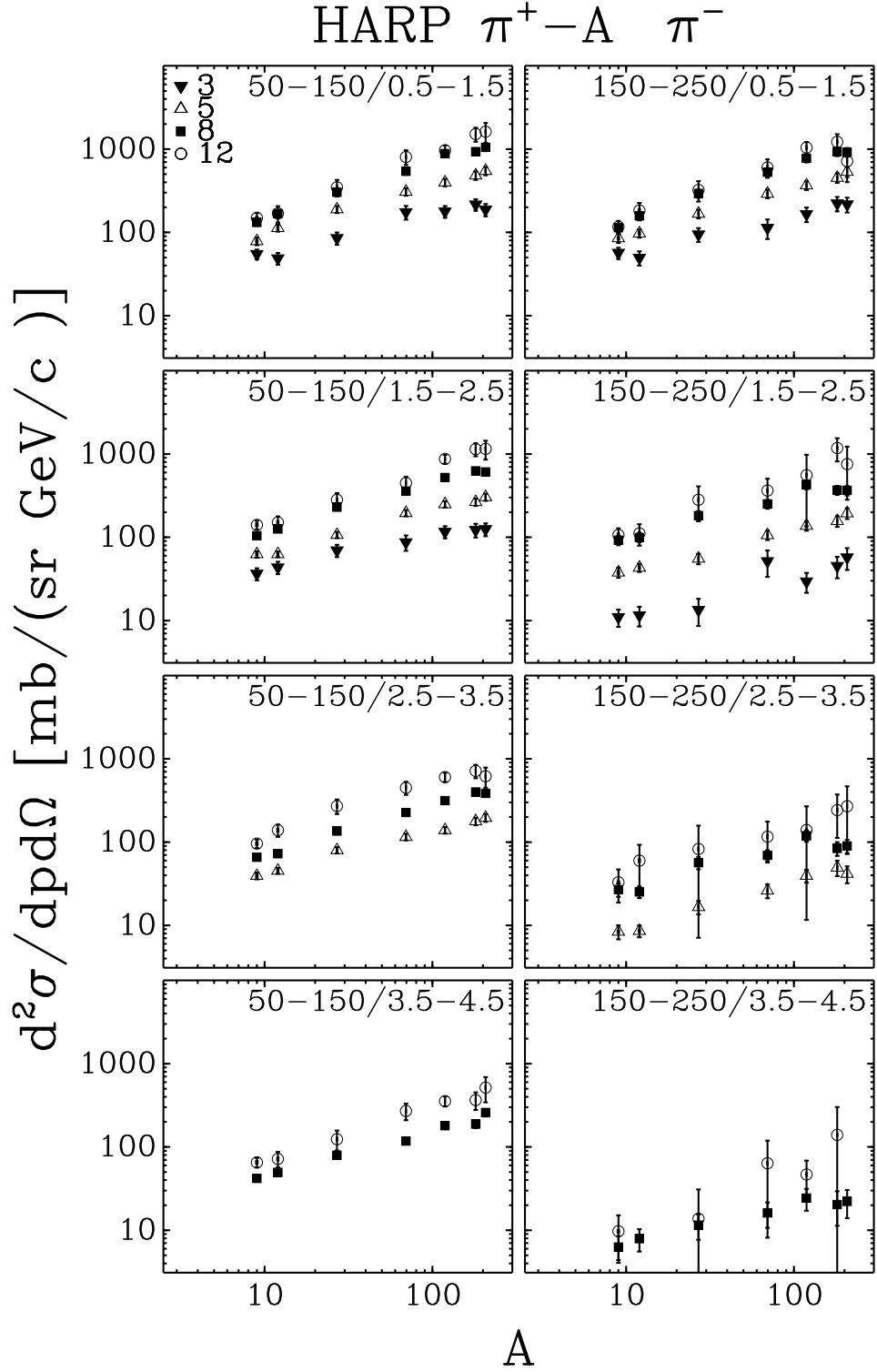}
\end{center}
\caption{
 The dependence on the atomic number $A$ of the \pim  production yields
 in $\pi^+$--Be, $\pi^+$--C, $\pi^+$--Al, $\pi^+$--Cu, $\pi^+$--Sn, $\pi^+$--Ta, 
  $\pi^+$--Pb
 interactions averaged over two forward angular region
 ($0.05~\rad \leq \theta < 0.15~\rad$ and 
  $0.15~\rad \leq \theta < 0.25~\rad$)
 and four momentum regions 
  ($0.5~\GeVc \leq p < 1.5~\GeVc$,
   $1.5~\GeVc \leq p < 2.5~\GeVc$,
   $2.5~\GeVc \leq p < 3.5~\GeVc$ and
   $3.5~\GeVc \leq p < 4.5~\GeVc$), for the four different
  incoming beam momenta (from 3~\GeVc to 12~\GeVc).
}
\label{fig:xs-a-dep-3}
\end{figure*}

\begin{figure*}[tbp]
\begin{center}
  \includegraphics[width=0.79\textwidth]{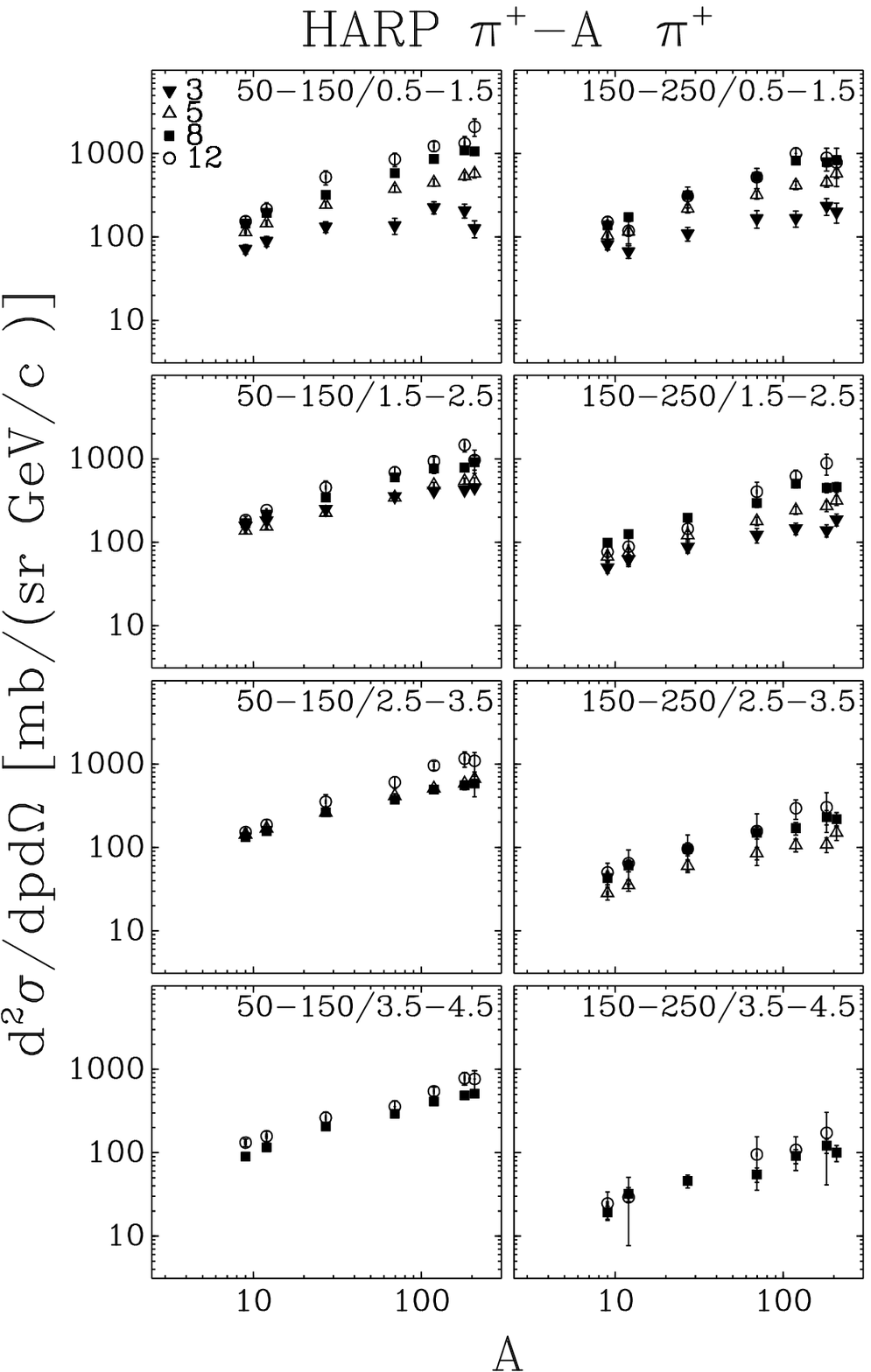}
\end{center}
\caption{
 The dependence on the atomic number $A$ of the \pip  production yields
 in $\pi^+$--Be, $\pi^+$--C, $\pi^+$--Al, $\pi^+$--Cu, $\pi^+$--Sn, $\pi^+$--Ta, 
  $\pi^+$--Pb
 interactions averaged over two forward angular region
 ($0.05~\rad \leq \theta < 0.15~\rad$ and 
  $0.15~\rad \leq \theta < 0.25~\rad$)
 and four momentum regions 
  ($0.5~\GeVc \leq p < 1.5~\GeVc$,
   $1.5~\GeVc \leq p < 2.5~\GeVc$,
   $2.5~\GeVc \leq p < 3.5~\GeVc$ and
   $3.5~\GeVc \leq p < 4.5~\GeVc$), for the four different
  incoming beam momenta (from 3~\GeVc to 12~\GeVc).
}
\label{fig:xs-a-dep-4}
\end{figure*}

In the following we will show only some comparisons 
of the HARP data with  publicly
available Monte Carlo simulations: GEANT4~\cite{ref:geant4}
(version 9.1p02)
and MARS~\cite{ref:mars}, not previously tuned to our
data-sets, by using different
models.
The comparison will be shown for a limited set of plots
and only for the Be (Figures \ref{fig:G43a} to \ref{fig:G46}) and Ta (Figures 
\ref{fig:G53} to \ref{fig:G56}) 
targets, as examples of a light and a heavy target.

At intermediate energies (up to 5--10 GeV),
GEANT4 uses two types of intra-nuclear cascade models: the Bertini
model~\cite{ref:bert,ref:bert1} (valid up to $\sim 10$ GeV) and the binary
model~\cite{ref:bin} (with a validity range up to $\sim 3$ GeV). 
Both models treat the target
nucleus in detail, taking into account density variations and tracking in the
nuclear field.
The binary model is based on hadron collisions with nucleons, giving
resonances that decay according to their quantum numbers. The Bertini
model is based on the cascade code reported in \cite{ref:bert2}
and hadron collisions are assumed to proceed according to free-space partial
cross-sections and final-state distributions measured for the incident
particle types.

At higher energies, instead, two parton string models,
the quark-gluon string (QGS)  model~\cite{ref:bert,ref:QGSP} and the Fritiof
(FTP) model~\cite{ref:QGSP} are used, in addition to a High Energy
Parametrized model (HEP)
derived from the high energy part of the GHEISHA code used inside
GEANT3~\cite{ref:gheisha}.

The parametrized models of GEANT4 (HEP and LEP) are intended to be fast,
but conserve energy and momentum on average and not event by
event.

A realistic GEANT4 simulation is built by combining models and physics processes
into what is called a ``physics list''. In high energy calorimetry the two
most commonly used are the QGSP physics list, based on the QGS model, the pre-compound
nucleus model and some of the Low Energy Parametrized (LEP) model~\footnote{
Also this model, at low energy, has its root in the Gheisha code inside
GEANT3.} and the LHEP physics list~\cite{ref:lhep} based on the parametrized
LEP model and HEP models.

Currently is also popular the
QGSP-BERT physics list in which the Bertini model is used at 3 and 5 GeV/c
and the QGSP model at 8 and 12 GeV/c.

\begin{figure}[tbp]
\begin{center}
\epsfig{figure=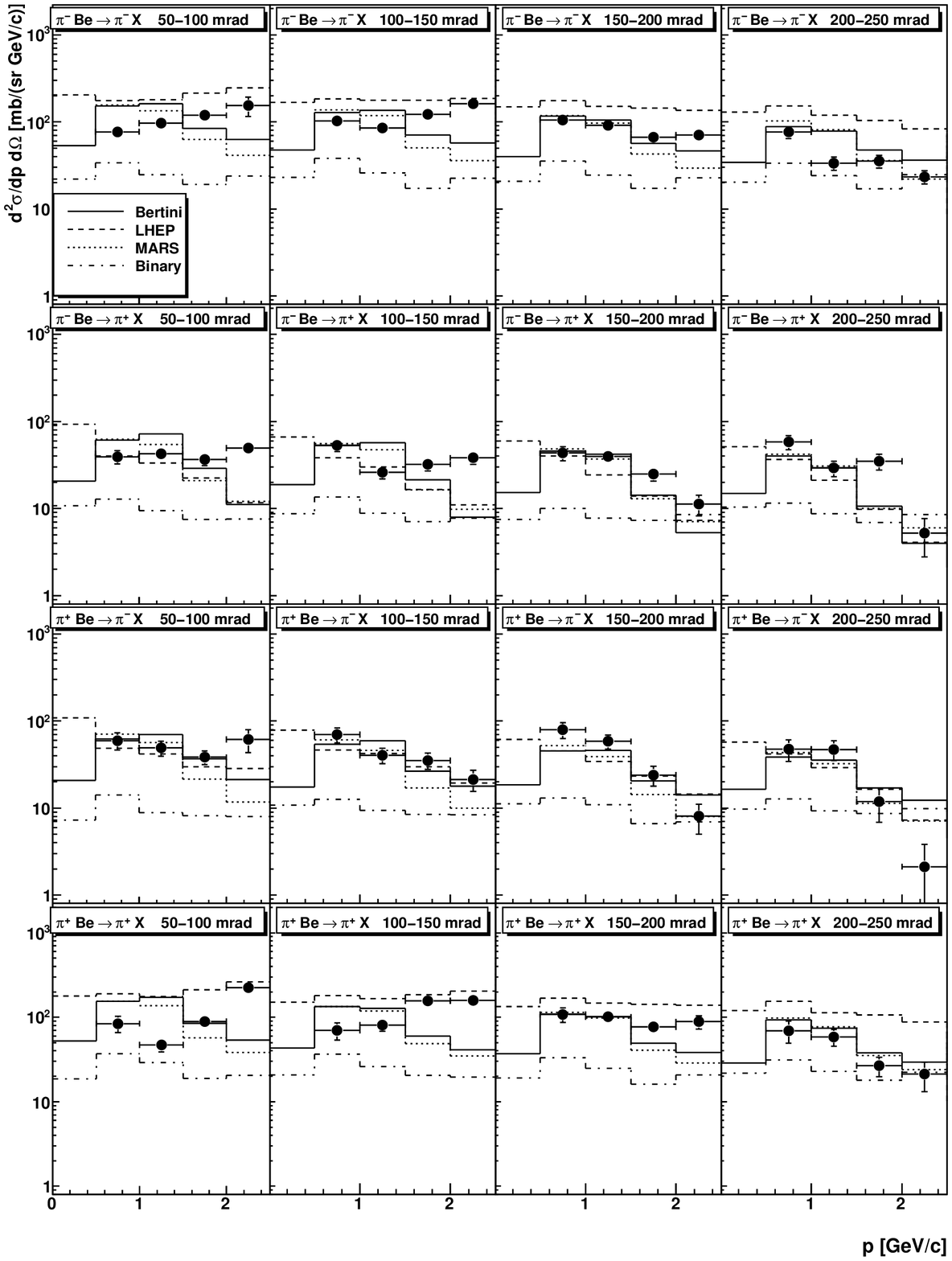,width=\textwidth}
\end{center}
\caption{
 Comparison of HARP double-differential $\pi^{\pm}$ cross-sections for $\pi$--Be at 3~\GeVc with
 GEANT4 and MARS MC predictions, using several generator models.}
\label{fig:G43a}
\end{figure}

\begin{figure}[tbp]
\begin{center}
\epsfig{figure=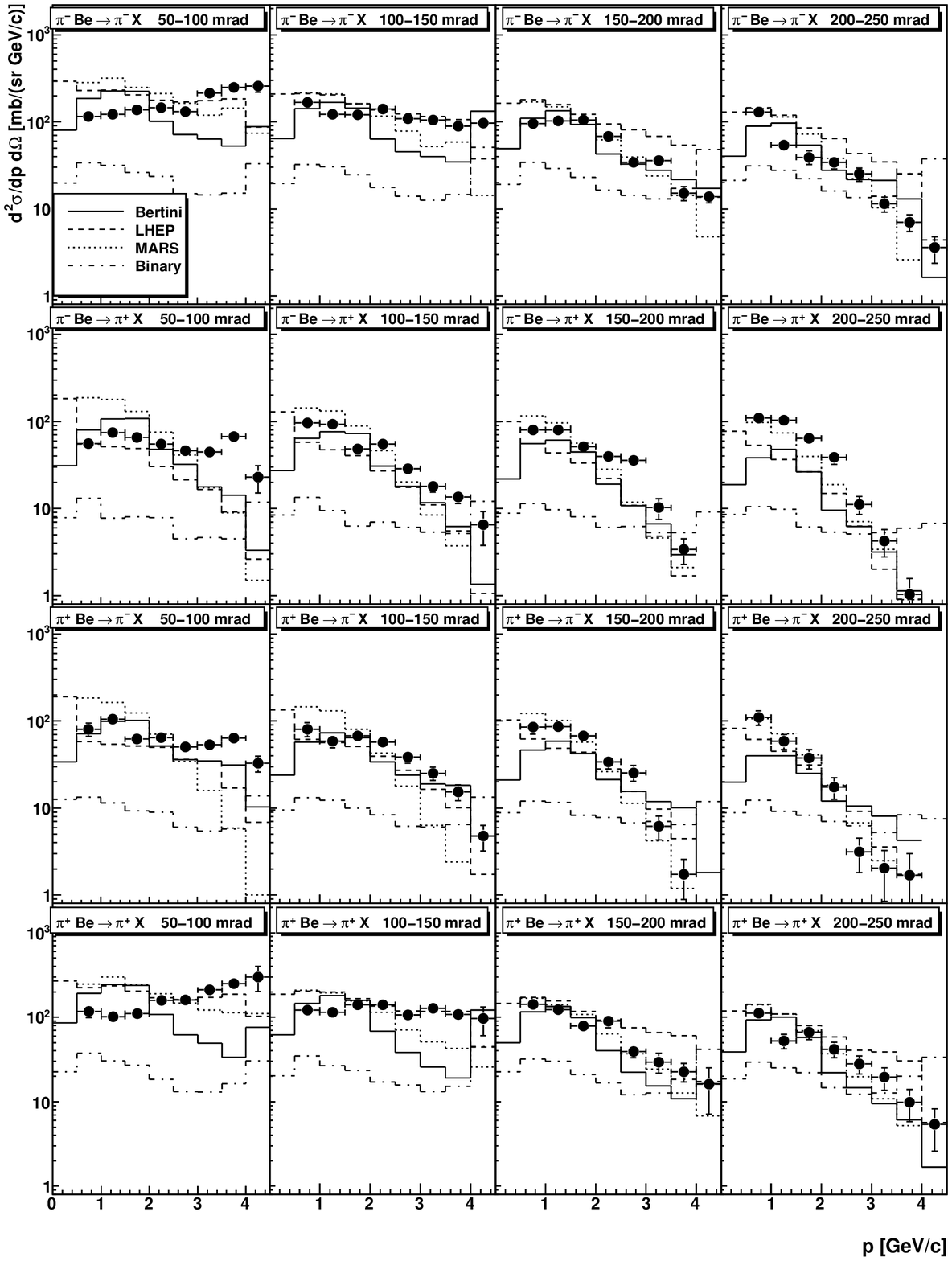,width=\textwidth}
\end{center}
\caption{
 Comparison of HARP double-differential $\pi^{\pm}$ cross-sections for $\pi$--Be at 5~\GeVc with
 GEANT4 and MARS MC predictions, using several generator models.}
\label{fig:G44a}
\end{figure}
\begin{figure}[tbp]
\begin{center}
\epsfig{figure=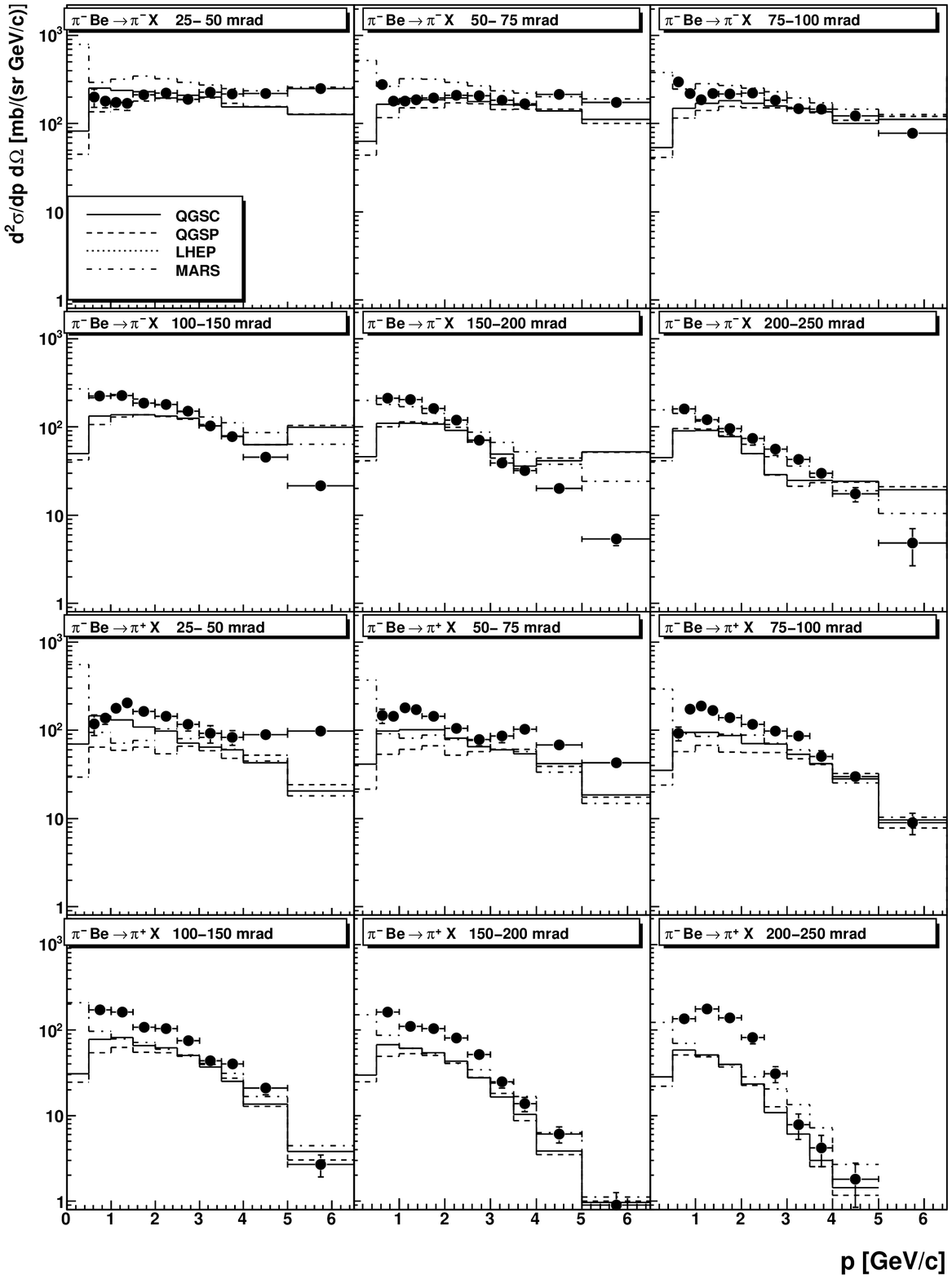,width=\textwidth}
\end{center}
\caption{
 Comparison of HARP double-differential $\pi^{\pm}$ cross-sections for $\pim$--Be at 8~\GeVc with
 GEANT4 and MARS MC predictions, using several generator models.
}
\label{fig:G45a}
\end{figure}
\begin{figure}[tbp]
\begin{center}
\epsfig{figure=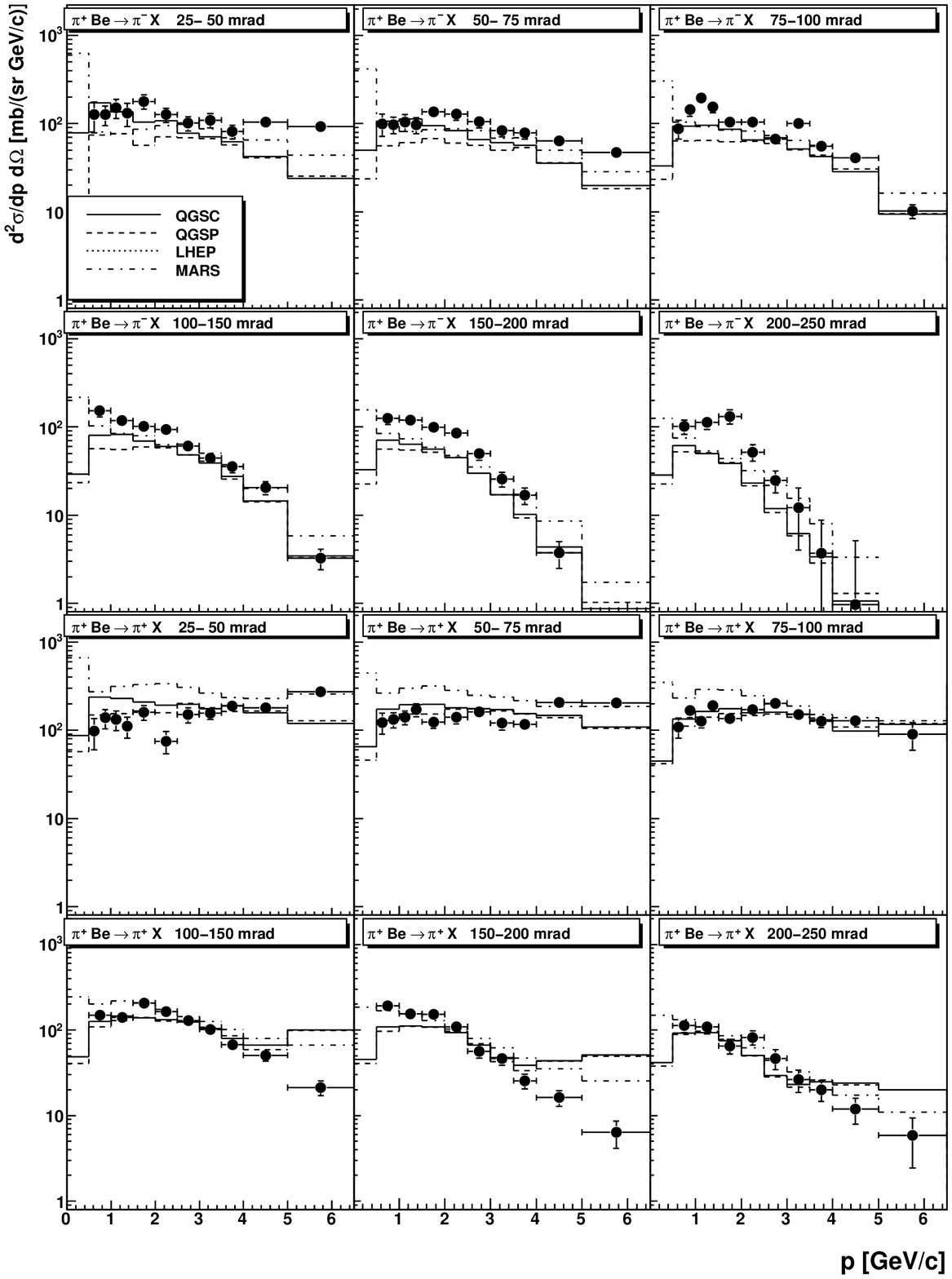,width=\textwidth}
\end{center}
\caption{
 Comparison of HARP double-differential $\pi^{\pm}$ cross-sections for $\pip$--Be at 8~\GeVc with
 GEANT4 and MARS MC predictions, using several generator models.
}
\label{fig:G45b}
\end{figure}

\begin{figure}[tbp]
\begin{center}
\epsfig{figure=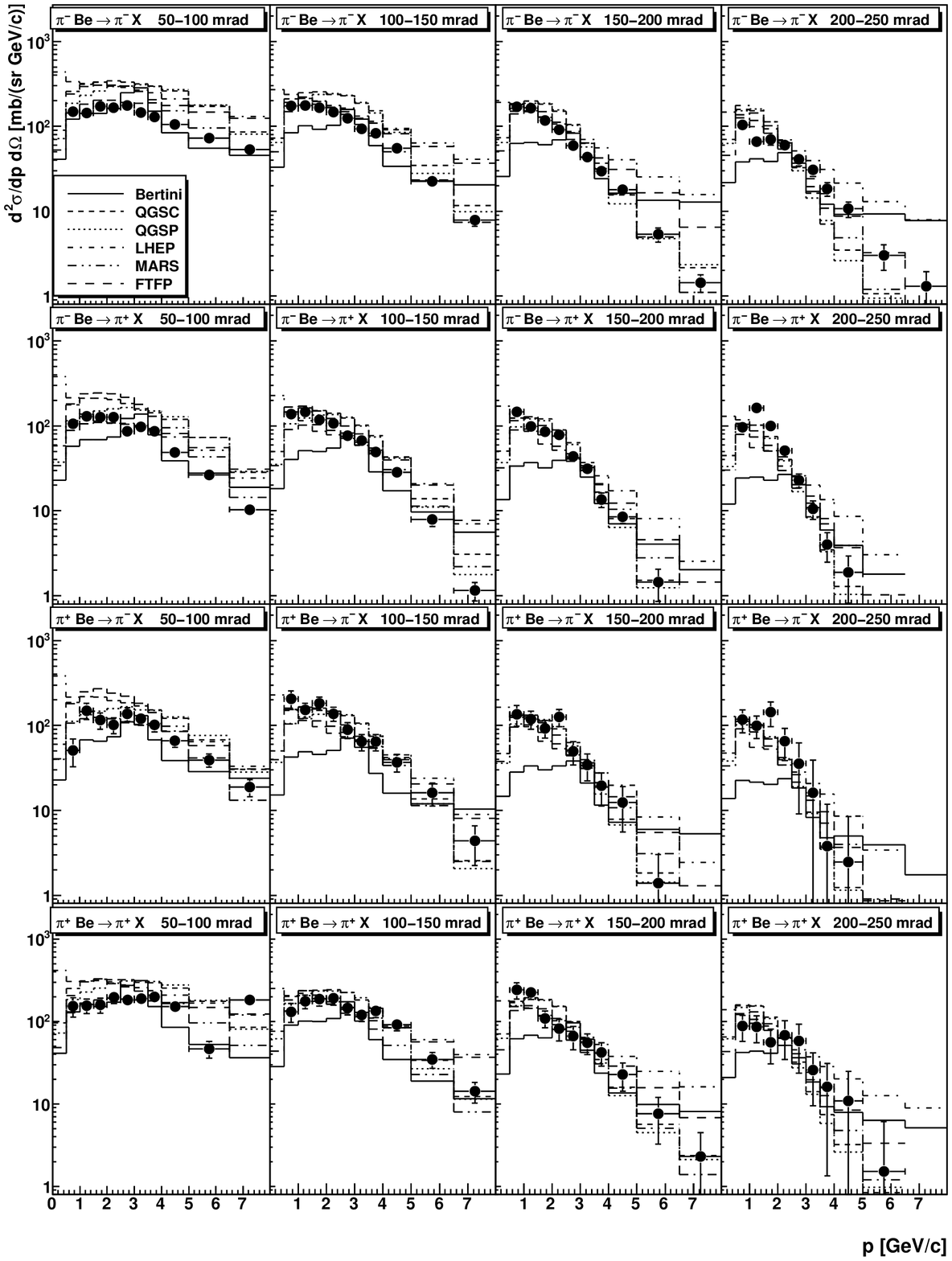,width=\textwidth}
\end{center}
\caption{
 Comparison of HARP double-differential $\pi^{\pm}$ cross-sections for $\pi$--Be at 12~\GeVc with
 GEANT4 and MARS MC predictions, using several generator models.
}
\label{fig:G46}
\end{figure}

\clearpage
\begin{figure}[tbp]
\begin{center}
\epsfig{figure=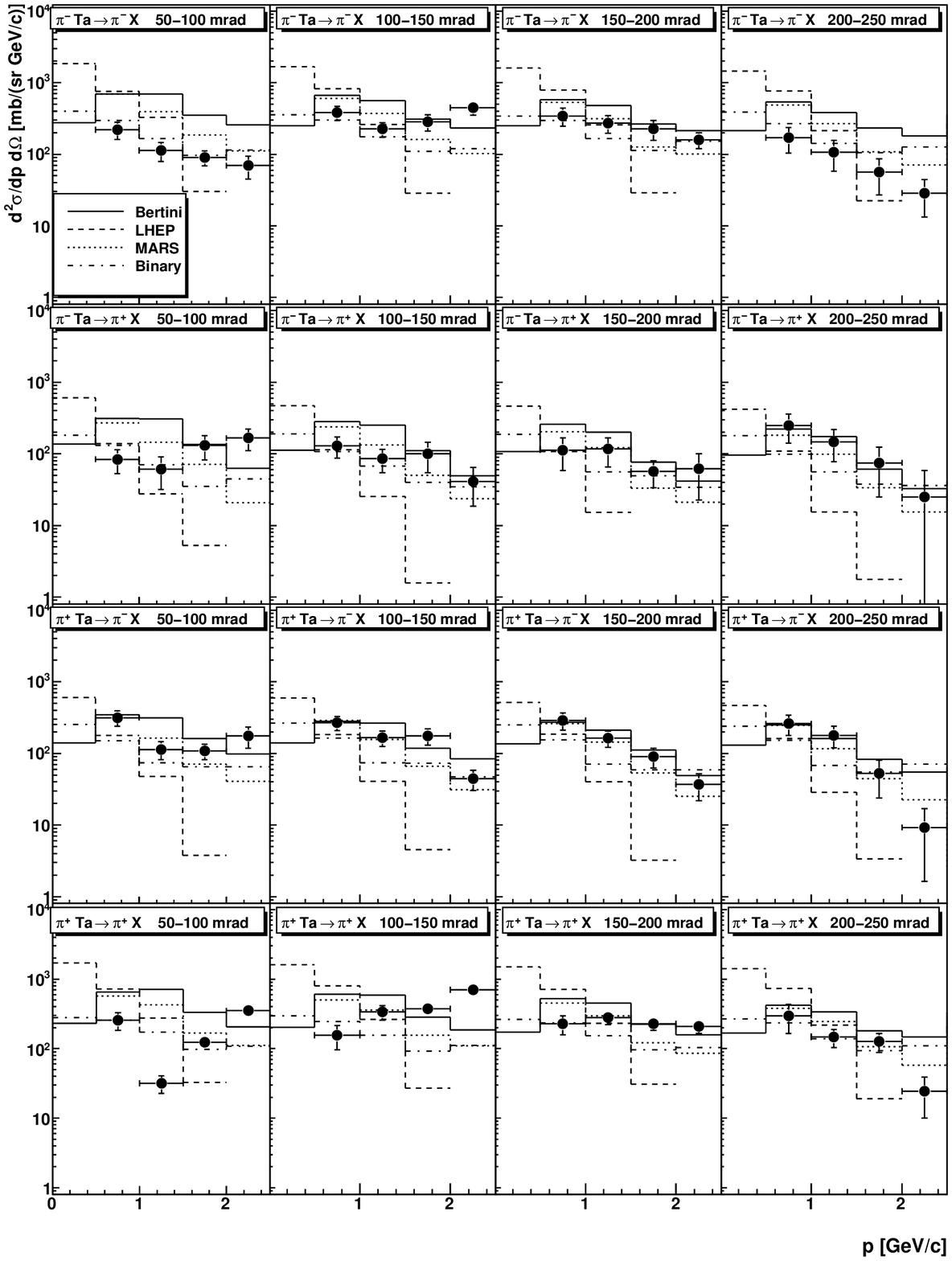,width=\textwidth}
\end{center}
\caption{
 Comparison of HARP double-differential $\pi^{\pm}$ cross-sections for $\pi$--Ta at 3~\GeVc with
 GEANT4 and MARS MC predictions, using several generator models. 
}
\label{fig:G53}
\end{figure}

\begin{figure}[tbp]
\begin{center}
\epsfig{figure=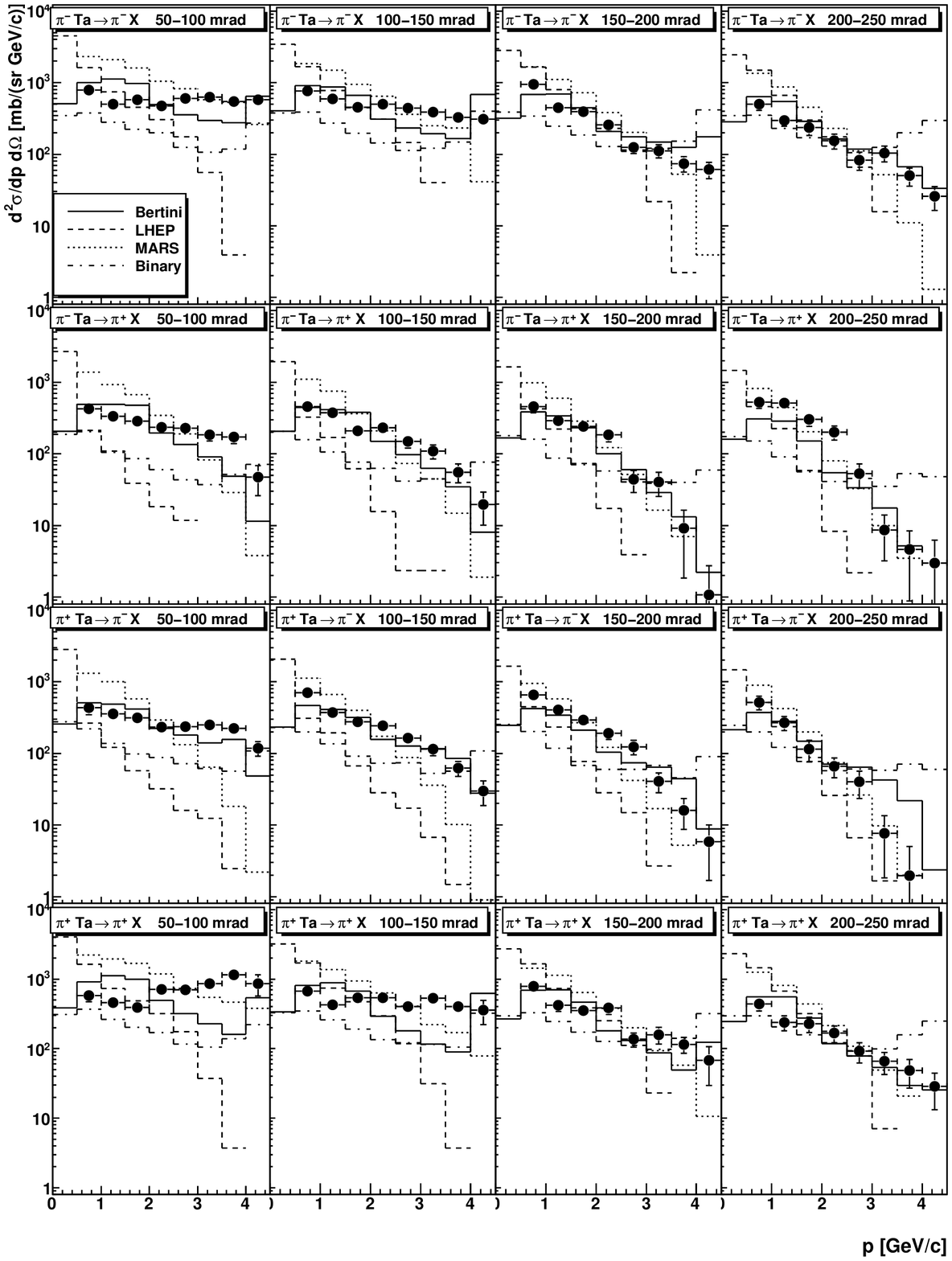,width=\textwidth}
\end{center}
\caption{
 Comparison of HARP double-differential $\pi^{\pm}$ cross-sections for $\pi$--Ta at 5~\GeVc with
 GEANT4 and MARS MC predictions, using several generator models.
}
\label{fig:G54a}
\end{figure}

\begin{figure}[tbp]
\begin{center}
\epsfig{figure=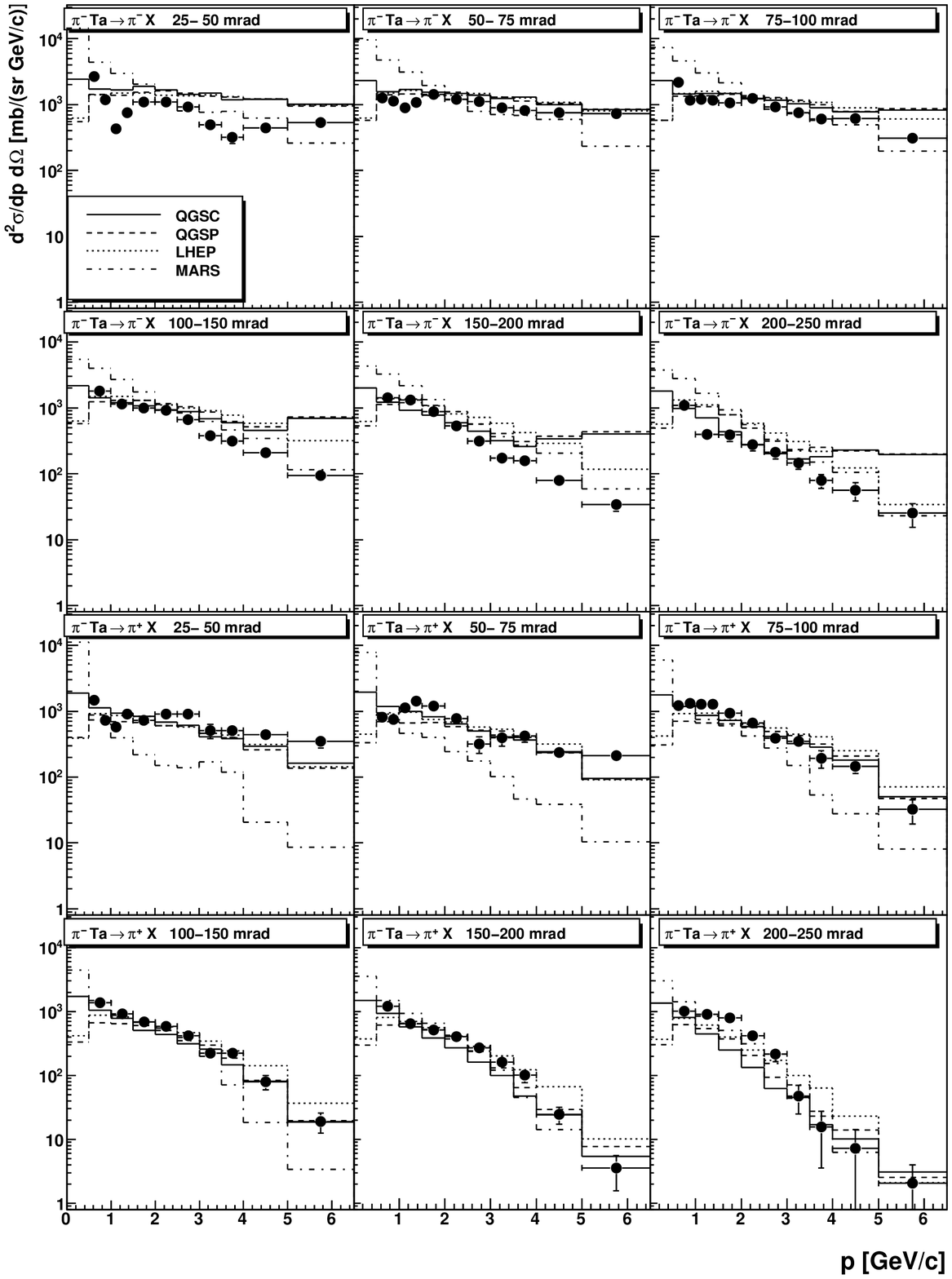,width=\textwidth}
\end{center}
\caption{
 Comparison of HARP double-differential $\pi^{\pm}$ cross-sections for $\pim$--Ta at 8~\GeVc with
 GEANT4 and MARS MC predictions, using several generator models. 
}
\label{fig:G55a}
\end{figure}
\begin{figure}[tbp]
\begin{center}
\epsfig{figure=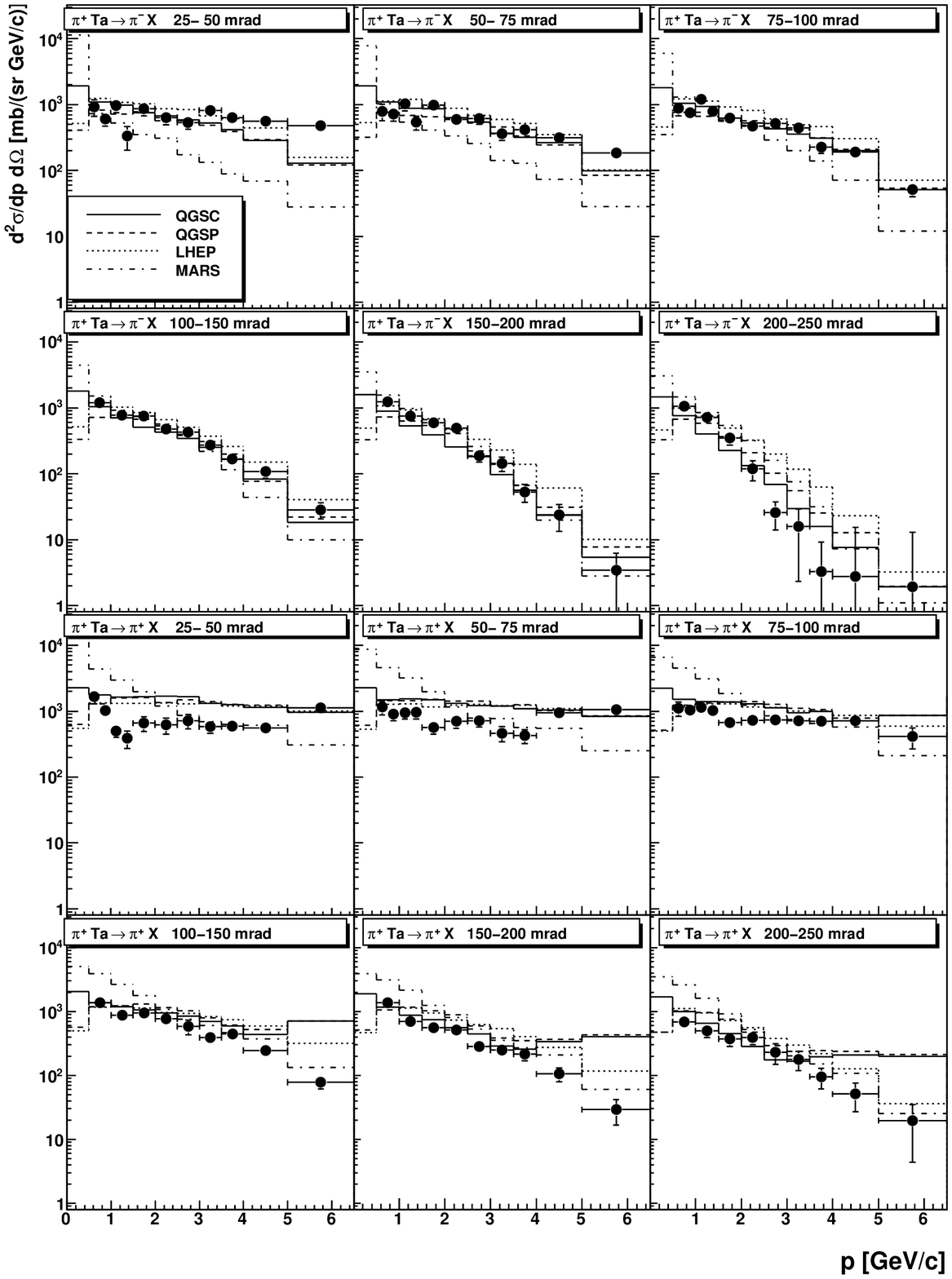,width=\textwidth}
\end{center}
\caption{
 Comparison of HARP double-differential $\pi^{\pm}$ cross-sections for $\pip$--Ta at 8~\GeVc with
 GEANT4 and MARS MC predictions, using several generator models. 
}
\label{fig:G55b}
\end{figure}

\begin{figure}[htbp]
\begin{center}
\epsfig{figure=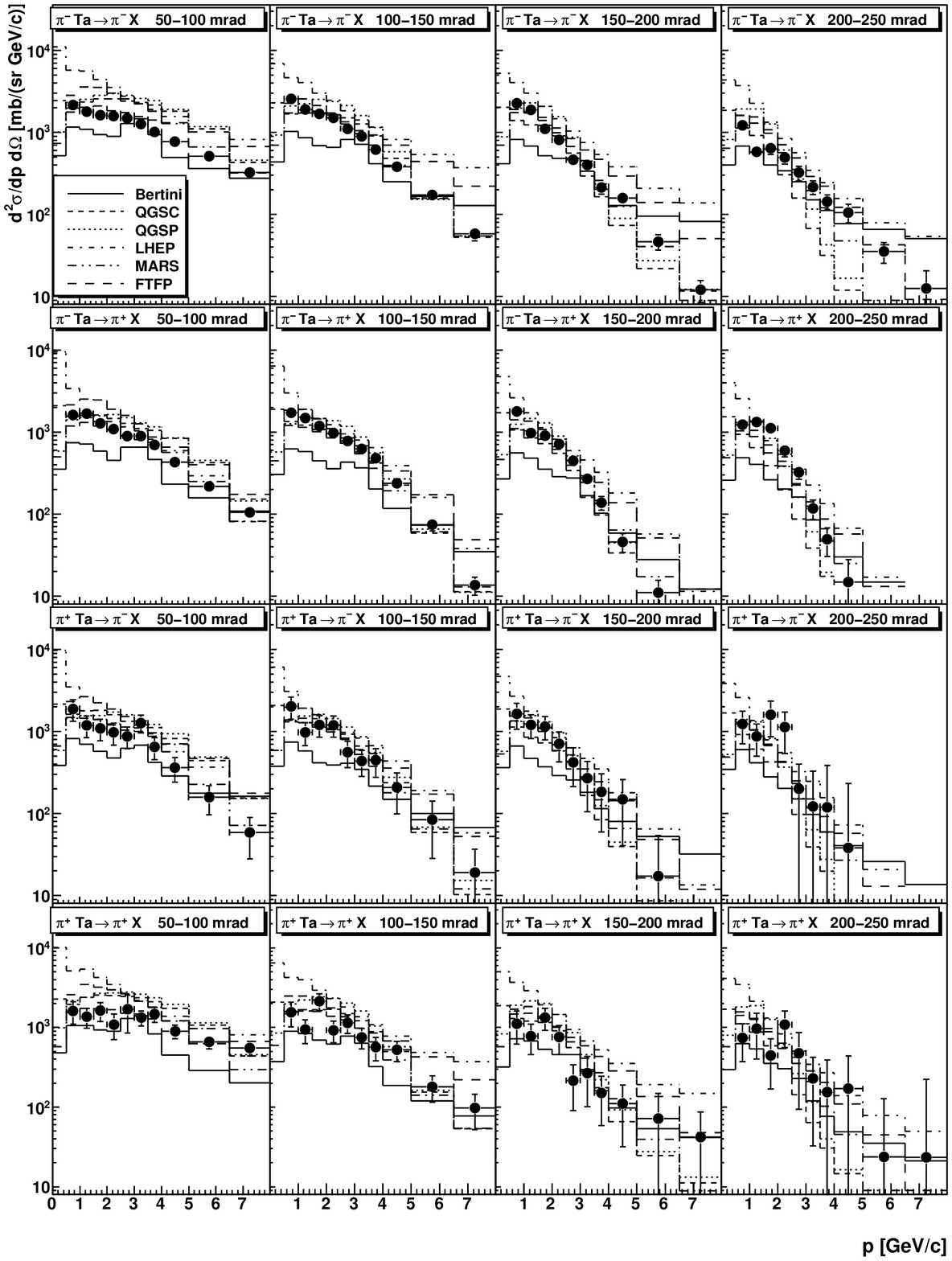,width=\textwidth}
\end{center}
\caption{
 Comparison of HARP double-differential $\pi^{\pm}$ cross-sections for $\pi$--Ta at 12~\GeVc with
 GEANT4 and MARS MC predictions, using several generator models.
}
\label{fig:G56}
\end{figure}

The MARS code system~\cite{ref:mars} uses as basic model an inclusive
approach to multiparticle production originated by R. Feynman in 1974. 
At each interaction vertex, a particle cascade tree is constructed by
using a fixed number of particles with weights that in the simplest case
are equal to the partial mean multiplicity of that event. 
Above 3~GeV
phenomenological particle production models are used, while below 5~GeV
a cascade-exciton model~\cite{ref:casca} combined with the Fermi
break-up model, the coalescence model, an evaporation model and a
multifragmentation extension are used instead.

HARP measurements have been compared to the available models, first renormalizing 
MC predictions to data and then computing the $\chi^2$ between models
and data themselves. Normalization factors are shown 
in Tables~\ref{tab-m} and~\ref{tab-n}.
An absolute comparison yields $\chi^2$ values that are too large to be meaningful.
Over the full energy range covered by the HARP experiment, the best
comparison is obtained with the MARS Monte Carlo which is on average 10\%
below the data, with a maximum shift of 30\%. 
This may be owing to the fact that MARS is using different models in
different energy regions, equivalent to use a collection of models as
implemented in the ``physics lists'' of GEANT4.
Unfortunately, all models have very bad $\chi^2$ compared with the data, showing
significant inconsistencies in the shape distribution. This is worse
than the situation with incident protons and may be explained by the 
fact that in the latter case some experimental data were available
for tuning of the Monte Carlo simulations. This was not the case
with incident pions in the momentum range studied here (below 15~\GeVc). 

At higher energies the FTP model (from GEANT4) and the MARS model
describe the data better, while at the lowest energies the Bertini
and binary cascade models (from GEANT4) seem more appropriate.
Parametrized models, such as LHEP from GEANT4, show important discrepancies.

\begin{table}[htbp!]
\caption{Normalization factors data over MC for some models, incident $\pi^+$.}
\label{tab-m}
{\small
\begin{center}
\begin{tabular}{c|cc|cc|cc|cc|cc|cc|cc|cc}
Generator & \multicolumn{2}{c}{Be } & \multicolumn{2}{c}{Ta } & \multicolumn{2}{c}{Be } & \multicolumn{2}{c}{Ta } &
 \multicolumn{2}{c}{Be } & \multicolumn{2}{c}{Ta } & \multicolumn{2}{c}{Be } & \multicolumn{2}{c}{Ta } \\
 & \multicolumn{2}{c}{ 3 \GeVc} & \multicolumn{2}{c}{ 3 \GeVc} & \multicolumn{2}{c}{ 5 \GeVc} & \multicolumn{2}{c}{ 5 \GeVc} &
 \multicolumn{2}{c}{ 8 \GeVc} & \multicolumn{2}{c}{ 8 \GeVc} & \multicolumn{2}{c}{ 12 \GeVc} & \multicolumn{2}{c}{ 12 \GeVc} \\
& \pip & \pim & \pip & \pim & \pip & \pim & \pip & \pim
& \pip & \pim & \pip & \pim & \pip & \pim & \pip & \pim \\
Bertini &     1.1 &     1.1 &     0.9 &     0.6 &     1.3 &     1.3 &     1.1 &     1.1 &     1.7 &     1.4 &     1.5 &     1.4 &     2.0 &     1.4 &     1.9 &     1.6 \\
LHEP &     1.3 &     0.6 &     2.8 &     1.0 &     1.5 &     0.8 &     2.5 &     1.1 &     1.3 &     0.8 &     1.0 &     0.5 &     1.0 &     0.8 &     0.7 &     0.5 \\
QGSC &-&-&-&-&-&-&-&-&     1.6 &     1.0 &     1.2 &     0.7 &     0.8 &     0.7 &     1.0 &     0.8 \\
QGSP &-&-&-&-&-&-&-&-&     1.9 &     1.0 &     1.2 &     0.6 &     0.9 &     0.8 &     0.8 &     0.6 \\
FTFP &-&-&-&-&-&-&-&-&     - & - &     0.8 &     0.7 &     1.1 &     1.0 &     0.9 &     0.8 \\
MARS &     1.3 &     1.2 &     1.2 &     0.9 &     0.9 &     1.0 &     0.7 &     0.6 &     1.0 &     1.1 &     0.8 &     0.6 &     0.8 &     1.0 &     0.8 &     0.7 \\
\end{tabular}
\end{center}
}
\end{table}
\begin{table}[htbp!]
\caption{Normalization factors data over MC for some models, incident $\pi^-$.}
\label{tab-n}
{\small
\begin{center}
\begin{tabular}{c|cc|cc|cc|cc|cc|cc|cc|cc}
Generator & \multicolumn{2}{c}{Be } & \multicolumn{2}{c}{Ta } & \multicolumn{2}{c}{Be } & \multicolumn{2}{c}{Ta } &
 \multicolumn{2}{c}{Be } & \multicolumn{2}{c}{Ta } & \multicolumn{2}{c}{Be } & \multicolumn{2}{c}{Ta  } \\
 & \multicolumn{2}{c}{ 3 \GeVc} & \multicolumn{2}{c}{ 3 \GeVc} & \multicolumn{2}{c}{ 5 \GeVc} & \multicolumn{2}{c}{ 5 \GeVc} &
 \multicolumn{2}{c}{ 8 \GeVc} & \multicolumn{2}{c}{ 8 \GeVc} & \multicolumn{2}{c}{ 12 \GeVc} & \multicolumn{2}{c}{ 12 \GeVc} \\
& \pip & \pim & \pip & \pim & \pip & \pim & \pip & \pim
& \pip & \pim & \pip & \pim & \pip & \pim & \pip & \pim \\
Bertini &     1.0 &     1.1 &     0.5 &     0.6 &     1.2 &     1.4 &     0.9 &     1.2 &     1.6 &     2.1 &     1.4 &     1.9 &     1.2 &     1.6 &     1.6 &     2.1 \\
LHEP &     0.5 &     1.6 &     0.7 &     2.9 &     0.8 &     1.9 &     1.0 &     2.6 &     0.9 &     1.7 &     0.6 &     1.2 &     0.6 &     1.0 &     0.5 &     0.7 \\
QGSC &-&-&-&-&-&-&-&-&     1.1 &     1.8 &     0.7 &     1.3 &     0.6 &     0.7 &     0.8 &     1.1 \\
QGSP &-&-&-&-&-&-&-&-&     1.2 &     2.2 &     0.7 &     1.4 &     0.6 &     0.8 &     0.6 &     0.9 \\
FTFP & - & - & - &-&-&-&-&-&-&-&     0.7 &     1.1 &     0.7 &     1.0 &     0.8 &     1.1 \\
MARS &     1.1 &     1.2 &     0.7 &     1.0 &     0.9 &     0.9 &     0.6 &     0.6 &     1.2 &     1.2 &     0.7 &     0.9 &     0.8 &     0.7 &     0.7 &     0.9 \\
\end{tabular}
\end{center}
}   
\end{table}
The full set of HARP data, taken with targets
spanning the full periodic table of elements, with small total errors and full
coverage of the solid angle in a single detector will definitely help the validation
of models used in hadronic simulations in the difficult energy range between
3~\GeVc and 15~\GeVc of incident momentum with incident pions, where data
were previously missing.

\section{Summary and conclusions}\label{sec:conclusions}
 

Double-differential cross-sections for the production
of positive and negative pions 
in the kinematic range 
0.5~\GeVc$\leq p_\pi \leq 8$~\GeVc 
and 0.025~rad $\leq \theta_\pi \leq$ 0.25~rad
from the interactions of 3, 5, 8 and 12~\GeVc
$\pi^\pm$ on beryllium, carbon, aluminium, copper,
tin, tantalum and lead targets of 5\% interaction length thickness have been reported.

We should stress that the HARP  data are the first 
precision measurements with incoming charged pion beams
in this kinematic region and may have a major impact on the
tuning of Monte Carlo generators at low energies.
Quantitative comparisons with a variety of generators
currently in use have been presented.
 
The pion yield averaged over different
momentum and angular ranges 
increase smoothly
with the atomic number $A$ of the target and with the
energy of the incoming pion beam.

\section{Acknowledgments}

We gratefully acknowledge the help and support of the PS beam staff
and of the numerous technical collaborators who contributed to the
detector design, construction, commissioning and operation.  
In particular, we would like to thank
G.~Barichello,
R.~Brocard,
K.~Burin,
V.~Carassiti,
F.~Chignoli,
D.~Conventi,
G.~Decreuse,
M.~Delattre,
C.~Detraz,  
A.~Domeniconi,
M.~Dwuznik,   
F.~Evangelisti,
B.~Friend,
A.~Iaciofano,
I.~Krasin, 
D.~Lacroix,
J.-C.~Legrand,
M.~Lobello, 
M.~Lollo,
J.~Loquet,
F.~Marinilli,
R.~Mazza,
J.~Mulon,
L.~Musa,
R.~Nicholson,
A.~Pepato,
P.~Petev, 
X.~Pons,
I.~Rusinov,
M.~Scandurra,
E.~Usenko,
R.~van der Vlugt,
for their support in the construction of the detector
and P. Dini for his contribution to Monte Carlo production. 
The collaboration acknowledges the major contributions and advice of
M.~Baldo-Ceolin, 
L.~Linssen, 
M.T.~Muciaccia and A. Pullia
during the construction of the experiment.
The collaboration is indebted to 
V.~Ableev,
F.~Bergsma,
P.~Binko,
E.~Boter,
M.~Calvi, 
C.~Cavion,
M.Chizov, 
A.~Chukanov,
A.~DeSanto, 
A.~DeMin, 
M.~Doucet,
D.~D\"{u}llmann,
V.~Ermilova, 
W.~Flegel,
Y.~Hayato,
A.~Ichikawa,
O.~Klimov,
T.~Kobayashi,
D.~Kustov, 
M.~Laveder, 
M.~Mass,
H.~Meinhard,
A.~Menegolli, 
T.~Nakaya,
K.~Nishikawa,
M.~Paganoni,
F.~Paleari,
M.~Pasquali,
M.~Placentino,
V.~Serdiouk,
S.~Simone,
P.J.~Soler,
S.~Troquereau,
S.~Ueda,
A.~Valassi and
R.~Veenhof
for their contributions to the experiment.

We acknowledge the contributions of 
V.~Ammosov,
G.~Chelkov,
D.~Dedovich,
F.~Dydak,
M.~Gostkin,
A.~Guskov,
D.~Khartchenko,
V.~Koreshev,
Z.~Kroumchtein,
I.~Nefedov,
A.~Semak,
J.~Wotschack,
V.~Zaets and
A.~Zhemchugov
to the work described in this paper.

 The experiment was made possible by grants from
the Institut Interuniversitaire des Sciences Nucl\'eair\-es and the
Interuniversitair Instituut voor Kernwetenschappen (Belgium), 
Ministerio de Educacion y Ciencia, Grant FPA2003-06921-c02-02 and
Generalitat Valenciana, grant GV00-054-1,
CERN (Geneva, Switzerland), 
the German Bundesministerium f\"ur Bildung und Forschung (Germany), 
the Istituto Na\-zio\-na\-le di Fisica Nucleare (Italy), 
INR RAS (Moscow), the Russian Foundation for Basic Research (grant 08-02-00018),
the Particle Physics and Astronomy Research Council (UK) and the Swiss National
Science Foundation, in the framework of the SCOPES programme.
We gratefully acknowledge their support.

\clearpage 
\begin{appendix}

\section{Cross-section data}\label{app:data}

The following tables report the measured differential cross-section
for positive and negative pion production in interactions
of 3, 5, 8 and 12~\GeVc momentum charged pions
on different types of nuclear targets.
The data are presented in the kinematic range of  
0.5~\GeVc$\leq p_\pi \leq 8$~\GeVc 
and 0.05~rad $\leq \theta_\pi \leq$ 0.25~rad.
The overall normalization uncertainty ( $ \leq 2 \%$) is not included 
in the reported errors.

Results at higher incoming beam momenta, from 8~\GeVc to 12.9~\GeVc,
are also presented 
extending to a lower value of the polar angle 
$\theta$ (0.025 rad), using a finer binning.
This second set of tables contains additional data for collisions
of 12.9~\GeVc $\pi^\pm$ on Al and 8.9~\GeVc $\pi^\pm$ on Be.

The cross-section values in some bins of the following tables are
omitted and replaced by the symbol $* \pm *$, due to instabilities
in the unfolding procedure coming from the low statistics available.
\clearpage
\input{xsec_results_Be_pim.tex}

\input{xsec_results_Be_pip_abs.tex}

\input{xsec_results_C_pim.tex}

\input{xsec_results_C_pip_abs.tex}

\input{xsec_results_Al_pim.tex}

\input{xsec_results_Al_pip_abs.tex}

\input{xsec_results_Cu_pim.tex}

\input{xsec_results_Cu_pip_abs.tex}

\input{xsec_results_Sn_pim.tex}

\input{xsec_results_Sn_pip_abs.tex}

\input{xsec_results_Ta_pim.tex}

\input{xsec_results_Ta_pip_abs.tex}

\input{xsec_results_Pb_pim.tex}

\input{xsec_results_Pb_pip_abs.tex}

\input{table_result_fw_036-pim_Be_OMEGA_fine.tex}
\input{table_result_fw_036-pip_Be_OMEGA_fine.tex}
\input{table_result_fw_036-pim_C_OMEGA_fine.tex}
\input{table_result_fw_036-pip_C_OMEGA_fine.tex}
\input{table_result_fw_036-pim_Al_OMEGA_fine.tex}
\input{table_result_fw_036-pip_Al_OMEGA_fine.tex}
\input{table_result_fw_036-pim_Cu_OMEGA_fine.tex}
\input{table_result_fw_036-pip_Cu_OMEGA_fine.tex}
\input{table_result_fw_036-pim_Sn_OMEGA_fine.tex}
\input{table_result_fw_036-pip_Sn_OMEGA_fine.tex}
\input{table_result_fw_036-pim_Ta_OMEGA_fine.tex}
\input{table_result_fw_036-pip_Ta_OMEGA_fine.tex}
\input{table_result_fw_036-pim_Pb_OMEGA_fine.tex}
\input{table_result_fw_036-pip_Pb_OMEGA_fine.tex}

\end{appendix}

\end{document}

%% file: All5_table1.tex
\begin{table}[bp!] 
\caption{Total number of events and tracks used in the various nuclear 
  5\%~$\lambda_{\mathrm{I}}$ target data sets and the number of
  incident pions on target as calculated from the pre-scaled incident beam triggers. Numbers are
for incident $\pi^{+}$ (in parenthesis for incident $\pi^{-}$) in units of $10^3$ events.} 
\label{tab:events}
{\small
\begin{center}
\begin{tabular}{llccccccc} \hline
\bf{Data set (\bfGeVc)}          &         &\bf{3}&\bf{5}&\bf{8}&\bf{8.9}&\bf{12} & \bf{12.9}\\ \hline
    Total DAQ events     & (Be)     & 1113 (2233) & 1296 (1798)  & 1935 (1585)    & 5868 &  1207 (1227)&                \\
                         &  (C)     & 1345 (1831) & 2628 (1279)      & 1846 (1399)&    &    1062 (646)&              \\
                         & (Al)     & 1159 (1523) & 1789 (920)     & 1707 (1059)   &  &  619 (741)  & 4713            \\
                         & (Cu)     & 624 (3325)  & 2079 (1805)      & 2089 (1615)    & &    745 (591)  &              \\
                         & (Sn)     & 1637 (1972) & 2828 (1625)      & 2404(1408)    & &   1803(937)  &        \\
                         & (Ta)     & 1783 (994) & 2084(1435)      & 1965(1505)    &  &  866(961)   &       \\   
                         & (Pb)     & 1911 (1282) & 2111(2074)      & 2266 (1496)    &  &  487(1706)  &        \\ 
\hline  
  Accepted  beam pions with     &  (Be)         & 246 (731)  &  384 (914)     &  341 (826)  & 1278  &    76 (693) & \\
  forward interaction               & (C)           & 257 (299)      &  754 (530)         &  358 (772)  &       &    41 (352)&  \\
                         &  (Al)    & 213(486)  &  523 (308)      & 335 (611)   &   &  27 (435)    & 332                 \\
                         &  (Cu)    & 168 (1185)  &  611 (850)     &  397 (966) &  &    33 (347)  &               \\
                         &  (Sn)    & 467 (778)  &  819 (732)     &  481 (804)   & &   79  (584)   &               \\
                         &  (Ta)    & 561 (426)  &  600 (671)     &  388 (893)   & &   37 (536)    &               \\
                         &  (Pb)    & 611 (473)  &  594 (997)     &  444 (896)   & &   20  (839)   &                 \\
\hline
  \bf{Final state $\bfpim$ } & (Be)  &  .9 (24.9)   &   3.5 (32.1)     &  6.3 (31.0) & 26.7    &  2.0 (27.5)&                \\
     {selected with PID}                             & (C)   & .7 (9.1)   &   5.9 (15.0)     &  5.9 (24.2)   &   &  1.0 (13.7)  &          \\
                                  & (Al)  & .5 (10.3)    &   4.0 (6.8)      &  5.6 (13.4)   &    &  .7 (13.4)   &  9.8               \\ 
                                  & (Cu)  & .3 (18.9)   &    4.3 (19.0)     &  6.0 (25.1) &     &  .9 (12.9) &             \\
                                  & (Sn)  & .7 (9.1)   &   4.8 (14.4)     &  6.9 (16.3)    &  &  1.9k (19.4) &              \\
                                  & (Ta)  & .6 (3.7)   &   3.1 (11.5)     &  4.9 (16.2)    &  &  .9k (16.6)  &               \\
                                  & (Pb)  & .6 (3.4)   &   3.0 (16.0)     &  5.2 (16.3)    &  &  .5k (25.5)  &                \\
\hline
  \bf{Final state $\bfpip$ } & (Be)  & 8.9 (1.3)  &    14 (4.8)     &  13.8 (10.6) & 50.0     &  3.4  (13.1) &             \\
     {selected with PID}                             & (C)   & 8.0 (.5)   &   23.8 (2.5)     &  13.3 (8.0)   &   &  1.7 (6.7) &            \\
                                  & (Al)  & 5.2 (.5)    &   15 (1.3)     &  12.1 (5.6)      &  & 1.2 (7.0) &  15.1               \\
                                  & (Cu)  & 2.8 (1.6)    &   14.8 (3.5)     &  12.2 (9.6) &    &  1.4 (6.6)  &               \\
                                  & (Sn)  & 5.9 (.6)  &   16.8 (2.8)     &  13.3 (6.1)   &  &  3.1 (10.2) &                \\
                                  & (Ta)  & 5.3 (.2)   &   10.8 (2.2)     &  9.4 (6.0)   &   &  1.4 (8.7) &               \\
                                  & (Pb)  & 5.0 (.2)   &   9.8 (2.9)     &  10.3 (6.1)   &   &   .7 (13.1)  &           \\
\hline
\end{tabular}
\end{center}
}
\end{table}

%% file: xsec_results_Be_pim.tex
\begin{table}[!ht]
  \caption{\label{tab:xsec_results_Be1}
    HARP results for the double-differential $\pi^+$  production
    cross-section in the laboratory system,
    $d^2\sigma^{\pi}/(dpd\Omega)$, for $\pi^{-}$--Be interactions at 3,5,8,12~\GeVc.
    Each row refers to a
    different $(p_{\hbox{\small min}} \le p<p_{\hbox{\small max}},
    \theta_{\hbox{\small min}} \le \theta<\theta_{\hbox{\small max}})$ bin,
    where $p$ and $\theta$ are the pion momentum and polar angle, respectively.
    The central value as well as the square-root of the diagonal elements
    of the covariance matrix are given.}

\small{
\begin{tabular}{rrrr|r@{$\pm$}lr@{$\pm$}lr@{$\pm$}lr@{$\pm$}l}
\hline
$\theta_{\hbox{\small min}}$ &
$\theta_{\hbox{\small max}}$ &
$p_{\hbox{\small min}}$ &
$p_{\hbox{\small max}}$ &
\multicolumn{8}{c}{$d^2\sigma^{\pi^+}/(dpd\Omega)$}
\\
(rad) & (rad) & (\GeVc) & (\GeVc) &
\multicolumn{8}{c}{(barn/(\GeVc sr))}
\\
  &  &  &
&\multicolumn{2}{c}{$ \bf{3 \ \GeVc}$}
&\multicolumn{2}{c}{$ \bf{5 \ \GeVc}$}
&\multicolumn{2}{c}{$ \bf{8 \ \GeVc}$}
&\multicolumn{2}{c}{$ \bf{12 \ \GeVc}$}
\\
\hline

0.05 & 0.10 & 0.50 & 1.00& 0.038 &  0.007& 0.054 &  0.007&  0.14 &   0.02& 0.106 &  0.012\\ 
      &      & 1.00 & 1.50& 0.042 &  0.006& 0.073 &  0.007& 0.177 &  0.011& 0.130 &  0.010\\ 
      &      & 1.50 & 2.00& 0.035 &  0.005& 0.064 &  0.006& 0.141 &  0.009& 0.127 &  0.010\\ 
      &      & 2.00 & 2.50& 0.048 &  0.007& 0.053 &  0.005& 0.112 &  0.008& 0.127 &  0.008\\ 
      &      & 2.50 & 3.00&       &       & 0.045 &  0.004& 0.090 &  0.008& 0.086 &  0.006\\ 
      &      & 3.00 & 3.50&       &       & 0.044 &  0.005& 0.086 &  0.008& 0.097 &  0.008\\ 
      &      & 3.50 & 4.00&       &       & 0.066 &  0.005& 0.072 &  0.007& 0.087 &  0.007\\ 
      &      & 4.00 & 5.00&       &       & 0.022 &  0.008& 0.046 &  0.004& 0.049 &  0.004\\ 
      &      & 5.00 & 6.50&       &       &       &       & 0.023 &  0.003& 0.027 &  0.002\\ 
      &      & 6.50 & 8.00&       &       &       &       &       &       & 0.010 &  0.001\\ 
 0.10 & 0.15 & 0.50 & 1.00& 0.052 &  0.008& 0.094 &  0.011&  0.17 &   0.02& 0.137 &  0.015\\ 
      &      & 1.00 & 1.50& 0.026 &  0.004& 0.091 &  0.009& 0.162 &  0.013& 0.145 &  0.012\\ 
      &      & 1.50 & 2.00& 0.032 &  0.005& 0.047 &  0.005& 0.107 &  0.010& 0.117 &  0.010\\ 
      &      & 2.00 & 2.50& 0.037 &  0.006& 0.054 &  0.006& 0.104 &  0.009& 0.107 &  0.010\\ 
      &      & 2.50 & 3.00&       &       & 0.028 &  0.004& 0.075 &  0.007& 0.077 &  0.007\\ 
      &      & 3.00 & 3.50&       &       & 0.018 &  0.003& 0.044 &  0.005& 0.068 &  0.006\\ 
      &      & 3.50 & 4.00&       &       & 0.013 &  0.002& 0.040 &  0.004& 0.049 &  0.005\\ 
      &      & 4.00 & 5.00&       &       & 0.006 &  0.003& 0.021 &  0.004& 0.028 &  0.003\\ 
      &      & 5.00 & 6.50&       &       &       &       & 0.003 &  0.001& 0.008 &  0.001\\ 
      &      & 6.50 & 8.00&       &       &       &       &       &       & 0.001 &  0.000\\ 
 0.15 & 0.20 & 0.50 & 1.00& 0.042 &  0.008& 0.078 &  0.010&  0.16 &   0.02&  0.15 &   0.02\\ 
      &      & 1.00 & 1.50& 0.039 &  0.005& 0.078 &  0.008& 0.110 &  0.010& 0.099 &  0.009\\ 
      &      & 1.50 & 2.00& 0.024 &  0.004& 0.050 &  0.005& 0.104 &  0.009& 0.085 &  0.008\\ 
      &      & 2.00 & 2.50& 0.011 &  0.003& 0.039 &  0.005& 0.081 &  0.008& 0.078 &  0.008\\ 
      &      & 2.50 & 3.00&       &       & 0.035 &  0.005& 0.052 &  0.007& 0.044 &  0.006\\ 
      &      & 3.00 & 3.50&       &       & 0.010 &  0.003& 0.025 &  0.004& 0.031 &  0.004\\ 
      &      & 3.50 & 4.00&       &       & 0.003 &  0.001& 0.014 &  0.003& 0.014 &  0.003\\ 
      &      & 4.00 & 5.00&       &       & 0.001 &  0.000& 0.006 &  0.001& 0.008 &  0.001\\ 
      &      & 5.00 & 6.50&       &       &       &       & 0.001 &  0.000& 0.001 &  0.001\\ 
      &      & 6.50 & 8.00&       &       &       &       &       &       &  &  \\ 
 0.20 & 0.25 & 0.50 & 1.00& 0.057 &  0.010& 0.107 &  0.014&  0.13 &   0.02& 0.097 &  0.013\\ 
      &      & 1.00 & 1.50& 0.028 &  0.006& 0.101 &  0.013&  0.18 &   0.02&  0.16 &   0.02\\ 
      &      & 1.50 & 2.00& 0.034 &  0.007& 0.063 &  0.009&  0.14 &   0.02& 0.099 &  0.014\\ 
      &      & 2.00 & 2.50& 0.005 &  0.002& 0.038 &  0.006& 0.082 &  0.013& 0.051 &  0.008\\ 
      &      & 2.50 & 3.00&       &       & 0.011 &  0.003& 0.031 &  0.007& 0.023 &  0.004\\ 
      &      & 3.00 & 3.50&       &       & 0.004 &  0.001& 0.008 &  0.003& 0.011 &  0.003\\ 
      &      & 3.50 & 4.00&       &       & 0.001 &  0.001& 0.004 &  0.002& 0.004 &  0.002\\ 
      &      & 4.00 & 5.00&       &       &  &  & 0.002 &  0.001& 0.002 &  0.001\\ 
      &      & 5.00 & 6.50&       &       &       &       & &  &  &  \\ 
      &      & 6.50 & 8.00&       &       &       &       &       &       & 0.000 &  0.001\\ 

\hline
\end{tabular}
}
\end{table}
\clearpage
\begin{table}[!ht]
  \caption{\label{tab:xsec_results_Be2}
    HARP results for the double-differential  $\pi^-$ production
    cross-section in the laboratory system,
    $d^2\sigma^{\pi}/(dpd\Omega)$, for $\pi^{-}$--Be interactions at 3,5,8,12~\GeVc.
    Each row refers to a
    different $(p_{\hbox{\small min}} \le p<p_{\hbox{\small max}},
    \theta_{\hbox{\small min}} \le \theta<\theta_{\hbox{\small max}})$ bin,
    where $p$ and $\theta$ are the pion momentum and polar angle, respectively.
    The central value as well as the square-root of the diagonal elements
    of the covariance matrix are given.}

\small{
\begin{tabular}{rrrr|r@{$\pm$}lr@{$\pm$}lr@{$\pm$}lr@{$\pm$}l}
\hline
$\theta_{\hbox{\small min}}$ &
$\theta_{\hbox{\small max}}$ &
$p_{\hbox{\small min}}$ &
$p_{\hbox{\small max}}$ &
\multicolumn{8}{c}{$d^2\sigma^{\pi^-}/(dpd\Omega)$}
\\
(rad) & (rad) & (\GeVc) & (\GeVc) &
\multicolumn{8}{c}{(barn/(\GeVc sr))}
\\
  &  &  &
&\multicolumn{2}{c}{$ \bf{3 \ \GeVc}$}
&\multicolumn{2}{c}{$ \bf{5 \ \GeVc}$}
&\multicolumn{2}{c}{$ \bf{8 \ \GeVc}$}
&\multicolumn{2}{c}{$ \bf{12 \ \GeVc}$}
\\
\hline

 0.05 & 0.10 & 0.50 & 1.00& 0.075 &  0.010& 0.113 &  0.012&  0.25 &   0.02&  0.15 &   0.02\\ 
      &      & 1.00 & 1.50& 0.094 &  0.010& 0.119 &  0.009& 0.195 &  0.014& 0.144 &  0.012\\ 
      &      & 1.50 & 2.00& 0.116 &  0.009& 0.134 &  0.009& 0.208 &  0.013& 0.172 &  0.011\\ 
      &      & 2.00 & 2.50&  0.15 &   0.04& 0.142 &  0.010& 0.217 &  0.014& 0.167 &  0.011\\ 
      &      & 2.50 & 3.00&       &       & 0.128 &  0.009& 0.194 &  0.011& 0.176 &  0.011\\ 
      &      & 3.00 & 3.50&       &       &  0.21 &   0.02& 0.163 &  0.011& 0.145 &  0.008\\ 
      &      & 3.50 & 4.00&       &       & 0.241 &  0.013& 0.155 &  0.011& 0.129 &  0.007\\ 
      &      & 4.00 & 5.00&       &       &  0.25 &   0.04& 0.161 &  0.009& 0.105 &  0.006\\ 
      &      & 5.00 & 6.50&       &       &       &       & 0.118 &  0.006& 0.073 &  0.004\\ 
      &      & 6.50 & 8.00&       &       &       &       &       &       & 0.053 &  0.003\\ 
 0.10 & 0.15 & 0.50 & 1.00& 0.100 &  0.012&  0.16 &   0.02&  0.22 &   0.02&  0.17 &   0.02\\ 
      &      & 1.00 & 1.50& 0.083 &  0.009& 0.119 &  0.010&  0.23 &   0.02& 0.177 &  0.014\\ 
      &      & 1.50 & 2.00& 0.119 &  0.012& 0.118 &  0.011& 0.186 &  0.014& 0.165 &  0.013\\ 
      &      & 2.00 & 2.50&  0.16 &   0.02& 0.137 &  0.011& 0.180 &  0.014& 0.147 &  0.012\\ 
      &      & 2.50 & 3.00&       &       & 0.107 &  0.009& 0.150 &  0.012& 0.124 &  0.010\\ 
      &      & 3.00 & 3.50&       &       & 0.103 &  0.008& 0.103 &  0.009& 0.093 &  0.007\\ 
      &      & 3.50 & 4.00&       &       & 0.087 &  0.007& 0.078 &  0.008& 0.083 &  0.006\\ 
      &      & 4.00 & 5.00&       &       & 0.094 &  0.009& 0.045 &  0.005& 0.055 &  0.005\\ 
      &      & 5.00 & 6.50&       &       &       &       & 0.022 &  0.003& 0.023 &  0.002\\ 
      &      & 6.50 & 8.00&       &       &       &       &       &       & 0.008 &  0.001\\ 
 0.15 & 0.20 & 0.50 & 1.00& 0.103 &  0.014& 0.093 &  0.012&  0.21 &   0.02&  0.17 &   0.02\\ 
      &      & 1.00 & 1.50& 0.089 &  0.010& 0.099 &  0.010&  0.20 &   0.02& 0.163 &  0.014\\ 
      &      & 1.50 & 2.00& 0.065 &  0.008& 0.103 &  0.010& 0.163 &  0.014& 0.118 &  0.011\\ 
      &      & 2.00 & 2.50& 0.069 &  0.009& 0.067 &  0.007& 0.120 &  0.011& 0.091 &  0.008\\ 
      &      & 2.50 & 3.00&       &       & 0.034 &  0.004& 0.071 &  0.007& 0.059 &  0.006\\ 
      &      & 3.00 & 3.50&       &       & 0.035 &  0.004& 0.039 &  0.005& 0.043 &  0.004\\ 
      &      & 3.50 & 4.00&       &       & 0.015 &  0.003& 0.032 &  0.003& 0.030 &  0.003\\ 
      &      & 4.00 & 5.00&       &       & 0.013 &  0.002& 0.020 &  0.003& 0.018 &  0.002\\ 
      &      & 5.00 & 6.50&       &       &       &       & 0.005 &  0.001& 0.005 &  0.001\\ 
      &      & 6.50 & 8.00&       &       &       &       &       &       & 0.001 &  0.000\\ 
 0.20 & 0.25 & 0.50 & 1.00& 0.075 &  0.012&  0.13 &   0.02&  0.16 &   0.02& 0.104 &  0.014\\ 
      &      & 1.00 & 1.50& 0.033 &  0.006& 0.053 &  0.008& 0.120 &  0.014& 0.066 &  0.009\\ 
      &      & 1.50 & 2.00& 0.035 &  0.006& 0.038 &  0.007& 0.096 &  0.014& 0.070 &  0.011\\ 
      &      & 2.00 & 2.50& 0.023 &  0.004& 0.033 &  0.005& 0.074 &  0.012& 0.060 &  0.009\\ 
      &      & 2.50 & 3.00&       &       & 0.025 &  0.004& 0.056 &  0.009& 0.041 &  0.006\\ 
      &      & 3.00 & 3.50&       &       & 0.011 &  0.002& 0.043 &  0.006& 0.031 &  0.005\\ 
      &      & 3.50 & 4.00&       &       & 0.007 &  0.002& 0.030 &  0.005& 0.018 &  0.003\\ 
      &      & 4.00 & 5.00&       &       & 0.003 &  0.001& 0.017 &  0.003& 0.011 &  0.002\\ 
      &      & 5.00 & 6.50&       &       &       &       & 0.005 &  0.002& 0.003 &  0.001\\ 
      &      & 6.50 & 8.00&       &       &       &       &       &       & 0.001 &  0.001\\ 
\hline
\end{tabular}
}
\end{table}

%% file: xsec_results_Be_pip_abs.tex
\begin{table}[!ht]
  \caption{\label{tab:xsec_results_Be3}
    HARP results for the double-differential $\pi^+$  production
    cross-section in the laboratory system,
    $d^2\sigma^{\pi}/(dpd\Omega)$, for $\pi^{+}$--Be interactions at 3,5,8,12~\GeVc.
    Each row refers to a
    different $(p_{\hbox{\small min}} \le p<p_{\hbox{\small max}},
    \theta_{\hbox{\small min}} \le \theta<\theta_{\hbox{\small max}})$ bin,
    where $p$ and $\theta$ are the pion momentum and polar angle, respectively.
    The central value as well as the square-root of the diagonal elements
    of the covariance matrix are given.}

\small{
\begin{tabular}{rrrr|r@{$\pm$}lr@{$\pm$}lr@{$\pm$}lr@{$\pm$}l}
\hline
$\theta_{\hbox{\small min}}$ &
$\theta_{\hbox{\small max}}$ &
$p_{\hbox{\small min}}$ &
$p_{\hbox{\small max}}$ &
\multicolumn{8}{c}{$d^2\sigma^{\pi^+}/(dpd\Omega)$}
\\
(rad) & (rad) & (\GeVc) & (\GeVc) &
\multicolumn{8}{c}{(barn/(\GeVc sr))}
\\
  &  &  &
&\multicolumn{2}{c}{$ \bf{3 \ \GeVc}$}
&\multicolumn{2}{c}{$ \bf{5 \ \GeVc}$}
&\multicolumn{2}{c}{$ \bf{8 \ \GeVc}$}
&\multicolumn{2}{c}{$ \bf{12 \ \GeVc}$}
\\
\hline
0.050 &0.100 & 0.50 & 1.00&  0.08 &   0.02&  0.12 &   0.02&  0.13 &   0.02&  0.15 &   0.04\\
      &      & 1.00 & 1.50& 0.047 &  0.008& 0.102 &  0.013&  0.16 &   0.02&  0.16 &   0.03\\
      &      & 1.50 & 2.00& 0.089 &  0.009& 0.110 &  0.012& 0.131 &  0.013&  0.16 &   0.03\\
      &      & 2.00 & 2.50&  0.22 &   0.02& 0.158 &  0.014& 0.158 &  0.015&  0.20 &   0.03\\
      &      & 2.50 & 3.00&       &       & 0.161 &  0.012&  0.18 &   0.02&  0.18 &   0.03\\
      &      & 3.00 & 3.50&       &       &  0.21 &   0.02& 0.138 &  0.014&  0.19 &   0.03\\
      &      & 3.50 & 4.00&       &       & 0.251 &  0.014& 0.122 &  0.012&  0.20 &   0.03\\
      &      & 4.00 & 5.00&       &       &  0.30 &   0.10& 0.161 &  0.010&  0.15 &   0.02\\
      &      & 5.00 & 6.50&       &       &       &       &  0.14 &   0.02& 0.047 &  0.011\\
      &      & 6.50 & 8.00&       &       &       &       &       &       &  0.18 &   0.02\\
0.100 &0.150 & 0.50 & 1.00&  0.07 &   0.02&  0.12 &   0.02&  0.15 &   0.02&  0.13 &   0.03\\
      &      & 1.00 & 1.50& 0.081 &  0.013& 0.114 &  0.014&  0.14 &   0.02&  0.18 &   0.03\\
      &      & 1.50 & 2.00&  0.16 &   0.02&  0.14 &   0.02&  0.21 &   0.02&  0.19 &   0.03\\
      &      & 2.00 & 2.50&  0.16 &   0.02&  0.14 &   0.02&  0.16 &   0.02&  0.19 &   0.03\\
      &      & 2.50 & 3.00&       &       & 0.107 &  0.012& 0.130 &  0.014&  0.15 &   0.02\\
      &      & 3.00 & 3.50&       &       & 0.128 &  0.013& 0.102 &  0.011&  0.12 &   0.02\\
      &      & 3.50 & 4.00&       &       & 0.108 &  0.011& 0.067 &  0.008&  0.13 &   0.02\\
      &      & 4.00 & 5.00&       &       &  0.10 &   0.04& 0.051 &  0.008& 0.091 &  0.014\\
      &      & 5.00 & 6.50&       &       &       &       & 0.021 &  0.004& 0.034 &  0.008\\
      &      & 6.50 & 8.00&       &       &       &       &       &       & 0.014 &  0.004\\
0.150 &0.200 & 0.50 & 1.00&  0.11 &   0.02&  0.14 &   0.02&  0.19 &   0.03&  0.24 &   0.05\\
      &      & 1.00 & 1.50& 0.101 &  0.015& 0.122 &  0.015&  0.16 &   0.02&  0.23 &   0.04\\
      &      & 1.50 & 2.00& 0.077 &  0.011& 0.079 &  0.010&  0.15 &   0.02&  0.11 &   0.02\\
      &      & 2.00 & 2.50&  0.09 &   0.02& 0.090 &  0.015& 0.109 &  0.013&  0.08 &   0.02\\
      &      & 2.50 & 3.00&       &       & 0.039 &  0.006& 0.056 &  0.009&  0.07 &   0.02\\
      &      & 3.00 & 3.50&       &       & 0.030 &  0.008& 0.046 &  0.008&  0.06 &   0.02\\
      &      & 3.50 & 4.00&       &       & 0.023 &  0.006& 0.026 &  0.005& 0.042 &  0.013\\
      &      & 4.00 & 5.00&       &       & 0.016 &  0.009& 0.016 &  0.003& 0.023 &  0.009\\
      &      & 5.00 & 6.50&       &       &       &       & 0.006 &  0.002& 0.008 &  0.004\\
      &      & 6.50 & 8.00&       &       &       &       &       &       & 0.002 &  0.002\\
0.200 &0.250 & 0.50 & 1.00&  0.07 &   0.02&  0.11 &   0.02&  0.11 &   0.02&  0.09 &   0.03\\
      &      & 1.00 & 1.50& 0.059 &  0.013& 0.053 &  0.010&  0.11 &   0.02&  0.09 &   0.03\\
      &      & 1.50 & 2.00& 0.027 &  0.007& 0.067 &  0.013& 0.065 &  0.013&  0.06 &   0.03\\
      &      & 2.00 & 2.50& 0.021 &  0.008& 0.042 &  0.009&  0.08 &   0.02&  0.07 &   0.03\\
      &      & 2.50 & 3.00&       &       & 0.028 &  0.007& 0.047 &  0.012&  0.06 &   0.03\\
      &      & 3.00 & 3.50&       &       & 0.019 &  0.006& 0.026 &  0.008&  0.03 &   0.02\\
      &      & 3.50 & 4.00&       &       & 0.010 &  0.004& 0.020 &  0.005& 0.016 &  0.015\\
      &      & 4.00 & 5.00&       &       & 0.005 &  0.003& 0.012 &  0.004& 0.011 &  0.014\\
      &      & 5.00 & 6.50&       &       &       &       & 0.006 &  0.003& 0.002 &  0.005\\
      &      & 6.50 & 8.00&       &       &       &       &       &       & 0.000 &  0.011\\

\hline
\end{tabular}
}
\end{table}
\clearpage
\begin{table}[!ht]
  \caption{\label{tab:xsec_results_Be4}
    HARP results for the double-differential  $\pi^-$ production
    cross-section in the laboratory system,
    $d^2\sigma^{\pi}/(dpd\Omega)$, for $\pi^{+}$--Be interactions at 3,5,8,12~\GeVc.
    Each row refers to a
    different $(p_{\hbox{\small min}} \le p<p_{\hbox{\small max}},
    \theta_{\hbox{\small min}} \le \theta<\theta_{\hbox{\small max}})$ bin,
    where $p$ and $\theta$ are the pion momentum and polar angle, respectively.
    The central value as well as the square-root of the diagonal elements
    of the covariance matrix are given.}

\small{
\begin{tabular}{rrrr|r@{$\pm$}lr@{$\pm$}lr@{$\pm$}lr@{$\pm$}l}
\hline
$\theta_{\hbox{\small min}}$ &
$\theta_{\hbox{\small max}}$ &
$p_{\hbox{\small min}}$ &
$p_{\hbox{\small max}}$ &
\multicolumn{8}{c}{$d^2\sigma^{\pi^-}/(dpd\Omega)$}
\\
(rad) & (rad) & (\GeVc) & (\GeVc) &
\multicolumn{8}{c}{(barn/(\GeVc sr))}
\\
  &  &  &
&\multicolumn{2}{c}{$ \bf{3 \ \GeVc}$}
&\multicolumn{2}{c}{$ \bf{5 \ \GeVc}$}
&\multicolumn{2}{c}{$ \bf{8 \ \GeVc}$}
&\multicolumn{2}{c}{$ \bf{12 \ \GeVc}$}
\\
\hline
0.050 &0.100 & 0.50 & 1.00& 0.060 &  0.013& 0.081 &  0.014&  0.11 &   0.02&  0.05 &   0.02\\
      &      & 1.00 & 1.50& 0.049 &  0.010& 0.106 &  0.013&  0.14 &   0.02&  0.15 &   0.03\\
      &      & 1.50 & 2.00& 0.038 &  0.007& 0.062 &  0.008& 0.118 &  0.013&  0.12 &   0.02\\
      &      & 2.00 & 2.50&  0.06 &   0.02& 0.064 &  0.008& 0.114 &  0.012&  0.10 &   0.02\\
      &      & 2.50 & 3.00&       &       & 0.050 &  0.007& 0.083 &  0.009&  0.14 &   0.02\\
      &      & 3.00 & 3.50&       &       & 0.054 &  0.008& 0.093 &  0.009&  0.12 &   0.02\\
      &      & 3.50 & 4.00&       &       & 0.063 &  0.007& 0.065 &  0.007&  0.10 &   0.02\\
      &      & 4.00 & 5.00&       &       & 0.033 &  0.007& 0.050 &  0.005& 0.066 &  0.011\\
      &      & 5.00 & 6.50&       &       &       &       & 0.026 &  0.003& 0.039 &  0.007\\
      &      & 6.50 & 8.00&       &       &       &       &       &       & 0.019 &  0.004\\
0.100 &0.150 & 0.50 & 1.00& 0.070 &  0.014& 0.081 &  0.015&  0.15 &   0.02&  0.20 &   0.05\\
      &      & 1.00 & 1.50& 0.041 &  0.008& 0.059 &  0.009& 0.118 &  0.014&  0.15 &   0.03\\
      &      & 1.50 & 2.00& 0.035 &  0.007& 0.067 &  0.010& 0.101 &  0.012&  0.18 &   0.03\\
      &      & 2.00 & 2.50& 0.021 &  0.006& 0.057 &  0.008& 0.093 &  0.011&  0.14 &   0.03\\
      &      & 2.50 & 3.00&       &       & 0.038 &  0.006& 0.061 &  0.008&  0.09 &   0.02\\
      &      & 3.00 & 3.50&       &       & 0.025 &  0.004& 0.044 &  0.006& 0.064 &  0.014\\
      &      & 3.50 & 4.00&       &       & 0.015 &  0.003& 0.036 &  0.005& 0.064 &  0.014\\
      &      & 4.00 & 5.00&       &       & 0.005 &  0.002& 0.020 &  0.004& 0.037 &  0.009\\
      &      & 5.00 & 6.50&       &       &       &       & 0.003 &  0.001& 0.016 &  0.005\\
      &      & 6.50 & 8.00&       &       &       &       &       &       & 0.004 &  0.002\\
0.150 &0.200 & 0.50 & 1.00&  0.08 &   0.02& 0.086 &  0.014&  0.13 &   0.02&  0.14 &   0.04\\
      &      & 1.00 & 1.50& 0.059 &  0.011& 0.086 &  0.012&  0.12 &   0.02&  0.12 &   0.03\\
      &      & 1.50 & 2.00& 0.024 &  0.006& 0.068 &  0.010& 0.099 &  0.013&  0.09 &   0.02\\
      &      & 2.00 & 2.50& 0.008 &  0.003& 0.034 &  0.006& 0.085 &  0.012&  0.13 &   0.03\\
      &      & 2.50 & 3.00&       &       & 0.025 &  0.005& 0.050 &  0.008&  0.05 &   0.02\\
      &      & 3.00 & 3.50&       &       & 0.006 &  0.002& 0.026 &  0.005& 0.034 &  0.012\\
      &      & 3.50 & 4.00&       &       & 0.002 &  0.001& 0.017 &  0.004& 0.019 &  0.008\\
      &      & 4.00 & 5.00&       &       & 0.001 &  0.001& 0.004 &  0.001& 0.012 &  0.007\\
      &      & 5.00 & 6.50&       &       &       &       & * &  *& 0.001 &  0.002\\
      &      & 6.50 & 8.00&       &       &       &       &       &       & 0.000 &  0.001\\
0.200 &0.250 & 0.50 & 1.00& 0.048 &  0.013&  0.11 &   0.02&  0.10 &   0.02&  0.12 &   0.04\\
      &      & 1.00 & 1.50& 0.047 &  0.012& 0.059 &  0.011&  0.11 &   0.02&  0.10 &   0.03\\
      &      & 1.50 & 2.00& 0.012 &  0.005& 0.038 &  0.010&  0.13 &   0.02&  0.14 &   0.05\\
      &      & 2.00 & 2.50& 0.002 &  0.002& 0.017 &  0.005& 0.052 &  0.011&  0.07 &   0.03\\
      &      & 2.50 & 3.00&       &       & 0.003 &  0.001& 0.025 &  0.007&  0.04 &   0.03\\
      &      & 3.00 & 3.50&       &       & 0.002 &  0.001& 0.012 &  0.008& 0.016 &  0.023\\
      &      & 3.50 & 4.00&       &       & 0.002 &  0.001& 0.004 &  0.005& 0.004 &  0.008\\
      &      & 4.00 & 5.00&       &       & 0.001 &  0.001& 0.001 &  0.004& 0.002 &  0.006\\
      &      & 5.00 & 6.50&       &       &       &       & 0.000 &  0.004& 0.000 &  0.001\\
      &      & 6.50 & 8.00&       &       &       &       &       &       & 0.000 &  0.006\\

\hline
\end{tabular}
}
\end{table}

%% file: xsec_results_C_pim.tex
\begin{table}[!ht]
  \caption{\label{tab:xsec_results_C1}
    HARP results for the double-differential $\pi^+$  production
    cross-section in the laboratory system,
    $d^2\sigma^{\pi}/(dpd\Omega)$, for $\pi^{-}$--C interactions at 3,5,8,12~\GeVc.
    Each row refers to a
    different $(p_{\hbox{\small min}} \le p<p_{\hbox{\small max}},
    \theta_{\hbox{\small min}} \le \theta<\theta_{\hbox{\small max}})$ bin,
    where $p$ and $\theta$ are the pion momentum and polar angle, respectively.
    The central value as well as the square-root of the diagonal elements
    of the covariance matrix are given.}

\small{
\begin{tabular}{rrrr|r@{$\pm$}lr@{$\pm$}lr@{$\pm$}lr@{$\pm$}l}
\hline
$\theta_{\hbox{\small min}}$ &
$\theta_{\hbox{\small max}}$ &
$p_{\hbox{\small min}}$ &
$p_{\hbox{\small max}}$ &
\multicolumn{8}{c}{$d^2\sigma^{\pi^+}/(dpd\Omega)$}
\\
(rad) & (rad) & (\GeVc) & (\GeVc) &
\multicolumn{8}{c}{(barn/(\GeVc sr))}
\\
  &  &  &
&\multicolumn{2}{c}{$ \bf{3 \ \GeVc}$}
&\multicolumn{2}{c}{$ \bf{5 \ \GeVc}$}
&\multicolumn{2}{c}{$ \bf{8 \ \GeVc}$}
&\multicolumn{2}{c}{$ \bf{12 \ \GeVc}$}
\\
\hline 
0.05 & 0.10 & 0.50 & 1.00& 0.044 &  0.012& 0.078 &  0.013&  0.14 &   0.02&  0.21 &   0.03\\ 
      &      & 1.00 & 1.50& 0.046 &  0.010& 0.099 &  0.012& 0.204 &  0.014&  0.26 &   0.02\\ 
      &      & 1.50 & 2.00& 0.057 &  0.011& 0.060 &  0.009& 0.174 &  0.011&  0.23 &   0.02\\ 
      &      & 2.00 & 2.50& 0.058 &  0.012& 0.070 &  0.009& 0.126 &  0.011&  0.20 &   0.02\\ 
      &      & 2.50 & 3.00&       &       & 0.049 &  0.007& 0.094 &  0.010& 0.175 &  0.014\\ 
      &      & 3.00 & 3.50&       &       & 0.050 &  0.007& 0.094 &  0.010& 0.171 &  0.014\\ 
      &      & 3.50 & 4.00&       &       & 0.059 &  0.008& 0.068 &  0.009& 0.150 &  0.013\\ 
      &      & 4.00 & 5.00&       &       & 0.030 &  0.009& 0.050 &  0.005& 0.082 &  0.008\\ 
      &      & 5.00 & 6.50&       &       &       &       & 0.023 &  0.003& 0.053 &  0.005\\ 
      &      & 6.50 & 8.00&       &       &       &       &       &       & 0.016 &  0.003\\ 
 0.10 & 0.15 & 0.50 & 1.00&  0.09 &   0.02& 0.090 &  0.014&  0.19 &   0.02&  0.20 &   0.03\\ 
      &      & 1.00 & 1.50& 0.054 &  0.011& 0.098 &  0.012& 0.167 &  0.015&  0.27 &   0.03\\ 
      &      & 1.50 & 2.00& 0.039 &  0.009& 0.073 &  0.010& 0.146 &  0.013&  0.21 &   0.02\\ 
      &      & 2.00 & 2.50& 0.035 &  0.009& 0.052 &  0.008& 0.130 &  0.012&  0.19 &   0.02\\ 
      &      & 2.50 & 3.00&       &       & 0.039 &  0.006& 0.086 &  0.010& 0.145 &  0.015\\ 
      &      & 3.00 & 3.50&       &       & 0.024 &  0.005& 0.053 &  0.007& 0.113 &  0.012\\ 
      &      & 3.50 & 4.00&       &       & 0.013 &  0.003& 0.050 &  0.006& 0.102 &  0.011\\ 
      &      & 4.00 & 5.00&       &       & 0.007 &  0.003& 0.019 &  0.004& 0.048 &  0.007\\ 
      &      & 5.00 & 6.50&       &       &       &       & 0.003 &  0.001& 0.014 &  0.003\\ 
      &      & 6.50 & 8.00&       &       &       &       &       &       & 0.002 &  0.000\\ 
 0.15 & 0.20 & 0.50 & 1.00&  0.06 &   0.02&  0.09 &   0.02&  0.19 &   0.02&  0.20 &   0.03\\ 
      &      & 1.00 & 1.50& 0.034 &  0.008& 0.066 &  0.010& 0.130 &  0.013&  0.19 &   0.02\\ 
      &      & 1.50 & 2.00& 0.028 &  0.008& 0.056 &  0.008& 0.115 &  0.011&  0.16 &   0.02\\ 
      &      & 2.00 & 2.50& 0.022 &  0.008& 0.042 &  0.007& 0.091 &  0.010&  0.16 &   0.02\\ 
      &      & 2.50 & 3.00&       &       & 0.022 &  0.005& 0.050 &  0.007& 0.073 &  0.010\\ 
      &      & 3.00 & 3.50&       &       & 0.010 &  0.004& 0.031 &  0.005& 0.068 &  0.010\\ 
      &      & 3.50 & 4.00&       &       & 0.001 &  0.001& 0.016 &  0.003& 0.026 &  0.005\\ 
      &      & 4.00 & 5.00&       &       & 0.001 &  0.001& 0.009 &  0.002& 0.011 &  0.003\\ 
      &      & 5.00 & 6.50&       &       &       &       & 0.001 &  0.000& 0.002 &  0.001\\ 
      &      & 6.50 & 8.00&       &       &       &       &       &       &  &  \\ 
 0.20 & 0.25 & 0.50 & 1.00&  0.06 &   0.02&  0.10 &   0.02&  0.14 &   0.02&  0.19 &   0.03\\ 
      &      & 1.00 & 1.50&  0.06 &   0.02&  0.08 &   0.02&  0.19 &   0.02&  0.25 &   0.03\\ 
      &      & 1.50 & 2.00& 0.027 &  0.010& 0.077 &  0.014&  0.12 &   0.02&  0.15 &   0.03\\ 
      &      & 2.00 & 2.50& 0.010 &  0.006& 0.039 &  0.008& 0.069 &  0.012&  0.10 &   0.02\\ 
      &      & 2.50 & 3.00&       &       & 0.018 &  0.005& 0.036 &  0.008& 0.044 &  0.011\\ 
      &      & 3.00 & 3.50&       &       & 0.004 &  0.002& 0.011 &  0.005& 0.015 &  0.005\\ 
      &      & 3.50 & 4.00&       &       & 0.001 &  0.001& 0.002 &  0.002& 0.008 &  0.004\\ 
      &      & 4.00 & 5.00&       &       &  &   & 0.001 &  0.001& 0.003 &  0.003\\ 
      &      & 5.00 & 6.50&       &       &       &       &  &   &   &   \\ 
      &      & 6.50 & 8.00&       &       &       &       &       &       & 0.000 &  0.003\\ 

\hline
\end{tabular}
}
\end{table}
\clearpage
\begin{table}[!ht]
  \caption{\label{tab:xsec_results_C2}
    HARP results for the double-differential  $\pi^-$ production
    cross-section in the laboratory system,
    $d^2\sigma^{\pi}/(dpd\Omega)$, for $\pi^{-}$--C interactions at 3,5,8,12~\GeVc.
    Each row refers to a
    different $(p_{\hbox{\small min}} \le p<p_{\hbox{\small max}},
    \theta_{\hbox{\small min}} \le \theta<\theta_{\hbox{\small max}})$ bin,
    where $p$ and $\theta$ are the pion momentum and polar angle, respectively.
    The central value as well as the square-root of the diagonal elements
    of the covariance matrix are given.}

\small{
\begin{tabular}{rrrr|r@{$\pm$}lr@{$\pm$}lr@{$\pm$}lr@{$\pm$}l}
\hline
$\theta_{\hbox{\small min}}$ &
$\theta_{\hbox{\small max}}$ &
$p_{\hbox{\small min}}$ &
$p_{\hbox{\small max}}$ &
\multicolumn{8}{c}{$d^2\sigma^{\pi^-}/(dpd\Omega)$}
\\
(rad) & (rad) & (\GeVc) & (\GeVc) &
\multicolumn{8}{c}{(barn/(\GeVc sr))}
\\
  &  &  &
&\multicolumn{2}{c}{$ \bf{3 \ \GeVc}$}
&\multicolumn{2}{c}{$ \bf{5 \ \GeVc}$}
&\multicolumn{2}{c}{$ \bf{8 \ \GeVc}$}
&\multicolumn{2}{c}{$ \bf{12 \ \GeVc}$}
\\
\hline

 0.05 & 0.10 & 0.50 & 1.00&  0.11 &   0.02&  0.14 &   0.02&  0.27 &   0.03&  0.28 &   0.03\\ 
      &      & 1.00 & 1.50& 0.080 &  0.014& 0.122 &  0.013&  0.23 &   0.02&  0.29 &   0.03\\ 
      &      & 1.50 & 2.00&  0.14 &   0.02& 0.123 &  0.013&  0.25 &   0.02&  0.32 &   0.02\\ 
      &      & 2.00 & 2.50&  0.15 &   0.04& 0.138 &  0.014&  0.24 &   0.02&  0.32 &   0.02\\ 
      &      & 2.50 & 3.00&       &       &  0.17 &   0.02& 0.222 &  0.013&  0.30 &   0.02\\ 
      &      & 3.00 & 3.50&       &       &  0.22 &   0.02& 0.183 &  0.014&  0.26 &   0.02\\ 
      &      & 3.50 & 4.00&       &       &  0.19 &   0.02& 0.177 &  0.014&  0.23 &   0.02\\ 
      &      & 4.00 & 5.00&       &       &  0.27 &   0.05& 0.172 &  0.011& 0.194 &  0.013\\ 
      &      & 5.00 & 6.50&       &       &       &       & 0.116 &  0.008& 0.134 &  0.007\\ 
      &      & 6.50 & 8.00&       &       &       &       &       &       & 0.089 &  0.007\\ 
 0.10 & 0.15 & 0.50 & 1.00&  0.16 &   0.03&  0.17 &   0.02&  0.25 &   0.03&  0.31 &   0.04\\ 
      &      & 1.00 & 1.50& 0.087 &  0.013& 0.133 &  0.015&  0.24 &   0.02&  0.30 &   0.03\\ 
      &      & 1.50 & 2.00&  0.15 &   0.02&  0.15 &   0.02&  0.21 &   0.02&  0.31 &   0.03\\ 
      &      & 2.00 & 2.50&  0.17 &   0.03& 0.140 &  0.014&  0.20 &   0.02&  0.26 &   0.02\\ 
      &      & 2.50 & 3.00&       &       & 0.126 &  0.013& 0.167 &  0.014&  0.23 &   0.02\\ 
      &      & 3.00 & 3.50&       &       & 0.110 &  0.012& 0.097 &  0.010&  0.18 &   0.02\\ 
      &      & 3.50 & 4.00&       &       & 0.100 &  0.011& 0.076 &  0.008& 0.140 &  0.014\\ 
      &      & 4.00 & 5.00&       &       &  0.10 &   0.02& 0.046 &  0.006& 0.090 &  0.009\\ 
      &      & 5.00 & 6.50&       &       &       &       & 0.020 &  0.003& 0.040 &  0.005\\ 
      &      & 6.50 & 8.00&       &       &       &       &       &       & 0.014 &  0.003\\ 
 0.15 & 0.20 & 0.50 & 1.00&  0.09 &   0.02&  0.16 &   0.02&  0.22 &   0.03&  0.31 &   0.04\\ 
      &      & 1.00 & 1.50&  0.14 &   0.02&  0.12 &   0.02&  0.23 &   0.02&  0.25 &   0.03\\ 
      &      & 1.50 & 2.00&  0.12 &   0.02& 0.112 &  0.013&  0.18 &   0.02&  0.23 &   0.02\\ 
      &      & 2.00 & 2.50& 0.095 &  0.014& 0.074 &  0.010& 0.120 &  0.013&  0.15 &   0.02\\ 
      &      & 2.50 & 3.00&       &       & 0.041 &  0.007& 0.077 &  0.008& 0.087 &  0.011\\ 
      &      & 3.00 & 3.50&       &       & 0.024 &  0.004& 0.042 &  0.006& 0.060 &  0.008\\ 
      &      & 3.50 & 4.00&       &       & 0.018 &  0.003& 0.029 &  0.004& 0.050 &  0.007\\ 
      &      & 4.00 & 5.00&       &       & 0.022 &  0.004& 0.017 &  0.003& 0.030 &  0.005\\ 
      &      & 5.00 & 6.50&       &       &       &       & 0.005 &  0.001& 0.009 &  0.002\\ 
      &      & 6.50 & 8.00&       &       &       &       &       &       & 0.003 &  0.001\\ 
 0.20 & 0.25 & 0.50 & 1.00&  0.08 &   0.02&  0.10 &   0.02&  0.15 &   0.02&  0.19 &   0.03\\ 
      &      & 1.00 & 1.50& 0.056 &  0.013& 0.046 &  0.009& 0.099 &  0.013&  0.12 &   0.02\\ 
      &      & 1.50 & 2.00&  0.07 &   0.02& 0.048 &  0.010& 0.076 &  0.014&  0.12 &   0.02\\ 
      &      & 2.00 & 2.50& 0.027 &  0.007& 0.040 &  0.008& 0.070 &  0.012& 0.075 &  0.015\\ 
      &      & 2.50 & 3.00&       &       & 0.024 &  0.006& 0.053 &  0.009& 0.069 &  0.013\\ 
      &      & 3.00 & 3.50&       &       & 0.014 &  0.003& 0.029 &  0.005& 0.052 &  0.011\\ 
      &      & 3.50 & 4.00&       &       & 0.009 &  0.003& 0.020 &  0.004& 0.023 &  0.006\\ 
      &      & 4.00 & 5.00&       &       & 0.005 &  0.002& 0.014 &  0.003& 0.011 &  0.004\\ 
      &      & 5.00 & 6.50&       &       &       &       & 0.003 &  0.001& 0.006 &  0.003\\ 
      &      & 6.50 & 8.00&       &       &       &       &       &       & 0.002 &  0.002\\ 
\hline
\end{tabular}
}
\end{table}

%% file: xsec_results_C_pip_abs.tex
\begin{table}[!ht]
  \caption{\label{tab:xsec_results_C3}
    HARP results for the double-differential $\pi^+$  production
    cross-section in the laboratory system,
    $d^2\sigma^{\pi}/(dpd\Omega)$, for $\pi^{+}$--C interactions at 3,5,8,12~\GeVc.
    Each row refers to a
    different $(p_{\hbox{\small min}} \le p<p_{\hbox{\small max}},
    \theta_{\hbox{\small min}} \le \theta<\theta_{\hbox{\small max}})$ bin,
    where $p$ and $\theta$ are the pion momentum and polar angle, respectively.
    The central value as well as the square-root of the diagonal elements
    of the covariance matrix are given.
}

\small{
\begin{tabular}{rrrr|r@{$\pm$}lr@{$\pm$}lr@{$\pm$}lr@{$\pm$}l}
\hline
$\theta_{\hbox{\small min}}$ &
$\theta_{\hbox{\small max}}$ &
$p_{\hbox{\small min}}$ &
$p_{\hbox{\small max}}$ &
\multicolumn{8}{c}{$d^2\sigma^{\pi^+}/(dpd\Omega)$}
\\
(rad) & (rad) & (\GeVc) & (\GeVc) &
\multicolumn{8}{c}{(barn/(\GeVc sr))}
\\
  &  &  &
&\multicolumn{2}{c}{$ \bf{3 \ \GeVc}$}
&\multicolumn{2}{c}{$ \bf{5 \ \GeVc}$}
&\multicolumn{2}{c}{$ \bf{8 \ \GeVc}$}
&\multicolumn{2}{c}{$ \bf{12 \ \GeVc}$}
\\
\hline
0.050 &0.100 & 0.50 & 1.00&  0.12 &   0.03&  0.14 &   0.02&  0.18 &   0.03&  0.25 &   0.08\\
      &      & 1.00 & 1.50& 0.065 &  0.012& 0.112 &  0.013&  0.21 &   0.02&  0.16 &   0.05\\
      &      & 1.50 & 2.00& 0.091 &  0.010& 0.137 &  0.013&  0.19 &   0.02& * &  * \\
      &      & 2.00 & 2.50&  0.25 &   0.02& 0.191 &  0.014&  0.19 &   0.02&  0.36 &   0.07\\
      &      & 2.50 & 3.00&       &       & 0.191 &  0.012&  0.20 &   0.02&  0.20 &   0.05\\
      &      & 3.00 & 3.50&       &       &  0.25 &   0.02&  0.19 &   0.02&  0.28 &   0.05\\
      &      & 3.50 & 4.00&       &       & 0.315 &  0.015&  0.14 &   0.02&  0.27 &   0.05\\
      &      & 4.00 & 5.00&       &       &  0.32 &   0.10& 0.196 &  0.014&  0.14 &   0.03\\
      &      & 5.00 & 6.50&       &       &       &       &  0.17 &   0.03&  0.12 &   0.02\\
      &      & 6.50 & 8.00&       &       &       &       &       &       &  0.09 &   0.02\\
0.100 &0.150 & 0.50 & 1.00&  0.08 &   0.02&  0.17 &   0.02&  0.18 &   0.03&  0.26 &   0.08\\
      &      & 1.00 & 1.50&  0.09 &   0.02&  0.15 &   0.02&  0.21 &   0.02&  0.18 &   0.05\\
      &      & 1.50 & 2.00&  0.18 &   0.02& 0.136 &  0.014&  0.25 &   0.03&  0.22 &   0.06\\
      &      & 2.00 & 2.50&  0.20 &   0.02& 0.160 &  0.015&  0.21 &   0.03&  0.19 &   0.06\\
      &      & 2.50 & 3.00&       &       & 0.138 &  0.013&  0.16 &   0.02&  0.16 &   0.06\\
      &      & 3.00 & 3.50&       &       & 0.138 &  0.013& 0.110 &  0.012&  0.15 &   0.04\\
      &      & 3.50 & 4.00&       &       & 0.115 &  0.010& 0.097 &  0.011&  0.15 &   0.04\\
      &      & 4.00 & 5.00&       &       &  0.11 &   0.04& 0.065 &  0.011&  0.07 &   0.02\\
      &      & 5.00 & 6.50&       &       &       &       & 0.022 &  0.005& 0.038 &  0.014\\
      &      & 6.50 & 8.00&       &       &       &       &       &       & 0.006 &  0.006\\
0.150 &0.200 & 0.50 & 1.00&  0.08 &   0.02&  0.16 &   0.02&  0.20 &   0.03&  0.09 &   0.05\\
      &      & 1.00 & 1.50&  0.10 &   0.02& 0.119 &  0.014&  0.18 &   0.02&  0.18 &   0.06\\
      &      & 1.50 & 2.00&  0.12 &   0.02& 0.105 &  0.011&  0.17 &   0.02&  0.13 &   0.05\\
      &      & 2.00 & 2.50&  0.09 &   0.02& 0.080 &  0.010&  0.13 &   0.02&  0.12 &   0.07\\
      &      & 2.50 & 3.00&       &       & 0.055 &  0.007& 0.080 &  0.013&  0.10 &   0.08\\
      &      & 3.00 & 3.50&       &       & 0.042 &  0.010& 0.076 &  0.012&  0.11 &   0.04\\
      &      & 3.50 & 4.00&       &       & 0.031 &  0.007& 0.043 &  0.008&  0.06 &   0.03\\
      &      & 4.00 & 5.00&       &       & 0.022 &  0.012& 0.025 &  0.005& 0.022 &  0.015\\
      &      & 5.00 & 6.50&       &       &       &       & 0.005 &  0.002& 0.015 &  0.012\\
      &      & 6.50 & 8.00&       &       &       &       &       &       & 0.002 &  0.004\\
0.200 &0.250 & 0.50 & 1.00&  0.05 &   0.02&  0.12 &   0.02&  0.19 &   0.03&  0.15 &   0.08\\
      &      & 1.00 & 1.50& 0.049 &  0.013& 0.070 &  0.012&  0.13 &   0.02&  0.05 &   0.04\\
      &      & 1.50 & 2.00& 0.043 &  0.011& 0.071 &  0.012&  0.13 &   0.02&  0.06 &   0.05\\
      &      & 2.00 & 2.50& 0.018 &  0.008& 0.053 &  0.009&  0.09 &   0.02&  0.06 &   0.05\\
      &      & 2.50 & 3.00&       &       & 0.034 &  0.007& 0.055 &  0.014&  0.05 &   0.05\\
      &      & 3.00 & 3.50&       &       & 0.016 &  0.004& 0.040 &  0.011& 0.016 &  0.016\\
      &      & 3.50 & 4.00&       &       & 0.012 &  0.005& 0.032 &  0.009& 0.010 &  0.030\\
      &      & 4.00 & 5.00&       &       & 0.008 &  0.004& 0.023 &  0.008& 0.025 &  0.050\\
      &      & 5.00 & 6.50&       &       &       &       & 0.007 &  0.004& 0.003 &  0.010\\
      &      & 6.50 & 8.00&       &       &       &       &       &       & 0.000 &  0.027\\

\hline
\end{tabular}
}
\end{table}
\clearpage
\begin{table}[!ht]
  \caption{\label{tab:xsec_results_C4}
    HARP results for the double-differential  $\pi^-$ production
    cross-section in the laboratory system,
    $d^2\sigma^{\pi}/(dpd\Omega)$, for $\pi^{+}$--C interactions at 3,5,8,12~\GeVc.
    Each row refers to a
    different $(p_{\hbox{\small min}} \le p<p_{\hbox{\small max}},
    \theta_{\hbox{\small min}} \le \theta<\theta_{\hbox{\small max}})$ bin,
    where $p$ and $\theta$ are the pion momentum and polar angle, respectively.
    The central value as well as the square-root of the diagonal elements
    of the covariance matrix are given.}

\small{
\begin{tabular}{rrrr|r@{$\pm$}lr@{$\pm$}lr@{$\pm$}lr@{$\pm$}l}
\hline
$\theta_{\hbox{\small min}}$ &
$\theta_{\hbox{\small max}}$ &
$p_{\hbox{\small min}}$ &
$p_{\hbox{\small max}}$ &
\multicolumn{8}{c}{$d^2\sigma^{\pi^-}/(dpd\Omega)$}
\\
(rad) & (rad) & (\GeVc) & (\GeVc) &
\multicolumn{8}{c}{(barn/(\GeVc sr))}
\\
  &  &  &
&\multicolumn{2}{c}{$ \bf{3 \ \GeVc}$}
&\multicolumn{2}{c}{$ \bf{5 \ \GeVc}$}
&\multicolumn{2}{c}{$ \bf{8 \ \GeVc}$}
&\multicolumn{2}{c}{$ \bf{12 \ \GeVc}$}
\\
\hline
0.050 &0.100 & 0.50 & 1.00& 0.031 &  0.011& 0.082 &  0.013&  0.14 &   0.02&  0.11 &   0.05\\
      &      & 1.00 & 1.50& 0.052 &  0.012& 0.111 &  0.012&  0.15 &   0.02&  0.18 &   0.06\\
      &      & 1.50 & 2.00& 0.041 &  0.008& 0.073 &  0.008&  0.15 &   0.02&  0.21 &   0.06\\
      &      & 2.00 & 2.50&  0.06 &   0.02& 0.067 &  0.007& 0.131 &  0.015&  0.11 &   0.03\\
      &      & 2.50 & 3.00&       &       & 0.064 &  0.007& 0.113 &  0.013&  0.16 &   0.05\\
      &      & 3.00 & 3.50&       &       & 0.067 &  0.008& 0.081 &  0.009&  0.20 &   0.04\\
      &      & 3.50 & 4.00&       &       & 0.062 &  0.006& 0.104 &  0.011&  0.08 &   0.02\\
      &      & 4.00 & 5.00&       &       & 0.030 &  0.006& 0.050 &  0.005&  0.09 &   0.02\\
      &      & 5.00 & 6.50&       &       &       &       & 0.026 &  0.003& 0.059 &  0.015\\
      &      & 6.50 & 8.00&       &       &       &       &       &       & 0.021 &  0.008\\
0.100 &0.150 & 0.50 & 1.00&  0.07 &   0.02&  0.16 &   0.02&  0.21 &   0.03&  0.25 &   0.08\\
      &      & 1.00 & 1.50& 0.041 &  0.010& 0.082 &  0.011&  0.15 &   0.02&  0.11 &   0.04\\
      &      & 1.50 & 2.00& 0.040 &  0.009& 0.061 &  0.008& 0.117 &  0.014&  0.15 &   0.05\\
      &      & 2.00 & 2.50& 0.036 &  0.009& 0.056 &  0.007& 0.120 &  0.015&  0.14 &   0.04\\
      &      & 2.50 & 3.00&       &       & 0.034 &  0.004& 0.056 &  0.008&  0.12 &   0.04\\
      &      & 3.00 & 3.50&       &       & 0.033 &  0.004& 0.060 &  0.009&  0.11 &   0.03\\
      &      & 3.50 & 4.00&       &       & 0.013 &  0.002& 0.037 &  0.006&  0.09 &   0.03\\
      &      & 4.00 & 5.00&       &       & 0.003 &  0.001& 0.022 &  0.004&  0.04 &   0.02\\
      &      & 5.00 & 6.50&       &       &       &       & 0.006 &  0.001& 0.013 &  0.008\\
      &      & 6.50 & 8.00&       &       &       &       &       &       & 0.005 &  0.004\\
0.150 &0.200 & 0.50 & 1.00&  0.07 &   0.02&  0.11 &   0.02&  0.20 &   0.03&  0.27 &   0.09\\
      &      & 1.00 & 1.50& 0.035 &  0.009& 0.073 &  0.010&  0.13 &   0.02&  0.20 &   0.06\\
      &      & 1.50 & 2.00& 0.021 &  0.006& 0.091 &  0.011&  0.12 &   0.02&  0.18 &   0.06\\
      &      & 2.00 & 2.50& 0.015 &  0.006& 0.046 &  0.006& 0.095 &  0.013&  0.12 &   0.04\\
      &      & 2.50 & 3.00&       &       & 0.023 &  0.004& 0.048 &  0.008&  0.07 &   0.03\\
      &      & 3.00 & 3.50&       &       & 0.007 &  0.002& 0.027 &  0.005& 0.032 &  0.035\\
      &      & 3.50 & 4.00&       &       & 0.004 &  0.001& 0.017 &  0.004& 0.005 &  0.009\\
      &      & 4.00 & 5.00&       &       &  &   & 0.005 &  0.002& 0.007 &  0.016\\
      &      & 5.00 & 6.50&       &       &       &       & 0.001 &  0.001& 0.003 &  0.009\\
      &      & 6.50 & 8.00&       &       &       &       &       &       & 0.000 &  0.001\\
0.200 &0.250 & 0.50 & 1.00&  0.07 &   0.02&  0.12 &   0.02&  0.15 &   0.03&  0.14 &   0.06\\
      &      & 1.00 & 1.50& 0.025 &  0.009& 0.080 &  0.013&  0.16 &   0.03&  0.15 &   0.07\\
      &      & 1.50 & 2.00& 0.005 &  0.003& 0.031 &  0.008&  0.13 &   0.02&  0.07 &   0.05\\
      &      & 2.00 & 2.50& 0.008 &  0.005& 0.015 &  0.004& 0.052 &  0.012&  0.08 &   0.06\\
      &      & 2.50 & 3.00&       &       & 0.005 &  0.002& 0.020 &  0.006&  0.08 &   0.06\\
      &      & 3.00 & 3.50&       &       & 0.002 &  0.001& 0.011 &  0.004& 0.051 &  0.059\\
      &      & 3.50 & 4.00&       &       & 0.000 &  0.000& 0.006 &  0.003& 0.021 &  0.038\\
      &      & 4.00 & 5.00&       &       &  &  & 0.002 &  0.003& 0.003 &  0.019\\
      &      & 5.00 & 6.50&       &       &       &       & 0.000 &  0.001& 0.002 &  0.015\\
      &      & 6.50 & 8.00&       &       &       &       &       &       & 0.001 &  0.018\\

\hline
\end{tabular}
}
\end{table}

%% file: xsec_results_Al_pim.tex
\begin{table}[!ht]
  \caption{\label{tab:xsec_results_Al_pim1}
    HARP results for the double-differential $\pi^+$  production
    cross-section in the laboratory system,
    $d^2\sigma^{\pi}/(dpd\Omega)$, for $\pi^{-}$--Al interactions at 3,5,8,12~\GeVc.
    Each row refers to a
    different $(p_{\hbox{\small min}} \le p<p_{\hbox{\small max}},
    \theta_{\hbox{\small min}} \le \theta<\theta_{\hbox{\small max}})$ bin,
    where $p$ and $\theta$ are the pion momentum and polar angle, respectively.
    The central value as well as the square-root of the diagonal elements
    of the covariance matrix are given.}

\small{
\begin{tabular}{rrrr|r@{$\pm$}lr@{$\pm$}lr@{$\pm$}lr@{$\pm$}l}
\hline
$\theta_{\hbox{\small min}}$ &
$\theta_{\hbox{\small max}}$ &
$p_{\hbox{\small min}}$ &
$p_{\hbox{\small max}}$ &
\multicolumn{8}{c}{$d^2\sigma^{\pi^+}/(dpd\Omega)$}
\\
(rad) & (rad) & (\GeVc) & (\GeVc) &
\multicolumn{8}{c}{(barn/(\GeVc sr))}
\\
  &  &  &
&\multicolumn{2}{c}{$ \bf{3 \ \GeVc}$}
&\multicolumn{2}{c}{$ \bf{5 \ \GeVc}$}
&\multicolumn{2}{c}{$ \bf{8 \ \GeVc}$}
&\multicolumn{2}{c}{$ \bf{12 \ \GeVc}$}
\\
\hline
0.05 & 0.10 & 0.50 & 1.00& 0.066 &  0.015&  0.14 &   0.03&  0.29 &   0.03&  0.32 &   0.04\\ 
      &      & 1.00 & 1.50& 0.086 &  0.015&  0.18 &   0.02&  0.36 &   0.03&  0.44 &   0.03\\ 
      &      & 1.50 & 2.00& 0.063 &  0.012&  0.13 &   0.02&  0.30 &   0.02&  0.37 &   0.03\\ 
      &      & 2.00 & 2.50& 0.090 &  0.015&  0.12 &   0.02&  0.22 &   0.02&  0.32 &   0.02\\ 
      &      & 2.50 & 3.00&       &       & 0.092 &  0.015&  0.14 &   0.02&  0.29 &   0.02\\ 
      &      & 3.00 & 3.50&       &       & 0.078 &  0.014&  0.17 &   0.02&  0.29 &   0.02\\ 
      &      & 3.50 & 4.00&       &       & 0.080 &  0.014&  0.13 &   0.02&  0.22 &   0.02\\ 
      &      & 4.00 & 5.00&       &       & 0.038 &  0.011& 0.083 &  0.009& 0.138 &  0.011\\ 
      &      & 5.00 & 6.50&       &       &       &       & 0.039 &  0.005& 0.071 &  0.006\\ 
      &      & 6.50 & 8.00&       &       &       &       &       &       & 0.030 &  0.004\\ 
 0.10 & 0.15 & 0.50 & 1.00&  0.12 &   0.02&  0.22 &   0.03&  0.32 &   0.04&  0.34 &   0.04\\ 
      &      & 1.00 & 1.50& 0.052 &  0.011&  0.17 &   0.02&  0.29 &   0.03&  0.44 &   0.04\\ 
      &      & 1.50 & 2.00& 0.047 &  0.010&  0.10 &   0.02&  0.24 &   0.02&  0.35 &   0.03\\ 
      &      & 2.00 & 2.50& 0.047 &  0.011& 0.074 &  0.014&  0.23 &   0.02&  0.33 &   0.03\\ 
      &      & 2.50 & 3.00&       &       &  0.08 &   0.02&  0.14 &   0.02&  0.25 &   0.02\\ 
      &      & 3.00 & 3.50&       &       & 0.048 &  0.013& 0.094 &  0.013&  0.21 &   0.02\\ 
      &      & 3.50 & 4.00&       &       & 0.011 &  0.005& 0.079 &  0.011& 0.150 &  0.015\\ 
      &      & 4.00 & 5.00&       &       & 0.002 &  0.002& 0.029 &  0.006& 0.084 &  0.011\\ 
      &      & 5.00 & 6.50&       &       &       &       & 0.006 &  0.002& 0.021 &  0.004\\ 
      &      & 6.50 & 8.00&       &       &       &       &       &       & 0.004 &  0.001\\ 
 0.15 & 0.20 & 0.50 & 1.00&  0.10 &   0.02&  0.17 &   0.03&  0.34 &   0.04&  0.44 &   0.05\\ 
      &      & 1.00 & 1.50& 0.082 &  0.014&  0.10 &   0.02&  0.25 &   0.03&  0.32 &   0.03\\ 
      &      & 1.50 & 2.00& 0.043 &  0.009&  0.10 &   0.02&  0.20 &   0.02&  0.27 &   0.03\\ 
      &      & 2.00 & 2.50& 0.024 &  0.008&  0.08 &   0.02&  0.16 &   0.02&  0.23 &   0.03\\ 
      &      & 2.50 & 3.00&       &       & 0.034 &  0.010& 0.094 &  0.013& 0.121 &  0.015\\ 
      &      & 3.00 & 3.50&       &       & 0.014 &  0.005& 0.055 &  0.010& 0.093 &  0.014\\ 
      &      & 3.50 & 4.00&       &       & 0.007 &  0.004& 0.028 &  0.006& 0.036 &  0.007\\ 
      &      & 4.00 & 5.00&       &       & 0.001 &  0.001& 0.010 &  0.003& 0.028 &  0.005\\ 
      &      & 5.00 & 6.50&       &       &       &       & 0.002 &  0.001& 0.004 &  0.001\\ 
      &      & 6.50 & 8.00&       &       &       &       &       &       & 0.001 &  0.000\\ 
 0.20 & 0.25 & 0.50 & 1.00&  0.10 &   0.02&  0.19 &   0.04&  0.29 &   0.04&  0.33 &   0.05\\ 
      &      & 1.00 & 1.50&  0.10 &   0.02&  0.19 &   0.04&  0.38 &   0.05&  0.36 &   0.05\\ 
      &      & 1.50 & 2.00& 0.039 &  0.012&  0.12 &   0.03&  0.27 &   0.04&  0.25 &   0.04\\ 
      &      & 2.00 & 2.50& 0.011 &  0.006&  0.06 &   0.02&  0.13 &   0.02&  0.18 &   0.03\\ 
      &      & 2.50 & 3.00&       &       & 0.019 &  0.008& 0.078 &  0.015&  0.08 &   0.02\\ 
      &      & 3.00 & 3.50&       &       & 0.009 &  0.005& 0.024 &  0.009& 0.038 &  0.009\\ 
      &      & 3.50 & 4.00&       &       & 0.003 &  0.002& 0.008 &  0.004& 0.017 &  0.006\\ 
      &      & 4.00 & 5.00&       &       & 0.002 &  0.002& 0.006 &  0.003& 0.007 &  0.004\\ 
      &      & 5.00 & 6.50&       &       &       &       & 0.001 &  0.001& 0.001 &  0.001\\ 
      &      & 6.50 & 8.00&       &       &       &       &       &       & 0.000 &  0.004\\ 

\hline
\end{tabular}
}
\end{table}
\clearpage
\begin{table}[!ht]
  \caption{\label{tab:xsec_results_Al_pim2}
    HARP results for the double-differential  $\pi^-$ production
    cross-section in the laboratory system,
    $d^2\sigma^{\pi}/(dpd\Omega)$, for $\pi^{-}$--Al interactions at 3,5,8,12~\GeVc.
    Each row refers to a
    different $(p_{\hbox{\small min}} \le p<p_{\hbox{\small max}},
    \theta_{\hbox{\small min}} \le \theta<\theta_{\hbox{\small max}})$ bin,
    where $p$ and $\theta$ are the pion momentum and polar angle, respectively.
    The central value as well as the square-root of the diagonal elements
    of the covariance matrix are given.}

\small{
\begin{tabular}{rrrr|r@{$\pm$}lr@{$\pm$}lr@{$\pm$}lr@{$\pm$}l}
\hline
$\theta_{\hbox{\small min}}$ &
$\theta_{\hbox{\small max}}$ &
$p_{\hbox{\small min}}$ &
$p_{\hbox{\small max}}$ &
\multicolumn{8}{c}{$d^2\sigma^{\pi^-}/(dpd\Omega)$}
\\
(rad) & (rad) & (\GeVc) & (\GeVc) &
\multicolumn{8}{c}{(barn/(\GeVc sr))}
\\
  &  &  &
&\multicolumn{2}{c}{$ \bf{3 \ \GeVc}$}
&\multicolumn{2}{c}{$ \bf{5 \ \GeVc}$}
&\multicolumn{2}{c}{$ \bf{8 \ \GeVc}$}
&\multicolumn{2}{c}{$ \bf{12 \ \GeVc}$}
\\
\hline

 0.05 & 0.10 & 0.50 & 1.00&  0.16 &   0.02&  0.27 &   0.04&  0.50 &   0.05&  0.49 &   0.05\\ 
      &      & 1.00 & 1.50&  0.14 &   0.02&  0.23 &   0.03&  0.41 &   0.03&  0.46 &   0.04\\ 
      &      & 1.50 & 2.00&  0.21 &   0.02&  0.25 &   0.03&  0.43 &   0.03&  0.52 &   0.03\\ 
      &      & 2.00 & 2.50&  0.21 &   0.05&  0.28 &   0.03&  0.42 &   0.03&  0.47 &   0.03\\ 
      &      & 2.50 & 3.00&       &       &  0.27 &   0.03&  0.39 &   0.02&  0.46 &   0.03\\ 
      &      & 3.00 & 3.50&       &       &  0.31 &   0.03&  0.30 &   0.02&  0.48 &   0.03\\ 
      &      & 3.50 & 4.00&       &       &  0.36 &   0.03&  0.28 &   0.02&  0.42 &   0.02\\ 
      &      & 4.00 & 5.00&       &       &  0.39 &   0.06&  0.27 &   0.02&  0.26 &   0.02\\ 
      &      & 5.00 & 6.50&       &       &       &       & 0.204 &  0.013& 0.199 &  0.010\\ 
      &      & 6.50 & 8.00&       &       &       &       &       &       & 0.136 &  0.009\\ 
 0.10 & 0.15 & 0.50 & 1.00&  0.18 &   0.03&  0.30 &   0.04&  0.49 &   0.05&  0.62 &   0.07\\ 
      &      & 1.00 & 1.50&  0.17 &   0.02&  0.23 &   0.03&  0.42 &   0.04&  0.59 &   0.05\\ 
      &      & 1.50 & 2.00&  0.22 &   0.03&  0.19 &   0.02&  0.39 &   0.03&  0.47 &   0.04\\ 
      &      & 2.00 & 2.50&  0.23 &   0.04&  0.22 &   0.03&  0.36 &   0.03&  0.46 &   0.04\\ 
      &      & 2.50 & 3.00&       &       &  0.21 &   0.03&  0.30 &   0.03&  0.35 &   0.03\\ 
      &      & 3.00 & 3.50&       &       &  0.17 &   0.02&  0.18 &   0.02&  0.26 &   0.02\\ 
      &      & 3.50 & 4.00&       &       &  0.15 &   0.02&  0.12 &   0.02&  0.20 &   0.02\\ 
      &      & 4.00 & 5.00&       &       &  0.14 &   0.02& 0.068 &  0.009& 0.140 &  0.013\\ 
      &      & 5.00 & 6.50&       &       &       &       & 0.040 &  0.005& 0.070 &  0.007\\ 
      &      & 6.50 & 8.00&       &       &       &       &       &       & 0.025 &  0.004\\ 
 0.15 & 0.20 & 0.50 & 1.00&  0.19 &   0.03&  0.26 &   0.04&  0.46 &   0.05&  0.55 &   0.06\\ 
      &      & 1.00 & 1.50&  0.15 &   0.02&  0.22 &   0.03&  0.42 &   0.04&  0.42 &   0.04\\ 
      &      & 1.50 & 2.00&  0.14 &   0.02&  0.14 &   0.02&  0.29 &   0.03&  0.35 &   0.03\\ 
      &      & 2.00 & 2.50&  0.12 &   0.02&  0.12 &   0.02&  0.21 &   0.02&  0.25 &   0.03\\ 
      &      & 2.50 & 3.00&       &       &  0.09 &   0.02&  0.12 &   0.02&  0.19 &   0.02\\ 
      &      & 3.00 & 3.50&       &       & 0.031 &  0.008& 0.079 &  0.011& 0.131 &  0.015\\ 
      &      & 3.50 & 4.00&       &       & 0.035 &  0.008& 0.053 &  0.010& 0.081 &  0.010\\ 
      &      & 4.00 & 5.00&       &       & 0.029 &  0.007& 0.025 &  0.004& 0.052 &  0.007\\ 
      &      & 5.00 & 6.50&       &       &       &       & 0.010 &  0.002& 0.018 &  0.003\\ 
      &      & 6.50 & 8.00&       &       &       &       &       &       & 0.004 &  0.001\\ 
 0.20 & 0.25 & 0.50 & 1.00&  0.10 &   0.02&  0.21 &   0.04&  0.30 &   0.04&  0.32 &   0.04\\ 
      &      & 1.00 & 1.50& 0.069 &  0.015&  0.14 &   0.03&  0.18 &   0.03&  0.18 &   0.03\\ 
      &      & 1.50 & 2.00&  0.08 &   0.02&  0.10 &   0.02&  0.14 &   0.03&  0.16 &   0.03\\ 
      &      & 2.00 & 2.50& 0.036 &  0.008&  0.07 &   0.02&  0.14 &   0.02&  0.15 &   0.03\\ 
      &      & 2.50 & 3.00&       &       &  0.06 &   0.02&  0.09 &   0.02&  0.09 &   0.02\\ 
      &      & 3.00 & 3.50&       &       & 0.025 &  0.007& 0.056 &  0.010& 0.067 &  0.013\\ 
      &      & 3.50 & 4.00&       &       & 0.022 &  0.007& 0.041 &  0.008& 0.046 &  0.010\\ 
      &      & 4.00 & 5.00&       &       & 0.011 &  0.006& 0.024 &  0.006& 0.023 &  0.006\\ 
      &      & 5.00 & 6.50&       &       &       &       & 0.009 &  0.004& 0.007 &  0.002\\ 
      &      & 6.50 & 8.00&       &       &       &       &       &       & 0.002 &  0.002\\

\hline
\end{tabular}
}
\end{table}

%% file: xsec_results_Al_pip_abs.tex
\begin{table}[!ht]
  \caption{\label{tab:xsec_results_Al_pip1}
    HARP results for the double-differential $\pi^+$  production
    cross-section in the laboratory system,
    $d^2\sigma^{\pi}/(dpd\Omega)$, for $\pi^{+}$--Al interactions at 3,5,8,12~\GeVc.
    Each row refers to a
    different $(p_{\hbox{\small min}} \le p<p_{\hbox{\small max}},
    \theta_{\hbox{\small min}} \le \theta<\theta_{\hbox{\small max}})$ bin,
    where $p$ and $\theta$ are the pion momentum and polar angle, respectively.
    The central value as well as the square-root of the diagonal elements
    of the covariance matrix are given.}

\small{
\begin{tabular}{rrrr|r@{$\pm$}lr@{$\pm$}lr@{$\pm$}lr@{$\pm$}l}
\hline
$\theta_{\hbox{\small min}}$ &
$\theta_{\hbox{\small max}}$ &
$p_{\hbox{\small min}}$ &
$p_{\hbox{\small max}}$ &
\multicolumn{8}{c}{$d^2\sigma^{\pi^+}/(dpd\Omega)$}
\\
(rad) & (rad) & (\GeVc) & (\GeVc) &
\multicolumn{8}{c}{(barn/(\GeVc sr))}
\\
  &  &  &
&\multicolumn{2}{c}{$ \bf{3 \ \GeVc}$}
&\multicolumn{2}{c}{$ \bf{5 \ \GeVc}$}
&\multicolumn{2}{c}{$ \bf{8 \ \GeVc}$}
&\multicolumn{2}{c}{$ \bf{12 \ \GeVc}$}
\\
\hline
0.050 &0.100 & 0.50 & 1.00&  0.14 &   0.04&  0.23 &   0.03&  0.30 &   0.05&  0.60 &   0.21\\
      &      & 1.00 & 1.50&  0.09 &   0.02&  0.22 &   0.03&  0.33 &   0.04&  0.47 &   0.14\\
      &      & 1.50 & 2.00& 0.118 &  0.014&  0.21 &   0.02&  0.32 &   0.03&  0.46 &   0.13\\
      &      & 2.00 & 2.50&  0.32 &   0.03&  0.24 &   0.02&  0.35 &   0.03&  0.59 &   0.18\\
      &      & 2.50 & 3.00&       &       &  0.29 &   0.02&  0.33 &   0.03&  0.36 &   0.21\\
      &      & 3.00 & 3.50&       &       &  0.38 &   0.04&  0.35 &   0.03&  0.39 &   0.09\\
      &      & 3.50 & 4.00&       &       &  0.40 &   0.02&  0.30 &   0.03&  0.52 &   0.12\\
      &      & 4.00 & 5.00&       &       &  0.43 &   0.14&  0.36 &   0.02&  0.29 &   0.06\\
      &      & 5.00 & 6.50&       &       &       &       &  0.26 &   0.04&  0.20 &   0.04\\
      &      & 6.50 & 8.00&       &       &       &       &       &       &  0.16 &   0.04\\
0.100 &0.150 & 0.50 & 1.00&  0.10 &   0.03&  0.27 &   0.04&  0.31 &   0.04&  0.53 &   0.18\\
      &      & 1.00 & 1.50&  0.19 &   0.03&  0.24 &   0.03&  0.33 &   0.04&  0.50 &   0.15\\
      &      & 1.50 & 2.00&  0.22 &   0.03&  0.22 &   0.03&  0.34 &   0.04&  0.40 &   0.12\\
      &      & 2.00 & 2.50&  0.31 &   0.04&  0.23 &   0.02&  0.35 &   0.04&  0.43 &   0.14\\
      &      & 2.50 & 3.00&       &       &  0.25 &   0.03&  0.24 &   0.03&  0.30 &   0.12\\
      &      & 3.00 & 3.50&       &       &  0.19 &   0.02&  0.20 &   0.02&  0.38 &   0.10\\
      &      & 3.50 & 4.00&       &       &  0.16 &   0.02&  0.14 &   0.02&  0.22 &   0.06\\
      &      & 4.00 & 5.00&       &       &  0.14 &   0.05&  0.11 &   0.02&  0.14 &   0.04\\
      &      & 5.00 & 6.50&       &       &       &       & 0.037 &  0.007&  0.07 &   0.03\\
      &      & 6.50 & 8.00&       &       &       &       &       &       & 0.023 &  0.014\\
0.150 &0.200 & 0.50 & 1.00&  0.17 &   0.04&  0.30 &   0.04&  0.47 &   0.06&  0.29 &   0.13\\
      &      & 1.00 & 1.50&  0.13 &   0.02&  0.21 &   0.03&  0.35 &   0.04&  0.50 &   0.16\\
      &      & 1.50 & 2.00&  0.14 &   0.02&  0.15 &   0.02&  0.26 &   0.03&  0.15 &   0.07\\
      &      & 2.00 & 2.50&  0.17 &   0.03&  0.13 &   0.02&  0.20 &   0.03&  0.24 &   0.11\\
      &      & 2.50 & 3.00&       &       & 0.069 &  0.011&  0.12 &   0.02&  0.25 &   0.10\\
      &      & 3.00 & 3.50&       &       &  0.06 &   0.02&  0.10 &   0.02&  0.11 &   0.06\\
      &      & 3.50 & 4.00&       &       & 0.055 &  0.013&  0.08 &   0.02&  0.02 &   0.02\\
      &      & 4.00 & 5.00&       &       &  0.04 &   0.02& 0.043 &  0.009&  0.03 &   0.03\\
      &      & 5.00 & 6.50&       &       &       &       & 0.014 &  0.005&  0.05 &   0.04\\
      &      & 6.50 & 8.00&       &       &       &       &       &       & 0.005 &  0.009\\
0.200 &0.250 & 0.50 & 1.00&  0.10 &   0.04&  0.22 &   0.03&  0.32 &   0.05&  0.35 &   0.18\\
      &      & 1.00 & 1.50&  0.06 &   0.02&  0.16 &   0.03&  0.17 &   0.03&  0.13 &   0.09\\
      &      & 1.50 & 2.00& 0.042 &  0.014&  0.10 &   0.02&  0.17 &   0.03&  0.17 &   0.19\\
      &      & 2.00 & 2.50& 0.031 &  0.015&  0.11 &   0.02&  0.16 &   0.03& 0.042 &  0.048\\
      &      & 2.50 & 3.00&       &       &  0.08 &   0.02&  0.11 &   0.03& 0.046 &  0.101\\
      &      & 3.00 & 3.50&       &       & 0.034 &  0.010&  0.07 &   0.02& 0.015 &  0.027\\
      &      & 3.50 & 4.00&       &       & 0.018 &  0.007& 0.035 &  0.012& 0.014 &  0.074\\
      &      & 4.00 & 5.00&       &       & 0.014 &  0.007& 0.021 &  0.008& 0.014 &  0.098\\
      &      & 5.00 & 6.50&       &       &       &       & 0.007 &  0.004& 0.007 &  0.038\\
      &      & 6.50 & 8.00&       &       &       &       &       &       & 0.003 &  0.077\\

\hline
\end{tabular}
}
\end{table}
\clearpage
\begin{table}[!ht]
  \caption{\label{tab:xsec_results_Al_pip2}
    HARP results for the double-differential  $\pi^-$ production
    cross-section in the laboratory system,
    $d^2\sigma^{\pi}/(dpd\Omega)$, for $\pi^{+}$--Al interactions at 3,5,8,12~\GeVc.
    Each row refers to a
    different $(p_{\hbox{\small min}} \le p<p_{\hbox{\small max}},
    \theta_{\hbox{\small min}} \le \theta<\theta_{\hbox{\small max}})$ bin,
    where $p$ and $\theta$ are the pion momentum and polar angle, respectively.
    The central value as well as the square-root of the diagonal elements
    of the covariance matrix are given.}

\small{
\begin{tabular}{rrrr|r@{$\pm$}lr@{$\pm$}lr@{$\pm$}lr@{$\pm$}l}
\hline
$\theta_{\hbox{\small min}}$ &
$\theta_{\hbox{\small max}}$ &
$p_{\hbox{\small min}}$ &
$p_{\hbox{\small max}}$ &
\multicolumn{8}{c}{$d^2\sigma^{\pi^-}/(dpd\Omega)$}
\\
(rad) & (rad) & (\GeVc) & (\GeVc) &
\multicolumn{8}{c}{(barn/(\GeVc sr))}
\\
  &  &  &
&\multicolumn{2}{c}{$ \bf{3 \ \GeVc}$}
&\multicolumn{2}{c}{$ \bf{5 \ \GeVc}$}
&\multicolumn{2}{c}{$ \bf{8 \ \GeVc}$}
&\multicolumn{2}{c}{$ \bf{12 \ \GeVc}$}
\\
\hline
0.050 &0.100 & 0.50 & 1.00&  0.05 &   0.02&  0.16 &   0.03&  0.25 &   0.04&  0.19 &   0.12\\
      &      & 1.00 & 1.50&  0.07 &   0.02&  0.16 &   0.02&  0.27 &   0.03&  0.38 &   0.13\\
      &      & 1.50 & 2.00&  0.09 &   0.02&  0.12 &   0.02&  0.25 &   0.03&  0.31 &   0.10\\
      &      & 2.00 & 2.50&  0.09 &   0.03& 0.120 &  0.014&  0.26 &   0.03&  0.27 &   0.09\\
      &      & 2.50 & 3.00&       &       & 0.107 &  0.013&  0.15 &   0.02&  0.39 &   0.11\\
      &      & 3.00 & 3.50&       &       & 0.102 &  0.014&  0.16 &   0.02&  0.26 &   0.08\\
      &      & 3.50 & 4.00&       &       & 0.095 &  0.011&  0.13 &   0.02&  0.25 &   0.08\\
      &      & 4.00 & 5.00&       &       & 0.050 &  0.010& 0.081 &  0.009&  0.13 &   0.04\\
      &      & 5.00 & 6.50&       &       &       &       & 0.048 &  0.006&  0.06 &   0.02\\
      &      & 6.50 & 8.00&       &       &       &       &       &       & 0.024 &  0.013\\
0.100 &0.150 & 0.50 & 1.00&  0.13 &   0.03&  0.24 &   0.04&  0.35 &   0.05&  0.28 &   0.12\\
      &      & 1.00 & 1.50&  0.06 &   0.02&  0.17 &   0.02&  0.31 &   0.04&  0.49 &   0.14\\
      &      & 1.50 & 2.00&  0.08 &   0.02&  0.11 &   0.02&  0.26 &   0.03&  0.31 &   0.10\\
      &      & 2.00 & 2.50& 0.025 &  0.009& 0.090 &  0.012&  0.17 &   0.02&  0.25 &   0.09\\
      &      & 2.50 & 3.00&       &       & 0.084 &  0.012&  0.14 &   0.02&  0.20 &   0.08\\
      &      & 3.00 & 3.50&       &       & 0.047 &  0.007& 0.109 &  0.015&  0.27 &   0.09\\
      &      & 3.50 & 4.00&       &       & 0.023 &  0.005& 0.077 &  0.012&  0.12 &   0.06\\
      &      & 4.00 & 5.00&       &       & 0.013 &  0.004& 0.038 &  0.007&  0.03 &   0.02\\
      &      & 5.00 & 6.50&       &       &       &       & 0.006 &  0.002&  0.02 &   0.02\\
      &      & 6.50 & 8.00&       &       &       &       &       &       & 0.004 &  0.005\\
0.150 &0.200 & 0.50 & 1.00&  0.15 &   0.04&  0.21 &   0.03&  0.38 &   0.06&  0.61 &   0.22\\
      &      & 1.00 & 1.50&  0.06 &   0.02&  0.12 &   0.02&  0.26 &   0.03&  0.33 &   0.12\\
      &      & 1.50 & 2.00& 0.016 &  0.007& 0.087 &  0.014&  0.19 &   0.03&  0.34 &   0.12\\
      &      & 2.00 & 2.50& 0.012 &  0.007& 0.065 &  0.011&  0.18 &   0.02&  0.23 &   0.10\\
      &      & 2.50 & 3.00&       &       & 0.037 &  0.008& 0.085 &  0.015&  0.09 &   0.05\\
      &      & 3.00 & 3.50&       &       & 0.022 &  0.006& 0.043 &  0.009&  0.14 &   0.10\\
      &      & 3.50 & 4.00&       &       & 0.005 &  0.002& 0.020 &  0.005&  0.02 &   0.02\\
      &      & 4.00 & 5.00&       &       & 0.004 &  0.002& 0.011 &  0.004&  0.03 &   0.03\\
      &      & 5.00 & 6.50&       &       &       &       & 0.002 &  0.001& * & *\\
      &      & 6.50 & 8.00&       &       &       &       &       &       & 0.001 &  0.005\\
0.200 &0.250 & 0.50 & 1.00&  0.06 &   0.02&  0.23 &   0.04&  0.31 &   0.05&  0.15 &   0.09\\
      &      & 1.00 & 1.50&  0.11 &   0.03&  0.11 &   0.02&  0.23 &   0.04&  0.27 &   0.15\\
      &      & 1.50 & 2.00& 0.018 &  0.010&  0.06 &   0.02&  0.24 &   0.04&  0.30 &   0.19\\
      &      & 2.00 & 2.50& 0.007 &  0.006& 0.018 &  0.005&  0.13 &   0.03& 0.257 &  0.275\\
      &      & 2.50 & 3.00&       &       & 0.010 &  0.004&  0.07 &   0.02&  0.10 &   0.19\\
      &      & 3.00 & 3.50&       &       & 0.003 &  0.002& 0.027 &  0.010& 0.016 &  0.075\\
      &      & 3.50 & 4.00&       &       & 0.002 &  0.002& 0.009 &  0.005& 0.000 &  0.014\\
      &      & 4.00 & 5.00&       &       &  &  & 0.003 &  0.005& 0.000 &  0.016\\
      &      & 5.00 & 6.50&       &       &       &       & 0.002 &  0.003& 0.000 &  0.003\\
      &      & 6.50 & 8.00&       &       &       &       &       &       &0.000 &  0.037\\

\hline
\end{tabular}
}
\end{table}

%% file: xsec_results_Cu_pim.tex
\begin{table}[!ht]
  \caption{\label{tab:xsec_results_Cu1}
    HARP results for the double-differential $\pi^+$  production
    cross-section in the laboratory system,
    $d^2\sigma^{\pi}/(dpd\Omega)$, for $\pi^{-}$--Cu interactions at 3,5,8,12~\GeVc.
    Each row refers to a
    different $(p_{\hbox{\small min}} \le p<p_{\hbox{\small max}},
    \theta_{\hbox{\small min}} \le \theta<\theta_{\hbox{\small max}})$ bin,
    where $p$ and $\theta$ are the pion momentum and polar angle, respectively.
    The central value as well as the square-root of the diagonal elements
    of the covariance matrix are given.}

\small{
\begin{tabular}{rrrr|r@{$\pm$}lr@{$\pm$}lr@{$\pm$}lr@{$\pm$}l}
\hline
$\theta_{\hbox{\small min}}$ &
$\theta_{\hbox{\small max}}$ &
$p_{\hbox{\small min}}$ &
$p_{\hbox{\small max}}$ &
\multicolumn{8}{c}{$d^2\sigma^{\pi^+}/(dpd\Omega)$}
\\
(rad) & (rad) & (\GeVc) & (\GeVc) &
\multicolumn{8}{c}{(barn/(\GeVc sr))}
\\
  &  &  &
&\multicolumn{2}{c}{$ \bf{3 \ \GeVc}$}
&\multicolumn{2}{c}{$ \bf{5 \ \GeVc}$}
&\multicolumn{2}{c}{$ \bf{8 \ \GeVc}$}
&\multicolumn{2}{c}{$ \bf{12 \ \GeVc}$}
\\
\hline
 
0.05 & 0.10 & 0.50 & 1.00&  0.11 &   0.02&  0.24 &   0.03&  0.63 &   0.06&  0.70 &   0.08\\ 
      &      & 1.00 & 1.50&  0.09 &   0.02&  0.26 &   0.03&  0.62 &   0.04&  0.79 &   0.07\\ 
      &      & 1.50 & 2.00&  0.10 &   0.02&  0.20 &   0.02&  0.53 &   0.03&  0.69 &   0.06\\ 
      &      & 2.00 & 2.50&  0.13 &   0.02&  0.16 &   0.02&  0.39 &   0.03&  0.69 &   0.05\\ 
      &      & 2.50 & 3.00&       &       &  0.13 &   0.02&  0.27 &   0.03&  0.54 &   0.04\\ 
      &      & 3.00 & 3.50&       &       &  0.11 &   0.02&  0.29 &   0.03&  0.53 &   0.04\\ 
      &      & 3.50 & 4.00&       &       &  0.13 &   0.02&  0.19 &   0.03&  0.40 &   0.04\\ 
      &      & 4.00 & 5.00&       &       &  0.06 &   0.02& 0.143 &  0.014&  0.26 &   0.02\\ 
      &      & 5.00 & 6.50&       &       &       &       & 0.066 &  0.010& 0.124 &  0.014\\ 
      &      & 6.50 & 8.00&       &       &       &       &       &       & 0.062 &  0.008\\ 
 0.10 & 0.15 & 0.50 & 1.00&  0.19 &   0.03&  0.35 &   0.04&  0.70 &   0.07&  0.91 &   0.10\\ 
      &      & 1.00 & 1.50&  0.11 &   0.02&  0.27 &   0.03&  0.55 &   0.05&  0.79 &   0.08\\ 
      &      & 1.50 & 2.00&  0.09 &   0.02&  0.21 &   0.02&  0.45 &   0.04&  0.69 &   0.07\\ 
      &      & 2.00 & 2.50& 0.081 &  0.014&  0.15 &   0.02&  0.37 &   0.03&  0.54 &   0.06\\ 
      &      & 2.50 & 3.00&       &       & 0.095 &  0.014&  0.27 &   0.03&  0.42 &   0.05\\ 
      &      & 3.00 & 3.50&       &       & 0.073 &  0.013&  0.15 &   0.02&  0.41 &   0.04\\ 
      &      & 3.50 & 4.00&       &       & 0.040 &  0.008&  0.11 &   0.02&  0.25 &   0.03\\ 
      &      & 4.00 & 5.00&       &       & 0.010 &  0.006& 0.051 &  0.010&  0.13 &   0.02\\ 
      &      & 5.00 & 6.50&       &       &       &       & 0.010 &  0.003& 0.043 &  0.008\\ 
      &      & 6.50 & 8.00&       &       &       &       &       &       & 0.006 &  0.002\\ 
 0.15 & 0.20 & 0.50 & 1.00&  0.12 &   0.02&  0.32 &   0.04&  0.70 &   0.07&  0.95 &   0.11\\ 
      &      & 1.00 & 1.50&  0.11 &   0.02&  0.22 &   0.03&  0.45 &   0.04&  0.57 &   0.06\\ 
      &      & 1.50 & 2.00& 0.055 &  0.010&  0.13 &   0.02&  0.36 &   0.03&  0.46 &   0.05\\ 
      &      & 2.00 & 2.50& 0.052 &  0.011&  0.13 &   0.02&  0.29 &   0.03&  0.39 &   0.05\\ 
      &      & 2.50 & 3.00&       &       & 0.051 &  0.011&  0.12 &   0.02&  0.27 &   0.04\\ 
      &      & 3.00 & 3.50&       &       & 0.028 &  0.007& 0.067 &  0.012&  0.16 &   0.02\\ 
      &      & 3.50 & 4.00&       &       & 0.005 &  0.004& 0.050 &  0.009& 0.068 &  0.015\\ 
      &      & 4.00 & 5.00&       &       &  &  & 0.021 &  0.005& 0.045 &  0.010\\ 
      &      & 5.00 & 6.50&       &       &       &       & 0.002 &  0.001& 0.006 &  0.003\\ 
      &      & 6.50 & 8.00&       &       &       &       &       &       & 0.000 &  0.001\\ 
 0.20 & 0.25 & 0.50 & 1.00&  0.22 &   0.04&  0.27 &   0.04&  0.51 &   0.07&  0.76 &   0.11\\ 
      &      & 1.00 & 1.50&  0.14 &   0.03&  0.26 &   0.04&  0.71 &   0.08&  0.72 &   0.10\\ 
      &      & 1.50 & 2.00&  0.07 &   0.02&  0.17 &   0.03&  0.46 &   0.06&  0.65 &   0.10\\ 
      &      & 2.00 & 2.50& 0.035 &  0.012&  0.09 &   0.02&  0.27 &   0.04&  0.35 &   0.06\\ 
      &      & 2.50 & 3.00&       &       & 0.039 &  0.011&  0.11 &   0.02&  0.18 &   0.04\\ 
      &      & 3.00 & 3.50&       &       & 0.003 &  0.002& 0.028 &  0.010&  0.06 &   0.02\\ 
      &      & 3.50 & 4.00&       &       &  &   & 0.011 &  0.006& 0.028 &  0.013\\ 
      &      & 4.00 & 5.00&       &       &   &   & 0.005 &  0.003& 0.008 &  0.009\\ 
      &      & 5.00 & 6.50&       &       &       &       &  &   & 0.000 &  0.001\\ 
      &      & 6.50 & 8.00&       &       &       &       &       &       & 0.000 &  0.009\\ 

\hline
\end{tabular}
}
\end{table}
\clearpage
\begin{table}[!ht]
  \caption{\label{tab:xsec_results_Cu2}
    HARP results for the double-differential  $\pi^-$ production
    cross-section in the laboratory system,
    $d^2\sigma^{\pi}/(dpd\Omega)$, for $\pi^{-}$--Cu interactions at 3,5,8,12~\GeVc.
    Each row refers to a
    different $(p_{\hbox{\small min}} \le p<p_{\hbox{\small max}},
    \theta_{\hbox{\small min}} \le \theta<\theta_{\hbox{\small max}})$ bin,
    where $p$ and $\theta$ are the pion momentum and polar angle, respectively.
    The central value as well as the square-root of the diagonal elements
    of the covariance matrix are given.}

\small{
\begin{tabular}{rrrr|r@{$\pm$}lr@{$\pm$}lr@{$\pm$}lr@{$\pm$}l}
\hline
$\theta_{\hbox{\small min}}$ &
$\theta_{\hbox{\small max}}$ &
$p_{\hbox{\small min}}$ &
$p_{\hbox{\small max}}$ &
\multicolumn{8}{c}{$d^2\sigma^{\pi^-}/(dpd\Omega)$}
\\
(rad) & (rad) & (\GeVc) & (\GeVc) &
\multicolumn{8}{c}{(barn/(\GeVc sr))}
\\
  &  &  &
&\multicolumn{2}{c}{$ \bf{3 \ \GeVc}$}
&\multicolumn{2}{c}{$ \bf{5 \ \GeVc}$}
&\multicolumn{2}{c}{$ \bf{8 \ \GeVc}$}
&\multicolumn{2}{c}{$ \bf{12 \ \GeVc}$}
\\
\hline

 0.05 & 0.10 & 0.50 & 1.00&  0.24 &   0.03&  0.45 &   0.05&  0.91 &   0.08&  1.12 &   0.12\\ 
      &      & 1.00 & 1.50&  0.21 &   0.03&  0.32 &   0.03&  0.71 &   0.05&  0.96 &   0.08\\ 
      &      & 1.50 & 2.00&  0.26 &   0.03&  0.34 &   0.03&  0.75 &   0.04&  0.94 &   0.07\\ 
      &      & 2.00 & 2.50&  0.15 &   0.04&  0.36 &   0.03&  0.68 &   0.04&  1.03 &   0.07\\ 
      &      & 2.50 & 3.00&       &       &  0.41 &   0.03&  0.63 &   0.04&  0.76 &   0.06\\ 
      &      & 3.00 & 3.50&       &       &  0.43 &   0.04&  0.47 &   0.04&  0.78 &   0.06\\ 
      &      & 3.50 & 4.00&       &       &  0.53 &   0.04&  0.41 &   0.03&  0.68 &   0.05\\ 
      &      & 4.00 & 5.00&       &       &  0.51 &   0.08&  0.42 &   0.03&  0.48 &   0.03\\ 
      &      & 5.00 & 6.50&       &       &       &       &  0.30 &   0.02&  0.34 &   0.02\\ 
      &      & 6.50 & 8.00&       &       &       &       &       &       &  0.21 &   0.02\\ 
 0.10 & 0.15 & 0.50 & 1.00&  0.33 &   0.04&  0.58 &   0.06&  1.01 &   0.11&  1.10 &   0.12\\ 
      &      & 1.00 & 1.50&  0.24 &   0.03&  0.40 &   0.04&  0.80 &   0.06&  1.00 &   0.09\\ 
      &      & 1.50 & 2.00&  0.26 &   0.03&  0.34 &   0.03&  0.67 &   0.05&  0.85 &   0.08\\ 
      &      & 2.00 & 2.50&  0.22 &   0.04&  0.31 &   0.03&  0.63 &   0.05&  0.82 &   0.07\\ 
      &      & 2.50 & 3.00&       &       &  0.31 &   0.03&  0.40 &   0.03&  0.63 &   0.06\\ 
      &      & 3.00 & 3.50&       &       &  0.34 &   0.03&  0.31 &   0.03&  0.49 &   0.05\\ 
      &      & 3.50 & 4.00&       &       &  0.21 &   0.03&  0.18 &   0.03&  0.33 &   0.03\\ 
      &      & 4.00 & 5.00&       &       &  0.21 &   0.04& 0.118 &  0.014&  0.25 &   0.03\\ 
      &      & 5.00 & 6.50&       &       &       &       & 0.059 &  0.007& 0.100 &  0.013\\ 
      &      & 6.50 & 8.00&       &       &       &       &       &       & 0.035 &  0.006\\ 
 0.15 & 0.20 & 0.50 & 1.00&  0.22 &   0.04&  0.42 &   0.05&  0.88 &   0.09&  1.05 &   0.13\\ 
      &      & 1.00 & 1.50&  0.21 &   0.03&  0.33 &   0.04&  0.76 &   0.07&  0.86 &   0.09\\ 
      &      & 1.50 & 2.00&  0.15 &   0.02&  0.27 &   0.03&  0.51 &   0.05&  0.65 &   0.07\\ 
      &      & 2.00 & 2.50&  0.17 &   0.02&  0.19 &   0.02&  0.35 &   0.03&  0.50 &   0.05\\ 
      &      & 2.50 & 3.00&       &       &  0.12 &   0.02&  0.20 &   0.02&  0.32 &   0.04\\ 
      &      & 3.00 & 3.50&       &       & 0.073 &  0.012& 0.114 &  0.015&  0.19 &   0.03\\ 
      &      & 3.50 & 4.00&       &       & 0.039 &  0.008& 0.097 &  0.012&  0.15 &   0.02\\ 
      &      & 4.00 & 5.00&       &       & 0.036 &  0.007& 0.049 &  0.007& 0.088 &  0.013\\ 
      &      & 5.00 & 6.50&       &       &       &       & 0.019 &  0.003& 0.037 &  0.007\\ 
      &      & 6.50 & 8.00&       &       &       &       &       &       & 0.009 &  0.003\\ 
 0.20 & 0.25 & 0.50 & 1.00&  0.17 &   0.03&  0.39 &   0.06&  0.59 &   0.07&  0.72 &   0.10\\ 
      &      & 1.00 & 1.50&  0.11 &   0.02&  0.18 &   0.03&  0.39 &   0.05&  0.38 &   0.06\\ 
      &      & 1.50 & 2.00&  0.11 &   0.02&  0.14 &   0.03&  0.31 &   0.05&  0.42 &   0.07\\ 
      &      & 2.00 & 2.50& 0.056 &  0.011&  0.07 &   0.02&  0.35 &   0.05&  0.31 &   0.06\\ 
      &      & 2.50 & 3.00&       &       & 0.047 &  0.011&  0.23 &   0.04&  0.16 &   0.04\\ 
      &      & 3.00 & 3.50&       &       & 0.040 &  0.009&  0.13 &   0.02&  0.13 &   0.03\\ 
      &      & 3.50 & 4.00&       &       & 0.032 &  0.007& 0.070 &  0.013&  0.11 &   0.02\\ 
      &      & 4.00 & 5.00&       &       & 0.026 &  0.006& 0.041 &  0.008&  0.08 &   0.02\\ 
      &      & 5.00 & 6.50&       &       &       &       & 0.013 &  0.004& 0.029 &  0.010\\ 
      &      & 6.50 & 8.00&       &       &       &       &       &       & 0.006 &  0.006\\ 
\hline
\end{tabular}
}
\end{table}

%% file: xsec_results_Cu_pip_abs.tex
\begin{table}[!ht]
  \caption{\label{tab:xsec_results_Cu3}
    HARP results for the double-differential $\pi^+$  production
    cross-section in the laboratory system,
    $d^2\sigma^{\pi}/(dpd\Omega)$, for $\pi^{+}$--Cu interactions at 3,5,8,12~\GeVc.
    Each row refers to a
    different $(p_{\hbox{\small min}} \le p<p_{\hbox{\small max}},
    \theta_{\hbox{\small min}} \le \theta<\theta_{\hbox{\small max}})$ bin,
    where $p$ and $\theta$ are the pion momentum and polar angle, respectively.
    The central value as well as the square-root of the diagonal elements
    of the covariance matrix are given.}

\small{
\begin{tabular}{rrrr|r@{$\pm$}lr@{$\pm$}lr@{$\pm$}lr@{$\pm$}l}
\hline
$\theta_{\hbox{\small min}}$ &
$\theta_{\hbox{\small max}}$ &
$p_{\hbox{\small min}}$ &
$p_{\hbox{\small max}}$ &
\multicolumn{8}{c}{$d^2\sigma^{\pi^+}/(dpd\Omega)$}
\\
(rad) & (rad) & (\GeVc) & (\GeVc) &
\multicolumn{8}{c}{(barn/(\GeVc sr))}
\\
  &  &  &
&\multicolumn{2}{c}{$ \bf{3 \ \GeVc}$}
&\multicolumn{2}{c}{$ \bf{5 \ \GeVc}$}
&\multicolumn{2}{c}{$ \bf{8 \ \GeVc}$}
&\multicolumn{2}{c}{$ \bf{12 \ \GeVc}$}
\\
\hline
0.050 &0.100 & 0.50 & 1.00&  0.16 &   0.06&  0.32 &   0.05&  0.58 &   0.09&  0.74 &   0.25\\
      &      & 1.00 & 1.50&  0.08 &   0.02&  0.32 &   0.04&  0.60 &   0.06&  0.81 &   0.21\\
      &      & 1.50 & 2.00&  0.15 &   0.02&  0.27 &   0.03&  0.57 &   0.06&  0.75 &   0.20\\
      &      & 2.00 & 2.50&  0.43 &   0.05&  0.34 &   0.03&  0.56 &   0.06&  0.68 &   0.18\\
      &      & 2.50 & 3.00&       &       &  0.49 &   0.04&  0.54 &   0.05&  0.77 &   0.19\\
      &      & 3.00 & 3.50&       &       &  0.60 &   0.06&  0.43 &   0.05&  0.68 &   0.15\\
      &      & 3.50 & 4.00&       &       &  0.73 &   0.04&  0.46 &   0.05&  0.71 &   0.16\\
      &      & 4.00 & 5.00&       &       &  0.68 &   0.22&  0.48 &   0.03&  0.55 &   0.10\\
      &      & 5.00 & 6.50&       &       &       &       &  0.41 &   0.06&  0.41 &   0.07\\
      &      & 6.50 & 8.00&       &       &       &       &       &       &  0.18 &   0.05\\
0.100 &0.150 & 0.50 & 1.00&  0.12 &   0.05&  0.45 &   0.06&  0.64 &   0.08&  0.86 &   0.27\\
      &      & 1.00 & 1.50&  0.18 &   0.05&  0.38 &   0.05&  0.51 &   0.06&  0.95 &   0.25\\
      &      & 1.50 & 2.00&  0.28 &   0.05&  0.40 &   0.05&  0.68 &   0.07&  0.79 &   0.20\\
      &      & 2.00 & 2.50&  0.51 &   0.07&  0.34 &   0.04&  0.56 &   0.06&  0.56 &   0.16\\
      &      & 2.50 & 3.00&       &       &  0.34 &   0.04&  0.31 &   0.04&  0.65 &   0.16\\
      &      & 3.00 & 3.50&       &       &  0.33 &   0.03&  0.31 &   0.04&  0.42 &   0.13\\
      &      & 3.50 & 4.00&       &       &  0.27 &   0.03&  0.21 &   0.03&  0.23 &   0.07\\
      &      & 4.00 & 5.00&       &       &  0.27 &   0.10&  0.16 &   0.03&  0.23 &   0.07\\
      &      & 5.00 & 6.50&       &       &       &       & 0.066 &  0.014&  0.07 &   0.03\\
      &      & 6.50 & 8.00&       &       &       &       &       &       &  0.03 &   0.02\\
0.150 &0.200 & 0.50 & 1.00&  0.21 &   0.07&  0.50 &   0.07&  0.74 &   0.10&  0.85 &   0.30\\
      &      & 1.00 & 1.50&  0.22 &   0.05&  0.32 &   0.04&  0.55 &   0.06&  0.58 &   0.19\\
      &      & 1.50 & 2.00&  0.23 &   0.05&  0.26 &   0.03&  0.39 &   0.05&  0.68 &   0.19\\
      &      & 2.00 & 2.50&  0.14 &   0.04&  0.19 &   0.03&  0.30 &   0.04&  0.72 &   0.21\\
      &      & 2.50 & 3.00&       &       &  0.13 &   0.02&  0.24 &   0.04&  0.28 &   0.11\\
      &      & 3.00 & 3.50&       &       &  0.09 &   0.02&  0.15 &   0.03&  0.23 &   0.11\\
      &      & 3.50 & 4.00&       &       &  0.07 &   0.02& 0.069 &  0.015&  0.16 &   0.08\\
      &      & 4.00 & 5.00&       &       &  0.05 &   0.03& 0.043 &  0.010&  0.14 &   0.07\\
      &      & 5.00 & 6.50&       &       &       &       & 0.017 &  0.007& 0.021 &  0.024\\
      &      & 6.50 & 8.00&       &       &       &       &       &       & 0.007 &  0.010\\
0.200 &0.250 & 0.50 & 1.00&  0.18 &   0.09&  0.28 &   0.05&  0.53 &   0.08&  0.41 &   0.22\\
      &      & 1.00 & 1.50&  0.07 &   0.03&  0.23 &   0.04&  0.33 &   0.06&  0.34 &   0.26\\
      &      & 1.50 & 2.00&  0.10 &   0.04&  0.14 &   0.03&  0.26 &   0.05&  0.21 &   0.17\\
      &      & 2.00 & 2.50&  0.05 &   0.03&  0.15 &   0.03&  0.25 &   0.05&  0.14 &   0.20\\
      &      & 2.50 & 3.00&       &       &  0.09 &   0.02&  0.14 &   0.04& 0.124 &  0.253\\
      &      & 3.00 & 3.50&       &       & 0.041 &  0.012&  0.10 &   0.03& 0.037 &  0.087\\
      &      & 3.50 & 4.00&       &       & 0.021 &  0.009&  0.06 &   0.02& 0.034 &  0.108\\
      &      & 4.00 & 5.00&       &       & 0.021 &  0.011& 0.038 &  0.014& 0.037 &  0.124\\
      &      & 5.00 & 6.50&       &       &       &       & 0.018 &  0.011& 0.012 &  0.036\\
      &      & 6.50 & 8.00&       &       &       &       &       &       & 0.003 &  0.102\\

\hline
\end{tabular}
}
\end{table}
\clearpage
\begin{table}[!ht]
  \caption{\label{tab:xsec_results_cu4}
    HARP results for the double-differential  $\pi^-$ production
    cross-section in the laboratory system,
    $d^2\sigma^{\pi}/(dpd\Omega)$, for $\pi^{+}$--Cu interactions at 3,5,8,12~\GeVc.
    Each row refers to a
    different $(p_{\hbox{\small min}} \le p<p_{\hbox{\small max}},
    \theta_{\hbox{\small min}} \le \theta<\theta_{\hbox{\small max}})$ bin,
    where $p$ and $\theta$ are the pion momentum and polar angle, respectively.
    The central value as well as the square-root of the diagonal elements
    of the covariance matrix are given.}

\small{
\begin{tabular}{rrrr|r@{$\pm$}lr@{$\pm$}lr@{$\pm$}lr@{$\pm$}l}
\hline
$\theta_{\hbox{\small min}}$ &
$\theta_{\hbox{\small max}}$ &
$p_{\hbox{\small min}}$ &
$p_{\hbox{\small max}}$ &
\multicolumn{8}{c}{$d^2\sigma^{\pi^+}/(dpd\Omega)$}
\\
(rad) & (rad) & (\GeVc) & (\GeVc) &
\multicolumn{8}{c}{(barn/(\GeVc sr))}
\\
  &  &  &
&\multicolumn{2}{c}{$ \bf{3 \ \GeVc}$}
&\multicolumn{2}{c}{$ \bf{5 \ \GeVc}$}
&\multicolumn{2}{c}{$ \bf{8 \ \GeVc}$}
&\multicolumn{2}{c}{$ \bf{12 \ \GeVc}$}
\\
\hline
0.050 &0.100 & 0.50 & 1.00&  0.11 &   0.04&  0.27 &   0.04&  0.51 &   0.07&  0.33 &   0.16\\
      &      & 1.00 & 1.50&  0.13 &   0.04&  0.24 &   0.03&  0.54 &   0.06&  0.72 &   0.21\\
      &      & 1.50 & 2.00&  0.13 &   0.03&  0.20 &   0.03&  0.41 &   0.05&  0.68 &   0.19\\
      &      & 2.00 & 2.50&  0.11 &   0.04&  0.19 &   0.02&  0.38 &   0.04&  0.47 &   0.14\\
      &      & 2.50 & 3.00&       &       &  0.16 &   0.02&  0.27 &   0.03&  0.83 &   0.20\\
      &      & 3.00 & 3.50&       &       &  0.16 &   0.02&  0.25 &   0.03&  0.29 &   0.09\\
      &      & 3.50 & 4.00&       &       &  0.18 &   0.02&  0.18 &   0.02&  0.41 &   0.12\\
      &      & 4.00 & 5.00&       &       &  0.07 &   0.02&  0.17 &   0.02&  0.29 &   0.08\\
      &      & 5.00 & 6.50&       &       &       &       & 0.065 &  0.008&  0.12 &   0.04\\
      &      & 6.50 & 8.00&       &       &       &       &       &       &  0.09 &   0.04\\
0.100 &0.150 & 0.50 & 1.00&  0.27 &   0.07&  0.45 &   0.06&  0.65 &   0.09&  1.10 &   0.35\\
      &      & 1.00 & 1.50&  0.14 &   0.04&  0.23 &   0.03&  0.46 &   0.05&  0.85 &   0.22\\
      &      & 1.50 & 2.00&  0.10 &   0.03&  0.21 &   0.03&  0.38 &   0.04&  0.31 &   0.12\\
      &      & 2.00 & 2.50&  0.04 &   0.02&  0.18 &   0.02&  0.29 &   0.04&  0.43 &   0.15\\
      &      & 2.50 & 3.00&       &       & 0.100 &  0.015&  0.24 &   0.03&  0.44 &   0.14\\
      &      & 3.00 & 3.50&       &       & 0.077 &  0.013&  0.18 &   0.02&  0.34 &   0.11\\
      &      & 3.50 & 4.00&       &       & 0.031 &  0.006& 0.082 &  0.014&  0.36 &   0.12\\
      &      & 4.00 & 5.00&       &       & 0.022 &  0.007& 0.073 &  0.012&  0.13 &   0.06\\
      &      & 5.00 & 6.50&       &       &       &       & 0.018 &  0.004&  0.04 &   0.03\\
      &      & 6.50 & 8.00&       &       &       &       &       &       &  0.002 &   0.003\\
0.150 &0.200 & 0.50 & 1.00&  0.11 &   0.04&  0.42 &   0.06&  0.58 &   0.08&  0.79 &   0.27\\
      &      & 1.00 & 1.50&  0.10 &   0.04&  0.23 &   0.03&  0.42 &   0.06&  0.66 &   0.23\\
      &      & 1.50 & 2.00&  0.11 &   0.04&  0.23 &   0.03&  0.33 &   0.04&  0.31 &   0.13\\
      &      & 2.00 & 2.50& 0.019 &  0.013&  0.11 &   0.02&  0.28 &   0.04&  0.23 &   0.11\\
      &      & 2.50 & 3.00&       &       & 0.064 &  0.013&  0.16 &   0.03&  0.16 &   0.09\\
      &      & 3.00 & 3.50&       &       & 0.016 &  0.004&  0.08 &   0.02&  0.22 &   0.10\\
      &      & 3.50 & 4.00&       &       & 0.012 &  0.005& 0.037 &  0.010&  0.18 &   0.09\\
      &      & 4.00 & 5.00&       &       & 0.005 &  0.003& 0.020 &  0.006&  0.11 &   0.07\\
      &      & 5.00 & 6.50&       &       &       &       & 0.004 &  0.002&  0.03 &   0.03\\
      &      & 6.50 & 8.00&       &       &       &       &       &       &  &  \\
0.200 &0.250 & 0.50 & 1.00&  0.17 &   0.07&  0.33 &   0.06&  0.60 &   0.09&  0.58 &   0.26\\
      &      & 1.00 & 1.50&  0.06 &   0.03&  0.20 &   0.04&  0.51 &   0.08&  0.44 &   0.23\\
      &      & 1.50 & 2.00&  0.08 &   0.04&  0.09 &   0.02&  0.29 &   0.05&  0.72 &   0.37\\
      &      & 2.00 & 2.50& 0.009 &  0.011& 0.024 &  0.007&  0.13 &   0.03&  0.16 &   0.18\\
      &      & 2.50 & 3.00&       &       & 0.014 &  0.005& 0.043 &  0.012&  0.07 &   0.11\\
      &      & 3.00 & 3.50&       &       & 0.017 &  0.007& 0.020 &  0.009& 0.050 &  0.065\\
      &      & 3.50 & 4.00&       &       & 0.008 &  0.004& 0.008 &  0.006& 0.010 &  0.031\\
      &      & 4.00 & 5.00&       &       & 0.005 &  0.004& 0.005 &  0.007& 0.019 &  0.109\\
      &      & 5.00 & 6.50&       &       &       &       & 0.002 &  0.004& 0.016 &  0.090\\
      &      & 6.50 & 8.00&       &       &       &       &       &       & 0.003 &  0.068\\

\hline
\end{tabular}
}
\end{table}

%% file: xsec_results_Sn_pim.tex
\begin{table}[!ht]
  \caption{\label{tab:xsec_results_Sn1}
    HARP results for the double-differential $\pi^+$  production
    cross-section in the laboratory system,
    $d^2\sigma^{\pi}/(dpd\Omega)$, for $\pi^{-}$--Sn interactions at 3,5,8,12~\GeVc.
    Each row refers to a
    different $(p_{\hbox{\small min}} \le p<p_{\hbox{\small max}},
    \theta_{\hbox{\small min}} \le \theta<\theta_{\hbox{\small max}})$ bin,
    where $p$ and $\theta$ are the pion momentum and polar angle, respectively.
    The central value as well as the square-root of the diagonal elements
    of the covariance matrix are given.}

\small{
\begin{tabular}{rrrr|r@{$\pm$}lr@{$\pm$}lr@{$\pm$}lr@{$\pm$}l}
\hline
$\theta_{\hbox{\small min}}$ &
$\theta_{\hbox{\small max}}$ &
$p_{\hbox{\small min}}$ &
$p_{\hbox{\small max}}$ &
\multicolumn{8}{c}{$d^2\sigma^{\pi^+}/(dpd\Omega)$}
\\
(rad) & (rad) & (\GeVc) & (\GeVc) &
\multicolumn{8}{c}{(barn/(\GeVc sr))}
\\
  &  &  &
&\multicolumn{2}{c}{$ \bf{3 \ \GeVc}$}
&\multicolumn{2}{c}{$ \bf{5 \ \GeVc}$}
&\multicolumn{2}{c}{$ \bf{8 \ \GeVc}$}
&\multicolumn{2}{c}{$ \bf{12 \ \GeVc}$}
\\
\hline
 
 0.05 & 0.10 & 0.50 & 1.00&  0.11 &   0.03&  0.32 &   0.05&  0.68 &   0.08&  1.15 &   0.12\\ 
      &      & 1.00 & 1.50&  0.07 &   0.02&  0.34 &   0.04&  0.89 &   0.07&  1.25 &   0.09\\ 
      &      & 1.50 & 2.00&  0.14 &   0.03&  0.29 &   0.04&  0.72 &   0.05&  1.02 &   0.08\\ 
      &      & 2.00 & 2.50&  0.11 &   0.03&  0.13 &   0.02&  0.52 &   0.05&  0.93 &   0.07\\ 
      &      & 2.50 & 3.00&       &       &  0.16 &   0.03&  0.37 &   0.04&  0.75 &   0.05\\ 
      &      & 3.00 & 3.50&       &       &  0.17 &   0.02&  0.36 &   0.05&  0.69 &   0.06\\ 
      &      & 3.50 & 4.00&       &       &  0.17 &   0.02&  0.21 &   0.04&  0.50 &   0.04\\ 
      &      & 4.00 & 5.00&       &       &  0.06 &   0.03&  0.15 &   0.02&  0.36 &   0.03\\ 
      &      & 5.00 & 6.50&       &       &       &       & 0.080 &  0.013&  0.19 &   0.02\\ 
      &      & 6.50 & 8.00&       &       &       &       &       &       & 0.081 &  0.011\\ 
 0.10 & 0.15 & 0.50 & 1.00&  0.18 &   0.03&  0.42 &   0.06&  0.94 &   0.10&  1.21 &   0.12\\ 
      &      & 1.00 & 1.50&  0.07 &   0.02&  0.35 &   0.04&  0.68 &   0.07&  1.23 &   0.11\\ 
      &      & 1.50 & 2.00&  0.08 &   0.02&  0.23 &   0.03&  0.61 &   0.06&  1.07 &   0.09\\ 
      &      & 2.00 & 2.50&  0.12 &   0.03&  0.21 &   0.03&  0.50 &   0.05&  0.87 &   0.08\\ 
      &      & 2.50 & 3.00&       &       &  0.14 &   0.02&  0.34 &   0.04&  0.65 &   0.06\\ 
      &      & 3.00 & 3.50&       &       &  0.09 &   0.02&  0.20 &   0.03&  0.55 &   0.05\\ 
      &      & 3.50 & 4.00&       &       & 0.036 &  0.009&  0.17 &   0.03&  0.41 &   0.04\\ 
      &      & 4.00 & 5.00&       &       & 0.022 &  0.009&  0.07 &   0.02&  0.21 &   0.02\\ 
      &      & 5.00 & 6.50&       &       &       &       & 0.013 &  0.005& 0.067 &  0.011\\ 
      &      & 6.50 & 8.00&       &       &       &       &       &       & 0.008 &  0.003\\ 
 0.15 & 0.20 & 0.50 & 1.00&  0.22 &   0.05&  0.45 &   0.07&  0.92 &   0.11&  1.22 &   0.13\\ 
      &      & 1.00 & 1.50&  0.10 &   0.03&  0.24 &   0.03&  0.57 &   0.06&  0.86 &   0.08\\ 
      &      & 1.50 & 2.00& 0.040 &  0.012&  0.24 &   0.03&  0.50 &   0.05&  0.82 &   0.07\\ 
      &      & 2.00 & 2.50&  0.05 &   0.02&  0.14 &   0.03&  0.34 &   0.04&  0.52 &   0.06\\ 
      &      & 2.50 & 3.00&       &       &  0.06 &   0.02&  0.19 &   0.03&  0.33 &   0.04\\ 
      &      & 3.00 & 3.50&       &       & 0.031 &  0.010&  0.12 &   0.02&  0.20 &   0.03\\ 
      &      & 3.50 & 4.00&       &       & 0.021 &  0.008&  0.07 &   0.02&  0.09 &   0.02\\ 
      &      & 4.00 & 5.00&       &       & 0.004 &  0.004& 0.027 &  0.007& 0.053 &  0.010\\ 
      &      & 5.00 & 6.50&       &       &       &       & 0.005 &  0.002& 0.013 &  0.004\\ 
      &      & 6.50 & 8.00&       &       &       &       &       &       & 0.002 &  0.001\\ 
 0.20 & 0.25 & 0.50 & 1.00&  0.23 &   0.06&  0.45 &   0.08&  0.73 &   0.10&  1.14 &   0.15\\ 
      &      & 1.00 & 1.50&  0.17 &   0.04&  0.36 &   0.06&  0.75 &   0.10&  1.14 &   0.14\\ 
      &      & 1.50 & 2.00&  0.08 &   0.03&  0.29 &   0.05&  0.56 &   0.09&  0.75 &   0.11\\ 
      &      & 2.00 & 2.50&  0.002 &   0.005&  0.15 &   0.03&  0.23 &   0.05&  0.56 &   0.09\\ 
      &      & 2.50 & 3.00&       &       &  0.05 &   0.02&  0.13 &   0.03&  0.25 &   0.05\\ 
      &      & 3.00 & 3.50&       &       & 0.013 &  0.007&  0.04 &   0.02&  0.09 &   0.02\\ 
      &      & 3.50 & 4.00&       &       & 0.003 &  0.002& 0.014 &  0.010& 0.047 &  0.015\\ 
      &      & 4.00 & 5.00&       &       & 0.001 &  0.001& 0.010 &  0.009& 0.018 &  0.011\\ 
      &      & 5.00 & 6.50&       &       &       &       & 0.001 &  0.001& 0.003 &  0.002\\ 
      &      & 6.50 & 8.00&       &       &       &       &       &       & 0.001 &  0.009\\ 

\hline
\end{tabular}
}
\end{table}
\clearpage
\begin{table}[!ht]
  \caption{\label{tab:xsec_results_Sn2}
    HARP results for the double-differential  $\pi^-$ production
    cross-section in the laboratory system,
    $d^2\sigma^{\pi}/(dpd\Omega)$, for $\pi^{-}$--Sn interactions at 3,5,8,12~\GeVc.
    Each row refers to a
    different $(p_{\hbox{\small min}} \le p<p_{\hbox{\small max}},
    \theta_{\hbox{\small min}} \le \theta<\theta_{\hbox{\small max}})$ bin,
    where $p$ and $\theta$ are the pion momentum and polar angle, respectively.
    The central value as well as the square-root of the diagonal elements
    of the covariance matrix are given.}

\small{
\begin{tabular}{rrrr|r@{$\pm$}lr@{$\pm$}lr@{$\pm$}lr@{$\pm$}l}
\hline
$\theta_{\hbox{\small min}}$ &
$\theta_{\hbox{\small max}}$ &
$p_{\hbox{\small min}}$ &
$p_{\hbox{\small max}}$ &
\multicolumn{8}{c}{$d^2\sigma^{\pi^-}/(dpd\Omega)$}
\\
(rad) & (rad) & (\GeVc) & (\GeVc) &
\multicolumn{8}{c}{(barn/(\GeVc sr))}
\\
  &  &  &
&\multicolumn{2}{c}{$ \bf{3 \ \GeVc}$}
&\multicolumn{2}{c}{$ \bf{5 \ \GeVc}$}
&\multicolumn{2}{c}{$ \bf{8 \ \GeVc}$}
&\multicolumn{2}{c}{$ \bf{12 \ \GeVc}$}
\\
\hline

 0.05 & 0.10 & 0.50 & 1.00&  0.30 &   0.05&  0.62 &   0.07&  1.26 &   0.11&  1.72 &   0.16\\ 
      &      & 1.00 & 1.50&  0.19 &   0.04&  0.41 &   0.04&  0.90 &   0.07&  1.29 &   0.10\\ 
      &      & 1.50 & 2.00&  0.17 &   0.03&  0.46 &   0.05&  0.94 &   0.06&  1.38 &   0.09\\ 
      &      & 2.00 & 2.50&  0.12 &   0.03&  0.51 &   0.05&  0.91 &   0.07&  1.34 &   0.09\\ 
      &      & 2.50 & 3.00&       &       &  0.46 &   0.04&  0.83 &   0.06&  1.24 &   0.08\\ 
      &      & 3.00 & 3.50&       &       &  0.61 &   0.06&  0.66 &   0.06&  1.02 &   0.07\\ 
      &      & 3.50 & 4.00&       &       &  0.62 &   0.06&  0.54 &   0.05&  0.94 &   0.07\\ 
      &      & 4.00 & 5.00&       &       &  0.66 &   0.10&  0.56 &   0.05&  0.67 &   0.05\\ 
      &      & 5.00 & 6.50&       &       &       &       &  0.38 &   0.03&  0.46 &   0.03\\ 
      &      & 6.50 & 8.00&       &       &       &       &       &       &  0.29 &   0.02\\ 
 0.10 & 0.15 & 0.50 & 1.00&  0.27 &   0.05&  0.76 &   0.09&  1.39 &   0.15&  1.91 &   0.20\\ 
      &      & 1.00 & 1.50&  0.25 &   0.04&  0.41 &   0.05&  1.00 &   0.09&  1.35 &   0.11\\ 
      &      & 1.50 & 2.00&  0.23 &   0.04&  0.49 &   0.05&  0.81 &   0.07&  1.35 &   0.11\\ 
      &      & 2.00 & 2.50&  0.44 &   0.08&  0.39 &   0.04&  0.73 &   0.06&  0.96 &   0.08\\ 
      &      & 2.50 & 3.00&       &       &  0.33 &   0.04&  0.64 &   0.06&  0.91 &   0.08\\ 
      &      & 3.00 & 3.50&       &       &  0.40 &   0.05&  0.37 &   0.04&  0.64 &   0.06\\ 
      &      & 3.50 & 4.00&       &       &  0.25 &   0.04&  0.28 &   0.04&  0.57 &   0.05\\ 
      &      & 4.00 & 5.00&       &       &  0.28 &   0.06&  0.17 &   0.02&  0.34 &   0.03\\ 
      &      & 5.00 & 6.50&       &       &       &       & 0.100 &  0.013& 0.141 &  0.015\\ 
      &      & 6.50 & 8.00&       &       &       &       &       &       & 0.053 &  0.009\\ 
 0.15 & 0.20 & 0.50 & 1.00&  0.25 &   0.05&  0.65 &   0.09&  1.07 &   0.13&  1.69 &   0.18\\ 
      &      & 1.00 & 1.50&  0.36 &   0.06&  0.43 &   0.05&  0.99 &   0.09&  1.33 &   0.12\\ 
      &      & 1.50 & 2.00&  0.22 &   0.04&  0.30 &   0.04&  0.66 &   0.07&  0.95 &   0.09\\ 
      &      & 2.00 & 2.50&  0.14 &   0.02&  0.22 &   0.03&  0.38 &   0.05&  0.65 &   0.06\\ 
      &      & 2.50 & 3.00&       &       &  0.19 &   0.03&  0.27 &   0.03&  0.46 &   0.05\\ 
      &      & 3.00 & 3.50&       &       &  0.10 &   0.02&  0.15 &   0.02&  0.28 &   0.03\\ 
      &      & 3.50 & 4.00&       &       & 0.062 &  0.012&  0.11 &   0.02&  0.18 &   0.03\\ 
      &      & 4.00 & 5.00&       &       & 0.077 &  0.015& 0.065 &  0.011&  0.11 &   0.02\\ 
      &      & 5.00 & 6.50&       &       &       &       & 0.026 &  0.005& 0.047 &  0.008\\ 
      &      & 6.50 & 8.00&       &       &       &       &       &       & 0.012 &  0.003\\ 
 0.20 & 0.25 & 0.50 & 1.00&  0.17 &   0.04&  0.55 &   0.09&  0.65 &   0.09&  1.08 &   0.14\\ 
      &      & 1.00 & 1.50&  0.12 &   0.03&  0.24 &   0.04&  0.37 &   0.05&  0.49 &   0.07\\ 
      &      & 1.50 & 2.00&  0.12 &   0.03&  0.24 &   0.05&  0.35 &   0.07&  0.43 &   0.07\\ 
      &      & 2.00 & 2.50&  0.08 &   0.02&  0.15 &   0.03&  0.26 &   0.05&  0.42 &   0.07\\ 
      &      & 2.50 & 3.00&       &       &  0.09 &   0.02&  0.18 &   0.03&  0.30 &   0.06\\ 
      &      & 3.00 & 3.50&       &       & 0.061 &  0.014&  0.12 &   0.02&  0.24 &   0.04\\ 
      &      & 3.50 & 4.00&       &       & 0.040 &  0.010&  0.08 &   0.02&  0.13 &   0.03\\ 
      &      & 4.00 & 5.00&       &       & 0.019 &  0.007& 0.052 &  0.015&  0.08 &   0.02\\ 
      &      & 5.00 & 6.50&       &       &       &       & 0.017 &  0.007& 0.019 &  0.006\\ 
      &      & 6.50 & 8.00&       &       &       &       &       &       & 0.004 &  0.005\\ 
\hline
\end{tabular}
}
\end{table}

%% file: xsec_results_Sn_pip_abs.tex
 \begin{table}[!ht]
  \caption{\label{tab:xsec_results_Sn3}
    HARP results for the double-differential $\pi^+$  production
    cross-section in the laboratory system,
    $d^2\sigma^{\pi}/(dpd\Omega)$, for $\pi^{+}$--Sn interactions at 3,5,8,12~\GeVc.
    Each row refers to a
    different $(p_{\hbox{\small min}} \le p<p_{\hbox{\small max}},
    \theta_{\hbox{\small min}} \le \theta<\theta_{\hbox{\small max}})$ bin,
    where $p$ and $\theta$ are the pion momentum and polar angle, respectively.
    The central value as well as the square-root of the diagonal elements
    of the covariance matrix are given.}

\small{
\begin{tabular}{rrrr|r@{$\pm$}lr@{$\pm$}lr@{$\pm$}lr@{$\pm$}l}
\hline
$\theta_{\hbox{\small min}}$ &
$\theta_{\hbox{\small max}}$ &
$p_{\hbox{\small min}}$ &
$p_{\hbox{\small max}}$ &
\multicolumn{8}{c}{$d^2\sigma^{\pi^+}/(dpd\Omega)$}
\\
(rad) & (rad) & (\GeVc) & (\GeVc) &
\multicolumn{8}{c}{(barn/(\GeVc sr))}
\\
  &  &  &
&\multicolumn{2}{c}{$ \bf{3 \ \GeVc}$}
&\multicolumn{2}{c}{$ \bf{5 \ \GeVc}$}
&\multicolumn{2}{c}{$ \bf{8 \ \GeVc}$}
&\multicolumn{2}{c}{$ \bf{12 \ \GeVc}$}
\\
\hline
0.050 &0.100 & 0.50 & 1.00&  0.28 &   0.07&  0.48 &   0.08&  0.99 &   0.14&  1.11 &   0.27\\
      &      & 1.00 & 1.50&  0.09 &   0.02&  0.35 &   0.05&  0.80 &   0.08&  1.32 &   0.23\\
      &      & 1.50 & 2.00&  0.17 &   0.02&  0.35 &   0.05&  0.76 &   0.15&  1.14 &   0.21\\
      &      & 2.00 & 2.50&  0.46 &   0.04&  0.55 &   0.05&  0.68 &   0.21&  1.05 &   0.21\\
      &      & 2.50 & 3.00&       &       &  0.57 &   0.04&  0.64 &   0.17&  1.12 &   0.28\\
      &      & 3.00 & 3.50&       &       &  0.74 &   0.07&  0.55 &   0.06&  0.96 &   0.15\\
      &      & 3.50 & 4.00&       &       &  0.93 &   0.05&  0.55 &   0.07&  0.78 &   0.13\\
      &      & 4.00 & 5.00&       &       &  0.85 &   0.28&  0.68 &   0.06&  0.73 &   0.10\\
      &      & 5.00 & 6.50&       &       &       &       &  0.61 &   0.10&  0.49 &   0.07\\
      &      & 6.50 & 8.00&       &       &       &       &       &       &  0.41 &   0.06\\
0.100 &0.150 & 0.50 & 1.00&  0.19 &   0.06&  0.55 &   0.08&  0.94 &   0.12&  1.16 &   0.26\\
      &      & 1.00 & 1.50&  0.31 &   0.07&  0.40 &   0.06&  0.75 &   0.08&  1.30 &   0.24\\
      &      & 1.50 & 2.00&  0.43 &   0.06&  0.48 &   0.06&  0.90 &   0.09&  0.71 &   0.15\\
      &      & 2.00 & 2.50&  0.51 &   0.06&  0.53 &   0.05&  0.69 &   0.09&  0.98 &   0.18\\
      &      & 2.50 & 3.00&       &       &  0.43 &   0.05&  0.47 &   0.06&  0.94 &   0.17\\
      &      & 3.00 & 3.50&       &       &  0.42 &   0.04&  0.41 &   0.05&  0.87 &   0.15\\
      &      & 3.50 & 4.00&       &       &  0.37 &   0.04&  0.38 &   0.04&  0.47 &   0.09\\
      &      & 4.00 & 5.00&       &       &  0.35 &   0.13&  0.22 &   0.03&  0.39 &   0.07\\
      &      & 5.00 & 6.50&       &       &       &       & 0.060 &  0.013&  0.18 &   0.04\\
      &      & 6.50 & 8.00&       &       &       &       &       &       &  0.06 &   0.02\\
0.150 &0.200 & 0.50 & 1.00&  0.31 &   0.08&  0.54 &   0.09&  0.99 &   0.13&  1.29 &   0.31\\
      &      & 1.00 & 1.50&  0.23 &   0.05&  0.41 &   0.05&  0.85 &   0.09&  0.97 &   0.20\\
      &      & 1.50 & 2.00&  0.22 &   0.04&  0.35 &   0.04&  0.64 &   0.07&  0.93 &   0.18\\
      &      & 2.00 & 2.50&  0.29 &   0.05&  0.31 &   0.04&  0.50 &   0.06&  0.62 &   0.15\\
      &      & 2.50 & 3.00&       &       &  0.15 &   0.02&  0.23 &   0.04&  0.40 &   0.11\\
      &      & 3.00 & 3.50&       &       &  0.12 &   0.03&  0.18 &   0.03&  0.42 &   0.11\\
      &      & 3.50 & 4.00&       &       &  0.08 &   0.02&  0.11 &   0.02&  0.20 &   0.07\\
      &      & 4.00 & 5.00&       &       &  0.06 &   0.03&  0.08 &   0.02&  0.11 &   0.05\\
      &      & 5.00 & 6.50&       &       &       &       & 0.019 &  0.008&  0.04 &   0.03\\
      &      & 6.50 & 8.00&       &       &       &       &       &       & 0.012 &  0.014\\
0.200 &0.250 & 0.50 & 1.00&  0.09 &   0.07&  0.52 &   0.08&  0.91 &   0.14&  1.15 &   0.31\\
      &      & 1.00 & 1.50&  0.09 &   0.03&  0.23 &   0.05&  0.59 &   0.10&  0.66 &   0.20\\
      &      & 1.50 & 2.00&  0.08 &   0.03&  0.19 &   0.04&  0.55 &   0.10&  0.74 &   0.23\\
      &      & 2.00 & 2.50&  0.04 &   0.02&  0.15 &   0.03&  0.36 &   0.07&  0.25 &   0.11\\
      &      & 2.50 & 3.00&       &       &  0.11 &   0.03&  0.14 &   0.04&  0.27 &   0.16\\
      &      & 3.00 & 3.50&       &       &  0.06 &   0.02&  0.14 &   0.04&  0.15 &   0.09\\
      &      & 3.50 & 4.00&       &       &  0.04 &   0.02&  0.10 &   0.03& 0.084 &  0.093\\
      &      & 4.00 & 5.00&       &       & 0.022 &  0.011&  0.06 &   0.02& 0.039 &  0.072\\
      &      & 5.00 & 6.50&       &       &       &       & 0.020 &  0.012& 0.034 &  0.049\\
      &      & 6.50 & 8.00&       &       &       &       &       &       & 0.008 &  0.087\\
 
\hline
\end{tabular}
}
\end{table}
\clearpage
\begin{table}[!ht]
  \caption{\label{tab:xsec_results_Sn4}
    HARP results for the double-differential  $\pi^-$ production
    cross-section in the laboratory system,
    $d^2\sigma^{\pi}/(dpd\Omega)$, for $\pi^{+}$--Sn interactions at 3,5,8,12~\GeVc.
    Each row refers to a
    different $(p_{\hbox{\small min}} \le p<p_{\hbox{\small max}},
    \theta_{\hbox{\small min}} \le \theta<\theta_{\hbox{\small max}})$ bin,
    where $p$ and $\theta$ are the pion momentum and polar angle, respectively.
    The central value as well as the square-root of the diagonal elements
    of the covariance matrix are given.}

\small{
\begin{tabular}{rrrr|r@{$\pm$}lr@{$\pm$}lr@{$\pm$}lr@{$\pm$}l}
\hline
$\theta_{\hbox{\small min}}$ &
$\theta_{\hbox{\small max}}$ &
$p_{\hbox{\small min}}$ &
$p_{\hbox{\small max}}$ &
\multicolumn{8}{c}{$d^2\sigma^{\pi^-}/(dpd\Omega)$}
\\
(rad) & (rad) & (\GeVc) & (\GeVc) &
\multicolumn{8}{c}{(barn/(\GeVc sr))}
\\
  &  &  &
&\multicolumn{2}{c}{$ \bf{3 \ \GeVc}$}
&\multicolumn{2}{c}{$ \bf{5 \ \GeVc}$}
&\multicolumn{2}{c}{$ \bf{8 \ \GeVc}$}
&\multicolumn{2}{c}{$ \bf{12 \ \GeVc}$}
\\
\hline
0.050 &0.100 & 0.50 & 1.00&  0.17 &   0.05&  0.37 &   0.06&  0.78 &   0.10&  1.18 &   0.27\\
      &      & 1.00 & 1.50&  0.14 &   0.03&  0.34 &   0.04&  0.70 &   0.08&  0.78 &   0.17\\
      &      & 1.50 & 2.00&  0.18 &   0.03&  0.30 &   0.04&  0.60 &   0.06&  1.03 &   0.19\\
      &      & 2.00 & 2.50&  0.17 &   0.05&  0.25 &   0.03&  0.51 &   0.06&  0.80 &   0.16\\
      &      & 2.50 & 3.00&       &       &  0.20 &   0.03&  0.40 &   0.04&  0.80 &   0.15\\
      &      & 3.00 & 3.50&       &       &  0.18 &   0.02&  0.36 &   0.04&  0.62 &   0.11\\
      &      & 3.50 & 4.00&       &       &  0.20 &   0.02&  0.29 &   0.03&  0.60 &   0.11\\
      &      & 4.00 & 5.00&       &       &  0.09 &   0.02&  0.19 &   0.02&  0.36 &   0.06\\
      &      & 5.00 & 6.50&       &       &       &       & 0.084 &  0.010&  0.21 &   0.04\\
      &      & 6.50 & 8.00&       &       &       &       &       &       &  0.10 &   0.02\\
0.100 &0.150 & 0.50 & 1.00&  0.25 &   0.06&  0.50 &   0.08&  1.22 &   0.16&  1.18 &   0.26\\
      &      & 1.00 & 1.50&  0.14 &   0.03&  0.34 &   0.05&  0.71 &   0.08&  0.74 &   0.16\\
      &      & 1.50 & 2.00&  0.12 &   0.03&  0.28 &   0.04&  0.58 &   0.07&  0.98 &   0.19\\
      &      & 2.00 & 2.50& 0.045 &  0.014&  0.19 &   0.03&  0.42 &   0.05&  0.71 &   0.15\\
      &      & 2.50 & 3.00&       &       &  0.12 &   0.02&  0.30 &   0.04&  0.51 &   0.11\\
      &      & 3.00 & 3.50&       &       & 0.094 &  0.015&  0.24 &   0.03&  0.58 &   0.12\\
      &      & 3.50 & 4.00&       &       & 0.055 &  0.010&  0.16 &   0.03&  0.36 &   0.08\\
      &      & 4.00 & 5.00&       &       & 0.022 &  0.007&  0.09 &   0.02&  0.20 &   0.05\\
      &      & 5.00 & 6.50&       &       &       &       & 0.017 &  0.004&  0.07 &   0.03\\
      &      & 6.50 & 8.00&       &       &       &       &       &       & 0.006 &  0.004\\
0.150 &0.200 & 0.50 & 1.00&  0.14 &   0.05&  0.45 &   0.07&  0.95 &   0.13&  1.41 &   0.32\\
      &      & 1.00 & 1.50&  0.18 &   0.04&  0.32 &   0.04&  0.69 &   0.08&  0.86 &   0.20\\
      &      & 1.50 & 2.00&  0.09 &   0.03&  0.26 &   0.04&  0.54 &   0.07&  0.52 &   0.14\\
      &      & 2.00 & 2.50& 0.025 &  0.011&  0.15 &   0.02&  0.38 &   0.05&  0.40 &   0.12\\
      &      & 2.50 & 3.00&       &       &  0.09 &   0.02&  0.20 &   0.03&  0.20 &   0.08\\
      &      & 3.00 & 3.50&       &       & 0.036 &  0.009&  0.10 &   0.02&  0.23 &   0.08\\
      &      & 3.50 & 4.00&       &       & 0.021 &  0.007& 0.041 &  0.010&  0.14 &   0.06\\
      &      & 4.00 & 5.00&       &       & 0.006 &  0.003& 0.030 &  0.009&  0.06 &   0.04\\
      &      & 5.00 & 6.50&       &       &       &       & 0.005 &  0.003& 0.007 &  0.008\\
      &      & 6.50 & 8.00&       &       &       &       &       &       & 0.004 &  0.007\\
0.200 &0.250 & 0.50 & 1.00&  0.19 &   0.06&  0.40 &   0.08&  0.78 &   0.12&  1.17 &   0.30\\
      &      & 1.00 & 1.50&  0.15 &   0.05&  0.31 &   0.05&  0.70 &   0.11&  0.78 &   0.24\\
      &      & 1.50 & 2.00& 0.015 &  0.010&  0.12 &   0.03&  0.42 &   0.08&  0.72 &   0.46\\
      &      & 2.00 & 2.50& 0.002 &  0.002& 0.051 &  0.013&  0.39 &   0.08&  0.54 &   1.07\\
      &      & 2.50 & 3.00&       &       & 0.023 &  0.007&  0.15 &   0.04&  0.14 &   0.35\\
      &      & 3.00 & 3.50&       &       & 0.023 &  0.009&  0.04 &   0.02&  0.02 &   0.09\\
      &      & 3.50 & 4.00&       &       & 0.009 &  0.005& 0.016 &  0.009&  0.01 &   0.03\\
      &      & 4.00 & 5.00&       &       & 0.004 &  0.003& 0.004 &  0.007& 0.000 &  0.003\\
      &      & 5.00 & 6.50&       &       &       &       & 0.000 &  0.002& 0.000 &  0.002\\
      &      & 6.50 & 8.00&       &       &       &       &       &       &0.000 &  0.036\\

\hline
\end{tabular}
}
\end{table}

%% file: xsec_results_Ta_pim.tex
 \begin{table}[!ht]
  \caption{\label{tab:xsec_results_Ta1}
    HARP results for the double-differential $\pi^+$  production
    cross-section in the laboratory system,
    $d^2\sigma^{\pi}/(dpd\Omega)$, for $\pi^{-}$--Ta interactions at 3,5,8,12~\GeVc.
    Each row refers to a
    different $(p_{\hbox{\small min}} \le p<p_{\hbox{\small max}},
    \theta_{\hbox{\small min}} \le \theta<\theta_{\hbox{\small max}})$ bin,
    where $p$ and $\theta$ are the pion momentum and polar angle, respectively.
    The central value as well as the square-root of the diagonal elements
    of the covariance matrix are given.}

\small{
\begin{tabular}{rrrr|r@{$\pm$}lr@{$\pm$}lr@{$\pm$}lr@{$\pm$}l}
\hline
$\theta_{\hbox{\small min}}$ &
$\theta_{\hbox{\small max}}$ &
$p_{\hbox{\small min}}$ &
$p_{\hbox{\small max}}$ &
\multicolumn{8}{c}{$d^2\sigma^{\pi^+}/(dpd\Omega)$}
\\
(rad) & (rad) & (\GeVc) & (\GeVc) &
\multicolumn{8}{c}{(barn/(\GeVc sr))}
\\
  &  &  &
&\multicolumn{2}{c}{$ \bf{3 \ \GeVc}$}
&\multicolumn{2}{c}{$ \bf{5 \ \GeVc}$}
&\multicolumn{2}{c}{$ \bf{8 \ \GeVc}$}
&\multicolumn{2}{c}{$ \bf{12 \ \GeVc}$}
\\
\hline
 
 0.05 & 0.10 & 0.50 & 1.00&  0.08 &   0.03&  0.42 &   0.06&  1.06 &   0.11&  1.60 &   0.17\\ 
      &      & 1.00 & 1.50&  0.06 &   0.03&  0.33 &   0.05&  1.27 &   0.09&  1.67 &   0.13\\ 
      &      & 1.50 & 2.00&  0.13 &   0.05&  0.29 &   0.04&  1.04 &   0.07&  1.27 &   0.09\\ 
      &      & 2.00 & 2.50&  0.17 &   0.06&  0.24 &   0.04&  0.70 &   0.07&  1.08 &   0.08\\ 
      &      & 2.50 & 3.00&       &       &  0.23 &   0.04&  0.36 &   0.06&  0.89 &   0.07\\ 
      &      & 3.00 & 3.50&       &       &  0.19 &   0.03&  0.37 &   0.06&  0.89 &   0.07\\ 
      &      & 3.50 & 4.00&       &       &  0.17 &   0.03&  0.29 &   0.05&  0.69 &   0.07\\ 
      &      & 4.00 & 5.00&       &       &  0.05 &   0.02&  0.18 &   0.03&  0.43 &   0.04\\ 
      &      & 5.00 & 6.50&       &       &       &       &  0.11 &   0.02&  0.22 &   0.03\\ 
      &      & 6.50 & 8.00&       &       &       &       &       &       & 0.104 &  0.014\\ 
 0.10 & 0.15 & 0.50 & 1.00&  0.13 &   0.04&  0.46 &   0.07&  1.37 &   0.14&  1.72 &   0.17\\ 
      &      & 1.00 & 1.50&  0.09 &   0.03&  0.38 &   0.05&  0.92 &   0.09&  1.48 &   0.13\\ 
      &      & 1.50 & 2.00&  0.10 &   0.05&  0.21 &   0.04&  0.68 &   0.07&  1.19 &   0.10\\ 
      &      & 2.00 & 2.50&  0.04 &   0.02&  0.23 &   0.04&  0.58 &   0.06&  0.97 &   0.09\\ 
      &      & 2.50 & 3.00&       &       &  0.15 &   0.03&  0.42 &   0.05&  0.78 &   0.07\\ 
      &      & 3.00 & 3.50&       &       &  0.11 &   0.02&  0.22 &   0.04&  0.62 &   0.06\\ 
      &      & 3.50 & 4.00&       &       &  0.06 &   0.02&  0.22 &   0.03&  0.48 &   0.05\\ 
      &      & 4.00 & 5.00&       &       & 0.020 &  0.010&  0.08 &   0.02&  0.24 &   0.03\\ 
      &      & 5.00 & 6.50&       &       &       &       & 0.019 &  0.007& 0.074 &  0.012\\ 
      &      & 6.50 & 8.00&       &       &       &       &       &       & 0.014 &  0.003\\ 
 0.15 & 0.20 & 0.50 & 1.00&  0.11 &   0.05&  0.45 &   0.08&  1.22 &   0.14&  1.78 &   0.19\\ 
      &      & 1.00 & 1.50&  0.12 &   0.05&  0.29 &   0.05&  0.64 &   0.07&  0.98 &   0.10\\ 
      &      & 1.50 & 2.00&  0.06 &   0.02&  0.24 &   0.04&  0.51 &   0.06&  0.90 &   0.09\\ 
      &      & 2.00 & 2.50&  0.06 &   0.04&  0.18 &   0.04&  0.41 &   0.05&  0.71 &   0.08\\ 
      &      & 2.50 & 3.00&       &       &  0.04 &   0.02&  0.27 &   0.04&  0.45 &   0.06\\ 
      &      & 3.00 & 3.50&       &       & 0.040 &  0.015&  0.16 &   0.03&  0.27 &   0.04\\ 
      &      & 3.50 & 4.00&       &       & 0.009 &  0.007&  0.10 &   0.02&  0.14 &   0.03\\ 
      &      & 4.00 & 5.00&       &       & 0.001 &  0.002& 0.024 &  0.007& 0.046 &  0.012\\ 
      &      & 5.00 & 6.50&       &       &       &       & 0.004 &  0.002& 0.011 &  0.005\\ 
      &      & 6.50 & 8.00&       &       &       &       &       &       & 0.002 &  0.001\\ 
 0.20 & 0.25 & 0.50 & 1.00&  0.25 &   0.11&  0.53 &   0.10&  1.01 &   0.14&  1.24 &   0.17\\ 
      &      & 1.00 & 1.50&  0.15 &   0.07&  0.51 &   0.09&  0.90 &   0.12&  1.32 &   0.17\\ 
      &      & 1.50 & 2.00&  0.07 &   0.05&  0.30 &   0.06&  0.79 &   0.12&  1.11 &   0.16\\ 
      &      & 2.00 & 2.50&  0.02 &   0.03&  0.20 &   0.04&  0.42 &   0.08&  0.59 &   0.10\\ 
      &      & 2.50 & 3.00&       &       &  0.05 &   0.02&  0.22 &   0.05&  0.32 &   0.06\\ 
      &      & 3.00 & 3.50&       &       & 0.009 &  0.005&  0.05 &   0.02&  0.12 &   0.03\\ 
      &      & 3.50 & 4.00&       &       & 0.005 &  0.004& 0.016 &  0.012&  0.05 &   0.02\\ 
      &      & 4.00 & 5.00&       &       & 0.003 &  0.003& 0.007 &  0.007& 0.015 &  0.013\\ 
      &      & 5.00 & 6.50&       &       &       &       & 0.002 &  0.002& 0.001 &  0.001\\ 
      &      & 6.50 & 8.00&       &       &       &       &       &       & 0.000 &  0.013\\ 

\hline
\end{tabular}
}
\end{table}
\clearpage
\begin{table}[!ht]
  \caption{\label{tab:xsec_results_Ta2}
    HARP results for the double-differential  $\pi^-$ production
    cross-section in the laboratory system,
    $d^2\sigma^{\pi}/(dpd\Omega)$, for $\pi^{-}$--Ta interactions at 3,5,8,12~\GeVc.
    Each row refers to a
    different $(p_{\hbox{\small min}} \le p<p_{\hbox{\small max}},
    \theta_{\hbox{\small min}} \le \theta<\theta_{\hbox{\small max}})$ bin,
    where $p$ and $\theta$ are the pion momentum and polar angle, respectively.
    The central value as well as the square-root of the diagonal elements
    of the covariance matrix are given.}

\small{
\begin{tabular}{rrrr|r@{$\pm$}lr@{$\pm$}lr@{$\pm$}lr@{$\pm$}l}
\hline
$\theta_{\hbox{\small min}}$ &
$\theta_{\hbox{\small max}}$ &
$p_{\hbox{\small min}}$ &
$p_{\hbox{\small max}}$ &
\multicolumn{8}{c}{$d^2\sigma^{\pi^-}/(dpd\Omega)$}
\\
(rad) & (rad) & (\GeVc) & (\GeVc) &
\multicolumn{8}{c}{(barn/(\GeVc sr))}
\\
  &  &  &
&\multicolumn{2}{c}{$ \bf{3 \ \GeVc}$}
&\multicolumn{2}{c}{$ \bf{5 \ \GeVc}$}
&\multicolumn{2}{c}{$ \bf{8 \ \GeVc}$}
&\multicolumn{2}{c}{$ \bf{12 \ \GeVc}$}
\\
\hline

 0.05 & 0.10 & 0.50 & 1.00&  0.22 &   0.06&  0.79 &   0.09&  1.48 &   0.13&  2.16 &   0.20\\ 
      &      & 1.00 & 1.50&  0.11 &   0.03&  0.50 &   0.06&  1.10 &   0.09&  1.80 &   0.14\\ 
      &      & 1.50 & 2.00&  0.09 &   0.02&  0.58 &   0.06&  1.22 &   0.08&  1.63 &   0.11\\ 
      &      & 2.00 & 2.50&  0.07 &   0.02&  0.48 &   0.06&  1.23 &   0.10&  1.59 &   0.11\\ 
      &      & 2.50 & 3.00&       &       &  0.61 &   0.07&  1.00 &   0.07&  1.49 &   0.10\\ 
      &      & 3.00 & 3.50&       &       &  0.63 &   0.07&  0.81 &   0.08&  1.27 &   0.08\\ 
      &      & 3.50 & 4.00&       &       &  0.55 &   0.07&  0.69 &   0.07&  1.02 &   0.07\\ 
      &      & 4.00 & 5.00&       &       &  0.57 &   0.10&  0.67 &   0.06&  0.77 &   0.05\\ 
      &      & 5.00 & 6.50&       &       &       &       &  0.49 &   0.05&  0.51 &   0.03\\ 
      &      & 6.50 & 8.00&       &       &       &       &       &       &  0.32 &   0.03\\ 
 0.10 & 0.15 & 0.50 & 1.00&  0.38 &   0.08&  0.77 &   0.10&  1.80 &   0.20&  2.56 &   0.27\\ 
      &      & 1.00 & 1.50&  0.22 &   0.05&  0.59 &   0.07&  1.14 &   0.10&  1.91 &   0.16\\ 
      &      & 1.50 & 2.00&  0.28 &   0.07&  0.45 &   0.06&  0.99 &   0.09&  1.68 &   0.13\\ 
      &      & 2.00 & 2.50&  0.45 &   0.09&  0.50 &   0.06&  0.91 &   0.08&  1.51 &   0.13\\ 
      &      & 2.50 & 3.00&       &       &  0.44 &   0.05&  0.67 &   0.06&  1.11 &   0.10\\ 
      &      & 3.00 & 3.50&       &       &  0.39 &   0.05&  0.38 &   0.05&  0.89 &   0.08\\ 
      &      & 3.50 & 4.00&       &       &  0.33 &   0.04&  0.31 &   0.04&  0.62 &   0.06\\ 
      &      & 4.00 & 5.00&       &       &  0.31 &   0.05&  0.21 &   0.03&  0.38 &   0.04\\ 
      &      & 5.00 & 6.50&       &       &       &       & 0.095 &  0.015&  0.17 &   0.02\\ 
      &      & 6.50 & 8.00&       &       &       &       &       &       & 0.058 &  0.010\\ 
 0.15 & 0.20 & 0.50 & 1.00&  0.34 &   0.10&  0.95 &   0.13&  1.43 &   0.17&  2.25 &   0.25\\ 
      &      & 1.00 & 1.50&  0.27 &   0.07&  0.45 &   0.06&  1.31 &   0.12&  1.89 &   0.17\\ 
      &      & 1.50 & 2.00&  0.22 &   0.07&  0.39 &   0.06&  0.88 &   0.09&  1.10 &   0.11\\ 
      &      & 2.00 & 2.50&  0.16 &   0.04&  0.26 &   0.05&  0.53 &   0.06&  0.81 &   0.08\\ 
      &      & 2.50 & 3.00&       &       &  0.12 &   0.02&  0.31 &   0.04&  0.46 &   0.05\\ 
      &      & 3.00 & 3.50&       &       &  0.11 &   0.02&  0.17 &   0.03&  0.40 &   0.05\\ 
      &      & 3.50 & 4.00&       &       &  0.07 &   0.02&  0.16 &   0.02&  0.21 &   0.03\\ 
      &      & 4.00 & 5.00&       &       &  0.06 &   0.02& 0.079 &  0.014&  0.16 &   0.02\\ 
      &      & 5.00 & 6.50&       &       &       &       & 0.034 &  0.007& 0.047 &  0.010\\ 
      &      & 6.50 & 8.00&       &       &       &       &       &       & 0.012 &  0.004\\ 
 0.20 & 0.25 & 0.50 & 1.00&  0.17 &   0.07&  0.50 &   0.09&  1.10 &   0.14&  1.22 &   0.16\\ 
      &      & 1.00 & 1.50&  0.11 &   0.05&  0.30 &   0.05&  0.40 &   0.06&  0.58 &   0.08\\ 
      &      & 1.50 & 2.00&  0.06 &   0.03&  0.23 &   0.05&  0.39 &   0.08&  0.64 &   0.10\\ 
      &      & 2.00 & 2.50&  0.03 &   0.02&  0.15 &   0.04&  0.28 &   0.05&  0.49 &   0.08\\ 
      &      & 2.50 & 3.00&       &       &  0.08 &   0.02&  0.21 &   0.04&  0.32 &   0.07\\ 
      &      & 3.00 & 3.50&       &       &  0.10 &   0.03&  0.15 &   0.03&  0.21 &   0.04\\ 
      &      & 3.50 & 4.00&       &       & 0.050 &  0.014&  0.08 &   0.02&  0.14 &   0.03\\ 
      &      & 4.00 & 5.00&       &       & 0.026 &  0.009&  0.06 &   0.02&  0.11 &   0.03\\ 
      &      & 5.00 & 6.50&       &       &       &       & 0.025 &  0.010& 0.035 &  0.010\\ 
      &      & 6.50 & 8.00&       &       &       &       &       &       & 0.012 &  0.008\\ 
\hline
\end{tabular}
}
\end{table}

%% file: xsec_results_Ta_pip_abs.tex
\begin{table}[!ht]
  \caption{\label{tab:xsec_results_Ta3}
    HARP results for the double-differential $\pi^+$  production
    cross-section in the laboratory system,
    $d^2\sigma^{\pi}/(dpd\Omega)$, for $\pi^{+}$--Ta interactions at 3,5,8,12~\GeVc.
    Each row refers to a
    different $(p_{\hbox{\small min}} \le p<p_{\hbox{\small max}},
    \theta_{\hbox{\small min}} \le \theta<\theta_{\hbox{\small max}})$ bin,
    where $p$ and $\theta$ are the pion momentum and polar angle, respectively.
    The central value as well as the square-root of the diagonal elements
    of the covariance matrix are given.}

\small{
\begin{tabular}{rrrr|r@{$\pm$}lr@{$\pm$}lr@{$\pm$}lr@{$\pm$}l}
\hline
$\theta_{\hbox{\small min}}$ &
$\theta_{\hbox{\small max}}$ &
$p_{\hbox{\small min}}$ &
$p_{\hbox{\small max}}$ &
\multicolumn{8}{c}{$d^2\sigma^{\pi^+}/(dpd\Omega)$}
\\
(rad) & (rad) & (\GeVc) & (\GeVc) &
\multicolumn{8}{c}{(barn/(\GeVc sr))}
\\
  &  &  &
&\multicolumn{2}{c}{$ \bf{3 \ \GeVc}$}
&\multicolumn{2}{c}{$ \bf{5 \ \GeVc}$}
&\multicolumn{2}{c}{$ \bf{8 \ \GeVc}$}
&\multicolumn{2}{c}{$ \bf{12 \ \GeVc}$}
\\
\hline
 
0.050 &0.100 & 0.50 & 1.00&  0.26 &   0.07&  0.58 &   0.10&  1.06 &   0.17&  1.60 &   0.51\\
      &      & 1.00 & 1.50& 0.032 &  0.009&  0.46 &   0.07&  1.03 &   0.12&  1.36 &   0.38\\
      &      & 1.50 & 2.00& 0.123 &  0.015&  0.39 &   0.06&  0.63 &   0.09&  1.62 &   0.44\\
      &      & 2.00 & 2.50&  0.35 &   0.03&  0.71 &   0.07&  0.71 &   0.10&  1.09 &   0.38\\
      &      & 2.50 & 3.00&       &       &  0.70 &   0.06&  0.73 &   0.09&  1.69 &   0.84\\
      &      & 3.00 & 3.50&       &       &  0.87 &   0.09&  0.61 &   0.08&  1.33 &   0.29\\
      &      & 3.50 & 4.00&       &       &  1.15 &   0.07&  0.59 &   0.08&  1.46 &   0.31\\
      &      & 4.00 & 5.00&       &       &  0.86 &   0.29&  0.81 &   0.06&  0.90 &   0.17\\
      &      & 5.00 & 6.50&       &       &       &       &  0.68 &   0.10&  0.66 &   0.12\\
      &      & 6.50 & 8.00&       &       &       &       &       &       &  0.55 &   0.11\\
0.100 &0.150 & 0.50 & 1.00&  0.16 &   0.06&  0.67 &   0.11&  1.38 &   0.18&  1.55 &   0.53\\
      &      & 1.00 & 1.50&  0.34 &   0.08&  0.43 &   0.08&  0.88 &   0.11&  0.94 &   0.31\\
      &      & 1.50 & 2.00&  0.37 &   0.06&  0.54 &   0.07&  0.94 &   0.12&  2.13 &   0.51\\
      &      & 2.00 & 2.50&  0.70 &   0.09&  0.54 &   0.07&  0.77 &   0.14&  0.92 &   0.28\\
      &      & 2.50 & 3.00&       &       &  0.40 &   0.05&  0.58 &   0.15&  1.14 &   0.30\\
      &      & 3.00 & 3.50&       &       &  0.53 &   0.06&  0.39 &   0.06&  0.75 &   0.22\\
      &      & 3.50 & 4.00&       &       &  0.40 &   0.05&  0.45 &   0.06&  0.56 &   0.18\\
      &      & 4.00 & 5.00&       &       &  0.36 &   0.14&  0.25 &   0.04&  0.52 &   0.15\\
      &      & 5.00 & 6.50&       &       &       &       &  0.08 &   0.02&  0.18 &   0.07\\
      &      & 6.50 & 8.00&       &       &       &       &       &       &  0.10 &   0.05\\
0.150 &0.200 & 0.50 & 1.00&  0.23 &   0.07&  0.79 &   0.13&  1.38 &   0.20&  1.11 &   0.49\\
      &      & 1.00 & 1.50&  0.28 &   0.06&  0.42 &   0.08&  0.70 &   0.10&  0.78 &   0.32\\
      &      & 1.50 & 2.00&  0.23 &   0.04&  0.36 &   0.06&  0.56 &   0.08&  1.33 &   0.41\\
      &      & 2.00 & 2.50&  0.21 &   0.04&  0.38 &   0.07&  0.51 &   0.08&  0.75 &   0.30\\
      &      & 2.50 & 3.00&       &       &  0.14 &   0.03&  0.28 &   0.06&  0.22 &   0.13\\
      &      & 3.00 & 3.50&       &       &  0.16 &   0.04&  0.25 &   0.04&  0.27 &   0.16\\
      &      & 3.50 & 4.00&       &       &  0.12 &   0.03&  0.21 &   0.04&  0.15 &   0.09\\
      &      & 4.00 & 5.00&       &       &  0.07 &   0.04&  0.11 &   0.03&  0.11 &   0.08\\
      &      & 5.00 & 6.50&       &       &       &       & 0.029 &  0.012& 0.072 &  0.076\\
      &      & 6.50 & 8.00&       &       &       &       &       &       & 0.042 &  0.045\\
0.200 &0.250 & 0.50 & 1.00&  0.30 &   0.13&  0.44 &   0.09&  0.69 &   0.13&  0.74 &   0.37\\
      &      & 1.00 & 1.50&  0.15 &   0.04&  0.24 &   0.06&  0.50 &   0.11&  0.96 &   0.56\\
      &      & 1.50 & 2.00&  0.13 &   0.04&  0.23 &   0.06&  0.37 &   0.09&  0.45 &   0.28\\
      &      & 2.00 & 2.50& 0.024 &  0.014&  0.17 &   0.05&  0.39 &   0.09&  1.09 &   0.53\\
      &      & 2.50 & 3.00&       &       &  0.09 &   0.03&  0.23 &   0.08&  0.48 &   0.38\\
      &      & 3.00 & 3.50&       &       &  0.07 &   0.02&  0.18 &   0.06&  0.23 &   0.20\\
      &      & 3.50 & 4.00&       &       &  0.05 &   0.02&  0.10 &   0.03& 0.154 &  0.237\\
      &      & 4.00 & 5.00&       &       &  0.03 &   0.02&  0.05 &   0.02&  0.17 &   0.27\\
      &      & 5.00 & 6.50&       &       &       &       &  0.02 &   0.02&  0.02 &   0.10\\
      &      & 6.50 & 8.00&       &       &       &       &       &       &  0.02 &   0.20\\

\hline
\end{tabular}
}
\end{table}
\clearpage
\begin{table}[!ht]
  \caption{\label{tab:xsec_results_Ta4}
    HARP results for the double-differential  $\pi^-$ production
    cross-section in the laboratory system,
    $d^2\sigma^{\pi}/(dpd\Omega)$, for $\pi^{+}$--Ta interactions at 3,5,8,12~\GeVc.
    Each row refers to a
    different $(p_{\hbox{\small min}} \le p<p_{\hbox{\small max}},
    \theta_{\hbox{\small min}} \le \theta<\theta_{\hbox{\small max}})$ bin,
    where $p$ and $\theta$ are the pion momentum and polar angle, respectively.
    The central value as well as the square-root of the diagonal elements
    of the covariance matrix are given.}

\small{
\begin{tabular}{rrrr|r@{$\pm$}lr@{$\pm$}lr@{$\pm$}lr@{$\pm$}l}
\hline
$\theta_{\hbox{\small min}}$ &
$\theta_{\hbox{\small max}}$ &
$p_{\hbox{\small min}}$ &
$p_{\hbox{\small max}}$ &
\multicolumn{8}{c}{$d^2\sigma^{\pi^-}/(dpd\Omega)$}
\\
(rad) & (rad) & (\GeVc) & (\GeVc) &
\multicolumn{8}{c}{(barn/(\GeVc sr))}
\\
  &  &  &
&\multicolumn{2}{c}{$ \bf{3 \ \GeVc}$}
&\multicolumn{2}{c}{$ \bf{5 \ \GeVc}$}
&\multicolumn{2}{c}{$ \bf{8 \ \GeVc}$}
&\multicolumn{2}{c}{$ \bf{12 \ \GeVc}$}
\\
\hline
0.050 &0.100 & 0.50 & 1.00&  0.32 &   0.08&  0.43 &   0.09&  0.79 &   0.13&  1.89 &   0.55\\
      &      & 1.00 & 1.50&  0.11 &   0.03&  0.36 &   0.06&  0.90 &   0.11&  1.19 &   0.35\\
      &      & 1.50 & 2.00&  0.11 &   0.03&  0.31 &   0.05&  0.77 &   0.09&  1.10 &   0.33\\
      &      & 2.00 & 2.50&  0.18 &   0.06&  0.23 &   0.03&  0.53 &   0.07&  0.98 &   0.30\\
      &      & 2.50 & 3.00&       &       &  0.24 &   0.04&  0.56 &   0.07&  0.88 &   0.26\\
      &      & 3.00 & 3.50&       &       &  0.25 &   0.04&  0.41 &   0.05&  1.27 &   0.31\\
      &      & 3.50 & 4.00&       &       &  0.22 &   0.03&  0.30 &   0.04&  0.65 &   0.20\\
      &      & 4.00 & 5.00&       &       &  0.12 &   0.03&  0.24 &   0.03&  0.36 &   0.12\\
      &      & 5.00 & 6.50&       &       &       &       & 0.107 &  0.015&  0.16 &   0.06\\
      &      & 6.50 & 8.00&       &       &       &       &       &       &  0.06 &   0.03\\
0.100 &0.150 & 0.50 & 1.00&  0.27 &   0.06&  0.70 &   0.12&  1.19 &   0.18&  2.02 &   0.60\\
      &      & 1.00 & 1.50&  0.17 &   0.04&  0.37 &   0.06&  0.78 &   0.10&  0.98 &   0.31\\
      &      & 1.50 & 2.00&  0.18 &   0.04&  0.28 &   0.05&  0.75 &   0.09&  1.21 &   0.36\\
      &      & 2.00 & 2.50& 0.044 &  0.014&  0.24 &   0.04&  0.48 &   0.07&  1.20 &   0.35\\
      &      & 2.50 & 3.00&       &       &  0.16 &   0.03&  0.43 &   0.06&  0.56 &   0.20\\
      &      & 3.00 & 3.50&       &       &  0.11 &   0.02&  0.27 &   0.04&  0.44 &   0.15\\
      &      & 3.50 & 4.00&       &       & 0.062 &  0.015&  0.17 &   0.03&  0.45 &   0.17\\
      &      & 4.00 & 5.00&       &       & 0.030 &  0.011&  0.11 &   0.02&  0.21 &   0.11\\
      &      & 5.00 & 6.50&       &       &       &       & 0.028 &  0.008&  0.08 &   0.06\\
      &      & 6.50 & 8.00&       &       &       &       &       &       &  0.02 &   0.02\\
0.150 &0.200 & 0.50 & 1.00&  0.29 &   0.08&  0.65 &   0.11&  1.23 &   0.18&  1.65 &   0.57\\
      &      & 1.00 & 1.50&  0.16 &   0.04&  0.41 &   0.06&  0.75 &   0.11&  1.21 &   0.39\\
      &      & 1.50 & 2.00&  0.09 &   0.03&  0.29 &   0.05&  0.60 &   0.09&  1.15 &   0.37\\
      &      & 2.00 & 2.50&  0.04 &   0.02&  0.19 &   0.03&  0.49 &   0.08&  0.70 &   0.28\\
      &      & 2.50 & 3.00&       &       &  0.12 &   0.03&  0.19 &   0.04&  0.42 &   0.21\\
      &      & 3.00 & 3.50&       &       & 0.041 &  0.013&  0.14 &   0.03&  0.27 &   0.16\\
      &      & 3.50 & 4.00&       &       & 0.016 &  0.007&  0.05 &   0.02&  0.18 &   0.12\\
      &      & 4.00 & 5.00&       &       & 0.006 &  0.004& 0.024 &  0.010&  0.15 &   0.11\\
      &      & 5.00 & 6.50&       &       &       &       & 0.003 &  0.003& 0.017 &  0.037\\
      &      & 6.50 & 8.00&       &       &       &       &       &       & 0.000 &  0.001\\
0.200 &0.250 & 0.50 & 1.00&  0.26 &   0.08&  0.52 &   0.11&  1.07 &   0.18&  1.24 &   0.54\\
      &      & 1.00 & 1.50&  0.18 &   0.06&  0.27 &   0.06&  0.72 &   0.13&  0.87 &   0.37\\
      &      & 1.50 & 2.00&  0.05 &   0.03&  0.11 &   0.04&  0.35 &   0.07&  1.61 &   0.74\\
      &      & 2.00 & 2.50& 0.009 &  0.008&  0.07 &   0.02&  0.12 &   0.04&  1.14 &   0.60\\
      &      & 2.50 & 3.00&       &       &  0.04 &   0.02& 0.026 &  0.012&  0.20 &   0.20\\
      &      & 3.00 & 3.50&       &       & 0.008 &  0.006& 0.016 &  0.013&  0.12 &   0.21\\
      &      & 3.50 & 4.00&       &       &  0.00 &   0.00&  0.00 &   0.01&  0.12 &   0.27\\
      &      & 4.00 & 5.00&       &       &  0.00 &   0.00&  0.00 &   0.01&  0.04 &   0.19\\
      &      & 5.00 & 6.50&       &       &       &       &  0.00 &   0.01&  0.00 &   0.04\\
      &      & 6.50 & 8.00&       &       &       &       &       &       &  0.00 &   0.10\\

\hline
\end{tabular}
}
\end{table}

%% file: xsec_results_Pb_pim.tex
\begin{table}[!ht]
  \caption{\label{tab:xsec_results_Pb1}
    HARP results for the double-differential $\pi^+$  production
    cross-section in the laboratory system,
    $d^2\sigma^{\pi}/(dpd\Omega)$, for $\pi^{-}$--Pb interactions at 3,5,8,12~\GeVc.
    Each row refers to a
    different $(p_{\hbox{\small min}} \le p<p_{\hbox{\small max}},
    \theta_{\hbox{\small min}} \le \theta<\theta_{\hbox{\small max}})$ bin,
    where $p$ and $\theta$ are the pion momentum and polar angle, respectively.
    The central value as well as the square-root of the diagonal elements
    of the covariance matrix are given.}

\small{
\begin{tabular}{rrrr|r@{$\pm$}lr@{$\pm$}lr@{$\pm$}lr@{$\pm$}l}
\hline
$\theta_{\hbox{\small min}}$ &
$\theta_{\hbox{\small max}}$ &
$p_{\hbox{\small min}}$ &
$p_{\hbox{\small max}}$ &
\multicolumn{8}{c}{$d^2\sigma^{\pi^+}/(dpd\Omega)$}
\\
(rad) & (rad) & (\GeVc) & (\GeVc) &
\multicolumn{8}{c}{(barn/(\GeVc sr))}
\\
  &  &  &
&\multicolumn{2}{c}{$ \bf{3 \ \GeVc}$}
&\multicolumn{2}{c}{$ \bf{5 \ \GeVc}$}
&\multicolumn{2}{c}{$ \bf{8 \ \GeVc}$}
&\multicolumn{2}{c}{$ \bf{12 \ \GeVc}$}
\\
\hline
 0.05 & 0.10 & 0.50 & 1.00&  0.04 &   0.02&  0.39 &   0.06&  0.99 &   0.11&  1.64 &   0.16\\ 
      &      & 1.00 & 1.50&  0.04 &   0.02&  0.42 &   0.05&  1.46 &   0.10&  1.75 &   0.12\\ 
      &      & 1.50 & 2.00&  0.10 &   0.05&  0.32 &   0.04&  1.08 &   0.07&  1.41 &   0.10\\ 
      &      & 2.00 & 2.50&  0.10 &   0.04&  0.20 &   0.03&  0.65 &   0.07&  1.19 &   0.08\\ 
      &      & 2.50 & 3.00&       &       &  0.14 &   0.03&  0.41 &   0.06&  0.94 &   0.07\\ 
      &      & 3.00 & 3.50&       &       &  0.19 &   0.03&  0.42 &   0.06&  0.96 &   0.07\\ 
      &      & 3.50 & 4.00&       &       &  0.15 &   0.03&  0.31 &   0.05&  0.64 &   0.07\\ 
      &      & 4.00 & 5.00&       &       &  0.06 &   0.02&  0.22 &   0.03&  0.37 &   0.03\\ 
      &      & 5.00 & 6.50&       &       &       &       &  0.10 &   0.02&  0.20 &   0.02\\ 
      &      & 6.50 & 8.00&       &       &       &       &       &       & 0.074 &  0.011\\ 
 0.10 & 0.15 & 0.50 & 1.00&  0.18 &   0.06&  0.46 &   0.07&  1.45 &   0.15&  1.89 &   0.18\\ 
      &      & 1.00 & 1.50&  0.05 &   0.02&  0.39 &   0.05&  0.95 &   0.09&  1.59 &   0.13\\ 
      &      & 1.50 & 2.00&  0.09 &   0.05&  0.27 &   0.04&  0.75 &   0.08&  1.21 &   0.10\\ 
      &      & 2.00 & 2.50&  0.06 &   0.03&  0.24 &   0.04&  0.65 &   0.07&  1.00 &   0.09\\ 
      &      & 2.50 & 3.00&       &       &  0.16 &   0.03&  0.37 &   0.05&  0.72 &   0.07\\ 
      &      & 3.00 & 3.50&       &       &  0.11 &   0.02&  0.25 &   0.04&  0.57 &   0.06\\ 
      &      & 3.50 & 4.00&       &       &  0.09 &   0.02&  0.17 &   0.03&  0.45 &   0.04\\ 
      &      & 4.00 & 5.00&       &       & 0.018 &  0.012&  0.08 &   0.02&  0.24 &   0.03\\ 
      &      & 5.00 & 6.50&       &       &       &       & 0.016 &  0.007& 0.085 &  0.014\\ 
      &      & 6.50 & 8.00&       &       &       &       &       &       & 0.018 &  0.004\\ 
 0.15 & 0.20 & 0.50 & 1.00&  0.02 &   0.02&  0.48 &   0.08&  1.29 &   0.15&  1.86 &   0.19\\ 
      &      & 1.00 & 1.50&  0.14 &   0.07&  0.26 &   0.04&  0.80 &   0.08&  0.99 &   0.09\\ 
      &      & 1.50 & 2.00& 0.020 &  0.011&  0.22 &   0.03&  0.68 &   0.07&  0.94 &   0.09\\ 
      &      & 2.00 & 2.50&  0.04 &   0.04&  0.14 &   0.03&  0.45 &   0.06&  0.73 &   0.07\\ 
      &      & 2.50 & 3.00&       &       &  0.07 &   0.02&  0.31 &   0.05&  0.43 &   0.05\\ 
      &      & 3.00 & 3.50&       &       &  0.07 &   0.02&  0.15 &   0.03&  0.37 &   0.04\\ 
      &      & 3.50 & 4.00&       &       & 0.011 &  0.009&  0.07 &   0.02&  0.17 &   0.03\\ 
      &      & 4.00 & 5.00&       &       & 0.001 &  0.001& 0.042 &  0.010&  0.08 &   0.02\\ 
      &      & 5.00 & 6.50&       &       &       &       & 0.008 &  0.003& 0.012 &  0.005\\ 
      &      & 6.50 & 8.00&       &       &       &       &       &       & 0.002 &  0.001\\ 
 0.20 & 0.25 & 0.50 & 1.00&  0.20 &   0.11&  0.63 &   0.10&  1.01 &   0.14&  1.44 &   0.18\\ 
      &      & 1.00 & 1.50&  0.10 &   0.06&  0.54 &   0.08&  0.97 &   0.13&  1.53 &   0.18\\ 
      &      & 1.50 & 2.00&  0.12 &   0.07&  0.33 &   0.06&  0.75 &   0.12&  1.20 &   0.16\\ 
      &      & 2.00 & 2.50&  0.02 &   0.03&  0.18 &   0.04&  0.36 &   0.07&  0.73 &   0.11\\ 
      &      & 2.50 & 3.00&       &       &  0.05 &   0.02&  0.21 &   0.05&  0.32 &   0.05\\ 
      &      & 3.00 & 3.50&       &       & 0.016 &  0.011&  0.06 &   0.03&  0.15 &   0.03\\ 
      &      & 3.50 & 4.00&       &       & 0.002 &  0.002& 0.019 &  0.014&  0.07 &   0.02\\ 
      &      & 4.00 & 5.00&       &       & 0.000 &  0.001& 0.012 &  0.010& 0.024 &  0.012\\ 
      &      & 5.00 & 6.50&       &       &       &       & 0.002 &  0.002& 0.003 &  0.002\\ 
      &      & 6.50 & 8.00&       &       &       &       &       &       & 0.001 &  0.009\\ 

\hline
\end{tabular}
}
\end{table}
\clearpage
\begin{table}[!ht]
  \caption{\label{tab:xsec_results_Pb2}
    HARP results for the double-differential  $\pi^-$ production
    cross-section in the laboratory system,
    $d^2\sigma^{\pi}/(dpd\Omega)$, for $\pi^{-}$--Pb interactions at 3,5,8,12~\GeVc.
    Each row refers to a
    different $(p_{\hbox{\small min}} \le p<p_{\hbox{\small max}},
    \theta_{\hbox{\small min}} \le \theta<\theta_{\hbox{\small max}})$ bin,
    where $p$ and $\theta$ are the pion momentum and polar angle, respectively.
    The central value as well as the square-root of the diagonal elements
    of the covariance matrix are given.}

\small{
\begin{tabular}{rrrr|r@{$\pm$}lr@{$\pm$}lr@{$\pm$}lr@{$\pm$}l}
\hline
$\theta_{\hbox{\small min}}$ &
$\theta_{\hbox{\small max}}$ &
$p_{\hbox{\small min}}$ &
$p_{\hbox{\small max}}$ &
\multicolumn{8}{c}{$d^2\sigma^{\pi^-}/(dpd\Omega)$}
\\
(rad) & (rad) & (\GeVc) & (\GeVc) &
\multicolumn{8}{c}{(barn/(\GeVc sr))}
\\
  &  &  &
&\multicolumn{2}{c}{$ \bf{3 \ \GeVc}$}
&\multicolumn{2}{c}{$ \bf{5 \ \GeVc}$}
&\multicolumn{2}{c}{$ \bf{8 \ \GeVc}$}
&\multicolumn{2}{c}{$ \bf{12 \ \GeVc}$}
\\
\hline

 0.05 & 0.10 & 0.50 & 1.00&  0.23 &   0.06&  0.73 &   0.08&  1.71 &   0.15&  2.20 &   0.19\\ 
      &      & 1.00 & 1.50&  0.07 &   0.02&  0.46 &   0.05&  1.29 &   0.10&  1.60 &   0.13\\ 
      &      & 1.50 & 2.00&  0.06 &   0.02&  0.49 &   0.05&  1.40 &   0.10&  1.70 &   0.11\\ 
      &      & 2.00 & 2.50&  0.06 &   0.02&  0.60 &   0.06&  1.18 &   0.08&  1.61 &   0.11\\ 
      &      & 2.50 & 3.00&       &       &  0.59 &   0.06&  1.00 &   0.08&  1.34 &   0.09\\ 
      &      & 3.00 & 3.50&       &       &  0.69 &   0.07&  0.81 &   0.08&  1.19 &   0.08\\ 
      &      & 3.50 & 4.00&       &       &  0.59 &   0.08&  0.72 &   0.08&  0.96 &   0.07\\ 
      &      & 4.00 & 5.00&       &       &  0.54 &   0.09&  0.61 &   0.06&  0.63 &   0.05\\ 
      &      & 5.00 & 6.50&       &       &       &       &  0.47 &   0.04&  0.47 &   0.03\\ 
      &      & 6.50 & 8.00&       &       &       &       &       &       &  0.32 &   0.03\\ 
 0.10 & 0.15 & 0.50 & 1.00&  0.25 &   0.06&  0.70 &   0.09&  1.95 &   0.21&  2.55 &   0.26\\ 
      &      & 1.00 & 1.50&  0.22 &   0.06&  0.61 &   0.07&  1.22 &   0.10&  2.08 &   0.16\\ 
      &      & 1.50 & 2.00&  0.25 &   0.07&  0.46 &   0.05&  1.09 &   0.10&  1.67 &   0.13\\ 
      &      & 2.00 & 2.50&  0.38 &   0.09&  0.49 &   0.06&  0.89 &   0.08&  1.37 &   0.11\\ 
      &      & 2.50 & 3.00&       &       &  0.37 &   0.05&  0.72 &   0.07&  0.99 &   0.09\\ 
      &      & 3.00 & 3.50&       &       &  0.39 &   0.05&  0.44 &   0.05&  0.83 &   0.07\\ 
      &      & 3.50 & 4.00&       &       &  0.34 &   0.05&  0.29 &   0.04&  0.55 &   0.05\\ 
      &      & 4.00 & 5.00&       &       &  0.29 &   0.04&  0.19 &   0.03&  0.40 &   0.04\\ 
      &      & 5.00 & 6.50&       &       &       &       & 0.083 &  0.013&  0.15 &   0.02\\ 
      &      & 6.50 & 8.00&       &       &       &       &       &       & 0.054 &  0.009\\ 
 0.15 & 0.20 & 0.50 & 1.00&  0.43 &   0.12&  0.81 &   0.11&  1.49 &   0.17&  2.30 &   0.24\\ 
      &      & 1.00 & 1.50&  0.34 &   0.10&  0.49 &   0.06&  1.23 &   0.12&  1.71 &   0.15\\ 
      &      & 1.50 & 2.00&  0.06 &   0.04&  0.44 &   0.05&  0.92 &   0.09&  1.15 &   0.11\\ 
      &      & 2.00 & 2.50& 0.038 &  0.014&  0.32 &   0.05&  0.54 &   0.06&  0.82 &   0.09\\ 
      &      & 2.50 & 3.00&       &       &  0.19 &   0.03&  0.38 &   0.04&  0.44 &   0.05\\ 
      &      & 3.00 & 3.50&       &       &  0.13 &   0.03&  0.22 &   0.03&  0.38 &   0.04\\ 
      &      & 3.50 & 4.00&       &       &  0.07 &   0.02&  0.16 &   0.02&  0.27 &   0.03\\ 
      &      & 4.00 & 5.00&       &       & 0.061 &  0.013&  0.10 &   0.02&  0.15 &   0.02\\ 
      &      & 5.00 & 6.50&       &       &       &       & 0.026 &  0.006& 0.044 &  0.008\\ 
      &      & 6.50 & 8.00&       &       &       &       &       &       & 0.015 &  0.004\\ 
 0.20 & 0.25 & 0.50 & 1.00&  0.18 &   0.07&  0.52 &   0.09&  1.01 &   0.14&  1.48 &   0.18\\ 
      &      & 1.00 & 1.50&  0.05 &   0.03&  0.25 &   0.04&  0.73 &   0.10&  0.68 &   0.09\\ 
      &      & 1.50 & 2.00&  0.04 &   0.02&  0.17 &   0.04&  0.56 &   0.10&  0.71 &   0.10\\ 
      &      & 2.00 & 2.50&  0.04 &   0.02&  0.09 &   0.02&  0.40 &   0.09&  0.62 &   0.10\\ 
      &      & 2.50 & 3.00&       &       &  0.06 &   0.02&  0.27 &   0.06&  0.37 &   0.06\\ 
      &      & 3.00 & 3.50&       &       &  0.07 &   0.02&  0.14 &   0.03&  0.30 &   0.05\\ 
      &      & 3.50 & 4.00&       &       & 0.058 &  0.014&  0.09 &   0.02&  0.22 &   0.04\\ 
      &      & 4.00 & 5.00&       &       & 0.031 &  0.010&  0.09 &   0.02&  0.12 &   0.02\\ 
      &      & 5.00 & 6.50&       &       &       &       & 0.030 &  0.010& 0.035 &  0.011\\ 
      &      & 6.50 & 8.00&       &       &       &       &       &       & 0.009 &  0.006\\ 
\hline
\end{tabular}
}
\end{table}

%% file: xsec_results_Pb_pip_abs.tex
\begin{table}[!ht]
  \caption{\label{tab:xsec_results_Pb3}
    HARP results for the double-differential $\pi^+$  production
    cross-section in the laboratory system,
    $d^2\sigma^{\pi}/(dpd\Omega)$, for $\pi^{+}$--Pb interactions at 3,5,8,12~\GeVc.
    Each row refers to a
    different $(p_{\hbox{\small min}} \le p<p_{\hbox{\small max}},
    \theta_{\hbox{\small min}} \le \theta<\theta_{\hbox{\small max}})$ bin,
    where $p$ and $\theta$ are the pion momentum and polar angle, respectively.
    The central value as well as the square-root of the diagonal elements
    of the covariance matrix are given.}

\small{
\begin{tabular}{rrrr|r@{$\pm$}lr@{$\pm$}lr@{$\pm$}lr@{$\pm$}l}
\hline
$\theta_{\hbox{\small min}}$ &
$\theta_{\hbox{\small max}}$ &
$p_{\hbox{\small min}}$ &
$p_{\hbox{\small max}}$ &
\multicolumn{8}{c}{$d^2\sigma^{\pi^+}/(dpd\Omega)$}
\\
(rad) & (rad) & (\GeVc) & (\GeVc) &
\multicolumn{8}{c}{(barn/(\GeVc sr))}
\\
  &  &  &
&\multicolumn{2}{c}{$ \bf{3 \ \GeVc}$}
&\multicolumn{2}{c}{$ \bf{5 \ \GeVc}$}
&\multicolumn{2}{c}{$ \bf{8 \ \GeVc}$}
&\multicolumn{2}{c}{$ \bf{12 \ \GeVc}$}
\\
\hline
0.050 &0.100 & 0.50 & 1.00&  0.20 &   0.06&  0.64 &   0.12&  0.90 &   0.16&  2.89 &   1.10\\
      &      & 1.00 & 1.50& 0.042 &  0.011&  0.38 &   0.07&  0.78 &   0.10&  0.82 &   0.42\\
      &      & 1.50 & 2.00& 0.093 &  0.011&  0.43 &   0.07&  0.77 &   0.10&  1.54 &   0.65\\
      &      & 2.00 & 2.50&  0.29 &   0.03&  0.72 &   0.08&  0.82 &   0.11&  0.73 &   0.36\\
      &      & 2.50 & 3.00&       &       &  0.75 &   0.07&  0.74 &   0.10&  1.64 &   0.53\\
      &      & 3.00 & 3.50&       &       &  0.89 &   0.09&  0.61 &   0.08&  1.55 &   0.50\\
      &      & 3.50 & 4.00&       &       &  1.04 &   0.07&  0.68 &   0.09&  1.38 &   0.45\\
      &      & 4.00 & 5.00&       &       &  0.78 &   0.26&  0.89 &   0.07&  0.73 &   0.23\\
      &      & 5.00 & 6.50&       &       &       &       &  0.72 &   0.11& * &   *\\
      &      & 6.50 & 8.00&       &       &       &       &       &       &  0.14 &   0.15\\
0.100 &0.150 & 0.50 & 1.00&  0.08 &   0.04&  0.87 &   0.14&  1.40 &   0.19&  2.46 &   0.91\\
      &      & 1.00 & 1.50&  0.18 &   0.06&  0.36 &   0.07&  1.00 &   0.15&  2.03 &   0.73\\
      &      & 1.50 & 2.00&  0.52 &   0.08&  0.55 &   0.08&  1.07 &   0.21&  0.54 &   0.31\\
      &      & 2.00 & 2.50&  0.69 &   0.09&  0.52 &   0.07&  0.89 &   0.36&  1.20 &   0.65\\
      &      & 2.50 & 3.00&       &       &  0.49 &   0.06&  0.49 &   0.43&  0.88 &   0.53\\
      &      & 3.00 & 3.50&       &       &  0.66 &   0.08&  0.57 &   0.12&  0.70 &   0.37\\
      &      & 3.50 & 4.00&       &       &  0.57 &   0.06&  0.40 &   0.06&  0.53 &   0.26\\
      &      & 4.00 & 5.00&       &       &  0.35 &   0.13&  0.29 &   0.05&  0.39 &   0.19\\
      &      & 5.00 & 6.50&       &       &       &       &  0.12 &   0.03&  0.24 &   0.16\\
      &      & 6.50 & 8.00&       &       &       &       &       &       & 0.027 &  0.035\\
0.150 &0.200 & 0.50 & 1.00&  0.18 &   0.06&  0.79 &   0.14&  1.35 &   0.19&  2.46 &   1.08\\
      &      & 1.00 & 1.50&  0.26 &   0.06&  0.64 &   0.10&  0.83 &   0.11&  0.97 &   0.60\\
      &      & 1.50 & 2.00&  0.31 &   0.06&  0.37 &   0.06&  0.61 &   0.09& * &   *\\
      &      & 2.00 & 2.50&  0.26 &   0.05&  0.40 &   0.07&  0.53 &   0.09& * &   *\\
      &      & 2.50 & 3.00&       &       &  0.18 &   0.04&  0.30 &   0.07&  0.19 &   0.29\\
      &      & 3.00 & 3.50&       &       &  0.20 &   0.06&  0.26 &   0.05&  0.45 &   0.41\\
      &      & 3.50 & 4.00&       &       &  0.15 &   0.04&  0.12 &   0.03& 0.210 &  0.250\\
      &      & 4.00 & 5.00&       &       &  0.09 &   0.05&  0.08 &   0.02& 0.127 &  0.185\\
      &      & 5.00 & 6.50&       &       &       &       & 0.031 &  0.013& 0.145 &  0.162\\
      &      & 6.50 & 8.00&       &       &       &       &       &       & 0.020 &  0.049\\
0.200 &0.250 & 0.50 & 1.00&  0.26 &   0.14&  0.55 &   0.11&  0.86 &   0.15& 0.057 &  0.212\\
      &      & 1.00 & 1.50&  0.11 &   0.04&  0.39 &   0.08&  0.42 &   0.09&  0.06 &   0.73\\
      &      & 1.50 & 2.00&  0.16 &   0.05&  0.31 &   0.07&  0.43 &   0.10&  2.18 &   0.64\\
      &      & 2.00 & 2.50&  0.06 &   0.03&  0.22 &   0.06&  0.31 &   0.08&  2.25 &   1.17\\
      &      & 2.50 & 3.00&       &       &  0.16 &   0.05&  0.23 &   0.08&  0.13 &   0.65\\
      &      & 3.00 & 3.50&       &       &  0.07 &   0.03&  0.11 &   0.04& 0.075 &  0.359\\
      &      & 3.50 & 4.00&       &       &  0.03 &   0.02&  0.10 &   0.03& 0.026 &  0.488\\
      &      & 4.00 & 5.00&       &       &  0.03 &   0.02&  0.08 &   0.03&* &  *\\
      &      & 5.00 & 6.50&       &       &       &       &  0.02 &   0.02&  0.08 &   0.24\\

\hline
\end{tabular}
}
\end{table}
\clearpage
\begin{table}[!ht]
  \caption{\label{tab:xsec_results_Pb4}
    HARP results for the double-differential  $\pi^-$ production
    cross-section in the laboratory system,
    $d^2\sigma^{\pi}/(dpd\Omega)$, for $\pi^{+}$--Pb interactions at 3,5,8,12~\GeVc.
    Each row refers to a
    different $(p_{\hbox{\small min}} \le p<p_{\hbox{\small max}},
    \theta_{\hbox{\small min}} \le \theta<\theta_{\hbox{\small max}})$ bin,
    where $p$ and $\theta$ are the pion momentum and polar angle, respectively.
    The central value as well as the square-root of the diagonal elements
    of the covariance matrix are given.}

\small{
\begin{tabular}{rrrr|r@{$\pm$}lr@{$\pm$}lr@{$\pm$}lr@{$\pm$}l}
\hline
$\theta_{\hbox{\small min}}$ &
$\theta_{\hbox{\small max}}$ &
$p_{\hbox{\small min}}$ &
$p_{\hbox{\small max}}$ &
\multicolumn{8}{c}{$d^2\sigma^{\pi^-}/(dpd\theta)$}
\\
(rad) & (rad) & (\GeVc) & (\GeVc) &
\multicolumn{8}{c}{(barn/(\GeVc rad))}
\\
  &  &  &
&\multicolumn{2}{c}{$ \bf{3 \ \GeVc}$}
&\multicolumn{2}{c}{$ \bf{5 \ \GeVc}$}
&\multicolumn{2}{c}{$ \bf{8 \ \GeVc}$}
&\multicolumn{2}{c}{$ \bf{12 \ \GeVc}$}
\\
\hline
0.050 &0.100 & 0.50 & 1.00&  0.26 &   0.07&  0.60 &   0.11&  0.98 &   0.14&  0.93 &   0.62\\
      &      & 1.00 & 1.50&  0.17 &   0.04&  0.53 &   0.07&  1.00 &   0.12&  1.92 &   0.70\\
      &      & 1.50 & 2.00&  0.28 &   0.05&  0.39 &   0.06&  0.79 &   0.10&  1.34 &   0.53\\
      &      & 2.00 & 2.50&  0.20 &   0.06&  0.31 &   0.04&  0.48 &   0.07&  0.79 &   0.36\\
      &      & 2.50 & 3.00&       &       &  0.31 &   0.05&  0.51 &   0.07&  0.64 &   0.31\\
      &      & 3.00 & 3.50&       &       &  0.23 &   0.04&  0.41 &   0.05&  1.38 &   0.48\\
      &      & 3.50 & 4.00&       &       &  0.26 &   0.04&  0.39 &   0.05&  0.96 &   0.36\\
      &      & 4.00 & 5.00&       &       &  0.13 &   0.03&  0.29 &   0.03&  0.40 &   0.18\\
      &      & 5.00 & 6.50&       &       &       &       &  0.13 &   0.02&  0.27 &   0.13\\
      &      & 6.50 & 8.00&       &       &       &       &       &       &  0.06 &   0.05\\
0.100 &0.150 & 0.50 & 1.00&  0.27 &   0.06&  0.66 &   0.12&  1.36 &   0.20&  2.21 &   0.90\\
      &      & 1.00 & 1.50&  0.07 &   0.02&  0.42 &   0.07&  0.81 &   0.10&  1.27 &   0.58\\
      &      & 1.50 & 2.00&  0.07 &   0.03&  0.30 &   0.05&  0.79 &   0.10&  0.70 &   0.37\\
      &      & 2.00 & 2.50& 0.043 &  0.014&  0.25 &   0.04&  0.39 &   0.06&  1.70 &   0.65\\
      &      & 2.50 & 3.00&       &       &  0.20 &   0.03&  0.39 &   0.06&  0.18 &   0.14\\
      &      & 3.00 & 3.50&       &       &  0.10 &   0.02&  0.28 &   0.05&  0.58 &   0.31\\
      &      & 3.50 & 4.00&       &       & 0.046 &  0.012&  0.27 &   0.05&  0.66 &   0.35\\
      &      & 4.00 & 5.00&       &       &  0.05 &   0.02&  0.10 &   0.02&  0.26 &   0.20\\
      &      & 5.00 & 6.50&       &       &       &       & 0.015 &  0.005&  0.04 &   0.07\\
      &      & 6.50 & 8.00&       &       &       &       &       &       & 0.023 &  0.039\\
0.150 &0.200 & 0.50 & 1.00&  0.34 &   0.09&  0.79 &   0.13&  1.21 &   0.19&  0.68 &   0.55\\
      &      & 1.00 & 1.50&  0.20 &   0.05&  0.43 &   0.07&  0.76 &   0.11&  1.56 &   0.76\\
      &      & 1.50 & 2.00&  0.11 &   0.03&  0.31 &   0.05&  0.65 &   0.09&  0.82 &   0.48\\
      &      & 2.00 & 2.50&  0.06 &   0.02&  0.28 &   0.05&  0.40 &   0.07&  0.84 &   0.43\\
      &      & 2.50 & 3.00&       &       &  0.09 &   0.02&  0.18 &   0.04&  0.66 &   0.38\\
      &      & 3.00 & 3.50&       &       & 0.036 &  0.013&  0.13 &   0.03&  0.30 &   0.23\\
      &      & 3.50 & 4.00&       &       & 0.015 &  0.008&  0.07 &   0.02&  0.52 &   0.42\\
      &      & 4.00 & 5.00&       &       & 0.002 &  0.002& 0.019 &  0.009&  0.13 &   0.19\\
      &      & 5.00 & 6.50&       &       &       &       & 0.004 &  0.003& 0.001 &  0.006\\
      &      & 6.50 & 8.00&       &       &       &       &       &       & 0.005 &  0.023\\
0.200 &0.250 & 0.50 & 1.00&  0.21 &   0.08&  0.59 &   0.13&  0.90 &   0.16&  0.78 &   0.65\\
      &      & 1.00 & 1.50&  0.14 &   0.05&  0.36 &   0.08&  0.83 &   0.15&  0.03 &   0.06\\
      &      & 1.50 & 2.00&  0.07 &   0.04&  0.14 &   0.05&  0.38 &   0.09&  0.67 &   1.08\\
      &      & 2.00 & 2.50&   &   &  0.09 &   0.03&  0.11 &   0.04&  0.72 &   0.77\\
      &      & 2.50 & 3.00&       &       & 0.033 &  0.014&  0.06 &   0.02&  0.11 &   0.16\\
      &      & 3.00 & 3.50&       &       & 0.013 &  0.008&  0.02 &   0.02&  0.10 &   0.44\\
      &      & 3.50 & 4.00&       &       & 0.013 &  0.010&  0.00 &   0.01&  0.09 &   0.93\\
      &      & 4.00 & 5.00&       &       &  0.00 &   0.01&  0.00 &   0.01&  0.02 &   0.35\\
      &      & 5.00 & 6.50&       &       &       &       &   &   &  0.00 &   0.05\\
      &      & 6.50 & 8.00&       &       &       &       &       &       & 0.00 &   0.20\\

\hline
\end{tabular}
}
\end{table}

%% file: table_result_fw_036-pim_Be_OMEGA_fine.tex
\begin{table}[!ht]
  \caption{\label{tab:xsec_results_
Be7} 
    HARP results for the double-differential $\pi^+$  and $\pi^-$ production
    cross-section in the laboratory system,
    $d^2\sigma^{\pi}/(dpd\Omega)$, for $\pi^{-}$--
Be 
    interactions at 8 and 12~\GeVc.
    Each row refers to a
    different $(p_{\hbox{\small min}} \le p<p_{\hbox{\small max}},
    \theta_{\hbox{\small min}} \le \theta<\theta_{\hbox{\small max}})$ bin,
    where $p$ and $\theta$ are the pion momentum and polar angle, respectively.
    A finer angular binning than in the previous set of tables is used.
    The central value as well as the square-root of the diagonal elements
    of the covariance matrix are given.}

\small{
\begin{tabular}{rrrr|r@{$\pm$}lr@{$\pm$}l|r@{$\pm$}lr@{$\pm$}l}
\hline
$\theta_{\hbox{\small min}}$ &
$\theta_{\hbox{\small max}}$ &
$p_{\hbox{\small min}}$ &
$p_{\hbox{\small max}}$ &
\multicolumn{4}{c}{$d^2\sigma^{\pi^+}/(dpd\Omega)$} &
\multicolumn{4}{c}{$d^2\sigma^{\pi^-}/(dpd\Omega)$}
\\
(rad) & (rad) & (\GeVc) & (\GeVc) &
\multicolumn{4}{c}{(barn/(\GeVc rad))} &
\multicolumn{4}{c}{(barn/(\GeVc rad))}
\\
  &  &  &
&\multicolumn{2}{c}{$ \bf{8 \ \GeVc}$}
&\multicolumn{2}{c}{$ \bf{12 \ \GeVc}$}
&\multicolumn{2}{c}{$ \bf{8 \ \GeVc}$}
&\multicolumn{2}{c}{$ \bf{12 \ \GeVc}$}
\\
\hline
0.025 &0.050 & 0.50 & 0.75&  0.12 &   0.03&  0.07 &   0.02&  0.20 &   0.05&  0.13 &   0.03\\ 
      &      & 0.75 & 1.00&  0.14 &   0.02&  0.11 &   0.02&  0.18 &   0.03&  0.15 &   0.02\\ 
      &      & 1.00 & 1.25&  0.18 &   0.02&  0.09 &   0.02&  0.17 &   0.03&  0.12 &   0.02\\ 
      &      & 1.25 & 1.50&  0.20 &   0.02&  0.13 &   0.03&  0.17 &   0.03&  0.14 &   0.02\\ 
      &      & 1.50 & 2.00&  0.16 &   0.02&  0.13 &   0.02&  0.21 &   0.02&  0.17 &   0.02\\ 
      &      & 2.00 & 2.50&  0.14 &   0.02&  0.13 &   0.02&  0.22 &   0.02&  0.17 &   0.02\\ 
      &      & 2.50 & 3.00&  0.12 &   0.02&  0.15 &   0.02&  0.19 &   0.02&  0.14 &   0.02\\ 
      &      & 3.00 & 3.50&  0.09 &   0.02&  0.12 &   0.02&  0.23 &   0.02&  0.16 &   0.02\\ 
      &      & 3.50 & 4.00&  0.08 &   0.02& 0.094 &  0.014&  0.22 &   0.02& 0.132 &  0.014\\ 
      &      & 4.00 & 5.00& 0.089 &  0.013& 0.064 &  0.009&  0.22 &   0.02& 0.129 &  0.012\\ 
      &      & 5.00 & 6.50& 0.098 &  0.011& 0.065 &  0.007&  0.25 &   0.02& 0.117 &  0.010\\ 
      &      & 6.50 & 8.00&       &       & 0.053 &  0.006&       &       & 0.142 &  0.010\\ 
0.050 &0.075 & 0.50 & 0.75&  0.15 &   0.03&  0.08 &   0.02&  0.28 &   0.04&  0.17 &   0.03\\ 
      &      & 0.75 & 1.00&  0.14 &   0.02&  0.10 &   0.02&  0.18 &   0.02&  0.11 &   0.02\\ 
      &      & 1.00 & 1.25&  0.18 &   0.02&  0.12 &   0.02&  0.18 &   0.02&  0.13 &   0.02\\ 
      &      & 1.25 & 1.50&  0.17 &   0.02&  0.13 &   0.02&  0.19 &   0.02&  0.16 &   0.02\\ 
      &      & 1.50 & 2.00& 0.145 &  0.012& 0.119 &  0.012&  0.20 &   0.02&  0.15 &   0.02\\ 
      &      & 2.00 & 2.50& 0.105 &  0.012& 0.133 &  0.012&  0.21 &   0.02& 0.159 &  0.014\\ 
      &      & 2.50 & 3.00& 0.079 &  0.012& 0.088 &  0.009&  0.21 &   0.02& 0.158 &  0.014\\ 
      &      & 3.00 & 3.50& 0.086 &  0.014& 0.106 &  0.011&  0.18 &   0.02& 0.163 &  0.014\\ 
      &      & 3.50 & 4.00& 0.103 &  0.012& 0.096 &  0.010&  0.17 &   0.02& 0.135 &  0.011\\ 
      &      & 4.00 & 5.00& 0.068 &  0.007& 0.056 &  0.006&  0.21 &   0.02& 0.131 &  0.009\\ 
      &      & 5.00 & 6.50& 0.043 &  0.005& 0.037 &  0.004& 0.174 &  0.012& 0.100 &  0.006\\ 
      &      & 6.50 & 8.00&       &       & 0.018 &  0.003&       &       & 0.086 &  0.005\\ 
0.075 &0.100 & 0.50 & 0.75&  0.09 &   0.02&  0.10 &   0.02&  0.30 &   0.04&  0.20 &   0.03\\ 
      &      & 0.75 & 1.00&  0.17 &   0.02& 0.128 &  0.015&  0.22 &   0.02& 0.109 &  0.013\\ 
      &      & 1.00 & 1.25&  0.19 &   0.02& 0.124 &  0.013&  0.19 &   0.02&  0.15 &   0.02\\ 
      &      & 1.25 & 1.50& 0.167 &  0.014& 0.141 &  0.013&  0.22 &   0.02&  0.14 &   0.02\\ 
      &      & 1.50 & 2.00& 0.139 &  0.010& 0.133 &  0.012&  0.22 &   0.02& 0.185 &  0.013\\ 
      &      & 2.00 & 2.50& 0.117 &  0.010& 0.122 &  0.010& 0.222 &  0.015& 0.172 &  0.013\\ 
      &      & 2.50 & 3.00& 0.098 &  0.009& 0.085 &  0.007& 0.184 &  0.012& 0.189 &  0.013\\ 
      &      & 3.00 & 3.50& 0.086 &  0.008& 0.091 &  0.009& 0.147 &  0.012& 0.133 &  0.009\\ 
      &      & 3.50 & 4.00& 0.050 &  0.009& 0.080 &  0.009& 0.146 &  0.013& 0.124 &  0.008\\ 
      &      & 4.00 & 5.00& 0.030 &  0.004& 0.043 &  0.004& 0.123 &  0.009& 0.086 &  0.006\\ 
      &      & 5.00 & 6.50& 0.009 &  0.002& 0.019 &  0.002& 0.078 &  0.005& 0.053 &  0.004\\ 
      &      & 6.50 & 8.00&       &       & 0.005 &  0.001&       &       & 0.030 &  0.002\\ 
\hline
\end{tabular}
}
\end{table}

%% file: table_result_fw_036-pip_Be_OMEGA_fine.tex
\begin{table}[!ht]
  \caption{\label{tab:xsec_results_
Be7} 
    HARP results for the double-differential $\pi^+$  and $\pi^-$ production
    cross-section in the laboratory system,
    $d^2\sigma^{\pi}/(dpd\Omega)$, for $\pi^{+}$--
Be 
    interactions at 8 and 8.9~\GeVc.
    Each row refers to a
    different $(p_{\hbox{\small min}} \le p<p_{\hbox{\small max}},
    \theta_{\hbox{\small min}} \le \theta<\theta_{\hbox{\small max}})$ bin,
    where $p$ and $\theta$ are the pion momentum and polar angle, respectively.
    A finer angular binning than in the previous set of tables is used.
    The central value as well as the square-root of the diagonal elements
    of the covariance matrix are given.}

\small{
\begin{tabular}{rrrr|r@{$\pm$}lr@{$\pm$}l|r@{$\pm$}lr@{$\pm$}l}
\hline
$\theta_{\hbox{\small min}}$ &
$\theta_{\hbox{\small max}}$ &
$p_{\hbox{\small min}}$ &
$p_{\hbox{\small max}}$ &
\multicolumn{4}{c}{$d^2\sigma^{\pi^+}/(dpd\Omega)$} &
\multicolumn{4}{c}{$d^2\sigma^{\pi^-}/(dpd\Omega)$}
\\
(rad) & (rad) & (\GeVc) & (\GeVc) &
\multicolumn{4}{c}{(barn/(\GeVc rad))} &
\multicolumn{4}{c}{(barn/(\GeVc rad))}
\\
  &  &  &
&\multicolumn{2}{c}{$ \bf{8 \ \GeVc}$}
&\multicolumn{2}{c}{$ \bf{8.9 \ \GeVc}$}
&\multicolumn{2}{c}{$ \bf{8 \ \GeVc}$}
&\multicolumn{2}{c}{$ \bf{8.9 \ \GeVc}$}
\\
\hline
0.025 &0.050 & 0.50 & 0.75&  0.10 &   0.04&  0.13 &   0.04&  0.13 &   0.05&  0.10 &   0.03\\ 
      &      & 0.75 & 1.00&  0.14 &   0.03&  0.09 &   0.03&  0.13 &   0.03&  0.12 &   0.03\\ 
      &      & 1.00 & 1.25&  0.13 &   0.03&  0.14 &   0.03&  0.15 &   0.04&  0.11 &   0.03\\ 
      &      & 1.25 & 1.50&  0.11 &   0.03&  0.13 &   0.03&  0.13 &   0.04&  0.16 &   0.03\\ 
      &      & 1.50 & 2.00&  0.16 &   0.03&  0.15 &   0.03&  0.18 &   0.03&  0.13 &   0.02\\ 
      &      & 2.00 & 2.50&  0.08 &   0.02&  0.15 &   0.02&  0.13 &   0.02&  0.12 &   0.02\\ 
      &      & 2.50 & 3.00&  0.15 &   0.03&  0.19 &   0.02&  0.10 &   0.02&  0.07 &   0.02\\ 
      &      & 3.00 & 3.50&  0.16 &   0.02&  0.15 &   0.02&  0.11 &   0.02&  0.10 &   0.02\\ 
      &      & 3.50 & 4.00&  0.19 &   0.03&  0.21 &   0.02& 0.081 &  0.014& 0.081 &  0.015\\ 
      &      & 4.00 & 5.00&  0.18 &   0.02&  0.21 &   0.02& 0.104 &  0.014& 0.085 &  0.012\\ 
      &      & 5.00 & 6.50&  0.27 &   0.02&  0.22 &   0.02& 0.092 &  0.012& 0.079 &  0.009\\ 
0.050 &0.075 & 0.50 & 0.75&  0.12 &   0.03&  0.12 &   0.03&  0.10 &   0.03&  0.09 &   0.02\\ 
      &      & 0.75 & 1.00&  0.13 &   0.03&  0.10 &   0.02&  0.10 &   0.02&  0.11 &   0.02\\ 
      &      & 1.00 & 1.25&  0.14 &   0.03&  0.11 &   0.02&  0.10 &   0.02&  0.12 &   0.02\\ 
      &      & 1.25 & 1.50&  0.17 &   0.03&  0.16 &   0.02&  0.10 &   0.02&  0.12 &   0.02\\ 
      &      & 1.50 & 2.00&  0.12 &   0.02&  0.15 &   0.02&  0.14 &   0.02&  0.14 &   0.02\\ 
      &      & 2.00 & 2.50&  0.14 &   0.02&  0.14 &   0.02&  0.13 &   0.02& 0.129 &  0.014\\ 
      &      & 2.50 & 3.00&  0.16 &   0.02&  0.14 &   0.02&  0.11 &   0.02& 0.096 &  0.013\\ 
      &      & 3.00 & 3.50&  0.12 &   0.02&  0.17 &   0.02& 0.083 &  0.013& 0.106 &  0.011\\ 
      &      & 3.50 & 4.00&  0.12 &   0.02&  0.17 &   0.02& 0.079 &  0.012& 0.084 &  0.009\\ 
      &      & 4.00 & 5.00&  0.21 &   0.02& 0.173 &  0.012& 0.064 &  0.009& 0.062 &  0.007\\ 
      &      & 5.00 & 6.50& 0.203 &  0.013& 0.136 &  0.010& 0.047 &  0.006& 0.035 &  0.004\\ 
0.075 &0.100 & 0.50 & 0.75&  0.11 &   0.03&  0.15 &   0.03&  0.09 &   0.02&  0.12 &   0.02\\ 
      &      & 0.75 & 1.00&  0.17 &   0.03&  0.15 &   0.02&  0.14 &   0.02&  0.08 &   0.02\\ 
      &      & 1.00 & 1.25&  0.13 &   0.02&  0.18 &   0.02&  0.20 &   0.03&  0.14 &   0.02\\ 
      &      & 1.25 & 1.50&  0.19 &   0.03&  0.18 &   0.02&  0.16 &   0.02&  0.16 &   0.02\\ 
      &      & 1.50 & 2.00&  0.14 &   0.02&  0.19 &   0.02& 0.104 &  0.014& 0.132 &  0.012\\ 
      &      & 2.00 & 2.50&  0.17 &   0.02& 0.188 &  0.014& 0.105 &  0.014& 0.118 &  0.010\\ 
      &      & 2.50 & 3.00&  0.20 &   0.02& 0.150 &  0.011& 0.067 &  0.009& 0.088 &  0.008\\ 
      &      & 3.00 & 3.50&  0.15 &   0.02& 0.168 &  0.013& 0.100 &  0.012& 0.078 &  0.008\\ 
      &      & 3.50 & 4.00&  0.13 &   0.02& 0.143 &  0.011& 0.055 &  0.008& 0.061 &  0.006\\ 
      &      & 4.00 & 5.00& 0.128 &  0.010& 0.109 &  0.009& 0.041 &  0.005& 0.036 &  0.004\\ 
      &      & 5.00 & 6.50&  0.09 &   0.03& 0.074 &  0.005& 0.010 &  0.002& 0.011 &  0.002\\ 
\hline
\end{tabular}
}
\end{table}

%% file: table_result_fw_036-pim_C_OMEGA_fine.tex
\begin{table}[!ht]
  \caption{\label{tab:xsec_results_
C7} 
    HARP results for the double-differential $\pi^+$  and $\pi^-$ production
    cross-section in the laboratory system,
    $d^2\sigma^{\pi}/(dpd\Omega)$, for $\pi^{-}$--
C 
    interactions at 8 and 12~\GeVc.
    Each row refers to a
    different $(p_{\hbox{\small min}} \le p<p_{\hbox{\small max}},
    \theta_{\hbox{\small min}} \le \theta<\theta_{\hbox{\small max}})$ bin,
    where $p$ and $\theta$ are the pion momentum and polar angle, respectively.
    A finer angular binning than in the previous set of tables is used.
    The central value as well as the square-root of the diagonal elements
    of the covariance matrix are given.}

\small{
\begin{tabular}{rrrr|r@{$\pm$}lr@{$\pm$}l|r@{$\pm$}lr@{$\pm$}l}
\hline
$\theta_{\hbox{\small min}}$ &
$\theta_{\hbox{\small max}}$ &
$p_{\hbox{\small min}}$ &
$p_{\hbox{\small max}}$ &
\multicolumn{4}{c}{$d^2\sigma^{\pi^+}/(dpd\Omega)$} &
\multicolumn{4}{c}{$d^2\sigma^{\pi^-}/(dpd\Omega)$}
\\
(rad) & (rad) & (\GeVc) & (\GeVc) &
\multicolumn{4}{c}{(barn/(\GeVc rad))} &
\multicolumn{4}{c}{(barn/(\GeVc rad))}
\\
  &  &  &
&\multicolumn{2}{c}{$ \bf{8 \ \GeVc}$}
&\multicolumn{2}{c}{$ \bf{12 \ \GeVc}$}
&\multicolumn{2}{c}{$ \bf{8 \ \GeVc}$}
&\multicolumn{2}{c}{$ \bf{12 \ \GeVc}$}
\\
\hline
0.025 &0.050 & 0.50 & 0.75&  0.06 &   0.02&  0.13 &   0.04&  0.22 &   0.06&  0.21 &   0.07\\ 
      &      & 0.75 & 1.00&  0.15 &   0.03&  0.25 &   0.05&  0.24 &   0.04&  0.30 &   0.05\\ 
      &      & 1.00 & 1.25&  0.20 &   0.03&  0.12 &   0.03&  0.18 &   0.03&  0.24 &   0.05\\ 
      &      & 1.25 & 1.50&  0.24 &   0.03&  0.22 &   0.05&  0.23 &   0.04&  0.22 &   0.04\\ 
      &      & 1.50 & 2.00&  0.17 &   0.03&  0.25 &   0.04&  0.29 &   0.03&  0.31 &   0.04\\ 
      &      & 2.00 & 2.50&  0.16 &   0.02&  0.26 &   0.04&  0.24 &   0.02&  0.34 &   0.05\\ 
      &      & 2.50 & 3.00&  0.15 &   0.03&  0.24 &   0.03&  0.21 &   0.03&  0.25 &   0.03\\ 
      &      & 3.00 & 3.50&  0.05 &   0.02&  0.22 &   0.03&  0.23 &   0.03&  0.33 &   0.04\\ 
      &      & 3.50 & 4.00&  0.11 &   0.02&  0.17 &   0.03&  0.17 &   0.03&  0.24 &   0.03\\ 
      &      & 4.00 & 5.00&  0.08 &   0.02&  0.13 &   0.02&  0.23 &   0.03&  0.24 &   0.03\\ 
      &      & 5.00 & 6.50& 0.098 &  0.015& 0.119 &  0.014&  0.24 &   0.03&  0.27 &   0.03\\ 
      &      & 6.50 & 8.00&       &       & 0.108 &  0.015&       &       &  0.30 &   0.02\\ 
0.050 &0.075 & 0.50 & 0.75&  0.15 &   0.03&  0.24 &   0.05&  0.30 &   0.05&  0.29 &   0.06\\ 
      &      & 0.75 & 1.00&  0.13 &   0.02&  0.22 &   0.04&  0.25 &   0.03&  0.21 &   0.03\\ 
      &      & 1.00 & 1.25&  0.23 &   0.03&  0.26 &   0.04&  0.24 &   0.03&  0.29 &   0.04\\ 
      &      & 1.25 & 1.50&  0.19 &   0.02&  0.20 &   0.03&  0.18 &   0.02&  0.23 &   0.03\\ 
      &      & 1.50 & 2.00&  0.18 &   0.02&  0.20 &   0.03&  0.25 &   0.02&  0.34 &   0.04\\ 
      &      & 2.00 & 2.50&  0.14 &   0.02&  0.17 &   0.02&  0.25 &   0.02&  0.29 &   0.03\\ 
      &      & 2.50 & 3.00&  0.10 &   0.02&  0.17 &   0.02&  0.22 &   0.02&  0.26 &   0.03\\ 
      &      & 3.00 & 3.50&  0.09 &   0.02&  0.18 &   0.02&  0.22 &   0.02&  0.28 &   0.03\\ 
      &      & 3.50 & 4.00& 0.090 &  0.013&  0.17 &   0.02&  0.20 &   0.02&  0.24 &   0.03\\ 
      &      & 4.00 & 5.00& 0.083 &  0.010& 0.096 &  0.012&  0.21 &   0.02&  0.23 &   0.02\\ 
      &      & 5.00 & 6.50& 0.045 &  0.006& 0.073 &  0.009&  0.17 &   0.02& 0.173 &  0.013\\ 
      &      & 6.50 & 8.00&       &       & 0.026 &  0.005&       &       & 0.140 &  0.012\\ 
0.075 &0.100 & 0.50 & 0.75&  0.12 &   0.02&  0.14 &   0.03&  0.30 &   0.04&  0.37 &   0.06\\ 
      &      & 0.75 & 1.00&  0.16 &   0.02&  0.23 &   0.03&  0.22 &   0.02&  0.22 &   0.03\\ 
      &      & 1.00 & 1.25&  0.21 &   0.02&  0.27 &   0.03&  0.25 &   0.02&  0.29 &   0.03\\ 
      &      & 1.25 & 1.50&  0.18 &   0.02&  0.30 &   0.03&  0.25 &   0.03&  0.33 &   0.04\\ 
      &      & 1.50 & 2.00& 0.166 &  0.013&  0.25 &   0.02&  0.25 &   0.02&  0.30 &   0.03\\ 
      &      & 2.00 & 2.50& 0.114 &  0.012&  0.21 &   0.02&  0.24 &   0.02&  0.34 &   0.03\\ 
      &      & 2.50 & 3.00& 0.092 &  0.010&  0.18 &   0.02&  0.23 &   0.02&  0.32 &   0.03\\ 
      &      & 3.00 & 3.50& 0.096 &  0.010&  0.17 &   0.02&  0.16 &   0.02&  0.25 &   0.02\\ 
      &      & 3.50 & 4.00& 0.053 &  0.012& 0.136 &  0.014&  0.16 &   0.02&  0.23 &   0.02\\ 
      &      & 4.00 & 5.00& 0.026 &  0.005& 0.072 &  0.008& 0.144 &  0.013& 0.171 &  0.014\\ 
      &      & 5.00 & 6.50& 0.008 &  0.003& 0.039 &  0.006& 0.078 &  0.006& 0.105 &  0.008\\ 
      &      & 6.50 & 8.00&       &       & 0.008 &  0.002&       &       & 0.052 &  0.006\\ 
\hline
\end{tabular}
}
\end{table}

%% file: table_result_fw_036-pip_C_OMEGA_fine.tex
\begin{table}[!ht]
  \caption{\label{tab:xsec_results_
C7} 
    HARP results for the double-differential $\pi^+$  and $\pi^-$ production
    cross-section in the laboratory system,
    $d^2\sigma^{\pi}/(dpd\Omega)$, for $\pi^{+}$--
C 
    interactions at 8~\GeVc.
    Each row refers to a
    different $(p_{\hbox{\small min}} \le p<p_{\hbox{\small max}},
    \theta_{\hbox{\small min}} \le \theta<\theta_{\hbox{\small max}})$ bin,
    where $p$ and $\theta$ are the pion momentum and polar angle, respectively.
    A finer angular binning than in the previous set of tables is used.
    The central value as well as the square-root of the diagonal elements
    of the covariance matrix are given.}

\small{
\begin{tabular}{rrrr|r@{$\pm$}l|r@{$\pm$}l}
\hline
$\theta_{\hbox{\small min}}$ &
$\theta_{\hbox{\small max}}$ &
$p_{\hbox{\small min}}$ &
$p_{\hbox{\small max}}$ &
\multicolumn{2}{c}{$d^2\sigma^{\pi^+}/(dpd\Omega)$} &
\multicolumn{2}{c}{$d^2\sigma^{\pi^-}/(dpd\Omega)$}
\\
(rad) & (rad) & (\GeVc) & (\GeVc) &
\multicolumn{2}{c}{(barn/(\GeVc rad))} &
\multicolumn{2}{c}{(barn/(\GeVc rad))}
\\
  &  &  &
&\multicolumn{2}{c}{$ \bf{8 \ \GeVc}$}
&\multicolumn{2}{c}{$ \bf{8 \ \GeVc}$}
\\
\hline
0.025 &0.050 & 0.50 & 0.75&  0.15 &   0.06&  0.27 &   0.08\\ 
      &      & 0.75 & 1.00&  0.19 &   0.05&  0.19 &   0.05\\ 
      &      & 1.00 & 1.25&  0.05 &   0.02&  0.12 &   0.03\\ 
      &      & 1.25 & 1.50&  0.07 &   0.03&  0.18 &   0.06\\ 
      &      & 1.50 & 2.00&  0.11 &   0.03&  0.17 &   0.04\\ 
      &      & 2.00 & 2.50&  0.15 &   0.04&  0.12 &   0.03\\ 
      &      & 2.50 & 3.00&  0.18 &   0.04&  0.09 &   0.02\\ 
      &      & 3.00 & 3.50&  0.23 &   0.03&  0.15 &   0.05\\ 
      &      & 3.50 & 4.00&  0.17 &   0.02&  0.10 &   0.02\\ 
      &      & 4.00 & 5.00&  0.21 &   0.02&  0.11 &   0.02\\ 
      &      & 5.00 & 6.50&  0.36 &   0.02& 0.097 &  0.014\\ 
0.050 &0.075 & 0.50 & 0.75&  0.18 &   0.05&  0.16 &   0.04\\ 
      &      & 0.75 & 1.00&  0.17 &   0.03&  0.12 &   0.03\\ 
      &      & 1.00 & 1.25&  0.14 &   0.03&  0.17 &   0.04\\ 
      &      & 1.25 & 1.50&  0.25 &   0.04&  0.15 &   0.03\\ 
      &      & 1.50 & 2.00&  0.17 &   0.03&  0.15 &   0.02\\ 
      &      & 2.00 & 2.50&  0.16 &   0.03&  0.16 &   0.02\\ 
      &      & 2.50 & 3.00&  0.22 &   0.03&  0.11 &   0.02\\ 
      &      & 3.00 & 3.50&  0.17 &   0.03& 0.081 &  0.014\\ 
      &      & 3.50 & 4.00&  0.14 &   0.02&  0.13 &   0.02\\ 
      &      & 4.00 & 5.00&  0.26 &   0.02& 0.065 &  0.010\\ 
      &      & 5.00 & 6.50&  0.25 &   0.02& 0.046 &  0.006\\ 
0.075 &0.100 & 0.50 & 0.75&  0.13 &   0.03&  0.17 &   0.04\\ 
      &      & 0.75 & 1.00&  0.22 &   0.04&  0.13 &   0.02\\ 
      &      & 1.00 & 1.25&  0.23 &   0.03&  0.14 &   0.03\\ 
      &      & 1.25 & 1.50&  0.21 &   0.03&  0.15 &   0.02\\ 
      &      & 1.50 & 2.00&  0.20 &   0.02&  0.15 &   0.02\\ 
      &      & 2.00 & 2.50&  0.20 &   0.03&  0.11 &   0.02\\ 
      &      & 2.50 & 3.00&  0.19 &   0.02&  0.12 &   0.02\\ 
      &      & 3.00 & 3.50&  0.20 &   0.02& 0.081 &  0.011\\ 
      &      & 3.50 & 4.00&  0.14 &   0.02& 0.088 &  0.013\\ 
      &      & 4.00 & 5.00& 0.153 &  0.014& 0.039 &  0.005\\ 
      &      & 5.00 & 6.50&  0.12 &   0.04& 0.011 &  0.002\\ 
\hline
\end{tabular}
}
\end{table}

%% file: table_result_fw_036-pim_Al_OMEGA_fine.tex
chop 
\begin{table}[!ht]
  \caption{\label{tab:xsec_results_
Al7} 
    HARP results for the double-differential $\pi^+$  and $\pi^-$ production
    cross-section in the laboratory system,
    $d^2\sigma^{\pi}/(dpd\Omega)$, for $\pi^{-}$--
Al 
    interactions at 8 and 12~\GeVc.
    Each row refers to a
    different $(p_{\hbox{\small min}} \le p<p_{\hbox{\small max}},
    \theta_{\hbox{\small min}} \le \theta<\theta_{\hbox{\small max}})$ bin,
    where $p$ and $\theta$ are the pion momentum and polar angle, respectively.
    A finer angular binning than in the previous set of tables is used.
    The central value as well as the square-root of the diagonal elements
    of the covariance matrix are given.}

\small{
\begin{tabular}{rrrr|r@{$\pm$}lr@{$\pm$}l|r@{$\pm$}lr@{$\pm$}l}
\hline
$\theta_{\hbox{\small min}}$ &
$\theta_{\hbox{\small max}}$ &
$p_{\hbox{\small min}}$ &
$p_{\hbox{\small max}}$ &
\multicolumn{4}{c}{$d^2\sigma^{\pi^+}/(dpd\Omega)$} &
\multicolumn{4}{c}{$d^2\sigma^{\pi^-}/(dpd\Omega)$}
\\
(rad) & (rad) & (\GeVc) & (\GeVc) &
\multicolumn{4}{c}{(barn/(\GeVc rad))} &
\multicolumn{4}{c}{(barn/(\GeVc rad))}
\\
  &  &  &
&\multicolumn{2}{c}{$ \bf{8 \ \GeVc}$}
&\multicolumn{2}{c}{$ \bf{12 \ \GeVc}$}
&\multicolumn{2}{c}{$ \bf{8 \ \GeVc}$}
&\multicolumn{2}{c}{$ \bf{12 \ \GeVc}$}
\\
\hline
0.025 &0.050 & 0.50 & 0.75&  0.12 &   0.04&  0.32 &   0.09&  0.49 &   0.11&  0.50 &   0.12\\ 
      &      & 0.75 & 1.00&  0.27 &   0.05&  0.37 &   0.06&  0.41 &   0.06&  0.42 &   0.07\\ 
      &      & 1.00 & 1.25&  0.22 &   0.04&  0.33 &   0.06&  0.36 &   0.06&  0.38 &   0.06\\ 
      &      & 1.25 & 1.50&  0.42 &   0.07&  0.38 &   0.06&  0.34 &   0.06&  0.50 &   0.09\\ 
      &      & 1.50 & 2.00&  0.33 &   0.05&  0.36 &   0.06&  0.45 &   0.05&  0.51 &   0.06\\ 
      &      & 2.00 & 2.50&  0.32 &   0.05&  0.48 &   0.06&  0.42 &   0.05&  0.49 &   0.07\\ 
      &      & 2.50 & 3.00&  0.19 &   0.04&  0.45 &   0.06&  0.40 &   0.05&  0.37 &   0.05\\ 
      &      & 3.00 & 3.50&  0.15 &   0.04&  0.30 &   0.04&  0.38 &   0.04&  0.43 &   0.05\\ 
      &      & 3.50 & 4.00&  0.17 &   0.04&  0.31 &   0.04&  0.31 &   0.04&  0.41 &   0.04\\ 
      &      & 4.00 & 5.00&  0.15 &   0.03&  0.26 &   0.03&  0.32 &   0.04&  0.41 &   0.04\\ 
      &      & 5.00 & 6.50&  0.14 &   0.02&  0.19 &   0.02&  0.35 &   0.04&  0.39 &   0.03\\ 
      &      & 6.50 & 8.00&       &       &  0.16 &   0.02&       &       &  0.40 &   0.03\\ 
0.050 &0.075 & 0.50 & 0.75&  0.28 &   0.06&  0.23 &   0.05&  0.55 &   0.09&  0.48 &   0.08\\ 
      &      & 0.75 & 1.00&  0.25 &   0.04&  0.38 &   0.05&  0.45 &   0.05&  0.38 &   0.05\\ 
      &      & 1.00 & 1.25&  0.34 &   0.04&  0.39 &   0.05&  0.32 &   0.04&  0.38 &   0.05\\ 
      &      & 1.25 & 1.50&  0.30 &   0.04&  0.43 &   0.06&  0.33 &   0.05&  0.43 &   0.06\\ 
      &      & 1.50 & 2.00&  0.33 &   0.03&  0.35 &   0.04&  0.46 &   0.04&  0.48 &   0.05\\ 
      &      & 2.00 & 2.50&  0.20 &   0.03&  0.34 &   0.04&  0.40 &   0.04&  0.47 &   0.05\\ 
      &      & 2.50 & 3.00&  0.13 &   0.03&  0.28 &   0.03&  0.39 &   0.04&  0.45 &   0.04\\ 
      &      & 3.00 & 3.50&  0.18 &   0.03&  0.35 &   0.04&  0.33 &   0.04&  0.48 &   0.05\\ 
      &      & 3.50 & 4.00&  0.17 &   0.02&  0.24 &   0.03&  0.33 &   0.04&  0.48 &   0.04\\ 
      &      & 4.00 & 5.00&  0.12 &   0.02&  0.16 &   0.02&  0.33 &   0.03&  0.31 &   0.03\\ 
      &      & 5.00 & 6.50& 0.073 &  0.010& 0.098 &  0.011&  0.29 &   0.02&  0.27 &   0.02\\ 
      &      & 6.50 & 8.00&       &       & 0.049 &  0.007&       &       &  0.23 &   0.02\\ 
0.075 &0.100 & 0.50 & 0.75&  0.26 &   0.05&  0.25 &   0.04&  0.57 &   0.08&  0.63 &   0.09\\ 
      &      & 0.75 & 1.00&  0.35 &   0.04&  0.40 &   0.05&  0.43 &   0.04&  0.43 &   0.05\\ 
      &      & 1.00 & 1.25&  0.36 &   0.04&  0.52 &   0.05&  0.44 &   0.04&  0.49 &   0.05\\ 
      &      & 1.25 & 1.50&  0.40 &   0.04&  0.42 &   0.04&  0.50 &   0.05&  0.52 &   0.06\\ 
      &      & 1.50 & 2.00&  0.28 &   0.02&  0.39 &   0.03&  0.41 &   0.03&  0.54 &   0.04\\ 
      &      & 2.00 & 2.50&  0.23 &   0.02&  0.31 &   0.03&  0.44 &   0.04&  0.48 &   0.04\\ 
      &      & 2.50 & 3.00&  0.14 &   0.02&  0.30 &   0.02&  0.40 &   0.03&  0.48 &   0.04\\ 
      &      & 3.00 & 3.50&  0.17 &   0.02&  0.25 &   0.02&  0.28 &   0.03&  0.47 &   0.03\\ 
      &      & 3.50 & 4.00&  0.10 &   0.02&  0.21 &   0.02&  0.24 &   0.03&  0.38 &   0.03\\ 
      &      & 4.00 & 5.00& 0.060 &  0.009& 0.124 &  0.012&  0.23 &   0.02&  0.23 &   0.02\\ 
      &      & 5.00 & 6.50& 0.015 &  0.005& 0.052 &  0.006& 0.140 &  0.011& 0.146 &  0.011\\ 
      &      & 6.50 & 8.00&       &       & 0.016 &  0.003&       &       & 0.071 &  0.007\\ 
\hline
\end{tabular}
}
\end{table}

%% file: table_result_fw_036-pip_Al_OMEGA_fine.tex
chop 
\begin{table}[!ht]
  \caption{\label{tab:xsec_results_
Al7} 
    HARP results for the double-differential $\pi^+$  and $\pi^-$ production
    cross-section in the laboratory system,
    $d^2\sigma^{\pi}/(dpd\Omega)$, for $\pi^{+}$--
Al 
    interactions at 8 and 12.9~\GeVc.
    Each row refers to a
    different $(p_{\hbox{\small min}} \le p<p_{\hbox{\small max}},
    \theta_{\hbox{\small min}} \le \theta<\theta_{\hbox{\small max}})$ bin,
    where $p$ and $\theta$ are the pion momentum and polar angle, respectively.
    A finer angular binning than in the previous set of tables is used.
    The central value as well as the square-root of the diagonal elements
    of the covariance matrix are given.}

\small{
\begin{tabular}{rrrr|r@{$\pm$}lr@{$\pm$}l|r@{$\pm$}lr@{$\pm$}l}
\hline
$\theta_{\hbox{\small min}}$ &
$\theta_{\hbox{\small max}}$ &
$p_{\hbox{\small min}}$ &
$p_{\hbox{\small max}}$ &
\multicolumn{4}{c}{$d^2\sigma^{\pi^+}/(dpd\Omega)$} &
\multicolumn{4}{c}{$d^2\sigma^{\pi^-}/(dpd\Omega)$}
\\
(rad) & (rad) & (\GeVc) & (\GeVc) &
\multicolumn{4}{c}{(barn/(\GeVc rad))} &
\multicolumn{4}{c}{(barn/(\GeVc rad))}
\\
  &  &  &
&\multicolumn{2}{c}{$ \bf{8 \ \GeVc}$}
&\multicolumn{2}{c}{$ \bf{12.9 \ \GeVc}$}
&\multicolumn{2}{c}{$ \bf{8 \ \GeVc}$}
&\multicolumn{2}{c}{$ \bf{12.9 \ \GeVc}$}
\\
\hline
0.025 &0.050 & 0.50 & 0.75&  0.37 &   0.11&  0.51 &   0.16&  0.15 &   0.06&  0.35 &   0.16\\  
      &      & 0.75 & 1.00&  0.38 &   0.08&  0.43 &   0.13&  0.37 &   0.09&  0.38 &   0.12\\  
      &      & 1.00 & 1.25&  0.24 &   0.07&  0.36 &   0.12&  0.27 &   0.07&  0.36 &   0.14\\  
      &      & 1.25 & 1.50&  0.11 &   0.04&  0.27 &   0.10&  0.48 &   0.12&  0.13 &   0.08\\  
      &      & 1.50 & 2.00&  0.23 &   0.05&  0.54 &   0.10&  0.26 &   0.06&  0.30 &   0.09\\  
      &      & 2.00 & 2.50&  0.27 &   0.06&  0.42 &   0.09&  0.24 &   0.05&  0.34 &   0.11\\  
      &      & 2.50 & 3.00&  0.35 &   0.06&  0.63 &   0.10&  0.21 &   0.04&  0.27 &   0.07\\  
      &      & 3.00 & 3.50&  0.31 &   0.05&  0.47 &   0.08&  0.25 &   0.04&  0.35 &   0.08\\  
      &      & 3.50 & 4.00&  0.38 &   0.05&  0.49 &   0.09&  0.21 &   0.04&  0.40 &   0.08\\  
      &      & 4.00 & 5.00&  0.37 &   0.04&  0.39 &   0.06&  0.19 &   0.03&  0.21 &   0.04\\  
      &      & 5.00 & 6.50&  0.50 &   0.03&  0.41 &   0.05&  0.19 &   0.03&  0.14 &   0.03\\  
      &      & 6.50 & 8.00&       &       &  0.42 &   0.05&       &       &  0.16 &   0.03\\  
0.050 &0.075 & 0.50 & 0.75&  0.38 &   0.09&  0.48 &   0.11&  0.22 &   0.06&  0.21 &   0.08\\  
      &      & 0.75 & 1.00&  0.33 &   0.06&  0.64 &   0.10&  0.15 &   0.04&  0.27 &   0.07\\  
      &      & 1.00 & 1.25&  0.33 &   0.06&  0.36 &   0.08&  0.24 &   0.05&  0.40 &   0.09\\  
      &      & 1.25 & 1.50&  0.21 &   0.05&  0.36 &   0.08&  0.18 &   0.04&  0.39 &   0.09\\  
      &      & 1.50 & 2.00&  0.28 &   0.04&  0.43 &   0.07&  0.23 &   0.04&  0.43 &   0.06\\  
      &      & 2.00 & 2.50&  0.37 &   0.05&  0.52 &   0.07&  0.24 &   0.04&  0.33 &   0.06\\  
      &      & 2.50 & 3.00&  0.34 &   0.05&  0.45 &   0.06&  0.16 &   0.03&  0.24 &   0.05\\  
      &      & 3.00 & 3.50&  0.33 &   0.05&  0.46 &   0.06&  0.21 &   0.03&  0.32 &   0.06\\  
      &      & 3.50 & 4.00&  0.29 &   0.04&  0.42 &   0.06&  0.18 &   0.03&  0.29 &   0.05\\  
      &      & 4.00 & 5.00&  0.42 &   0.04&  0.41 &   0.04&  0.13 &   0.02&  0.17 &   0.03\\  
      &      & 5.00 & 6.50&  0.40 &   0.03&  0.29 &   0.03& 0.080 &  0.011&  0.14 &   0.02\\  
      &      & 6.50 & 8.00&       &       &  0.21 &   0.03&       &       & 0.052 &  0.012\\  
0.075 &0.100 & 0.50 & 0.75&  0.26 &   0.06&  0.35 &   0.09&  0.32 &   0.07&  0.24 &   0.08\\  
      &      & 0.75 & 1.00&  0.26 &   0.05&  0.59 &   0.08&  0.26 &   0.04&  0.30 &   0.07\\  
      &      & 1.00 & 1.25&  0.34 &   0.05&  0.47 &   0.07&  0.33 &   0.05&  0.41 &   0.07\\  
      &      & 1.25 & 1.50&  0.40 &   0.06&  0.45 &   0.07&  0.28 &   0.04&  0.41 &   0.07\\  
      &      & 1.50 & 2.00&  0.35 &   0.04&  0.43 &   0.06&  0.26 &   0.03&  0.36 &   0.05\\  
      &      & 2.00 & 2.50&  0.34 &   0.04&  0.54 &   0.06&  0.27 &   0.03&  0.36 &   0.04\\  
      &      & 2.50 & 3.00&  0.33 &   0.04&  0.34 &   0.04&  0.15 &   0.02&  0.29 &   0.04\\  
      &      & 3.00 & 3.50&  0.36 &   0.04&  0.48 &   0.05&  0.12 &   0.02&  0.27 &   0.03\\  
      &      & 3.50 & 4.00&  0.31 &   0.05&  0.36 &   0.04&  0.09 &   0.02&  0.29 &   0.03\\  
      &      & 4.00 & 5.00&  0.31 &   0.02&  0.25 &   0.03& 0.043 &  0.007&  0.12 &   0.02\\  
      &      & 5.00 & 6.50&  0.16 &   0.05&  0.14 &   0.02& 0.025 &  0.005& 0.051 &  0.010\\  
      &      & 6.50 & 8.00&       &       & 0.080 &  0.012&       &       & 0.015 &  0.004\\  
\hline
\end{tabular}
}
\end{table}

%% file: table_result_fw_036-pim_Cu_OMEGA_fine.tex
\begin{table}[!ht]
  \caption{\label{tab:xsec_results_
Cu7} 
    HARP results for the double-differential $\pi^+$  and $\pi^-$ production
    cross-section in the laboratory system,
    $d^2\sigma^{\pi}/(dpd\Omega)$, for $\pi^{-}$--
Cu 
    interactions at 8 and 12~\GeVc.
    Each row refers to a
    different $(p_{\hbox{\small min}} \le p<p_{\hbox{\small max}},
    \theta_{\hbox{\small min}} \le \theta<\theta_{\hbox{\small max}})$ bin,
    where $p$ and $\theta$ are the pion momentum and polar angle, respectively.
    A finer angular binning than in the previous set of tables is used.
    The central value as well as the square-root of the diagonal elements
    of the covariance matrix are given.}

\small{
\begin{tabular}{rrrr|r@{$\pm$}lr@{$\pm$}l|r@{$\pm$}lr@{$\pm$}l}
\hline
$\theta_{\hbox{\small min}}$ &
$\theta_{\hbox{\small max}}$ &
$p_{\hbox{\small min}}$ &
$p_{\hbox{\small max}}$ &
\multicolumn{4}{c}{$d^2\sigma^{\pi^+}/(dpd\Omega)$} &
\multicolumn{4}{c}{$d^2\sigma^{\pi^-}/(dpd\Omega)$}
\\
(rad) & (rad) & (\GeVc) & (\GeVc) &
\multicolumn{4}{c}{(barn/(\GeVc rad))} &
\multicolumn{4}{c}{(barn/(\GeVc rad))}
\\
  &  &  &
&\multicolumn{2}{c}{$ \bf{8 \ \GeVc}$}
&\multicolumn{2}{c}{$ \bf{12 \ \GeVc}$}
&\multicolumn{2}{c}{$ \bf{8 \ \GeVc}$}
&\multicolumn{2}{c}{$ \bf{12 \ \GeVc}$}
\\
\hline
0.025 &0.050 & 0.50 & 0.75&  0.44 &   0.10&  0.73 &   0.19&  1.03 &   0.16&  1.23 &   0.27\\ 
      &      & 0.75 & 1.00&  0.47 &   0.06&  0.68 &   0.12&  0.55 &   0.07&  1.08 &   0.16\\ 
      &      & 1.00 & 1.25&  0.42 &   0.05&  0.64 &   0.11&  0.46 &   0.06&  0.54 &   0.10\\ 
      &      & 1.25 & 1.50&  0.70 &   0.09&  0.80 &   0.14&  0.55 &   0.09&  0.89 &   0.17\\ 
      &      & 1.50 & 2.00&  0.55 &   0.07&  0.71 &   0.11&  0.78 &   0.07&  0.93 &   0.12\\ 
      &      & 2.00 & 2.50&  0.50 &   0.07&  0.54 &   0.09&  0.68 &   0.06&  0.80 &   0.11\\ 
      &      & 2.50 & 3.00&  0.46 &   0.07&  0.73 &   0.11&  0.57 &   0.06&  0.73 &   0.10\\ 
      &      & 3.00 & 3.50&  0.24 &   0.06&  0.76 &   0.11&  0.42 &   0.05&  0.79 &   0.10\\ 
      &      & 3.50 & 4.00&  0.24 &   0.05&  0.61 &   0.09&  0.35 &   0.05&  0.63 &   0.08\\ 
      &      & 4.00 & 5.00&  0.29 &   0.05&  0.33 &   0.05&  0.37 &   0.05&  0.61 &   0.07\\ 
      &      & 5.00 & 6.50&  0.24 &   0.04&  0.34 &   0.04&  0.41 &   0.05&  0.57 &   0.06\\ 
      &      & 6.50 & 8.00&       &       &  0.27 &   0.04&       &       &  0.61 &   0.05\\ 
0.050 &0.075 & 0.50 & 0.75&  0.55 &   0.09&  0.81 &   0.15&  1.12 &   0.15&  1.25 &   0.22\\ 
      &      & 0.75 & 1.00&  0.63 &   0.07&  0.72 &   0.11&  0.77 &   0.07&  1.09 &   0.14\\ 
      &      & 1.00 & 1.25&  0.59 &   0.06&  0.85 &   0.12&  0.61 &   0.06&  0.99 &   0.13\\ 
      &      & 1.25 & 1.50&  0.67 &   0.07&  0.83 &   0.12&  0.62 &   0.07&  0.81 &   0.11\\ 
      &      & 1.50 & 2.00&  0.61 &   0.05&  0.63 &   0.08&  0.76 &   0.06&  0.92 &   0.10\\ 
      &      & 2.00 & 2.50&  0.37 &   0.05&  0.60 &   0.07&  0.70 &   0.06&  0.93 &   0.10\\ 
      &      & 2.50 & 3.00&  0.24 &   0.04&  0.49 &   0.06&  0.62 &   0.05&  0.67 &   0.08\\ 
      &      & 3.00 & 3.50&  0.29 &   0.05&  0.56 &   0.07&  0.50 &   0.06&  0.70 &   0.08\\ 
      &      & 3.50 & 4.00&  0.25 &   0.04&  0.47 &   0.07&  0.43 &   0.06&  0.63 &   0.09\\ 
      &      & 4.00 & 5.00&  0.23 &   0.03&  0.28 &   0.04&  0.55 &   0.06&  0.59 &   0.05\\ 
      &      & 5.00 & 6.50&  0.12 &   0.02&  0.17 &   0.02&  0.43 &   0.04&  0.48 &   0.04\\ 
      &      & 6.50 & 8.00&       &       &  0.12 &   0.02&       &       &  0.34 &   0.03\\ 
0.075 &0.100 & 0.50 & 0.75&  0.55 &   0.09&  0.72 &   0.12&  1.08 &   0.13&  1.42 &   0.19\\ 
      &      & 0.75 & 1.00&  0.75 &   0.07&  0.59 &   0.08&  0.70 &   0.06&  0.74 &   0.09\\ 
      &      & 1.00 & 1.25&  0.67 &   0.06&  0.87 &   0.10&  0.82 &   0.07&  1.10 &   0.12\\ 
      &      & 1.25 & 1.50&  0.58 &   0.05&  0.65 &   0.07&  0.75 &   0.07&  0.89 &   0.10\\ 
      &      & 1.50 & 2.00&  0.48 &   0.04&  0.73 &   0.08&  0.74 &   0.05&  0.96 &   0.08\\ 
      &      & 2.00 & 2.50&  0.40 &   0.04&  0.75 &   0.07&  0.67 &   0.05&  1.10 &   0.09\\ 
      &      & 2.50 & 3.00&  0.29 &   0.03&  0.58 &   0.05&  0.63 &   0.05&  0.83 &   0.07\\ 
      &      & 3.00 & 3.50&  0.30 &   0.03&  0.51 &   0.05&  0.45 &   0.05&  0.84 &   0.08\\ 
      &      & 3.50 & 4.00&  0.16 &   0.04&  0.35 &   0.04&  0.39 &   0.04&  0.71 &   0.08\\ 
      &      & 4.00 & 5.00& 0.082 &  0.014&  0.24 &   0.03&  0.33 &   0.03&  0.40 &   0.03\\ 
      &      & 5.00 & 6.50& 0.024 &  0.007& 0.089 &  0.014&  0.21 &   0.02&  0.24 &   0.02\\ 
      &      & 6.50 & 8.00&       &       & 0.018 &  0.005&       &       & 0.122 &  0.013\\ 
\hline
\end{tabular}
}
\end{table}

%% file: table_result_fw_036-pip_Cu_OMEGA_fine.tex
\begin{table}[!ht]
  \caption{\label{tab:xsec_results_
Cu7} 
    HARP results for the double-differential $\pi^+$  and $\pi^-$ production
    cross-section in the laboratory system,
    $d^2\sigma^{\pi}/(dpd\Omega)$, for $\pi^{+}$--
Cu 
    interactions at 8~\GeVc.
    Each row refers to a
    different $(p_{\hbox{\small min}} \le p<p_{\hbox{\small max}},
    \theta_{\hbox{\small min}} \le \theta<\theta_{\hbox{\small max}})$ bin,
    where $p$ and $\theta$ are the pion momentum and polar angle, respectively.
    A finer angular binning than in the previous set of tables is used.
    The central value as well as the square-root of the diagonal elements
    of the covariance matrix are given.}

\small{
\begin{tabular}{rrrr|r@{$\pm$}l|r@{$\pm$}l}
\hline
$\theta_{\hbox{\small min}}$ &
$\theta_{\hbox{\small max}}$ &
$p_{\hbox{\small min}}$ &
$p_{\hbox{\small max}}$ &
\multicolumn{2}{c}{$d^2\sigma^{\pi^+}/(dpd\Omega)$} &
\multicolumn{2}{c}{$d^2\sigma^{\pi^-}/(dpd\Omega)$}
\\
(rad) & (rad) & (\GeVc) & (\GeVc) &
\multicolumn{2}{c}{(barn/(\GeVc rad))} &
\multicolumn{2}{c}{(barn/(\GeVc rad))}
\\
  &  &  &
&\multicolumn{2}{c}{$ \bf{8 \ \GeVc}$}
&\multicolumn{2}{c}{$ \bf{8 \ \GeVc}$}
\\
\hline
0.025 &0.050 & 0.50 & 0.75&  0.22 &   0.08&  0.45 &   0.15\\ 
      &      & 0.75 & 1.00&  0.32 &   0.08&  0.45 &   0.11\\ 
      &      & 1.00 & 1.25&  0.26 &   0.07&  0.37 &   0.10\\ 
      &      & 1.25 & 1.50&  0.49 &   0.12&  0.31 &   0.11\\ 
      &      & 1.50 & 2.00&  0.32 &   0.08&  0.64 &   0.12\\ 
      &      & 2.00 & 2.50&  0.34 &   0.09&  0.36 &   0.08\\ 
      &      & 2.50 & 3.00&  0.61 &   0.12&  0.27 &   0.06\\ 
      &      & 3.00 & 3.50&  0.32 &   0.06&  0.40 &   0.07\\ 
      &      & 3.50 & 4.00&  0.47 &   0.07&  0.31 &   0.05\\ 
      &      & 4.00 & 5.00&  0.40 &   0.05&  0.32 &   0.04\\ 
      &      & 5.00 & 6.50&  0.76 &   0.05&  0.30 &   0.04\\ 
0.050 &0.075 & 0.50 & 0.75&  0.57 &   0.14&  0.44 &   0.11\\ 
      &      & 0.75 & 1.00&  0.59 &   0.11&  0.49 &   0.09\\ 
      &      & 1.00 & 1.25&  0.49 &   0.09&  0.48 &   0.10\\ 
      &      & 1.25 & 1.50&  0.66 &   0.11&  0.56 &   0.10\\ 
      &      & 1.50 & 2.00&  0.60 &   0.08&  0.44 &   0.07\\ 
      &      & 2.00 & 2.50&  0.59 &   0.09&  0.36 &   0.06\\ 
      &      & 2.50 & 3.00&  0.51 &   0.08&  0.31 &   0.05\\ 
      &      & 3.00 & 3.50&  0.45 &   0.08&  0.30 &   0.05\\ 
      &      & 3.50 & 4.00&  0.44 &   0.07&  0.22 &   0.04\\ 
      &      & 4.00 & 5.00&  0.58 &   0.06&  0.23 &   0.03\\ 
      &      & 5.00 & 6.50&  0.62 &   0.04&  0.12 &   0.02\\ 
0.075 &0.100 & 0.50 & 0.75&  0.64 &   0.15&  0.61 &   0.12\\ 
      &      & 0.75 & 1.00&  0.53 &   0.09&  0.47 &   0.08\\ 
      &      & 1.00 & 1.25&  0.66 &   0.10&  0.46 &   0.08\\ 
      &      & 1.25 & 1.50&  0.59 &   0.08&  0.64 &   0.09\\ 
      &      & 1.50 & 2.00&  0.55 &   0.07&  0.39 &   0.05\\ 
      &      & 2.00 & 2.50&  0.54 &   0.07&  0.40 &   0.05\\ 
      &      & 2.50 & 3.00&  0.56 &   0.07&  0.24 &   0.03\\ 
      &      & 3.00 & 3.50&  0.41 &   0.05&  0.22 &   0.03\\ 
      &      & 3.50 & 4.00&  0.47 &   0.07&  0.15 &   0.03\\ 
      &      & 4.00 & 5.00&  0.41 &   0.04&  0.13 &   0.02\\ 
      &      & 5.00 & 6.50&  0.27 &   0.09& 0.024 &  0.005\\ 
\hline
\end{tabular}
}
\end{table}

%% file: table_result_fw_036-pim_Sn_OMEGA_fine.tex
\begin{table}[!ht]
  \caption{\label{tab:xsec_results_
Sn7} 
    HARP results for the double-differential $\pi^+$  and $\pi^-$ production
    cross-section in the laboratory system,
    $d^2\sigma^{\pi}/(dpd\Omega)$, for $\pi^{-}$--
Sn 
    interactions at 8 and 12~\GeVc.
    Each row refers to a
    different $(p_{\hbox{\small min}} \le p<p_{\hbox{\small max}},
    \theta_{\hbox{\small min}} \le \theta<\theta_{\hbox{\small max}})$ bin,
    where $p$ and $\theta$ are the pion momentum and polar angle, respectively.
    A finer angular binning than in the previous set of tables is used.
    The central value as well as the square-root of the diagonal elements
    of the covariance matrix are given.}

\small{
\begin{tabular}{rrrr|r@{$\pm$}lr@{$\pm$}l|r@{$\pm$}lr@{$\pm$}l}
\hline
$\theta_{\hbox{\small min}}$ &
$\theta_{\hbox{\small max}}$ &
$p_{\hbox{\small min}}$ &
$p_{\hbox{\small max}}$ &
\multicolumn{4}{c}{$d^2\sigma^{\pi^+}/(dpd\Omega)$} &
\multicolumn{4}{c}{$d^2\sigma^{\pi^-}/(dpd\Omega)$}
\\
(rad) & (rad) & (\GeVc) & (\GeVc) &
\multicolumn{4}{c}{(barn/(\GeVc rad))} &
\multicolumn{4}{c}{(barn/(\GeVc rad))}
\\
  &  &  &
&\multicolumn{2}{c}{$ \bf{8 \ \GeVc}$}
&\multicolumn{2}{c}{$ \bf{12 \ \GeVc}$}
&\multicolumn{2}{c}{$ \bf{8 \ \GeVc}$}
&\multicolumn{2}{c}{$ \bf{12 \ \GeVc}$}
\\
\hline
0.025 &0.050 & 0.50 & 0.75&  0.64 &   0.14&  1.56 &   0.30&  1.39 &   0.21&  1.86 &   0.34\\ 
      &      & 0.75 & 1.00&  0.53 &   0.07&  1.16 &   0.19&  0.70 &   0.08&  1.53 &   0.17\\ 
      &      & 1.00 & 1.25&  0.36 &   0.05&  0.94 &   0.13&  0.29 &   0.04&  0.93 &   0.12\\ 
      &      & 1.25 & 1.50&  0.55 &   0.09&  1.15 &   0.17&  0.55 &   0.10&  1.34 &   0.20\\ 
      &      & 1.50 & 2.00&  0.63 &   0.10&  1.01 &   0.76&  0.80 &   0.08&  1.38 &   0.15\\ 
      &      & 2.00 & 2.50&  0.84 &   0.12&  1.07 &   0.36&  0.75 &   0.09&  1.19 &   0.16\\ 
      &      & 2.50 & 3.00&  0.53 &   0.11&  0.88 &   0.15&  0.70 &   0.09&  1.08 &   0.12\\ 
      &      & 3.00 & 3.50&  0.24 &   0.08&  0.92 &   0.16&  0.45 &   0.07&  1.04 &   0.11\\ 
      &      & 3.50 & 4.00&  0.37 &   0.09&  0.81 &   0.10&  0.27 &   0.05&  0.95 &   0.13\\ 
      &      & 4.00 & 5.00&  0.36 &   0.07&  0.58 &   0.07&  0.43 &   0.07&  0.91 &   0.09\\ 
      &      & 5.00 & 6.50&  0.25 &   0.05&  0.43 &   0.05&  0.63 &   0.09&  0.73 &   0.07\\ 
      &      & 6.50 & 8.00&       &       &  0.39 &   0.05&       &       &  0.84 &   0.07\\ 
0.050 &0.075 & 0.50 & 0.75&  0.55 &   0.11&  0.84 &   0.16&  1.19 &   0.17&  1.98 &   0.28\\ 
      &      & 0.75 & 1.00&  0.57 &   0.09&  0.99 &   0.13&  1.01 &   0.11&  1.13 &   0.14\\ 
      &      & 1.00 & 1.25&  0.79 &   0.10&  1.01 &   0.13&  0.82 &   0.10&  1.20 &   0.14\\ 
      &      & 1.25 & 1.50&  0.96 &   0.12&  1.24 &   0.17&  0.88 &   0.12&  1.17 &   0.14\\ 
      &      & 1.50 & 2.00&  0.77 &   0.07&  1.01 &   0.11&  0.93 &   0.09&  1.50 &   0.14\\ 
      &      & 2.00 & 2.50&  0.58 &   0.08&  0.95 &   0.11&  0.84 &   0.09&  1.20 &   0.11\\ 
      &      & 2.50 & 3.00&  0.32 &   0.07&  0.68 &   0.07&  0.69 &   0.08&  1.19 &   0.11\\ 
      &      & 3.00 & 3.50&  0.39 &   0.08&  0.87 &   0.10&  0.82 &   0.11&  1.10 &   0.11\\ 
      &      & 3.50 & 4.00&  0.31 &   0.06&  0.60 &   0.07&  0.58 &   0.09&  1.03 &   0.11\\ 
      &      & 4.00 & 5.00&  0.23 &   0.04&  0.41 &   0.05&  0.70 &   0.09&  0.79 &   0.07\\ 
      &      & 5.00 & 6.50&  0.16 &   0.03&  0.26 &   0.03&  0.56 &   0.06&  0.67 &   0.05\\ 
      &      & 6.50 & 8.00&       &       &  0.14 &   0.02&       &       &  0.45 &   0.04\\ 
0.075 &0.100 & 0.50 & 0.75&  0.73 &   0.12&  1.09 &   0.16&  1.61 &   0.20&  2.13 &   0.26\\ 
      &      & 0.75 & 1.00&  0.80 &   0.10&  1.54 &   0.16&  1.13 &   0.11&  1.56 &   0.15\\ 
      &      & 1.00 & 1.25&  1.08 &   0.11&  1.37 &   0.13&  0.98 &   0.09&  1.31 &   0.12\\ 
      &      & 1.25 & 1.50&  0.73 &   0.08&  1.30 &   0.11&  0.88 &   0.10&  1.43 &   0.15\\ 
      &      & 1.50 & 2.00&  0.68 &   0.06&  1.03 &   0.09&  0.94 &   0.08&  1.29 &   0.10\\ 
      &      & 2.00 & 2.50&  0.48 &   0.05&  0.92 &   0.07&  0.96 &   0.09&  1.43 &   0.11\\ 
      &      & 2.50 & 3.00&  0.41 &   0.05&  0.80 &   0.07&  0.93 &   0.08&  1.28 &   0.10\\ 
      &      & 3.00 & 3.50&  0.34 &   0.04&  0.56 &   0.05&  0.54 &   0.08&  0.97 &   0.07\\ 
      &      & 3.50 & 4.00&  0.14 &   0.05&  0.43 &   0.05&  0.51 &   0.07&  0.88 &   0.07\\ 
      &      & 4.00 & 5.00&  0.09 &   0.02&  0.33 &   0.03&  0.47 &   0.05&  0.57 &   0.06\\ 
      &      & 5.00 & 6.50& 0.021 &  0.008&  0.14 &   0.02&  0.26 &   0.03&  0.31 &   0.02\\ 
      &      & 6.50 & 8.00&       &       & 0.037 &  0.008&       &       &  0.17 &   0.02\\ 
\hline
\end{tabular}
}
\end{table}

%% file: table_result_fw_036-pip_Sn_OMEGA_fine.tex
\begin{table}[!ht]
  \caption{\label{tab:xsec_results_
Sn7} 
    HARP results for the double-differential $\pi^+$  and $\pi^-$ production
    cross-section in the laboratory system,
    $d^2\sigma^{\pi}/(dpd\Omega)$, for $\pi^{+}$--
Sn 
    interactions at 8~\GeVc.
    Each row refers to a
    different $(p_{\hbox{\small min}} \le p<p_{\hbox{\small max}},
    \theta_{\hbox{\small min}} \le \theta<\theta_{\hbox{\small max}})$ bin,
    where $p$ and $\theta$ are the pion momentum and polar angle, respectively.
    A finer angular binning than in the previous set of tables is used.
    The central value as well as the square-root of the diagonal elements
    of the covariance matrix are given.}

\small{
\begin{tabular}{rrrr|r@{$\pm$}l|r@{$\pm$}l}
\hline
$\theta_{\hbox{\small min}}$ &
$\theta_{\hbox{\small max}}$ &
$p_{\hbox{\small min}}$ &
$p_{\hbox{\small max}}$ &
\multicolumn{2}{c}{$d^2\sigma^{\pi^+}/(dpd\Omega)$} &
\multicolumn{2}{c}{$d^2\sigma^{\pi^-}/(dpd\Omega)$}
\\
(rad) & (rad) & (\GeVc) & (\GeVc) &
\multicolumn{2}{c}{(barn/(\GeVc rad))} &
\multicolumn{2}{c}{(barn/(\GeVc rad))}
\\
  &  &  &
&\multicolumn{2}{c}{$ \bf{8 \ \GeVc}$}
&\multicolumn{2}{c}{$ \bf{8 \ \GeVc}$}
\\
\hline
0.025 &0.050 & 0.50 & 0.75&  0.78 &   0.18&  0.58 &   0.14\\ 
      &      & 0.75 & 1.00&  0.63 &   0.10&  0.57 &   0.11\\ 
      &      & 1.00 & 1.25&  0.34 &   0.07&  0.68 &   0.14\\ 
      &      & 1.25 & 1.50&  0.27 &   0.08&  0.34 &   0.11\\ 
      &      & 1.50 & 2.00&  0.63 &   0.13&  0.82 &   0.16\\ 
      &      & 2.00 & 2.50&  0.55 &   0.13&  0.59 &   0.11\\ 
      &      & 2.50 & 3.00&  0.59 &   0.13&  0.67 &   0.11\\ 
      &      & 3.00 & 3.50&  0.52 &   0.09&  0.48 &   0.08\\ 
      &      & 3.50 & 4.00&  0.54 &   0.07&  0.45 &   0.07\\ 
      &      & 4.00 & 5.00&  0.61 &   0.06&  0.52 &   0.07\\ 
      &      & 5.00 & 6.50&  0.98 &   0.06&  0.39 &   0.06\\ 
0.050 &0.075 & 0.50 & 0.75&  0.85 &   0.19&  0.88 &   0.19\\ 
      &      & 0.75 & 1.00&  0.61 &   0.12&  0.54 &   0.11\\ 
      &      & 1.00 & 1.25&  0.60 &   0.11&  0.66 &   0.13\\ 
      &      & 1.25 & 1.50&  0.97 &   0.16&  0.81 &   0.15\\ 
      &      & 1.50 & 2.00&  0.85 &   0.11&  0.49 &   0.08\\ 
      &      & 2.00 & 2.50&  0.67 &   0.11&  0.61 &   0.10\\ 
      &      & 2.50 & 3.00&  0.54 &   0.09&  0.41 &   0.07\\ 
      &      & 3.00 & 3.50&  0.44 &   0.09&  0.42 &   0.07\\ 
      &      & 3.50 & 4.00&  0.39 &   0.08&  0.39 &   0.06\\ 
      &      & 4.00 & 5.00&  0.77 &   0.08&  0.28 &   0.04\\ 
      &      & 5.00 & 6.50&  0.88 &   0.06&  0.14 &   0.02\\ 
0.075 &0.100 & 0.50 & 0.75&  1.33 &   0.26&  0.91 &   0.17\\ 
      &      & 0.75 & 1.00&  1.01 &   0.15&  0.74 &   0.11\\ 
      &      & 1.00 & 1.25&  0.82 &   0.12&  0.67 &   0.11\\ 
      &      & 1.25 & 1.50&  0.79 &   0.12&  0.68 &   0.10\\ 
      &      & 1.50 & 2.00&  0.70 &   0.23&  0.69 &   0.08\\ 
      &      & 2.00 & 2.50&  0.68 &   0.35&  0.44 &   0.06\\ 
      &      & 2.50 & 3.00&  0.72 &   0.28&  0.40 &   0.05\\ 
      &      & 3.00 & 3.50&  0.62 &   0.08&  0.32 &   0.05\\ 
      &      & 3.50 & 4.00&  0.65 &   0.11&  0.22 &   0.03\\ 
      &      & 4.00 & 5.00&  0.62 &   0.08&  0.13 &   0.02\\ 
      &      & 5.00 & 6.50&  0.42 &   0.15& 0.044 &  0.009\\ 
\hline
\end{tabular}
}
\end{table}

%% file: table_result_fw_036-pim_Ta_OMEGA_fine.tex
\begin{table}[!ht]
  \caption{\label{tab:xsec_results_
Ta7} 
    HARP results for the double-differential $\pi^+$  and $\pi^-$ production
    cross-section in the laboratory system,
    $d^2\sigma^{\pi}/(dpd\Omega)$, for $\pi^{-}$--
Ta 
    interactions at 8 and 12~\GeVc.
    Each row refers to a
    different $(p_{\hbox{\small min}} \le p<p_{\hbox{\small max}},
    \theta_{\hbox{\small min}} \le \theta<\theta_{\hbox{\small max}})$ bin,
    where $p$ and $\theta$ are the pion momentum and polar angle, respectively.
    A finer angular binning than in the previous set of tables is used.
    The central value as well as the square-root of the diagonal elements
    of the covariance matrix are given.}

\small{
\begin{tabular}{rrrr|r@{$\pm$}lr@{$\pm$}l|r@{$\pm$}lr@{$\pm$}l}
\hline
$\theta_{\hbox{\small min}}$ &
$\theta_{\hbox{\small max}}$ &
$p_{\hbox{\small min}}$ &
$p_{\hbox{\small max}}$ &
\multicolumn{4}{c}{$d^2\sigma^{\pi^+}/(dpd\Omega)$} &
\multicolumn{4}{c}{$d^2\sigma^{\pi^-}/(dpd\Omega)$}
\\
(rad) & (rad) & (\GeVc) & (\GeVc) &
\multicolumn{4}{c}{(barn/(\GeVc rad))} &
\multicolumn{4}{c}{(barn/(\GeVc rad))}
\\
  &  &  &
&\multicolumn{2}{c}{$ \bf{8 \ \GeVc}$}
&\multicolumn{2}{c}{$ \bf{12 \ \GeVc}$}
&\multicolumn{2}{c}{$ \bf{8 \ \GeVc}$}
&\multicolumn{2}{c}{$ \bf{12 \ \GeVc}$}
\\
\hline
0.025 &0.050 & 0.50 & 0.75&  1.47 &   0.25&  1.65 &   0.32&  2.65 &   0.39&  2.44 &   0.39\\ 
      &      & 0.75 & 1.00&  0.72 &   0.08&  1.11 &   0.15&  1.18 &   0.12&  1.34 &   0.16\\ 
      &      & 1.00 & 1.25&  0.57 &   0.07&  1.02 &   0.15&  0.43 &   0.06&  0.89 &   0.13\\ 
      &      & 1.25 & 1.50&  0.90 &   0.12&  1.19 &   0.18&  0.76 &   0.13&  1.61 &   0.24\\ 
      &      & 1.50 & 2.00&  0.72 &   0.11&  1.21 &   0.17&  1.09 &   0.11&  1.51 &   0.19\\ 
      &      & 2.00 & 2.50&  0.90 &   0.15&  1.26 &   0.17&  1.10 &   0.12&  1.80 &   0.23\\ 
      &      & 2.50 & 3.00&  0.90 &   0.15&  1.26 &   0.16&  0.92 &   0.13&  1.34 &   0.18\\ 
      &      & 3.00 & 3.50&  0.51 &   0.12&  1.49 &   0.18&  0.50 &   0.08&  1.50 &   0.16\\ 
      &      & 3.50 & 4.00&  0.51 &   0.11&  0.77 &   0.14&  0.32 &   0.06&  1.08 &   0.13\\ 
      &      & 4.00 & 5.00&  0.44 &   0.08&  0.55 &   0.08&  0.44 &   0.08&  0.93 &   0.10\\ 
      &      & 5.00 & 6.50&  0.35 &   0.07&  0.55 &   0.07&  0.53 &   0.09&  0.92 &   0.11\\ 
      &      & 6.50 & 8.00&       &       &  0.36 &   0.05&       &       &  0.96 &   0.09\\ 
0.050 &0.075 & 0.50 & 0.75&  0.82 &   0.14&  1.39 &   0.25&  1.27 &   0.18&  2.13 &   0.31\\ 
      &      & 0.75 & 1.00&  0.75 &   0.11&  1.38 &   0.18&  1.12 &   0.12&  1.35 &   0.18\\ 
      &      & 1.00 & 1.25&  1.12 &   0.14&  1.27 &   0.17&  0.89 &   0.11&  1.62 &   0.20\\ 
      &      & 1.25 & 1.50&  1.41 &   0.16&  1.87 &   0.25&  1.08 &   0.15&  1.92 &   0.22\\ 
      &      & 1.50 & 2.00&  1.21 &   0.11&  1.23 &   0.13&  1.43 &   0.12&  1.64 &   0.16\\ 
      &      & 2.00 & 2.50&  0.77 &   0.11&  1.16 &   0.12&  1.21 &   0.13&  1.70 &   0.15\\ 
      &      & 2.50 & 3.00&  0.32 &   0.09&  1.02 &   0.11&  1.11 &   0.11&  1.43 &   0.14\\ 
      &      & 3.00 & 3.50&  0.40 &   0.10&  0.95 &   0.11&  0.89 &   0.13&  1.41 &   0.13\\ 
      &      & 3.50 & 4.00&  0.42 &   0.08&  0.87 &   0.12&  0.82 &   0.13&  1.06 &   0.11\\ 
      &      & 4.00 & 5.00&  0.24 &   0.04&  0.54 &   0.06&  0.75 &   0.10&  0.88 &   0.08\\ 
      &      & 5.00 & 6.50&  0.21 &   0.04&  0.31 &   0.04&  0.73 &   0.09&  0.73 &   0.06\\ 
      &      & 6.50 & 8.00&       &       &  0.19 &   0.03&       &       &  0.53 &   0.05\\ 
0.075 &0.100 & 0.50 & 0.75&  1.21 &   0.19&  1.42 &   0.22&  2.19 &   0.26&  2.67 &   0.34\\ 
      &      & 0.75 & 1.00&  1.31 &   0.14&  2.10 &   0.22&  1.17 &   0.12&  2.23 &   0.21\\ 
      &      & 1.00 & 1.25&  1.27 &   0.12&  1.77 &   0.17&  1.19 &   0.12&  1.78 &   0.17\\ 
      &      & 1.25 & 1.50&  1.27 &   0.12&  1.71 &   0.16&  1.16 &   0.13&  1.86 &   0.20\\ 
      &      & 1.50 & 2.00&  0.93 &   0.08&  1.30 &   0.11&  1.06 &   0.09&  1.62 &   0.13\\ 
      &      & 2.00 & 2.50&  0.66 &   0.07&  1.03 &   0.10&  1.24 &   0.11&  1.51 &   0.13\\ 
      &      & 2.50 & 3.00&  0.39 &   0.06&  0.79 &   0.08&  0.92 &   0.08&  1.53 &   0.12\\ 
      &      & 3.00 & 3.50&  0.35 &   0.05&  0.84 &   0.08&  0.75 &   0.09&  1.17 &   0.09\\ 
      &      & 3.50 & 4.00&  0.19 &   0.06&  0.57 &   0.06&  0.61 &   0.07&  0.98 &   0.08\\ 
      &      & 4.00 & 5.00&  0.14 &   0.03&  0.35 &   0.04&  0.61 &   0.07&  0.69 &   0.06\\ 
      &      & 5.00 & 6.50& 0.032 &  0.013&  0.15 &   0.02&  0.31 &   0.04&  0.36 &   0.03\\ 
      &      & 6.50 & 8.00&       &       & 0.040 &  0.009&       &       &  0.17 &   0.02\\ 
\hline
\end{tabular}
}
\end{table}

%% file: table_result_fw_036-pip_Ta_OMEGA_fine.tex
\begin{table}[!ht]
  \caption{\label{tab:xsec_results_
Ta7} 
    HARP results for the double-differential $\pi^+$  and $\pi^-$ production
    cross-section in the laboratory system,
    $d^2\sigma^{\pi}/(dpd\Omega)$, for $\pi^{+}$--
Ta 
    interactions at 8~\GeVc.
    Each row refers to a
    different $(p_{\hbox{\small min}} \le p<p_{\hbox{\small max}},
    \theta_{\hbox{\small min}} \le \theta<\theta_{\hbox{\small max}})$ bin,
    where $p$ and $\theta$ are the pion momentum and polar angle, respectively.
    A finer angular binning than in the previous set of tables is used.
    The central value as well as the square-root of the diagonal elements
    of the covariance matrix are given.}

\small{
\begin{tabular}{rrrr|r@{$\pm$}l|r@{$\pm$}l}
\hline
$\theta_{\hbox{\small min}}$ &
$\theta_{\hbox{\small max}}$ &
$p_{\hbox{\small min}}$ &
$p_{\hbox{\small max}}$ &
\multicolumn{2}{c}{$d^2\sigma^{\pi^+}/(dpd\Omega)$} &
\multicolumn{2}{c}{$d^2\sigma^{\pi^-}/(dpd\Omega)$}
\\
(rad) & (rad) & (\GeVc) & (\GeVc) &
\multicolumn{2}{c}{(barn/(\GeVc rad))} &
\multicolumn{2}{c}{(barn/(\GeVc rad))}
\\
  &  &  &
&\multicolumn{2}{c}{$ \bf{8 \ \GeVc}$}
&\multicolumn{2}{c}{$ \bf{8 \ \GeVc}$}
\\
\hline
0.025 &0.050 & 0.50 & 0.75&  1.66 &   0.37&  0.92 &   0.25\\ 
      &      & 0.75 & 1.00&  1.02 &   0.16&  0.60 &   0.13\\ 
      &      & 1.00 & 1.25&  0.50 &   0.10&  0.96 &   0.21\\ 
      &      & 1.25 & 1.50&  0.39 &   0.12&  0.33 &   0.13\\ 
      &      & 1.50 & 2.00&  0.66 &   0.16&  0.87 &   0.19\\ 
      &      & 2.00 & 2.50&  0.62 &   0.17&  0.63 &   0.14\\ 
      &      & 2.50 & 3.00&  0.72 &   0.18&  0.54 &   0.11\\ 
      &      & 3.00 & 3.50&  0.58 &   0.12&  0.82 &   0.15\\ 
      &      & 3.50 & 4.00&  0.59 &   0.11&  0.63 &   0.11\\ 
      &      & 4.00 & 5.00&  0.56 &   0.08&  0.56 &   0.09\\ 
      &      & 5.00 & 6.50&  1.12 &   0.08&  0.48 &   0.07\\ 
0.050 &0.075 & 0.50 & 0.75&  1.15 &   0.27&  0.79 &   0.22\\ 
      &      & 0.75 & 1.00&  0.90 &   0.19&  0.72 &   0.16\\ 
      &      & 1.00 & 1.25&  0.95 &   0.19&  1.02 &   0.22\\ 
      &      & 1.25 & 1.50&  0.96 &   0.20&  0.54 &   0.13\\ 
      &      & 1.50 & 2.00&  0.57 &   0.12&  0.98 &   0.16\\ 
      &      & 2.00 & 2.50&  0.70 &   0.15&  0.60 &   0.11\\ 
      &      & 2.50 & 3.00&  0.72 &   0.14&  0.62 &   0.12\\ 
      &      & 3.00 & 3.50&  0.46 &   0.12&  0.36 &   0.07\\ 
      &      & 3.50 & 4.00&  0.42 &   0.10&  0.41 &   0.08\\ 
      &      & 4.00 & 5.00&  0.95 &   0.11&  0.31 &   0.05\\ 
      &      & 5.00 & 6.50&  1.06 &   0.08&  0.18 &   0.03\\ 
0.075 &0.100 & 0.50 & 0.75&  1.11 &   0.27&  0.88 &   0.21\\ 
      &      & 0.75 & 1.00&  1.04 &   0.18&  0.75 &   0.14\\ 
      &      & 1.00 & 1.25&  1.15 &   0.19&  1.19 &   0.20\\ 
      &      & 1.25 & 1.50&  1.03 &   0.17&  0.79 &   0.14\\ 
      &      & 1.50 & 2.00&  0.67 &   0.11&  0.62 &   0.10\\ 
      &      & 2.00 & 2.50&  0.72 &   0.12&  0.47 &   0.08\\ 
      &      & 2.50 & 3.00&  0.74 &   0.12&  0.51 &   0.08\\ 
      &      & 3.00 & 3.50&  0.72 &   0.10&  0.44 &   0.07\\ 
      &      & 3.50 & 4.00&  0.70 &   0.11&  0.22 &   0.04\\ 
      &      & 4.00 & 5.00&  0.71 &   0.07&  0.19 &   0.03\\ 
      &      & 5.00 & 6.50&  0.41 &   0.15& 0.051 &  0.011\\ 
\hline
\end{tabular}
}
\end{table}

%% file: table_result_fw_036-pim_Pb_OMEGA_fine.tex
\begin{table}[!ht]
  \caption{\label{tab:xsec_results_
Pb7} 
    HARP results for the double-differential $\pi^+$  and $\pi^-$ production
    cross-section in the laboratory system,
    $d^2\sigma^{\pi}/(dpd\Omega)$, for $\pi^{-}$--
Pb 
    interactions at 8 and 12~\GeVc.
    Each row refers to a
    different $(p_{\hbox{\small min}} \le p<p_{\hbox{\small max}},
    \theta_{\hbox{\small min}} \le \theta<\theta_{\hbox{\small max}})$ bin,
    where $p$ and $\theta$ are the pion momentum and polar angle, respectively.
    A finer angular binning than in the previous set of tables is used.
    The central value as well as the square-root of the diagonal elements
    of the covariance matrix are given.}

\small{
\begin{tabular}{rrrr|r@{$\pm$}lr@{$\pm$}l|r@{$\pm$}lr@{$\pm$}l}
\hline
$\theta_{\hbox{\small min}}$ &
$\theta_{\hbox{\small max}}$ &
$p_{\hbox{\small min}}$ &
$p_{\hbox{\small max}}$ &
\multicolumn{4}{c}{$d^2\sigma^{\pi^+}/(dpd\Omega)$} &
\multicolumn{4}{c}{$d^2\sigma^{\pi^-}/(dpd\Omega)$}
\\
(rad) & (rad) & (\GeVc) & (\GeVc) &
\multicolumn{4}{c}{(barn/(\GeVc rad))} &
\multicolumn{4}{c}{(barn/(\GeVc rad))}
\\
  &  &  &
&\multicolumn{2}{c}{$ \bf{8 \ \GeVc}$}
&\multicolumn{2}{c}{$ \bf{12 \ \GeVc}$}
&\multicolumn{2}{c}{$ \bf{8 \ \GeVc}$}
&\multicolumn{2}{c}{$ \bf{12 \ \GeVc}$}
\\
\hline
0.025 &0.050 & 0.50 & 0.75&  1.60 &   0.27&  1.58 &   0.29&  3.24 &   0.43&  2.52 &   0.41\\ 
      &      & 0.75 & 1.00&  0.87 &   0.09&  1.37 &   0.16&  1.26 &   0.14&  1.74 &   0.18\\ 
      &      & 1.00 & 1.25&  0.56 &   0.07&  1.06 &   0.14&  0.54 &   0.07&  0.88 &   0.13\\ 
      &      & 1.25 & 1.50&  0.76 &   0.11&  1.25 &   0.18&  0.54 &   0.09&  0.84 &   0.14\\ 
      &      & 1.50 & 2.00&  0.82 &   0.13&  1.40 &   0.18&  1.27 &   0.14&  1.53 &   0.20\\ 
      &      & 2.00 & 2.50&  1.12 &   0.16&  1.24 &   0.17&  1.24 &   0.14&  1.66 &   0.20\\ 
      &      & 2.50 & 3.00&  0.76 &   0.15&  1.28 &   0.16&  1.02 &   0.13&  0.87 &   0.13\\ 
      &      & 3.00 & 3.50&  0.41 &   0.12&  0.92 &   0.14&  0.63 &   0.10&  1.27 &   0.16\\ 
      &      & 3.50 & 4.00&  0.53 &   0.12&  0.85 &   0.13&  0.42 &   0.09&  1.05 &   0.13\\ 
      &      & 4.00 & 5.00&  0.47 &   0.09&  0.56 &   0.09&  0.54 &   0.10&  0.73 &   0.10\\ 
      &      & 5.00 & 6.50&  0.40 &   0.08&  0.44 &   0.07&  0.65 &   0.11&  0.62 &   0.09\\ 
      &      & 6.50 & 8.00&       &       &  0.33 &   0.06&       &       &  0.77 &   0.10\\ 
0.050 &0.075 & 0.50 & 0.75&  0.92 &   0.15&  1.45 &   0.22&  1.91 &   0.24&  2.46 &   0.33\\ 
      &      & 0.75 & 1.00&  0.75 &   0.11&  1.33 &   0.17&  1.39 &   0.15&  1.34 &   0.18\\ 
      &      & 1.00 & 1.25&  1.00 &   0.13&  1.82 &   0.20&  1.12 &   0.14&  1.56 &   0.19\\ 
      &      & 1.25 & 1.50&  1.72 &   0.20&  1.70 &   0.20&  1.15 &   0.16&  1.49 &   0.18\\ 
      &      & 1.50 & 2.00&  1.19 &   0.10&  1.38 &   0.14&  1.43 &   0.14&  1.66 &   0.16\\ 
      &      & 2.00 & 2.50&  0.71 &   0.10&  1.21 &   0.12&  1.24 &   0.13&  1.52 &   0.15\\ 
      &      & 2.50 & 3.00&  0.36 &   0.10&  1.00 &   0.10&  0.95 &   0.11&  1.29 &   0.13\\ 
      &      & 3.00 & 3.50&  0.41 &   0.11&  1.04 &   0.11&  0.90 &   0.14&  1.16 &   0.12\\ 
      &      & 3.50 & 4.00&  0.42 &   0.08&  0.76 &   0.10&  0.84 &   0.13&  0.93 &   0.11\\ 
      &      & 4.00 & 5.00&  0.36 &   0.06&  0.43 &   0.06&  0.82 &   0.12&  0.76 &   0.08\\ 
      &      & 5.00 & 6.50&  0.20 &   0.04&  0.29 &   0.04&  0.67 &   0.08&  0.65 &   0.06\\ 
      &      & 6.50 & 8.00&       &       &  0.13 &   0.02&       &       &  0.51 &   0.05\\ 
0.075 &0.100 & 0.50 & 0.75&  0.96 &   0.16&  1.81 &   0.23&  2.21 &   0.26&  2.93 &   0.34\\ 
      &      & 0.75 & 1.00&  1.25 &   0.15&  1.83 &   0.18&  1.28 &   0.13&  1.92 &   0.19\\ 
      &      & 1.00 & 1.25&  1.75 &   0.16&  1.77 &   0.16&  1.27 &   0.12&  1.59 &   0.15\\ 
      &      & 1.25 & 1.50&  1.30 &   0.13&  1.71 &   0.14&  1.53 &   0.18&  1.71 &   0.19\\ 
      &      & 1.50 & 2.00&  1.00 &   0.09&  1.42 &   0.12&  1.38 &   0.11&  1.73 &   0.13\\ 
      &      & 2.00 & 2.50&  0.61 &   0.07&  1.17 &   0.10&  1.14 &   0.10&  1.67 &   0.12\\ 
      &      & 2.50 & 3.00&  0.45 &   0.06&  0.90 &   0.08&  1.04 &   0.10&  1.38 &   0.11\\ 
      &      & 3.00 & 3.50&  0.43 &   0.06&  0.91 &   0.08&  0.74 &   0.11&  1.22 &   0.09\\ 
      &      & 3.50 & 4.00&  0.24 &   0.06&  0.55 &   0.07&  0.63 &   0.09&  0.99 &   0.10\\ 
      &      & 4.00 & 5.00&  0.12 &   0.03&  0.32 &   0.04&  0.46 &   0.07&  0.54 &   0.06\\ 
      &      & 5.00 & 6.50& 0.033 &  0.014&  0.14 &   0.02&  0.32 &   0.04&  0.33 &   0.03\\ 
      &      & 6.50 & 8.00&       &       & 0.032 &  0.008&       &       &  0.19 &   0.02\\ 
\hline
\end{tabular}
}
\end{table}

%% file: table_result_fw_036-pip_Pb_OMEGA_fine.tex
\begin{table}[!ht]
  \caption{\label{tab:xsec_results_
Pb7} 
    HARP results for the double-differential $\pi^+$  and $\pi^-$ production
    cross-section in the laboratory system,
    $d^2\sigma^{\pi}/(dpd\Omega)$, for $\pi^{+}$--
Pb 
    interactions at 8~\GeVc.
    Each row refers to a
    different $(p_{\hbox{\small min}} \le p<p_{\hbox{\small max}},
    \theta_{\hbox{\small min}} \le \theta<\theta_{\hbox{\small max}})$ bin,
    where $p$ and $\theta$ are the pion momentum and polar angle, respectively.
    A finer angular binning than in the previous set of tables is used.
    The central value as well as the square-root of the diagonal elements
    of the covariance matrix are given.}

\small{
\begin{tabular}{rrrr|r@{$\pm$}l|r@{$\pm$}l}
\hline
$\theta_{\hbox{\small min}}$ &
$\theta_{\hbox{\small max}}$ &
$p_{\hbox{\small min}}$ &
$p_{\hbox{\small max}}$ &
\multicolumn{2}{c}{$d^2\sigma^{\pi^+}/(dpd\Omega)$} &
\multicolumn{2}{c}{$d^2\sigma^{\pi^-}/(dpd\Omega)$}
\\
(rad) & (rad) & (\GeVc) & (\GeVc) &
\multicolumn{2}{c}{(barn/(\GeVc rad))} &
\multicolumn{2}{c}{(barn/(\GeVc rad))}
\\
  &  &  &
&\multicolumn{2}{c}{$ \bf{8 \ \GeVc}$}
&\multicolumn{2}{c}{$ \bf{8 \ \GeVc}$}
\\
\hline
0.025 &0.050 & 0.50 & 0.75&  1.72 &   0.38&  0.75 &   0.21\\ 
      &      & 0.75 & 1.00&  1.04 &   0.16&  0.85 &   0.17\\ 
      &      & 1.00 & 1.25&  0.42 &   0.09&  0.73 &   0.17\\ 
      &      & 1.25 & 1.50&  0.58 &   0.15&  0.79 &   0.25\\ 
      &      & 1.50 & 2.00&  0.73 &   0.17&  0.66 &   0.17\\ 
      &      & 2.00 & 2.50&  0.63 &   0.20&  0.72 &   0.16\\ 
      &      & 2.50 & 3.00&  0.49 &   0.15&  0.60 &   0.13\\ 
      &      & 3.00 & 3.50&  0.65 &   0.13&  0.51 &   0.10\\ 
      &      & 3.50 & 4.00&  0.59 &   0.10&  0.67 &   0.12\\ 
      &      & 4.00 & 5.00&  0.63 &   0.08&  0.58 &   0.09\\ 
      &      & 5.00 & 6.50&  1.18 &   0.08&  0.49 &   0.08\\ 
0.050 &0.075 & 0.50 & 0.75&  0.66 &   0.18&  0.75 &   0.20\\ 
      &      & 0.75 & 1.00&  0.77 &   0.18&  0.73 &   0.16\\ 
      &      & 1.00 & 1.25&  0.43 &   0.12&  0.69 &   0.17\\ 
      &      & 1.25 & 1.50&  0.75 &   0.18&  1.00 &   0.20\\ 
      &      & 1.50 & 2.00&  0.69 &   0.13&  0.77 &   0.13\\ 
      &      & 2.00 & 2.50&  0.74 &   0.15&  0.69 &   0.13\\ 
      &      & 2.50 & 3.00&  0.85 &   0.16&  0.49 &   0.10\\ 
      &      & 3.00 & 3.50&  0.55 &   0.13&  0.51 &   0.10\\ 
      &      & 3.50 & 4.00&  0.50 &   0.11&  0.52 &   0.09\\ 
      &      & 4.00 & 5.00&  1.09 &   0.12&  0.40 &   0.06\\ 
      &      & 5.00 & 6.50&  1.11 &   0.09&  0.22 &   0.03\\ 
0.075 &0.100 & 0.50 & 0.75&  1.04 &   0.27&  1.05 &   0.23\\ 
      &      & 0.75 & 1.00&  1.02 &   0.19&  1.27 &   0.20\\ 
      &      & 1.00 & 1.25&  0.86 &   0.16&  1.30 &   0.21\\ 
      &      & 1.25 & 1.50&  0.99 &   0.17&  0.91 &   0.15\\ 
      &      & 1.50 & 2.00&  0.84 &   0.13&  0.81 &   0.12\\ 
      &      & 2.00 & 2.50&  0.88 &   0.15&  0.33 &   0.06\\ 
      &      & 2.50 & 3.00&  0.66 &   0.11&  0.53 &   0.08\\ 
      &      & 3.00 & 3.50&  0.66 &   0.09&  0.34 &   0.06\\ 
      &      & 3.50 & 4.00&  0.82 &   0.13&  0.30 &   0.06\\ 
      &      & 4.00 & 5.00&  0.75 &   0.07&  0.21 &   0.03\\ 
      &      & 5.00 & 6.50&  0.44 &   0.15& 0.065 &  0.013\\ 
\hline
\end{tabular}
}
\end{table}